\documentclass[a4paper,10pt]{article}
\pdfoutput=1

\usepackage{hyperref}
\usepackage[section]{placeins}
\usepackage{
graphicx,
hyperref,
amsmath,
amssymb,
charter,
xcolor,
ifluatex,
multirow}
        
\usepackage[skip=0.5ex]{subcaption}
\usepackage{lipsum}

\definecolor{Gray}{gray}{0.95}
\definecolor{RGray}{gray}{0.85}
\definecolor{CGray}{gray}{0.92}

\usepackage{booktabs}

\newcommand{\iab}{{\ensuremath\rm ab^{-1}}}

\definecolor{tit}{rgb}{0.1,0.2,0.4}
\definecolor{blus}{cmyk}{1,1,0,0.6}
\definecolor{verde}{cmyk}{0.92,0,0.59,0.25}
        
\newcommand{\e}[1]{\cdot 10^{#1}}
\usepackage{tipa}
\usepackage{amsmath,amssymb,amsfonts, bm}
\usepackage{epsfig}
\usepackage{graphicx}
\usepackage{slashed}
\usepackage{color}

\usepackage{caption}
\usepackage{subcaption}
\usepackage{slashed} 
\usepackage{float}
\usepackage{caption} 
\usepackage{booktabs} 

\usepackage{tcolorbox}

\newcommand{\eps}{\epsilon}

\hypersetup{pdfauthor={Name}}
\newcommand{\Heff}{{\cal H}_\text{NP}}

\newcommand{\meg}{\mu \to e \gamma}

\newcommand{\M}{{\cal M}}

\usepackage[normalem]{ulem}
\usepackage{calc}

\newcommand{\be}{\begin{equation}}
\newcommand{\ee}{\end{equation}}

\newcommand{\bea}{\begin{eqnarray}}
\newcommand{\eea}{\end{eqnarray}}

\newcommand{\bfig}{\begin{figure}}
\newcommand{\efig}{\end{figure}}

\newcommand{\meee}{\mu \to e \bar{e} e}
\newcommand{\lag}{\ensuremath{\mathcal{L}}}

\newcommand{\qqquad}{\qquad \qquad}

\newcommand{\br}{\text{BR}}

\usepackage{tabularx}

\usepackage{multirow}
\newcommand*{\myalign}[2]{\multicolumn{1}{#1}{#2}}

\makeatletter
\newcommand*{\rom}[1]{\expandafter\@slowromancap\romannumeral #1@}
\makeatother

\usepackage{xcolor}  
\usepackage{color}

\begin{document}
\allowdisplaybreaks
\vspace*{-2.5cm}
\begin{flushright}
{\small
IIT-BHU
}
\end{flushright}

\vspace{2cm}

\begin{center}
{\LARGE \bf \color{tit} Finding flavons at colliders}\\[1cm]

{\large\bf Gauhar Abbas$^{a}$\footnote{email: gauhar.phy@iitbhu.ac.in} \\ 
Ashutosh Kumar Alok$^{b}$\footnote{email: akalok@iitj.ac.in }   \\
Neetu Raj Singh Chundawat $^{b}$\footnote{email: chundawat.1@iitj.ac.in }   \\
Najimuddin Khan$^{c}$\footnote{email: nkhan.ph@amu.ac.in }\\
Neelam  Singh$^{a}$\footnote{email: neelamsingh.rs.phy19@itbhu.ac.in }   } \\[7mm]
{\it $^a$ } {\em Department of Physics, Indian Institute of Technology (BHU), Varanasi 221005, India}\\[3mm]
{\it $^b$  Indian Institute of Technology Jodhpur, Jodhpur 342037, India } {\em }\\[3mm]
{\it $^c$   Department of Physics, Aligarh Muslim University, Aligarh-202002, India } {\em }\\[3mm]

\vspace{1cm}
{\large\bf\color{blus} Abstract}
\begin{quote}
We conduct a comprehensive investigation into the flavor phenomenology and collider signatures of flavon of $\mathcal{Z}_{\rm N} \times \mathcal{Z}_{\rm M}$ flavor symmetries for the soft symmetry-breaking scenario and a new symmetry-conserving mechanism at the high-luminosity LHC, high energy LHC, and a 100 TeV hadron collider. The flavor physics of quark and leptonic observables places different bounds on the parameter space of flavons of $\mathcal{Z}_{\rm N} \times \mathcal{Z}_{\rm M}$ flavor symmetries.  
On the collider side, the decay $t \rightarrow c a$ can be probed by the high-luminosity LHC, high energy LHC, and a 100 TeV hadron collider for the $\mathcal{Z}_{\rm 8} \times \mathcal{Z}_{\rm 22}$ flavor symmetry.  The inclusive production signatures can be used to probe the flavon of all the  $\mathcal{Z}_{\rm N} \times \mathcal{Z}_{\rm M} $ flavor symmetries for the soft symmetry-breaking scenario for a heavy flavon  at a 100 TeV collider.  Flavons of all the  $\mathcal{Z}_{\rm N} \times \mathcal{Z}_{\rm M} $ flavor symmetries can be probed at high energy LHC and  a 100 TeV collider for a low mass in the case of soft symmetry-breaking.  The di-flavon production is within reach of the high-luminosity LHC, high energy LHC, and a 100 TeV collider only for a light flavon. The 14 TeV high-luminosity LHC can probe only the $\mathcal{Z}_{\rm 2} \times \mathcal{Z}_{\rm 5}$ and $\mathcal{Z}_{\rm 8} \times \mathcal{Z}_{\rm 22}$  flavor symmetries for a few specific  inclusive signatures. The symmetry-conserving scenario remains beyond the detection capabilities of any collider.
\end{quote}
\thispagestyle{empty}
\end{center}

\begin{quote}
{\large\noindent\color{blus} 
}

\end{quote}

\newpage
\setcounter{footnote}{0}
\section{Introduction}
 The enigma of fermion masses and quark mixing has persisted since the advent of the Standard Model \cite{Abbas:2023ivi,Abbas:2024wzp}. This challenge has been further exacerbated by the inclusion of neutrino masses and mixing. Potential resolutions to this issue may arise from various theoretical frameworks. Among these are the dark technicolor model, which addresses the problem through hierarchical vacuum expectation values (VEVs) manifested as sequential multi-fermion chiral condensates \cite{Abbas:2017vws,Abbas:2020frs,Abbas:2023dpf,Abbas:2023bmm},  continuous flavor symmetries \cite{Froggatt:1978nt,flavor_symm1,Chun:1996xv,flavor_symm2,flavor_symm3,Davidson:1983fy,Davidson:1987tr,Berezhiani:1990wn,Berezhiani:1990jj,Berezhiani:1989fp,Sakharov:1994pr,Shaikh:2024ufv,Yang:2024znv,Hinata:2020cdt}, discrete symmetries \cite{Abbas:2018lga,Abbas:2022zfb,Higaki:2019ojq}, and alternative frameworks \cite{higgs_coup,wf_local,partial_comp,Fuentes-Martin:2022xnb}.

The $\mathcal{Z}_{\rm N} \times \mathcal{Z}_{\rm M}$ flavor symmetry \cite{Abbas:2018lga} is capable of explaining the masses and mixing of fermions through the different realizations of the Froggatt-Nielsen (FN) mechanism \cite{Froggatt:1978nt}. This symmetry is a subset of a more generic $\mathcal{Z}_{\rm N} \times \mathcal{Z}_{\rm M} \times \mathcal{Z}_{\rm P}$ flavor symmetry, which naturally emerges in a dark technicolour framework \cite{Abbas:2020frs,Abbas:2023ivi}.  

The  FN mechanism is built upon the idea that an Abelian flavor symmetry $U(1)_F$ can differentiate among the different flavors among inter and intra- fermionic generations of the SM.  A flavon field, $\chi$,  is employed for this purpose, which couples with the top quark at tree-level, and the masses of other fermions originate from the higher dimensional non-renormalizable operators constructed using the flavon field and choosing the appropriate charges of the fields under the  flavor symmetry $U(1)_F$.  For instance,  if the charges of the fermions fields  $\psi_i^c$ and $\psi_j$ are $\theta_i$ and $\theta_j$, respectively, under the $U(1)_F$ flavor symmetry, and  the Higgs field is assigned zero charge, the Yukawa Lagrangian for the masses of fermions takes the form of  non-renormalizable operators,

\begin{eqnarray}
\label{flavon_ops}
\mathcal{O}  &=&  y (\dfrac{\chi}{\Lambda})^{(\theta_i + \theta_j)} \bar{\psi_i} \varphi \psi_j, \\ \nonumber
&=&  y \epsilon^{(\theta_i + \theta_j)} \bar{\psi_i} \varphi \psi_j = Y \bar{\psi_i} \varphi \psi_j
\end{eqnarray}
where $y$ denotes the dimensionless  coupling constant,  $\Lambda$ represents the scale of the flavor symmetry,  $\epsilon = \dfrac{\langle \chi \rangle}{\Lambda}$, $Y= y \epsilon^{(\theta_i + \theta_j)}$ stands for the effective Yukawa coupling,  and  the transformation of the  gauge singlet flavon scalar field  $\chi$  under the $SU(3)_c \times SU(2)_L \times U(1)_Y$ symmetry of the SM is given by,
\begin{eqnarray}
\chi :(1,1,0).
 \end{eqnarray} 

The VEV of the flavon field $\chi$ breaks the  $U(1)_F$ flavor symmetry  spontaneously.  The scale $\Lambda$,   where the higher dimension operators appearing in equation \ref{flavon_ops} get renormalized, is not predicted by the FN mechanism, and it can be anywhere between the weak and the Planck scale.  The essential requirement for the FN mechanism is that the flavor symmetry is broken in a way such that the ratio $ \dfrac{\langle \chi \rangle} { \Lambda}  <1 $.

 The flavor phenomenology of the flavon field is determined by the scale $\Lambda$. The flavon field can impact flavor observables such as neutral meson mixing and $CP$-violation, meson decays, and charged lepton flavor violating effects. These effects could be detectable if the scale $\Lambda$ is relatively low, closer to the weak scale. In such a scenario, it would also be possible to investigate the direct production of flavons in hadron colliders, like the Large Hadron Collider (LHC). Therefore, a critical question is how low the scale $\Lambda$ can be, while still satisfying constraints from flavor physics data. This question also hinges on the nature of the flavor symmetry implementing the FN mechanism. For a continuous symmetry such as the $U(1)_F$ flavor symmetry, it is essential to determine whether it is a global or gauged symmetry. If $U(1)_F$ is a gauged symmetry, the flavor and collider phenomenology of the flavon field will include contributions from the exchange of the corresponding gauge boson. Conversely, a global $U(1)_F$ flavor symmetry would face the issue of a massless Goldstone boson.

The $\mathcal{Z}_{\rm N} \times \mathcal{Z}_{\rm M}$ flavor symmetry creates the FN mechanism through the product of two discrete symmetries.  This results in a flavor structure of the SM, which is written in terms of the small parameter $\epsilon$.  The neutrino masses and mixing can also be recovered in this framework \cite{Abbas:2022zfb}.  The theoretical origin of the $\mathcal{Z}_{\rm N} \times \mathcal{Z}_{\rm M}$ flavor symmetry may be traced back to an underlying Abelian or non-Abelian continuous symmetry or their products.  For instance, in the dark technicolour framework, this symmetry is the result of the breaking of the two $U(1)$ axial symmetries  \cite{Abbas:2020frs,Abbas:2023ivi}.

The flavor phenomenology of two simple prototypes of the  $\mathcal{Z}_{\rm N} \times \mathcal{Z}_{\rm M}$ flavor symmetries is investigated in reference  \cite{Abbas:2022zfb} where $\rm N = 2$ and $\rm M= 5, 9$ are used.  The choice  $\rm N = 2$ is inspired by the use of the  discrete $\mathcal{Z}_2$ symmetry in different scenarios, such as the two-Higgs-doublet model (2HDM) and the minimal supersymmetric SM (MSSM), as well as the left-right symmetric mirror models \cite{Abbas:2017vle,Abbas:2017hzw,Abbas:2016qqc,Abbas:2016xgj}.  There are extensive phenomenological investigations of the flavon field in literature.  For instance, for flavor phenomenology, see references \cite{Abbas:2022zfb,Dorsner:2002wi,Bauer:2015kzy,Bauer:2016rxs,Calibbi:2015sfa,Fedele:2020fvh},  and for collider studies, see the references \cite{Tsumura:2009yf,Berger:2014gga,Huitu:2016pwk,Diaz-Cruz:2014pla,Arroyo-Urena:2018mvl,Arroyo-Urena:2019fyd,Higaki:2019ojq,Arroyo-Urena:2022oft,Alves:2023sdf,Chakraborty:2024fvn}.

 In this work, we conduct a comprehensive investigation into the flavor phenomenology and collider signatures of the flavon associated with $\mathcal{Z}_{\rm N} \times \mathcal{Z}_{\rm M}$ flavor symmetries, extending beyond established prototypes. We examine quark flavor constraints through neutral meson mixing and decays and derive constraints on the flavon of $\mathcal{Z}_{\rm N} \times \mathcal{Z}_{\rm M}$ flavor symmetries from a wide range of observables in diverse decay channels, specifically focusing on the quark-level transition $b \rightarrow s \mu^+ \mu^-$. Additionally, we investigate the muon forward-backward asymmetry in $B \rightarrow K \mu^+ \mu^-$ decays, which is predicted to be zero within the SM. Furthermore, we explore the bounds from future sensitivities of lepton flavor violation experiments such as MEG-II, Mu3e, PRISM/PRIME, DeeMe, Mu2e, and COMET. We demonstrate that these upcoming experiments have the potential to constrain the parameter space of the flavon associated with $\mathcal{Z}_{\rm N} \times \mathcal{Z}_{\rm M}$ flavor symmetries by orders of magnitude compared to existing limits.

In the conventional approach, flavor and collider studies of flavon physics involve imparting mass to the pseudoscalar component of the flavon field through a soft symmetry-breaking term \cite{Abbas:2022zfb,Dorsner:2002wi,Bauer:2015kzy,Calibbi:2015sfa,Fedele:2020fvh}. We investigate the flavor phenomenology and collider signatures of the flavon associated with $\mathcal{Z}_{\rm N} \times \mathcal{Z}_{\rm M}$ flavor symmetries using both the conventional approach and a novel symmetry-conserving mass mechanism for the pseudoscalar component of the flavon field. This novel scenario proves to be highly predictive.

For the collider signatures, our primary focus is on the High Luminosity LHC (HL-LHC), the High-Energy Large Hadron Collider (HE-LHC) \cite{FCC:2018bvk}, and a future high-luminosity 100 TeV hadron collider such as FCC-hh \cite{FCC:2018byv}. We classify the search for flavons into low and high-mass regions, with the low-mass region being particularly sensitive to the new symmetry-conserving mass mechanism of the pseudoscalar flavon.
We utilize inclusive decay channels such as $\ell \ell$, $b \bar{b}$, and $t \bar{t}$, as well as associative production channels like $t \bar{t} a \rightarrow t \bar{t} t \bar{t}$, $gg \rightarrow a b \bar{b} \rightarrow \tau \tau b \bar{b}$, and $gb \rightarrow a b \rightarrow \tau \tau b$ to identify the flavon of different $\mathcal{Z}_{\rm N} \times \mathcal{Z}_{\rm M}$ flavor symmetries. Additionally, we investigate di-flavon production for both light and heavy flavon to determine the sensitivities of flavon models associated with $\mathcal{Z}_{\rm N} \times \mathcal{Z}_{\rm M}$ flavor symmetries.
Throughout our investigation, we do not assume any specific ultraviolet (UV) completion of the $\mathcal{Z}_{\rm N} \times \mathcal{Z}_{\rm M}$ flavor symmetries, and we conduct our phenomenological studies in a model-independent manner.

This work will be presented along the following track:  In section \ref{sec:ZNXZM}, we discuss different $\mathcal{Z}_{\rm N} \times \mathcal{Z}_{\rm M}$ flavor symmetries and their motivations.  The scalar potential with the soft symmetry-breaking and the symmetry-conserving scenarios is presented in section \ref{potential}.  In section \ref{quark_flavor}, bounds on the parameter space of the flavon of different $\mathcal{Z}_{\rm N} \times \mathcal{Z}_{\rm M}$ flavor symmetries from quark flavor physics are derived.  We discuss constraints from lepton flavor violation experiments in section \ref{lepton_flavor}.  We discuss the  hadron collider physics of the flavon of the  $\mathcal{Z}_N \times \mathcal{Z}_M$ flavor symmetries in section \ref{collider}.  In section \ref{sum}, we summarize this work.

\section{ The $\mathcal{Z}_{\rm N} \times \mathcal{Z}_{\rm M} $ flavor symmetry  } 
\label{sec:ZNXZM}
In this section, we present different realizations of the $\mathcal{Z}_{\rm N} \times \mathcal{Z}_{\rm M} $ flavor symmetry, for which we  investigate the flavor phenomenology and collider signatures of the flavon field.  A minimal form of the  $\mathcal{Z}_{\rm N} \times \mathcal{Z}_{\rm M} $ flavor symmetry, $\mathcal{Z}_{\rm 2} \times \mathcal{Z}_{\rm 5} $,  was first proposed in reference \cite{Abbas:2018lga}, and flavor bounds on this minimal form, along with a non-minimal form, $\mathcal{Z}_{\rm 2} \times \mathcal{Z}_{\rm 9} $,  are derived in reference \cite{Abbas:2020frs}.  

The theoretical origin of the  $\mathcal{Z}_{\rm N} \times \mathcal{Z}_{\rm M}$  flavor symmetry may come from spontaneous breaking of a $U(1) \times U(1)$ symmetry, which could be a global or a local symmetry. It is also possible that this symmetry contains only one  $U(1)$ factor as a local symmetry.  The crucial difference between the FN mechanism based on the $\mathcal{Z}_{\rm N} \times \mathcal{Z}_{\rm M}$  flavor symmetry and the FN mechanism based on a $U(1)$ symmetry is that the  $\mathcal{Z}_{\rm N} \times \mathcal{Z}_{\rm M}$  flavor symmetry may also originate from the breaking of a global $U(1)_A \times U(1)_A$ symmetry.  This keeps apart the FN mechanism based on the  $\mathcal{Z}_{\rm N} \times \mathcal{Z}_{\rm M}$  flavor symmetry from that based on the conventional  $U(1)$ symmetry. Such a theoretical scenario is discussed in the appendix.

The FN mechanism  is produced  by imposing the $\mathcal{Z}_{\rm N} \times \mathcal{Z}_{\rm M} $ flavor symmetry on the SM.  This provides the following mass Lagrangian for fermions,
\bea
\label{mass1}
-{\mathcal{L}}_{\rm Yukawa} &=&    \left[  \dfrac{ \chi(\text{or} ~\chi^{\dagger})}{\Lambda} \right]^{n_{ij}^u}     y_{ij}^u \bar{ \psi}_{L_i}^q  \tilde{\varphi} \psi_{R_j}^{u}  +   \left[  \dfrac{ \chi(\text{or} ~\chi^{\dagger})}{\Lambda} \right]^{n_{ij}^d}      y_{ij}^d \bar{ \psi}_{L_i}^q  \varphi \psi_{R_j}^{d}  \nonumber \\
&+&    \left[  \dfrac{ \chi(\text{or} ~\chi^{\dagger})}{\Lambda} \right]^{n_{ij}^\ell}        y_{ij}^\ell \bar{ \psi}_{L_i}^\ell  \varphi \psi_{R_j}^{\ell} 
+  {\rm H.c.}, \\ \nonumber
\eea
The complex singlet scalar field $\chi$ acquires VEV $\langle \chi \rangle=f$, and the Lagrangian corresponding to the SM fermions (charged leptons and quarks) mass is given by
\bea
-{\mathcal{L}}_{\rm Yukawa} &=&  Y^u_{ij} \bar{ \psi}_{L_i}^q  \tilde{\varphi} \psi_{R_j}^{u}
+ Y^d_{ij} \bar{ \psi}_{L_i}^q  \varphi \psi_{R_j}^{d}
+ Y^\ell_{ij} \bar{ \psi}_{L_i}^\ell  \varphi \psi_{R_j}^{\ell}   + \text{H.c.}, 
\eea
where  $i$ and $j$   show the generation indices, $ \psi_{L}^q,  \psi_{L}^\ell  $ are left-handed doublets of the  quark and leptonic fields, $ \psi_{R}^u,  \psi_{R}^d, \psi_{R}^\ell$ are the right-handed up, down-type  quarks and  leptons, $ \tilde{\varphi}= -i \sigma_2 \varphi^*$, $\varphi$ represent the SM Higgs doublet,  and $\sigma_2$ is  the second Pauli matrix. The couplings $Y_{ij}$  stand for  the effective Yukawa couplings such that  $Y_{ij} = y_{ij} \epsilon^{n_{ij}}$, where $   \dfrac{\langle \chi \rangle} { \Lambda} = \dfrac{f}{\sqrt{2} \Lambda}= \epsilon <1$.

\subsection{$\mathcal{Z}_{\rm 2} \times \mathcal{Z}_{\rm 5} $ flavor symmetry}
The first minimal realization of the $\mathcal{Z}_{\rm N} \times \mathcal{Z}_{\rm M} $ flavor symmetry is the $\mathcal{Z}_{\rm 2} \times \mathcal{Z}_{\rm 5} $ flavor symmetry.  This can produce the charged fermion masses and mixing along with that of neutrinos.  This is an essential criteria for a solution to the flavor problem \cite{Abbas:2020frs}.    The transformations of the scalar and fermionic fields under this symmetry are provided in table \ref{tab_z25}.

 \begin{table}[h!]
\begin{center}
\begin{tabular}{|c|c|c|}
  \hline
  Fields             &        $\mathcal{Z}_2$                    & $\mathcal{Z}_5$        \\
  \hline
  $u_{R}, c_{R}, t_{R}$                 &   +  & $ \omega^2$                             \\
   $d_{R},  s_{R}, b_{R}, e_R, \mu_R, \tau_R$                 &   -  &     $\omega $                              \\
   $\psi_{L_1}^q$                 &   +  &    $\omega $                          \\
    $\psi_{L_2}^q$                 &   +  &     $\omega^4 $                         \\
     $\psi_{L_3}^q$                 &   +  &      $ \omega^2 $                         \\
  $\psi_{L_1}^\ell$                 &   +  &    $\omega $                          \\
    $\psi_{L_2}^\ell$                 &   +  &     $\omega^4 $                         \\
     $\psi_{L_3}^\ell$                 &   +  &      $ \omega^2 $                         \\     
    $\chi$                        & -  &       $ \omega$                                        \\
    $\varphi$              &   +        &     1 \\
  \hline
     \end{tabular}
\end{center}
\caption{The charges of left and  right-handed fermions of three families of the SM,  right-handed neutrinos,  Higgs,  and singlet scalar fields under $\mathcal{Z}_2$ and $\mathcal{Z}_5$  symmetries,  where $\omega$ is the fifth root of unity.}
 \label{tab_z25}
\end{table} 

The  $\mathcal{Z}_{\rm 2} \times \mathcal{Z}_{\rm 5} $ flavor symmetry enables us to write the mass Lagrangian for charged fermions as,
\bea
\label{massz5}
-{\mathcal{L}}_{\rm Yukawa} &=&    \left(  \dfrac{ \chi}{\Lambda} \right)^{4}  y_{11}^u \bar{ \psi}_{L_1}^q \tilde{\varphi} u_{R}+  \left(  \dfrac{ \chi}{\Lambda} \right)^{4}  y_{12}^u \bar{ \psi}_{L_1}^q \tilde{\varphi} c_{R} +  \left(  \dfrac{ \chi}{\Lambda} \right)^{4}  y_{13}^u \bar{ \psi}_{L_1}^q \tilde{\varphi}  t_{R} +  \left(  \dfrac{ \chi}{\Lambda} \right)^{2}  y_{21}^u \bar{ \psi}_{L_2}^q \tilde{\varphi} u_{R}\nonumber \\
&+& \left(  \dfrac{ \chi}{\Lambda} \right)^{2}  y_{22}^u \bar{ \psi}_{L_2}^q \tilde{\varphi} c_{R} + \left(  \dfrac{ \chi}{\Lambda} \right)^{2}  y_{23}^u \bar{ \psi}_{L_2}^q \tilde{\varphi} t_{R}+   y_{31}^u \bar{ \psi}_{L_3}^q \tilde{\varphi} u_{R} +   y_{32}^u \bar{ \psi}_{L_3}^q \tilde{\varphi} c_{R} +   y_{33}^u \bar{ \psi}_{L_3}^q \tilde{\varphi} t_{R} \nonumber \\
&+&  
 \left(  \dfrac{ \chi}{\Lambda} \right)^{5} y_{11}^d \bar{ \psi}_{L_1}^q  \varphi d_{R} + \left(  \dfrac{ \chi}{\Lambda} \right)^{5} y_{12}^d \bar{ \psi}_{L_1}^q  \varphi s_{R} + \left(  \dfrac{ \chi}{\Lambda} \right)^{5} y_{13}^d \bar{ \psi}_{L_1}^q  \varphi b_{R} + \left(  \dfrac{ \chi}{\Lambda} \right)^{3} y_{21}^d \bar{ \psi}_{L_2}^q  \varphi d_{R} \nonumber \\
&+& \left(  \dfrac{ \chi}{\Lambda} \right)^{3} y_{22}^d \bar{ \psi}_{L_2}^q  \varphi s_{R} + \left(  \dfrac{ \chi}{\Lambda} \right)^{3} y_{23}^d \bar{ \psi}_{L_2}^q  \varphi b_{R} + \left(  \dfrac{ \chi}{\Lambda} \right) y_{31}^d \bar{ \psi}_{L_3}^q  \varphi d_{R}+ \left(  \dfrac{ \chi}{\Lambda} \right) y_{32}^d \bar{ \psi}_{L_3}^q  \varphi s_{R}   \nonumber \\ 
&+& \left(  \dfrac{ \chi}{\Lambda} \right) y_{33}^d \bar{ \psi}_{L_3}^q  \varphi b_{R} + 
\left(  \dfrac{ \chi}{\Lambda} \right)^{5} y_{11}^\ell \bar{ \psi}_{L_1}^\ell  \varphi e_{R} + \left(  \dfrac{ \chi}{\Lambda} \right)^{5} y_{12}^\ell \bar{ \psi}_{L_1}^\ell  \varphi \mu_{R} + \left(  \dfrac{ \chi}{\Lambda} \right)^{5} y_{13}^\ell \bar{ \psi}_{L_1}^\ell  \varphi \tau_{R}  \nonumber \\
&+& \left(  \dfrac{ \chi}{\Lambda} \right)^{3} y_{21}^\ell \bar{ \psi}_{L_2}^\ell  \varphi e_{R} + \left(  \dfrac{ \chi}{\Lambda} \right)^{3} y_{22}^\ell \bar{ \psi}_{L_2}^\ell  \varphi \mu_{R} + \left(  \dfrac{ \chi}{\Lambda} \right)^{3} y_{23}^\ell \bar{ \psi}_{L_2}^\ell  \varphi \tau_{R} +  \left(  \dfrac{ \chi}{\Lambda} \right) y_{31}^\ell \bar{ \psi}_{L_3}^\ell  \varphi e_{R} \nonumber \\
&+& \left(  \dfrac{ \chi}{\Lambda} \right) y_{32}^\ell \bar{ \psi}_{L_3}^\ell  \varphi \mu_{R} \nonumber + \left(  \dfrac{ \chi}{\Lambda} \right) y_{33}^\ell \bar{ \psi}_{L_3}^\ell  \varphi \tau_{R} 
 + \text{H.c.}
\eea

The mass matrices for  up- and down-type quarks and charged leptons read,
\begin{equation}
\M_u = \dfrac{v}{\sqrt{2}}
\begin{pmatrix}
y_{11}^u  \epsilon^4 &  y_{12}^u \epsilon^4  & y_{13}^u \epsilon^4    \\
y_{21}^u  \epsilon^2    & y_{22}^u \epsilon^2  &  y_{23}^u \epsilon^2    \\
y_{31}^u     &  y_{32}^u      &  y_{33}^u
\end{pmatrix}, 
\M_d = \dfrac{v}{\sqrt{2}}
\begin{pmatrix}
y_{11}^d  \epsilon^5 &  y_{12}^d \epsilon^5 & y_{13}^d \epsilon^5   \\
y_{21}^d  \epsilon^3  & y_{22}^d \epsilon^3 &  y_{23}^d \epsilon^3  \\
 y_{31}^d \epsilon &  y_{32}^d \epsilon   &  y_{33}^d \epsilon
\end{pmatrix},
\M_\ell =  \dfrac{v}{\sqrt{2}}
\begin{pmatrix}
y_{11}^\ell  \epsilon^5 &  y_{12}^\ell \epsilon^5  & y_{13}^\ell \epsilon^5   \\
y_{21}^\ell  \epsilon^3  & y_{22}^\ell \epsilon^3  &  y_{23}^\ell \epsilon^3  \\
 y_{31}^\ell \epsilon   &  y_{32}^\ell \epsilon   &  y_{33}^\ell \epsilon
\end{pmatrix},
\end{equation}
where $\epsilon = 0.1$ can produce the required masses of the fermions.

The masses of the charged fermions can be written as\cite{Rasin:1998je},
\begin{align}
\label{eqn5}
\{m_t, m_c, m_u\} &\simeq \{|y_{33}^u| , ~ \left |y_{22}^u- \frac {y_{23}^u y_{32}^u} {|y_{33}^u|} \right| \epsilon^2,\\&
~ \left |y_{11}^u- \frac {y_{12}^u y_{21}^u}{|y_{22}^u-y_{23}^u y_{32}^u/y_{33}^u|}- \frac{y_{13}^u |y_{31}^u y_{22}^u-y_{21}^u y_{32}^u|-y_{31}^u y_{12}^u y_{23}^u}{|y_{22}^u- y_{23}^u y_{32}^u/y_{33}^u| |y_{33}^u|} \right| \epsilon^4\}v/\sqrt{2} ,\nonumber &\\ 
\label{eqn6}
\{m_b, m_s, m_d\} & \simeq \{|y_{33}^d| \epsilon, ~ \left |y_{22}^d- \frac {y_{23}^d y_{32}^d} {|y_{33}^d|} \right| \epsilon^3,\\&
~  \left |y_{11}^d- \frac {y_{12}^d y_{21}^d}{|y_{22}^d-y_{23}^d y_{32}^d/y_{33}^d|}- \frac{y_{13}^d |y_{31}^d y_{22}^d-y_{21}^d y_{32}^d|-y_{31}^d y_{12}^d y_{23}^d}{|y_{22}^d- y_{23}^d y_{32}^d/y_{33}^d| |y_{33}^d|} \right| \epsilon^5\}v/\sqrt{2} ,\nonumber &\\
\{m_\tau, m_\mu, m_e\} & \simeq \{|y_{33}^l| \epsilon, ~ \left|y_{22}^l- \frac {y_{23}^l y_{32}^l} {|y_{33}^l|} \right| \epsilon^3,\\& ~  \left |y_{11}^l- \frac {y_{12}^l y_{21}^l}{|y_{22}^l-y_{23}^l y_{32}^l/y_{33}^l|}- \frac{y_{13}^l |y_{31}^l y_{22}^l-y_{21}^l y_{32}^l|-y_{31}^l y_{12}^l y_{23}^l}{|y_{22}^l- y_{23}^l y_{32}^l/y_{33}^l| |y_{33}^l|} \right| \epsilon^5\}v/\sqrt{2}. \nonumber &\\
\end{align}

The quark mixing angles  are  \cite{tasi2000},
\begin{eqnarray}
\sin \theta_{12}  \simeq |V_{us}| &\simeq& \left|{y_{12}^d \over y_{22}^d}  -{y_{12}^u \over y_{22}^u}  \right| \epsilon^2, 
\sin \theta_{23}  \simeq |V_{cb}| \simeq  \left|{y_{23}^d \over y_{33}^d}  -{y_{23}^u \over y_{33}^u}  \right|  \epsilon^2,\nonumber \\
\sin \theta_{13}  \simeq |V_{ub}| &\simeq& \left|{y_{13}^d \over y_{33}^d}  -{y_{12}^u y_{23}^d \over y_{22}^u y_{33}^d} 
- {y_{13}^u \over y_{33}^u} \right|  \epsilon^4.
\end{eqnarray}

\subsection{$\mathcal{Z}_{\rm 2} \times \mathcal{Z}_{\rm 9} $ flavor symmetry}
A  non-minimal realization of the $\mathcal{Z}_{\rm N} \times \mathcal{Z}_{\rm M} $ flavor symmetry is the $\mathcal{Z}_{\rm 2} \times \mathcal{Z}_{\rm 9} $ flavor symmetry.  The scalar and fermionic fields transform under this symmetry, as given in table \ref{tab_z29}.

 \begin{table}[h!]
\begin{center}
\begin{tabular}{|c|c|c|}
  \hline
  Fields             &        $\mathcal{Z}_2$                    & $\mathcal{Z}_9$        \\
  \hline
  $u_{R}, t_{R}$                 &   +  & $ 1$                             \\
  $c_{R}$                 &   +  & $ \omega^4$                             \\
   $d_{R},  s_{R},  b_{R}, e_R, \mu_R, \tau_R $                 &   -  &     $\omega^3 $                              \\
   $\psi_{L_1}^q$                 &   +  &    $\omega $                          \\
    $\psi_{L_2}^q$                 &   +  &     $\omega^8 $                         \\
     $\psi_{L_3}^q$                 &   +  &      $ 1 $                         \\
     $\psi_{L_1}^\ell$                 &   +  &    $\omega $                          \\
    $\psi_{L_2}^\ell$                 &   +  &     $\omega^8 $                         \\
     $\psi_{L_3}^\ell$                 &   +  &      $ \omega^6 $                         \\     
    $\chi$                        & -  &       $ \omega$                                        \\
    $\varphi$              &   +        &     1 \\
  \hline
     \end{tabular}
\end{center}
\caption{The charges of left and  right-handed fermions of three families of the SM,  right-handed neutrinos,  Higgs,  and singlet scalar field under $\mathcal{Z}_2$ and $\mathcal{Z}_9$  symmetries,  where $\omega$ is the ninth root of unity. }
 \label{tab_z29}
\end{table} 

After imposing the $\mathcal{Z}_{\rm 2} \times \mathcal{Z}_{\rm 9} $ flavor symmetry on the SM, the mass Lagrangian for charged fermions reads as,
\bea
\label{massz9}
-{\mathcal{L}}_{\rm Yukawa} &=&    \left(  \dfrac{ \chi^\dag}{\Lambda} \right)^{8}  y_{11}^u \bar{ \psi}_{L_1}^q \tilde{\varphi} u_{R}+  \left(  \dfrac{ \chi}{\Lambda} \right)^{6}  y_{12}^u \bar{ \psi}_{L_1}^q \tilde{\varphi} c_{R} +  \left(  \dfrac{ \chi^\dag}{\Lambda} \right)^{8}  y_{13}^u \bar{ \psi}_{L_1}^q \tilde{\varphi}  t_{R} +  \left(  \dfrac{ \chi}{\Lambda} \right)^{8}  y_{21}^u \bar{ \psi}_{L_2}^q \tilde{\varphi} u_{R}\nonumber \\
&+& \left(  \dfrac{ \chi}{\Lambda} \right)^{4}  y_{22}^u \bar{ \psi}_{L_2}^q \tilde{\varphi} c_{R} + \left(  \dfrac{ \chi}{\Lambda} \right)^{8}  y_{23}^u \bar{ \psi}_{L_2}^q \tilde{\varphi} t_{R}+   y_{31}^u \bar{ \psi}_{L_3}^q \tilde{\varphi} u_{R} + \left(  \dfrac{ \chi^\dag}{\Lambda} \right)^{4}  y_{32}^u \bar{ \psi}_{L_3}^q \tilde{\varphi} c_{R} +   y_{33}^u \bar{ \psi}_{L_3}^q \tilde{\varphi} t_{R} \nonumber \\
&+&  
 \left(  \dfrac{ \chi}{\Lambda} \right)^{7} y_{11}^d \bar{ \psi}_{L_1}^q  \varphi d_{R} + \left(  \dfrac{ \chi}{\Lambda} \right)^{7} y_{12}^d \bar{ \psi}_{L_1}^q  \varphi s_{R} + \left(  \dfrac{ \chi}{\Lambda} \right)^{7} y_{13}^d \bar{ \psi}_{L_1}^q  \varphi b_{R} + \left(  \dfrac{ \chi}{\Lambda} \right)^{5} y_{21}^d \bar{ \psi}_{L_2}^q  \varphi d_{R} \nonumber \\
&+& \left(  \dfrac{ \chi}{\Lambda} \right)^{5} y_{22}^d \bar{ \psi}_{L_2}^q  \varphi s_{R} + \left(  \dfrac{ \chi}{\Lambda} \right)^{5} y_{23}^d \bar{ \psi}_{L_2}^q  \varphi b_{R} + \left(  \dfrac{ \chi^\dag}{\Lambda} \right)^{3} y_{31}^d \bar{ \psi}_{L_3}^q  \varphi d_{R}+ \left(  \dfrac{ \chi^\dag}{\Lambda} \right)^{3} y_{32}^d \bar{ \psi}_{L_3}^q  \varphi s_{R}   \nonumber \\ 
&+& \left(  \dfrac{ \chi^\dag}{\Lambda} \right)^{3} y_{33}^d \bar{ \psi}_{L_3}^q  \varphi b_{R} + 
\left(  \dfrac{ \chi}{\Lambda} \right)^{7} y_{11}^\ell \bar{ \psi}_{L_1}^\ell  \varphi e_{R} + \left(  \dfrac{ \chi}{\Lambda} \right)^{7} y_{12}^\ell \bar{ \psi}_{L_1}^\ell  \varphi \mu_{R} + \left(  \dfrac{ \chi}{\Lambda} \right)^{7} y_{13}^\ell \bar{ \psi}_{L_1}^\ell  \varphi \tau_{R}  \nonumber \\
&+& \left(  \dfrac{ \chi}{\Lambda} \right)^{5} y_{21}^\ell \bar{ \psi}_{L_2}^\ell  \varphi e_{R} + \left(  \dfrac{ \chi}{\Lambda} \right)^{5} y_{22}^\ell \bar{ \psi}_{L_2}^\ell  \varphi \mu_{R} + \left(  \dfrac{ \chi}{\Lambda} \right)^{5} y_{23}^\ell \bar{ \psi}_{L_2}^\ell  \varphi \tau_{R} +  \left(  \dfrac{ \chi}{\Lambda} \right)^{3} y_{31}^\ell \bar{ \psi}_{L_3}^\ell  \varphi e_{R} \nonumber \\
&+& \left(  \dfrac{ \chi}{\Lambda} \right)^{3} y_{32}^\ell \bar{ \psi}_{L_3}^\ell  \varphi \mu_{R} \nonumber + \left(  \dfrac{ \chi}{\Lambda} \right)^{3} y_{33}^\ell \bar{ \psi}_{L_3}^\ell  \varphi \tau_{R} 
 + \text{H.c.}
\eea

The mass matrices for  up and down-type quarks and charged leptons read,
\begin{equation}
\M_u = \dfrac{v}{\sqrt{2}}
\begin{pmatrix}
y_{11}^u  \epsilon^8 &  y_{12}^u \epsilon^{6}  & y_{13}^u \epsilon^{8}    \\
y_{21}^u  \epsilon^8    & y_{22}^u \epsilon^4  &  y_{23}^u \epsilon^8   \\
y_{31}^u      &  y_{32}^u  \epsilon^4     &  y_{33}^u  
\end{pmatrix}, 
\M_d = \dfrac{v}{\sqrt{2}}
\begin{pmatrix}
y_{11}^d  \epsilon^7 &  y_{12}^d \epsilon^7 & y_{13}^d \epsilon^7   \\
y_{21}^d  \epsilon^5  & y_{22}^d \epsilon^5 &  y_{23}^d \epsilon^5  \\
 y_{31}^d \epsilon^3 &  y_{32}^d \epsilon^3   &  y_{33}^d \epsilon^3
\end{pmatrix},
\M_\ell =  \dfrac{v}{\sqrt{2}}
\begin{pmatrix}
y_{11}^\ell  \epsilon^7 &  y_{12}^\ell \epsilon^7  & y_{13}^\ell \epsilon^7   \\
y_{21}^\ell  \epsilon^5  & y_{22}^\ell \epsilon^5  &  y_{23}^\ell \epsilon^5  \\
 y_{31}^\ell \epsilon^3   &  y_{32}^\ell \epsilon^3   &  y_{33}^\ell \epsilon^3
\end{pmatrix}.
\end{equation}
where $\epsilon = 0.225$ is chosen to produce the  masses of  fermions.

The masses of charged fermions are approximately \cite{Rasin:1998je},
\begin{align}
\label{eqn5}
\{m_t, m_c, m_u\} &\simeq \{|y_{33}^u| , ~ \left |y_{22}^u   \epsilon^4 - \frac {y_{23}^u y_{32}^u} {|y_{33}^u|   }  \epsilon^{12} \right|,\\&
~ \left |y_{11}^u \epsilon^8 - \frac {y_{12}^u y_{21}^u}{|y_{22}^u|} \epsilon^{10}- \frac{y_{13}^u |y_{31}^u y_{22}^u-y_{21}^u y_{32}^u|}{|y_{22}^u| |y_{33}^u|} \epsilon^8 \right| \}v/\sqrt{2} ,\nonumber &\\ 
\label{eqn6}
\{m_b, m_s, m_d\} & \simeq \{|y_{33}^d| \epsilon^3, ~ \left |y_{22}^d- \frac {y_{23}^d y_{32}^d} {|y_{33}^d|} \right| \epsilon^5,\\&
~  \left |y_{11}^d- \frac {y_{12}^d y_{21}^d}{|y_{22}^d-y_{23}^d y_{32}^d/y_{33}^d|}- \frac{y_{13}^d |y_{31}^d y_{22}^d-y_{21}^d y_{32}^d|-y_{31}^d y_{12}^d y_{23}^d}{|y_{22}^d- y_{23}^d y_{32}^d/y_{33}^d| |y_{33}^d|} \right| \epsilon^7\}v/\sqrt{2} ,\nonumber &\\
\{m_\tau, m_\mu, m_e\} & \simeq \{|y_{33}^l| \epsilon^3, ~ \left|y_{22}^l- \frac {y_{23}^l y_{32}^l} {|y_{33}^l|} \right| \epsilon^5,\\& ~  \left |y_{11}^l- \frac {y_{12}^l y_{21}^l}{|y_{22}^l-y_{23}^l y_{32}^l/y_{33}^l|}- \frac{y_{13}^l |y_{31}^l y_{22}^l-y_{21}^l y_{32}^l|-y_{31}^l y_{12}^l y_{23}^l}{|y_{22}^l- y_{23}^l y_{32}^l/y_{33}^l| |y_{33}^l|} \right| \epsilon^7\}v/\sqrt{2}. \nonumber &\\
\end{align}

The quark mixing angles are identical to that of the minimal  $\mathcal{Z}_{\rm 2} \times \mathcal{Z}_{\rm 5} $ flavor symmetry.

\subsection{$\mathcal{Z}_{\rm 2} \times \mathcal{Z}_{\rm 11} $ flavor symmetry}
The next non-minimal realization of the   $\mathcal{Z}_{\rm N} \times \mathcal{Z}_{\rm M} $ flavor symmetry is the $\mathcal{Z}_{\rm 2} \times \mathcal{Z}_{\rm 11} $ flavor symmetry.   The $\mathcal{Z}_{\rm 2} \times \mathcal{Z}_{\rm 11} $ flavor symmetry, in this work, is chosen to probe a relatively large value of the parameter $\epsilon$, which is $0.28$.  The  scalar and fermionic fields transform under this symmetry, as shown in table \ref{tab_z211}.

 \begin{table}[h!]
\begin{center}
\begin{tabular}{|c|c|c|}
  \hline
  Fields             &        $\mathcal{Z}_2$                    & $\mathcal{Z}_{11}$        \\
  \hline
  $u_{R}, c_{R}$                 &   +  & $ \omega^2$                             \\
  $t_{R}$                 &   +  & $ 1$                             \\
   $d_{R},  s_{R}, b_{R}, e_R, \mu_R, \tau_R$                 &   -  &     $\omega^3 $                              \\
   $\psi_{L}^1$                 &   +  &    $\omega $                          \\
    $\psi_{L}^2$                 &   +  &     $\omega^{7} $                         \\
     $\psi_{L}^3$                 &   +  &      $ 1 $                         \\
    $\chi$                        & -  &       $ \omega$                                        \\
    $\varphi$              &   +        &     1 \\
    $\sigma$              &   +        &     $ \omega^4 $  \\
  \hline
     \end{tabular}
\end{center}
\caption{The charges of left and  right-handed fermions of three families of the SM, right-handed neutrinos,  Higgs, and singlet scalar fields under $\mathcal{Z}_2$ and $\mathcal{Z}_{11}$  symmetries, where $\omega$ is the eleventh root of unity. }
 \label{tab_z211}
\end{table} 
The mass Lagrangian for charged fermions after imposing $\mathcal{Z}_{\rm 2} \times \mathcal{Z}_{\rm 11} $ flavor symmetry on the SM reads,
\bea
\label{massz11}
-{\mathcal{L}}_{\rm Yukawa} &=&    \left(  \dfrac{ \chi}{\Lambda} \right)^{10}  y_{11}^u \bar{ \psi}_{L_1}^q \tilde{\varphi} u_{R}+  \left(  \dfrac{ \chi}{\Lambda} \right)^{10}  y_{12}^u \bar{ \psi}_{L_1}^q \tilde{\varphi} c_{R} +  \left(  \dfrac{ \chi}{\Lambda} \right)^{12}  y_{13}^u \bar{ \psi}_{L_1}^q \tilde{\varphi}  t_{R} +  \left(  \dfrac{ \chi^\dag}{\Lambda} \right)^{6}  y_{21}^u \bar{ \psi}_{L_2}^q \tilde{\varphi} u_{R}\nonumber \\
&+& \left(  \dfrac{ \chi^\dag}{\Lambda} \right)^{6}  y_{22}^u \bar{ \psi}_{L_2}^q \tilde{\varphi} c_{R} + \left(  \dfrac{ \chi^\dag}{\Lambda} \right)^{4}  y_{23}^u \bar{ \psi}_{L_2}^q \tilde{\varphi} t_{R}+ \left(  \dfrac{ \chi^\dag}{\Lambda} \right)^{2}  y_{31}^u \bar{ \psi}_{L_3}^q \tilde{\varphi} u_{R} + \left(  \dfrac{ \chi^\dag}{\Lambda} \right)^{2}  y_{32}^u \bar{ \psi}_{L_3}^q \tilde{\varphi} c_{R} \nonumber \\
&+&   y_{33}^u \bar{ \psi}_{L_3}^q \tilde{\varphi} t_{R} 
+ \left(  \dfrac{ \chi}{\Lambda} \right)^{9} y_{11}^d \bar{ \psi}_{L_1}^q  \varphi d_{R} + \left(  \dfrac{ \chi}{\Lambda} \right)^{9} y_{12}^d \bar{ \psi}_{L_1}^q  \varphi s_{R} + \left(  \dfrac{ \chi}{\Lambda} \right)^{9} y_{13}^d \bar{ \psi}_{L_1}^q  \varphi b_{R} + \left(  \dfrac{ \chi^\dag}{\Lambda} \right)^{7} y_{21}^d \bar{ \psi}_{L_2}^q  \varphi d_{R} \nonumber \\
&+& \left(  \dfrac{ \chi^\dag}{\Lambda} \right)^{7} y_{22}^d \bar{ \psi}_{L_2}^q  \varphi s_{R} + \left(  \dfrac{ \chi^\dag}{\Lambda} \right)^{7} y_{23}^d \bar{ \psi}_{L_2}^q  \varphi b_{R} + \left(  \dfrac{ \chi^\dag}{\Lambda} \right)^{3} y_{31}^d \bar{ \psi}_{L_3}^q  \varphi d_{R}+ \left(  \dfrac{ \chi^\dag}{\Lambda} \right)^{3} y_{32}^d \bar{ \psi}_{L_3}^q  \varphi s_{R}   \nonumber \\ 
&+& \left(  \dfrac{ \chi^\dag}{\Lambda} \right)^{3} y_{33}^d \bar{ \psi}_{L_3}^q  \varphi b_{R} + 
\left(  \dfrac{ \chi}{\Lambda} \right)^{9} y_{11}^\ell \bar{ \psi}_{L_1}^\ell  \varphi e_{R} + \left(  \dfrac{ \chi}{\Lambda} \right)^{9} y_{12}^\ell \bar{ \psi}_{L_1}^\ell  \varphi \mu_{R} + \left(  \dfrac{ \chi}{\Lambda} \right)^{9} y_{13}^\ell \bar{ \psi}_{L_1}^\ell  \varphi \tau_{R}  \nonumber \\
&+& \left(  \dfrac{ \chi^\dag}{\Lambda} \right)^{7} y_{21}^\ell \bar{ \psi}_{L_2}^\ell  \varphi e_{R} + \left(  \dfrac{ \chi^\dag}{\Lambda} \right)^{7} y_{22}^\ell \bar{ \psi}_{L_2}^\ell  \varphi \mu_{R} + \left(  \dfrac{ \chi^\dag}{\Lambda} \right)^{7} y_{23}^\ell \bar{ \psi}_{L_2}^\ell  \varphi \tau_{R} +  \left(  \dfrac{ \chi^\dag}{\Lambda} \right)^{3} y_{31}^\ell \bar{ \psi}_{L_3}^\ell  \varphi e_{R} \nonumber \\
&+& \left(  \dfrac{\chi^\dag}{\Lambda} \right)^{3} y_{32}^\ell \bar{ \psi}_{L_3}^\ell  \varphi \mu_{R} \nonumber + \left(  \dfrac{ \chi^\dag}{\Lambda} \right)^{3} y_{33}^\ell \bar{ \psi}_{L_3}^\ell  \varphi \tau_{R} 
 + \text{H.c.}
\eea

The  mass matrices for  up- and down-type quarks and charged leptons  turn out to be,
\begin{equation}
\M_u = \dfrac{v}{\sqrt{2}}
\begin{pmatrix}
y_{11}^u  \epsilon^{10} &  y_{12}^u \epsilon^{10}  & y_{13}^u \epsilon^{12}    \\
y_{21}^u  \epsilon^6    & y_{22}^u \epsilon^6  &  y_{23}^u \epsilon^{4}    \\
y_{31}^u \epsilon^2    &  y_{32}^u  \epsilon^2    &  y_{33}^u
\end{pmatrix}, 
\M_d = \dfrac{v}{\sqrt{2}}
\begin{pmatrix}
y_{11}^d  \epsilon^9 &  y_{12}^d \epsilon^9 & y_{13}^d \epsilon^9   \\
y_{21}^d  \epsilon^7  & y_{22}^d \epsilon^7 &  y_{23}^d \epsilon^7  \\
 y_{31}^d \epsilon^3 &  y_{32}^d \epsilon^3   &  y_{33}^d \epsilon^3
\end{pmatrix},
\M_\ell =  \dfrac{v}{\sqrt{2}}
\begin{pmatrix}
y_{11}^\ell  \epsilon^9 &  y_{12}^\ell \epsilon^9  & y_{13}^\ell \epsilon^9   \\
y_{21}^\ell  \epsilon^7  & y_{22}^\ell \epsilon^7  &  y_{23}^\ell \epsilon^7  \\
 y_{31}^\ell \epsilon^3   &  y_{32}^\ell \epsilon^3   &  y_{33}^\ell \epsilon^3
\end{pmatrix}.
\end{equation}

The masses of quarks and charged leptons are approximately \cite{Rasin:1998je},
\begin{align}
\label{eqn5}
\{m_t, m_c, m_u\} &\simeq \{|y_{33}^u| , ~ \left |y_{22}^u- \frac {y_{23}^u y_{32}^u} {|y_{33}^u|} \right| \epsilon^6,\\&
~ \left |y_{11}^u- \frac {y_{12}^u y_{21}^u}{|y_{22}^u-y_{23}^u y_{32}^u/y_{33}^u|}- \frac{y_{13}^u |y_{31}^u y_{22}^u-y_{21}^u y_{32}^u|-y_{31}^u y_{12}^u y_{23}^u}{|y_{22}^u- y_{23}^u y_{32}^u/y_{33}^u| |y_{33}^u|} \right| \epsilon^{10}\}v/\sqrt{2} ,\nonumber &\\ 
\label{eqn6}
\{m_b, m_s, m_d\} & \simeq \{|y_{33}^d| \epsilon^3, ~ \left |y_{22}^d- \frac {y_{23}^d y_{32}^d} {|y_{33}^d|} \right| \epsilon^7,\\&
~  \left |y_{11}^d- \frac {y_{12}^d y_{21}^d}{|y_{22}^d-y_{23}^d y_{32}^d/y_{33}^d|}- \frac{y_{13}^d |y_{31}^d y_{22}^d-y_{21}^d y_{32}^d|-y_{31}^d y_{12}^d y_{23}^d}{|y_{22}^d- y_{23}^d y_{32}^d/y_{33}^d| |y_{33}^d|} \right| \epsilon^9\}v/\sqrt{2} ,\nonumber &\\
\{m_\tau, m_\mu, m_e\} & \simeq \{|y_{33}^l| \epsilon,^3 ~ \left|y_{22}^l- \frac {y_{23}^l y_{32}^l} {|y_{33}^l|} \right| \epsilon^7,\\& ~  \left |y_{11}^l- \frac {y_{12}^l y_{21}^l}{|y_{22}^l-y_{23}^l y_{32}^l/y_{33}^l|}- \frac{y_{13}^l |y_{31}^l y_{22}^l-y_{21}^l y_{32}^l|-y_{31}^l y_{12}^l y_{23}^l}{|y_{22}^l- y_{23}^l y_{32}^l/y_{33}^l| |y_{33}^l|} \right| \epsilon^9\}v/\sqrt{2} .\nonumber &\\
\end{align}

The mixing angles of quarks are found to be \cite{tasi2000},
\begin{eqnarray}
\sin \theta_{12}  \simeq |V_{us}| &\simeq& \left|{y_{12}^d \over y_{22}^d}  \epsilon^2 -{y_{12}^u \over y_{22}^u } \epsilon^4 \right| , 
\sin \theta_{23}  \simeq |V_{cb}| \simeq  \left|{y_{23}^d \over y_{33}^d}  -{y_{23}^u \over y_{33}^u }  \right|  \epsilon^4,\nonumber \\
\sin \theta_{13}  \simeq |V_{ub}| &\simeq& \left|{y_{13}^d \over y_{33}^d}  \epsilon^6  -{y_{12}^u y_{23}^d \over y_{22}^u y_{33}^d}  \epsilon^8 \right| .
\end{eqnarray}
The mixing angles in this have a consistent hierarchical pattern of the order ($\epsilon^2,~ \epsilon^4,~\epsilon^6$).  However, to produce the observed values of the fermionic masses and mixing angles, we need a relatively large value of the parameter $\epsilon$.

\subsection{The $\mathcal{Z}_{8} \times \mathcal{Z}_{22}$ flavor symmetry} 
This symmetry can provide the so-called flavonic dark matter \cite{Abbas:2023ion}, and is inspired by the hierarchical VEVs model\cite{Abbas:2017vws,Abbas:2020frs}, where the mass of the top quark originates from the dimension-5 operator.  The transformations of fermionic and scalar fields under this symmetry are given in table  \ref{tab_z822}.

   \begin{table}[h!]
 \small
\begin{center}
\begin{tabular}{|c|c|c|c|c|c|c|c|c| c|c|c| c|c|c|}
  \hline
  Fields             &        $\mathcal{Z}_8$                    & $\mathcal{Z}_{22}$ & Fields             &        $\mathcal{Z}_8$                    & $\mathcal{Z}_{22}$   & Fields             &        $\mathcal{Z}_8$                    & $\mathcal{Z}_{22}$    & Fields             &        $\mathcal{Z}_8$                    & $\mathcal{Z}_{22}$     & Fields             &        $\mathcal{Z}_8$                    & $\mathcal{Z}_{22}$        \\
  \hline
  $u_{R}$                 &   $ \omega^2$  &$ \omega^2$        &$c_{R}$                 &   $ \omega^5$  & $ \omega^5$    &$t_{R}$                 &   $ \omega^6$  & $ \omega^6$       & $d_{R}$                 &   $ \omega^3$  &     $\omega^{3} $           & $s_{R}$                 &   $ \omega^4$  &     $\omega^4 $           \\
  $b_{R}$                 &   $ \omega^4$  &     $\omega^4 $     &   $\psi_{L,1}^q$                 &    $ \omega^2$  &    $\omega^{10} $      & $\psi_{L,2}^q$                 &  $ \omega$  &     $\omega^{9} $       &  $\psi_{L,3}^q$                 &    $\omega^{7} $  &      $\omega^{7} $ & $\psi_{L,1}^\ell$                 &   $ \omega^3$  &    $\omega^3 $          \\
     $\psi_{L,2}^\ell$                  &   $ \omega^2$  &    $\omega^2 $    &   $\psi_{L,3}^\ell$                 &   $ \omega^2$  &    $\omega^2 $     &  $e_R$                 &   $\omega^{2} $  &     $\omega^{16} $          & $\mu_R$                 &  $\omega^5 $   &     $\omega^{19} $      &  $\tau_R $                 &   $ \omega^7$  &     $\omega^{21} $              \\
            $ \nu_{e_R} $                 &     $\omega^2 $    &     $1 $         & $   \nu_{\mu_R}$                 &     $\omega^5 $    &     $\omega^{3} $          &  $  \nu_{\tau_R} $                 &     $\omega^6 $    &     $\omega^{4} $        &   $\chi$                        & $ \omega$  &       $ \omega$       & $\varphi$              &   1        &     1                  \\          
  \hline
     \end{tabular}
\end{center}
\caption{The charges of the SM  and the flavon fields under the $\mathcal{Z}_8 \times \mathcal{Z}_{22}$  symmetry,  where $\omega$ is the 8th and  22nd root of unity. }
 \label{tab_z822}
\end{table} 

The $\mathcal{Z}_8 \times \mathcal{Z}_{22}$  symmetry allows us to write mass Lagrangian for the charged fermions as,
\bea
\label{massz11}
-{\mathcal{L}}_{\rm Yukawa} &=&    \left(  \dfrac{ \chi}{\Lambda} \right)^{8}  y_{11}^u \bar{ \psi}_{L_1}^q \tilde{\varphi} u_{R}+  \left(  \dfrac{ \chi}{\Lambda} \right)^{5}  y_{12}^u \bar{ \psi}_{L_1}^q \tilde{\varphi} c_{R} +  \left(  \dfrac{ \chi}{\Lambda} \right)^{4}  y_{13}^u \bar{ \psi}_{L_1}^q \tilde{\varphi}  t_{R} +  \left(  \dfrac{ \chi}{\Lambda} \right)^{7}  y_{21}^u \bar{ \psi}_{L_2}^q \tilde{\varphi} u_{R}\nonumber \\
&+& \left(  \dfrac{ \chi}{\Lambda} \right)^{4}  y_{22}^u \bar{ \psi}_{L_2}^q \tilde{\varphi} c_{R} + \left(  \dfrac{ \chi}{\Lambda} \right)^{3}  y_{23}^u \bar{ \psi}_{L_2}^q \tilde{\varphi} t_{R}+ \left(  \dfrac{ \chi}{\Lambda} \right)^{5}  y_{31}^u \bar{ \psi}_{L_3}^q \tilde{\varphi} u_{R} + \left(  \dfrac{ \chi}{\Lambda} \right)^{2}  y_{32}^u \bar{ \psi}_{L_3}^q \tilde{\varphi} c_{R} \nonumber \\
&+&  \left(  \dfrac{ \chi}{\Lambda} \right) y_{33}^u \bar{ \psi}_{L_3}^q \tilde{\varphi} t_{R} 
+ \left(  \dfrac{ \chi}{\Lambda} \right)^{7} y_{11}^d \bar{ \psi}_{L_1}^q  \varphi d_{R} + \left(  \dfrac{ \chi}{\Lambda} \right)^{6} y_{12}^d \bar{ \psi}_{L_1}^q  \varphi s_{R} + \left(  \dfrac{ \chi}{\Lambda} \right)^{6} y_{13}^d \bar{ \psi}_{L_1}^q  \varphi b_{R} + \left(  \dfrac{ \chi}{\Lambda} \right)^{6} y_{21}^d \bar{ \psi}_{L_2}^q  \varphi d_{R} \nonumber \\
&+& \left(  \dfrac{ \chi}{\Lambda} \right)^{5} y_{22}^d \bar{ \psi}_{L_2}^q  \varphi s_{R} + \left(  \dfrac{ \chi}{\Lambda} \right)^{5} y_{23}^d \bar{ \psi}_{L_2}^q  \varphi b_{R} + \left(  \dfrac{ \chi}{\Lambda} \right)^{4} y_{31}^d \bar{ \psi}_{L_3}^q  \varphi d_{R}+ \left(  \dfrac{ \chi}{\Lambda} \right)^{3} y_{32}^d \bar{ \psi}_{L_3}^q  \varphi s_{R}   \nonumber \\ 
&+& \left(  \dfrac{ \chi}{\Lambda} \right)^{3} y_{33}^d \bar{ \psi}_{L_3}^q  \varphi b_{R} + 
\left(  \dfrac{ \chi}{\Lambda} \right)^{9} y_{11}^\ell \bar{ \psi}_{L_1}^\ell  \varphi e_{R} + \left(  \dfrac{ \chi}{\Lambda} \right)^{6} y_{12}^\ell \bar{ \psi}_{L_1}^\ell  \varphi \mu_{R} + \left(  \dfrac{ \chi}{\Lambda} \right)^{4} y_{13}^\ell \bar{ \psi}_{L_1}^\ell  \varphi \tau_{R}  \nonumber \\
&+& \left(  \dfrac{ \chi}{\Lambda} \right)^{8} y_{21}^\ell \bar{ \psi}_{L_2}^\ell  \varphi e_{R} + \left(  \dfrac{ \chi}{\Lambda} \right)^{5} y_{22}^\ell \bar{ \psi}_{L_2}^\ell  \varphi \mu_{R} + \left(  \dfrac{ \chi}{\Lambda} \right)^{3} y_{23}^\ell \bar{ \psi}_{L_2}^\ell  \varphi \tau_{R} +  \left(  \dfrac{ \chi}{\Lambda} \right)^{8} y_{31}^\ell \bar{ \psi}_{L_3}^\ell  \varphi e_{R} \nonumber \\
&+& \left(  \dfrac{\chi}{\Lambda} \right)^{5} y_{32}^\ell \bar{ \psi}_{L_3}^\ell  \varphi \mu_{R} \nonumber + \left(  \dfrac{ \chi}{\Lambda} \right)^{3} y_{33}^\ell \bar{ \psi}_{L_3}^\ell  \varphi \tau_{R} 
 + \text{H.c.}
\eea

The mass matrices of the  up and down-type quarks and charged leptons take the forms,
\begin{align}
\M_u & = \dfrac{v}{\sqrt{2}}
\begin{pmatrix}
y_{11}^u  \epsilon^8 &  y_{12}^u \epsilon^{5}  & y_{13}^u \epsilon^{4}    \\
y_{21}^u \epsilon^7     & y_{22}^u \epsilon^4  &  y_{23}^u \epsilon^{3}  \\
y_{31}^u  \epsilon^{5}    &  y_{32}^u  \epsilon^2     &  y_{33}^u  \epsilon 
\end{pmatrix},  
\M_d   = \dfrac{v}{\sqrt{2}}
\begin{pmatrix}
y_{11}^d  \epsilon^7 &  y_{12}^d \epsilon^6 & y_{13}^d \epsilon^6   \\
y_{21}^d  \epsilon^6  & y_{22}^d \epsilon^5 &  y_{23}^d \epsilon^5  \\
 y_{31}^d \epsilon^4 &  y_{32}^d \epsilon^3   &  y_{33}^d \epsilon^3
\end{pmatrix}, 
\M_\ell =  \dfrac{v}{\sqrt{2}}
\begin{pmatrix}
y_{11}^\ell  \epsilon^9 &  y_{12}^\ell \epsilon^6  & y_{13}^\ell \epsilon^4   \\
y_{21}^\ell  \epsilon^{8}  & y_{22}^\ell \epsilon^5  &  y_{23}^\ell \epsilon^3  \\
 y_{31}^\ell \epsilon^{8}   &  y_{32}^\ell \epsilon^5   &  y_{33}^\ell \epsilon^3
\end{pmatrix},
\end{align}
where $\epsilon = 0.225$ produces the masses and mixing of fermions.

The masses of charged fermions are given by \cite{Rasin:1998je},
\begin{align}
\label{eqn5}
\{m_t, m_c, m_u\} &\simeq \{|y_{33}^u| \epsilon , ~ \left |y_{22}^u  - \frac {y_{23}^u y_{32}^u} {y_{33}^u  }   \right|  \epsilon^4 ,\\&
~ \left |y_{11}^u- \frac {y_{12}^u y_{21}^u}{y_{22}^u-y_{23}^u y_{32}^u/y_{33}^u}- \frac{y_{13}^u (y_{31}^u y_{22}^u-y_{21}^u y_{32}^u)-y_{31}^u y_{12}^u y_{23}^u}{(y_{22}^u- y_{23}^u y_{32}^u/y_{33}^u) y_{33}^u} \right| \epsilon^8\}v/\sqrt{2}  ,\nonumber \\ 
\{m_b, m_s, m_d\} & \simeq \{|y_{33}^d| \epsilon^3, ~ \left |y_{22}^d- \frac {y_{23}^d y_{32}^d} {y_{33}^d} \right| \epsilon^5,\\ \nonumber 
&  \left |y_{11}^d- \frac {y_{12}^d y_{21}^d}{y_{22}^d-y_{23}^d y_{32}^d/y_{33}^d}- \frac{y_{13}^d (y_{31}^d y_{22}^d-y_{21}^d y_{32}^d)-y_{31}^d y_{12}^d y_{23}^d}{(y_{22}^d- y_{23}^d y_{32}^d/y_{33}^d) y_{33}^d} \right| \epsilon^7\}v/\sqrt{2} ,\\ \nonumber 
\{m_\tau, m_\mu, m_e\} & \simeq \{|y_{33}^l| \epsilon^3, ~ \left|y_{22}^l- \frac {y_{23}^l y_{32}^l} {y_{33}^l} \right| \epsilon^5,\\& ~  \left |y_{11}^l- \frac {y_{12}^l y_{21}^l}{y_{22}^l-y_{23}^l y_{32}^l/y_{33}^l}- \frac{y_{13}^l \left( y_{31}^l y_{22}^l-y_{21}^l y_{32}^l \right) -y_{31}^l y_{12}^l y_{23}^l}{\left(  y_{22}^l- y_{23}^l y_{32}^l/y_{33}^l \right) y_{33}^l} \right| \epsilon^9\}v/\sqrt{2}.
\end{align}
The mixing angles of quarks can be written as\cite{Rasin:1998je},
\begin{eqnarray}
\sin \theta_{12}  \simeq |V_{us}| &\simeq& \left|{y_{12}^d \over y_{22}^d}  -{y_{12}^u \over y_{22}^u}  \right| \epsilon, ~
\sin \theta_{23}  \simeq |V_{cb}| \simeq  \left|{y_{23}^d \over y_{33}^d}   -{y_{23}^u \over y_{33}^u}   \right| \epsilon^2 , \\ \nonumber
\sin \theta_{13}  \simeq |V_{ub}| &\simeq& \left|{y_{13}^d \over y_{33}^d}    -{y_{12}^u y_{23}^d \over y_{22}^u y_{33}^d}      
- {y_{13}^u \over y_{33}^u}   \right|   \epsilon^3.
\end{eqnarray}

We show, in table \ref{tab:mod}, the values of the parameter $\epsilon$, and the minimum values of the Yukawa couplings $|y_{ij}|_{\text{min}}$ for all four $\mathcal{Z}_{\rm N} \times \mathcal{Z}_{\rm M}$ flavor symmetries, we discussed.  We notice that the models based on the symmetries $\mathcal{Z}_{\rm 2} \times \mathcal{Z}_{\rm 9}$ and $\mathcal{Z}_{\rm 8} \times \mathcal{Z}_{\rm 22}$ are described by ideally an order-one couplings providing the best description of fermion masses and mixings.

\begin{table}[H]
    \setlength{\tabcolsep}{7pt} 
\renewcommand{\arraystretch}{1.6} 
    \begin{center}
    \begin{tabular}{|c||c|c|}
\hline 
\bf{Model} & \bf{$ \boldsymbol{\epsilon}$}  & \bf{$\boldsymbol{|y_{ij}|_{\text{min}}}$} \\ \hline 
 \centering{$\mathcal{Z}_{\rm 2} \times \mathcal{Z}_{\rm 5}$} & 0.1 & $\approx 0.1$ \\
\hline
\centering{$\mathcal{Z}_{\rm 2} \times \mathcal{Z}_{\rm 9}$} & 0.23 & $\approx 1$ \\
\hline
\centering{$\mathcal{Z}_{\rm 2} \times \mathcal{Z}_{\rm 11}$} & 0.28 & $\approx 0.7 $ \\
\hline
\centering{$\mathcal{Z}_{\rm 8} \times \mathcal{Z}_{\rm 22}$} & 0.23 & $\approx 1$ \\
\hline
    \end{tabular}
    \caption{Values of $\epsilon$ and $|y_{ij}|_{\text{min}}$ for each $\mathcal{Z}_{\rm N} \times \mathcal{Z}_{\rm M}$ flavor symmetry. }
    \label{tab:mod}
    \end{center}
\end{table}

\section{The scalar potential}
\label{potential}
The scalar potential of the model  acquires the following form,
\begin{align}
- \lag_\text{potential}
=- \mu^2 \varphi^\dagger \varphi +\lambda (\varphi^\dagger \varphi)^2 - \mu_\chi^2\, \chi^*  \chi
  + \lambda_\chi\, (\chi^* \chi)^2 
  + \lambda_{\varphi  \chi}  (\chi^* \chi)  (\varphi^\dagger \varphi).
\label{eq:potential}
\end{align}
In the phenomenological investigation, we assume  $ \lambda_{\varphi  \chi}   =0$, i.e., no Higgs-flavon mixing.  The effects of this coupling are explored in reference \cite{Berger:2014gga}. We parametrize the scalar  fields  as,

\begin{align}
 \chi=\frac{f + s +i\, a}{\sqrt{2}}, ~ \varphi =\left( \begin{array}{c}
G^+ \\
\frac{v+h+i G^0}{\sqrt{2}} \\
\end{array} \right).
\label{chi}
\end{align}
We have the SM VEV given by $\langle\phi \rangle=v(\equiv v_{SM})\simeq  246$~GeV and $\langle \chi \rangle\!=f$. Here, $G^+, G^0$ are the Goldstone modes, which become the longitudinal components of the gauge bosons and give them mass after the electroweak symmetry is spontaneously broken.

\subsection{Softly broken scalar potential}
In the conventional approach \cite{Abbas:2022zfb,Dorsner:2002wi,Bauer:2015kzy,Calibbi:2015sfa,Fedele:2020fvh}, we can provide mass to the pseudoscalar component of the flavon field by adding  the following term to the scalar potential, which breaks the  $\mathcal{Z}_{\rm N} \times \mathcal{Z}_{\rm M} $ flavor symmetry  softly,

\begin{align}
V_{\rho}
= \rho  \ \chi^2 + \rm H.c. .
\label{eq:potential}
\end{align}
The parameter $\rho$ is complex, and its phase can be rotated away by redefining the flavon field $\chi$ resulting in a real value of the VEV of the field $\chi$.

The minimization conditions produce the masses of scalar and pseudoscalar flavons,
\begin{align}
m_s = \sqrt{\lambda_\chi} f 
\qqquad \text{and} \qqquad
m_a= \sqrt{2 \rho}.
\label{eq:masses}
\end{align}
The mass of the pseudoscalar flavon depends on the soft-breaking parameter $\rho$, and is a free parameter of the model.  The flavor and collider phenomenology of flavon based on the softly broken scalar potential can be found in references \cite{Abbas:2022zfb,Dorsner:2002wi,Bauer:2015kzy,Calibbi:2015sfa,Fedele:2020fvh}.

\subsection{Symmetry-conserving scalar potential}

The $\mathcal{Z}_N \times \mathcal{Z}_M$  flavor symmetry presents a novel scenario that simultaneously addresses the flavor problem and dark matter within a unified and generic framework \cite{Abbas:2023ion}. This is achieved by formulating the flavon potential as,
\begin{equation} 
\label{VN}
 V_{\lambda} = -\lambda {\chi^{\tilde N} \over \Lambda^{\tilde N-4}} + \text{H.c.}, 
\end{equation}
where $\tilde N $ is the least common multiple of $N$ and $M$.

Upon the $\mathcal{Z}_N \times \mathcal{Z}_M$ flavor symmetry  breaking by the VEV $\langle \chi \rangle$,   the flavonic Goldstone boson ($a $) receives the potential,
\begin{equation}
  V_{\rm \tilde N}= -{1\over4}|\lambda| {\epsilon^{\tilde N-4}} f^4  \cos\left( \tilde N {a \over f}+ \alpha\right),
\end{equation}
where $\lambda=|\lambda|e^{i\alpha}$. 

The axial flavon mass now turns out to be,
\begin{equation} 
\label{mphi1}
 m_a^2={1\over8} |\lambda| \tilde N^2 \epsilon^{\tilde N-4} f^2,
\end{equation}
which depends on the symmetry breaking scale and is no longer  an arbitrary free parameter of the model.  We notice that for a large value of the $\tilde N$, the axial flavon can be the so-called flavonic dark matter, a new class of dark matter particles \cite{Abbas:2023ion}.

\subsection{Couplings of flavon to fermions}
  Now, using equation \ref{chi},  we can write 
\begin{equation}
\label{eps}
\frac{\chi}{\Lambda} = \epsilon  [ 1 +  \frac{s + i a}{f} ].
\end{equation}

For obtaining the couplings of the scalar and pseudoscalar couplings to fermionic pair, we write the effective Yukawa couplings in the following form,


\begin{equation}
\label{coupling}
Y_{ij}^f  \left(\frac{v+ h }{\sqrt{2}} \right) = y_{ij}^f  \left(\frac{\chi}{\Lambda} \right)^{n^{f}_{ij}} \left(\frac{v+ h }{\sqrt{2}} \right) 
\cong  y_{ij}^f  \epsilon^{n^{f}_{ij}}  \frac{v}{\sqrt{2}} \left[1 + \frac{n^f_{ij} (s +  i  a )}{f} + \frac{h}{v}\right] = \M_f \left[1 + \frac{n^f_{ij} (s +  i  a )}{f} + \frac{h}{v}\right],
\end{equation}
where $f= u,d,\ell$, and $n^{f}_{ij}$ denote the power of the $\epsilon$, which appears in the mass matrices $\M_f$.

We keep only the linear terms of the flavon field components $s$ and $a$ in equation \ref{coupling} in our phenomenological analysis.   The higher-order terms are not relevant for the analysis given in this work.  The couplings of the scalar and pseudoscalar components $s$ and $a$  of the flavon field to the fermionic pair are derived from the matrix $n^f_{ij} \M_f$, which cannot be diagonalized.  This results in the non-diagonal flavor-changing and $CP$-violating interactions of the flavon field.

The couplings of the scalar component of the flavon field are obtained as,

\begin{equation}
y_{ij}= y_{sf_{iL} f_{iR}} = - i  y_{af_{iL} f_{iR}}. 
\end{equation}

\section{Quark flavor physics constraints on the  flavon of   the $\mathcal{Z}_N \times \mathcal{Z}_M$ flavor symmetries }
\label{quark_flavor}
In this section, we thoroughly investigate the bounds on the parameter space of the  $\mathcal{Z}_N \times \mathcal{Z}_M$ flavor symmetries using the quark-flavor physics data.  We shall present the bounds on the flavon VEV $f$ and mass of the pseudoscalar field $m_a$ in both soft symmetry-breaking as well as symmetry-preserving scenarios.

 We note that the bounds on the parameter space of flavons of different $\mathcal{Z}_N \times \mathcal{Z}_M$ flavor symmetries from  quark and leptonic flavor observables, in general, can be understood in terms of a generic behavior of the corresponding Wilson coefficients. The  Wilson coefficients, in general, behave as $y_{ij} \epsilon^{n_{ij}}  y_{ji} \epsilon^{n_{ji}} /f^2$, where $y_{ij}$ are the dimensionless couplings of flavons with a fermionic pair.  Thus, a small contribution of the Wilson coefficient from the flavon  to the corresponding observable, such as kaon mixing, can be accommodated by either increasing the value of the VEV $f$, or changing the values of the coupling  $y_{ij}$ and the order parameter $\epsilon$ while keeping a low value of the flavon  VEV $f$.  We shall analyze the constraints arising from quark and lepton flavor physics by employing the above observation.

\subsection{Neutral meson mixing}
The neutral meson-antimeson mixing receives a contribution from the FCNC interactions occurring at the tree-level due to the non-diagonal couplings of the flavon to fermions.  This contribution can be incorporated by writing the following   $\Delta F =2$  effective Hamiltonian,

\begin{align}
\Heff^{\Delta F=2}&=C_1^{ij} \,( \bar q^i_L\,\gamma_\mu \, q^j_L)^2+\widetilde C_1^{ij} \,( \bar q^i_R\,\gamma_\mu \, 
q^j_R)^2 +C_2^{ij} \,( \bar q^i_R \, q^j_L)^2+\widetilde C_2^{ij} \,( \bar q^i_L \, q^j_R)^2\notag\\
&+ C_4^{ij}\, ( \bar q^i_R \, q^j_L)\, ( \bar q^i_L \, q^j_R)\,+C_5^{ij}\, ( \bar q^i_L \,\gamma_\mu\, q^j_L)\, ( \bar q^i_R \,
\gamma^\mu q^j_R)\,+ \text{H.c.}, 
\label{eq:heffdf2}
\end{align}
where $q_{R,L} = \frac{1\pm \gamma_5}{2} q$,  and we do not show the colour indices  for simplicity. 

\begin{table}[h!]
\begin{center}
\begin{tabular}{|c|c || c|c|}\hline
$G_{F}$ & $1.166 \times 10^{-5}$ \,$~{\rm GeV}$ \cite{Zyla:2021} &$v$  & 246.22  \,GeV \cite{Zyla:2021}  \\  \cline{1-4}
     
     $\alpha_{s}[M_{Z}]$ & $0.1184$ \cite{Y.Aoki:2021} & $m_{u}$  & $ (2.16^{+0.49}_{-0.26} )\times 10^{-3}$~{\rm GeV} \cite{Zyla:2021} \\  \cline{1-2}
     
$M_{W}$  & $80.387 \pm 0.016$ \,GeV \cite{Zyla:2021} & $m_{d}$ & $ (4.67^{+0.48}_{-0.17}) \times 10^{-3}$~{\rm GeV} \cite{Zyla:2021} \\   \cline{1-2}     
      $f_{K}$  &  $159.8$ \,MeV \cite{Ciuchini:1998ix} & $m_{c}$ & $ 1.27 \pm 0.02 $~{\rm GeV} \cite{Zyla:2021}  \\  
     
      $m_{K}$  &  $497.611 \pm 0.013$ \,MeV \cite{Zyla:2021}  & $m_{s}$ & $ 93.4^{+8.6}_{-3.4}$~{\rm GeV}  \cite{Zyla:2021} \\ 
    $\hat{B_{K}}$  & $0.7625$  \cite{Y.Aoki:2021}  & $m_{t}$ & $ 172.69 \pm 0.30 $~{\rm GeV} \cite{Zyla:2021} \\ 
   $B_{1}^K$  & $0.60 (6)$  \cite{Ciuchini:1998ix}  & $m_{b}$ & $ 4.18^{+0.03}_{-0.02}$~{\rm GeV} \cite{Zyla:2021} \\ 
     
    $B_{2}^K$ & $0.66 (4)$ \cite{Ciuchini:1998ix} & $m_{c}(m_{c})$ & $ 1.275$~{\rm GeV}  \\ 
    $B_{3}^K$ & $1.05 (12)$ \cite{Ciuchini:1998ix}  &  $m_{b}(m_{b})$ & $ 4.18$~{\rm GeV} \\  
     $B_{4}^K$ & $ 1.03 (6)$ \cite{Ciuchini:1998ix}  &  $m_{t}(m_{t})$ & $ 162.883$~{\rm GeV}  \\ \cline{3-4}
     $B_{5}^K$ & $0.73 (10)$ \cite{Ciuchini:1998ix}  &$\alpha$ & $ 1/137.035$ \cite{Zyla:2021} \\ 
    
    $\eta_{1}$ & $1.87 \pm 0.76$ \cite{Brod:2012}& $e$ & $0.302862$~{\rm GeV} \\  
   $\eta_{2}$ & $0.574$ \cite{Buchalla:1996}& $m_{e}$ & $ 0.51099 $~{\rm MeV} \cite{Zyla:2021} \\
   $\eta_{3}$ & $0.496 \pm 0.047$ \cite{Brod:2010}& $m_{\mu}$ & $ 105.65837$~{\rm MeV} \cite{Zyla:2021} \\ \cline{1-2}
    
 $f_{B_{s}}$  &  $230.3 $ \,MeV \cite{Y.Aoki:2021} & $m_{\tau}$ & $ 1776.86 \pm 0.12$~{\rm MeV} \cite{Zyla:2021}  \\ 
 
$m_{B_{s}}$  &  $5366.88$ \,MeV \cite{Zyla:2021} & $\tau_{\mu}$ & $ 2.196811 \times 10^{-6}$~{\rm sec} \cite{Zyla:2021}\\ 

$\hat{B_{B_{s}}}$  & $1.232$\cite{Y.Aoki:2021}  & $\tau_{\tau}$ & $ (290.3 \pm 0.5) \times 10^{-15}$~{\rm sec} \cite{Zyla:2021}  \\

$B_{1}^{B_{s}}$ & $0.86 (2) (^{+5} _{-4})$ \cite{Becirevic:2001xt} & $m_{p}$ & $ 938.272$~{\rm MeV} \cite{Zyla:2021} \\
$B_{2}^{B_{s}}$ & $0.83 (2) (4)$ \cite{Becirevic:2001xt}& $m_{n}$ & $ 939.565$~{\rm MeV} \cite{Zyla:2021} \\ \cline{3-4}
$B_{3}^{B_{s}}$ & $1.03 (4) (9)$ \cite{Becirevic:2001xt}& $m_{D}$ & $ 1864.83$~{\rm MeV} \cite{Zyla:2021}\\
$B_{4}^{B_{s}}$ & $1.17 (2) (^{+5} _{-7})$ \cite{Becirevic:2001xt}& $f_{D}$ & $ 212$~{\rm MeV}\cite{Y.Aoki:2021}\\ 
$B_{5}^{B_{s}}$ & $1.94 (3) (^{+23} _{-7})$ \cite{Becirevic:2001xt}& $B_{1}^{D}$ & $ 0.861$ \cite{Bona:2007vi} \\ \cline{1-2}
$\eta_{2B}$ & $0.551$ \cite{Buchalla:1996}& $B_{2}^{D}$ & $ 0.82$ \cite{Bona:2007vi}\\ \cline{1-2}

$f_{B_{d}}$  &  $190.0 $ \,MeV \cite{Y.Aoki:2021} & $B_{3}^{D}$ & $ 1.07$ \cite{Bona:2007vi}\\ 
 
$m_{B_{d}}$  &  $5279.65$ \,MeV \cite{Zyla:2021}&  $B_{4}^{D}$ & $ 1.08$ \cite{Bona:2007vi}\\ 
$\hat{B_{B_{d}}}$  & $1.222$\cite{Y.Aoki:2021}& $B_{5}^{D}$ & $ 1.455$ \cite{Bona:2007vi}\\ \cline{3-4}
$B_{1}^{B_{d}}$ & $0.87 (4) (^{+5} _{-4})$ \cite{Becirevic:2001xt}& $\tau_{B_{d}}$ & $ (1.520 \pm 0.004)\times 10^{-12}$ ~{\rm sec} \cite{HFLAV:2016hnz} \\
$B_{2}^{B_{d}}$ & $0.82 (3) (4)$ \cite{Becirevic:2001xt}& $\tau_{B_{s}}$ & $ (1.505 \pm 0.005) \times 10^{-12}$ ~{\rm sec} \cite{HFLAV:2016hnz}\\
$B_{3}^{B_{d}}$ &$1.02 (6) (9)$ \cite{Becirevic:2001xt}& $\tau_{K_{L}}$ & $ (5.116 \pm 0.021) \times 10^{-8}$~{\rm sec} \cite{Zyla:2021}\\ 

$B_{4}^{B_{d}}$ & $1.16 (3) (^{+5} _{-7})$ \cite{Becirevic:2001xt} & $\tau_{D}$ & $ (410.1 \pm 1.5) \times 10^{-15}$~{\rm sec} \cite{Zyla:2021} \\
 
$B_{5}^{B_{d}}$ & $1.91 (4) (^{+22} _{-7})$ \cite{Becirevic:2001xt} &  &   \\   
\hline  
\end{tabular} \\
\end{center}

\caption{The numerical values of the  used  input parameters in this work.}
   \label{tab5}
   \end{table} 

The  Wilson coefficients corresponding to the tree-level contribution to neutral meson mixing due to the flavon exchange read as \cite{Buras:2013rqa,Crivellin:2013wna},
\begin{align}
C_2^{ij} &= -(y_{ji}^*)^2\left(\frac{1}{m_s^2}-\frac{1}{m_a^2}\right)\notag \\
\tilde C_2^{ij} &= -y_{ij}^2\left(\frac{1}{m_s^2}-\frac{1}{m_a^2}\right)\notag \\
C_4^{ij} &= -\frac{y_{ij}y_{ji}}{2}\left(\frac{1}{m_s^2}+\frac{1}{m_a^2}\right),
\label{eq:wilsons}
\end{align}
where $m_{s}$ and $m_{a}$ represent  the masses of scalar and pseudoscalar degrees of the flavon field.

We evolve down the Wilson coefficients $C_i$  from a scale $\Lambda$ to  the   hadronic scales, used in the lattice computations of the corresponding matrix elements
\cite{Ciuchini:1998ix, Becirevic:2001xt, Bona:2007vi},   $4.6$~GeV for bottom mesons,  $2.8$~GeV for charmed mesons,  and $2$~GeV for kaons.   The  renormalization group running of the matrix elements is performed using the procedure discussed in  reference~\cite{Bona:2007vi}, and matrix elements are adopted from reference~\cite{Ciuchini:1998ix, Becirevic:2001xt}.  Thus, the  $B_q -\bar B_q$ mixing amplitudes beyond  the SM can  be written as~\cite{Bona:2007vi},
\begin{equation}
\label{eq:magicbb}
\langle \bar B_q \vert {\cal H}_{\rm eff}^{\Delta B=2} \vert B_q \rangle_i = 
\sum_{j=1}^5 \sum_{r=1}^5
             \left(b^{(r,i)}_j + \eta \,c^{(r,i)}_j\right)
             \eta^{a_j} \,C_i(\Lambda)\, \langle \bar B_q \vert Q_r^{bq}
             \vert B_q \rangle\,,
\end{equation}
where $q=d,s$,    $\eta=\alpha_s(\Lambda)/\alpha_s(m_t)$,  $a_j$, $b^{(r,i)}_j$,  $c^{(r,i)}_j$  denote the so-called magic numbers  taken from reference~\cite{Becirevic:2001jj}, and $\alpha_s$ is the strong coupling constant.  A similar expression can be written for the  $D^0 - \bar D^0$ for which the  magic numbers are present in  reference~\cite{Bona:2007vi}.  The  corresponding expression for the  $K^0-\bar K^0$ mixing reads as  \cite{Bona:2007vi},

\begin{equation}
  \label{eq:kkbsm}
  \langle \bar K^0 \vert {\cal H}_{\rm eff}^{\Delta S=2} \vert K^0 \rangle_i = 
  \sum_{j=1}^5 \sum_{r=1}^5
  \left(b^{(r,i)}_j + \eta \,c^{(r,i)}_j\right)
  \eta^{a_j} \,C_i(\Lambda)\, R_r \, \langle \bar K^0 \vert Q_1^{sd}
  \vert K^0 \rangle,
\end{equation}
where $R_r$ denotes  the ratio of the matrix elements of NP operators over that of the SM \cite{Donini:1999nn}.  The   numerical values  of $R_r$  can be found in  reference  \cite{Bona:2007vi}.  For the $K^0-\bar K^0$ mixing, the magic numbers  are used  from reference \cite{Ciuchini:1998ix}.

For constraining the flavon mass and the VEV, we use the experimental measurements of the mixing observables of the  $K^0-\bar K^0$ mixing, which is given as \cite{Bona:2007vi},
\begin{align}
C_{\eps_K}&=\frac{ \text{Im} \langle K^0|\mathcal{H}_\text{eff}^{\Delta F=2}|\bar K^0\rangle}{\text{Im} \langle K^0| \mathcal{H}_\text{SM}^{\Delta F=2} |\bar K^0 \rangle} = 1.12_{-0.25}^{+0.27},
C_{\Delta m_K} =\frac{\text{Re}\langle K^0|\mathcal{H}_\text{eff}^{\Delta F=2}|\bar K^0\rangle}{\text{Re} \langle K^0| \mathcal{H}_\text{SM}^{\Delta F=2} |\bar K^0 \rangle} = 0.93_{-0.42}^{+1.14} ,
\end{align}
where  $\mathcal{H}_\text{eff}^{\Delta F=2}$ represents  the SM and flavon contributions,  and  $\mathcal{H}_\text{SM}^{\Delta F=2}$ denotes  only  the SM contribution. 

The   $B_q -\bar B_q$ mixing observables  are,
\begin{align*}
C_{B_{q}}e^{2i\phi_{B_{q}}}&=\frac{ \langle B_{q}^0|\mathcal{H}^{\Delta F=2}|\bar B_{q}^0\rangle}{ \langle B_{q}^0| \mathcal{H}_\text{SM}^{\Delta F=2} |\bar B_{q}^0 \rangle}, 
\end{align*}
where $q=s,d$ for the $B_{s}$ and $B_{d}$ mixing, respectively. We use  the following  measurements at 95 \% CL limits  in this work \cite{Bona:2007vi},
\begin{align*}
C_{B_{s}}= 1.110 \pm 0.090 \hspace{0.1cm} [0.942, 1.288] ,\hspace{0.5cm} \phi_{B_{s}}^o= 0.42 \pm 0.89 \hspace{0.1cm} [-1.35, 2.21]\\
C_{B_{d}}= 1.05 \pm 0.11 \hspace{0.1cm} [0.83, 1.29], \hspace{0.8cm} \phi_{B_{d}}^o= -2.0 \pm 1.8 \hspace{0.1cm} [ -6.0, 1.5].
\end{align*}

We write the  new physics contributions to neutral meson mixing in the following form,
 \be
 M_{12}^{d,s,K} =  (M_{12}^{d,s,K} )_{\rm SM} \left(  1 +  h_{d,s,K} e^{2 i \sigma_{d,s,K}}    \right).
  \ee

The following future sensitivity phases are adopted in this work \cite{Charles:2020dfl}:
\begin{enumerate}
\item
Phase \rom{1} which corresponds to  $50 fb^{-1}$ LHCb and $50 ab^{-1}$ Belle \rom{2} (late 2020s);
\item
Phase \rom{2} which corresponds to $300 fb^{-1}$ LHCb and $250 ab^{-1}$ Belle \rom{2} (late 2030s).
\end{enumerate}

 \begin{table}[tb]
\centering
\begin{tabular}{l|ccc}
\text{Observables} & \text{Phase \rom{1}} & \text{Phase \rom{2}} & \text{Ref.}  \\
\hline
$h_d$ & $0-0.04$    &   $0-0.028$    &  \cite{Charles:2020dfl}    \\
$h_s$  & $0-0.036$ &  $0-0.025$& \cite{Charles:2020dfl} \\
$h_K$  & $0-0.3$  & $-$  &    \cite{Charles:2013aka}\\
\hline
\end{tabular}
\caption{The future projected sensitivities of the neutral meson mixing. }
\label{hdsk}
\end{table}

In table  \ref{hdsk}, we show the expected sensitivities to the observables $C_{\Delta m_K} $ and $C_{B_{q}}$   in the future  phase \rom{1} and \rom{2} of the LHCb and the Belle \rom{2}.

\begin{figure}[h!]
	\centering
	\begin{subfigure}[]{0.4\linewidth}
    \includegraphics[width=\linewidth]{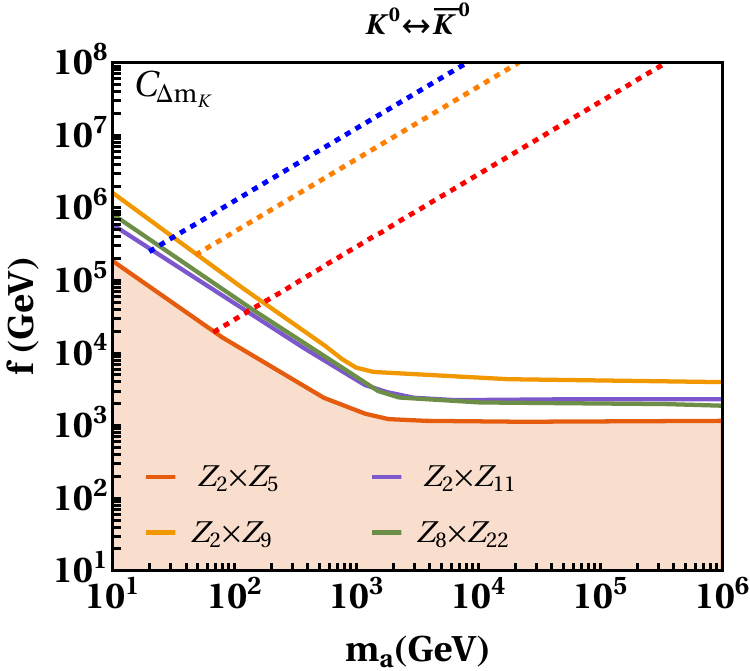}
    \caption{}
         \label{fig1a}	
\end{subfigure}
 \begin{subfigure}[]{0.4\linewidth}
 \includegraphics[width=\linewidth]{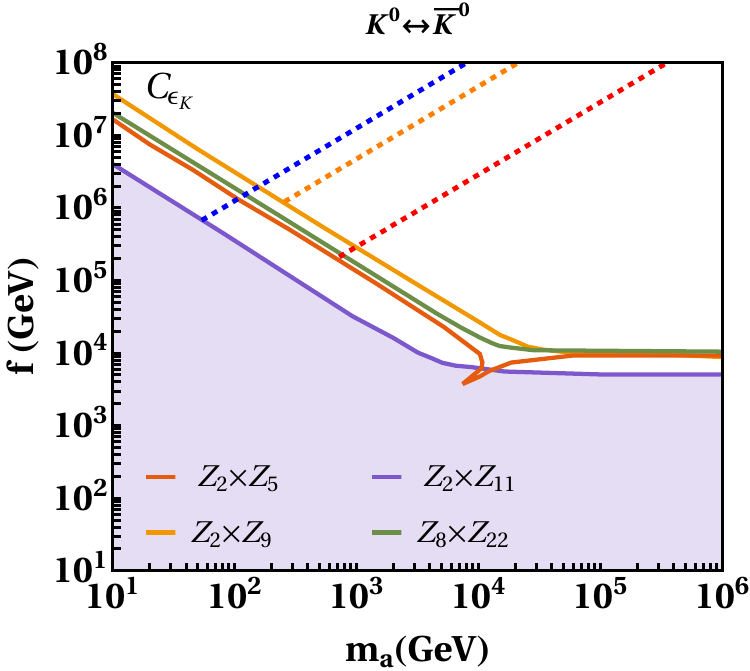}
 \caption{}
         \label{fig1b}
 \end{subfigure} 
 \begin{subfigure}[]{0.4\linewidth}
 \includegraphics[width=\linewidth]{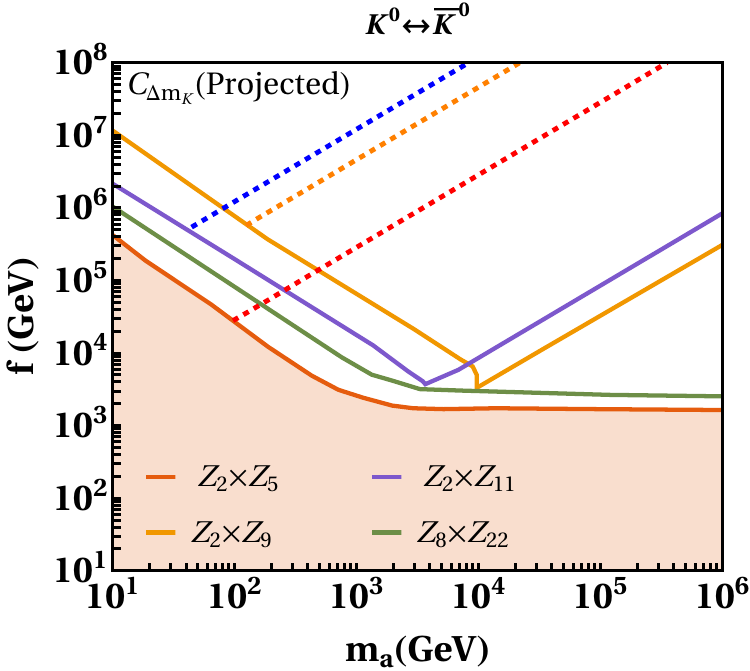}
 \caption{}
         \label{fig1c}
 \end{subfigure} 
 \caption{Bounds on the parameter space of the flavon of different $\mathcal{Z}_N \times \mathcal{Z}_M$ flavor symmetries  in $f-m_a$ plane from the neutral-kaon oscillations.  On the top left, we show the constraints arising from the observable $C_{\Delta m_K}$, and on the top right, the bounds from the observable $C_{\eps_K}$ are shown  for the value  $\lambda_\chi= 2$.  The figure \ref{fig1c} depicts the  allowed parameter space for projected sensitivities of $C_{\Delta m_K}$ in the LHCb  Phase-\rom{1}. The blank region above the continuous curves in figures shows the allowed $f-m_a$ parameter space in the soft symmetry-breaking framework, while the dashed curves are allowed for the symmetry-preserving scenario as defined by equation \ref{VN}. The colored region below the continuous curves indicates the $f-m_a$ parameter space excluded by all four $\mathcal{Z}_N \times \mathcal{Z}_{M}$ models in the soft symmetry-breaking scenario. }
  \label{fig1}
	\end{figure}

Now, we discuss the bounds on the parameter space of the flavon of the $\mathcal{Z}_N \times \mathcal{Z}_M$ flavor symmetries arising from the neutral meson mixing.  We discuss these bounds simultaneously for the soft symmetry-breaking as well as the symmetry-conserving scenarios. In figure \ref{fig1}, we show the bounds on the  parameter space of the flavon corresponding to different $\mathcal{Z}_N \times \mathcal{Z}_M$ flavor symmetries from neutral kaon mixing observable $C_{\Delta m_K}$ on the left,   and that from the observable   $C_{\eps_K}$ on the right for the quartic coupling $\lambda_\chi= 2$.  The region above the continuous curves show the allowed parameter space in the soft symmetry-breaking scenario for different $\mathcal{Z}_N \times \mathcal{Z}_M$ flavor symmetries.  On the other side, the dashed curves represent symmetry-conserving scenario defined by equation \ref{VN}.  In the case of the soft symmetry-breaking scenario, the shaded region of the parameter space below the continuous curves is excluded by the corresponding observables.  The symmetry-conserving scenario turns out to be extremely predictive in the sense that the allowed flavon parameter space is only along the dashed straight lines, and the rest of the region is excluded by the corresponding observables.

We observe that the bounds from the observable  $C_{\Delta m_K}$ are most stringent for the flavor symmetry $\mathcal{Z}_2 \times \mathcal{Z}_{9}$, shown by the region above yellow-coloured continuous curve in figure \ref{fig1a}. This is followed by the $\mathcal{Z}_2 \times \mathcal{Z}_{11}$ and $\mathcal{Z}_8 \times \mathcal{Z}_{22}$ flavor symmetries for which the allowed parameter space is almost similar, shown in figure \ref{fig1a} by the region above the violet and green coloured continuous curves, respectively. The bounds become more stringent when the observable $C_{\eps_K}$ is used as shown in figure \ref{fig1b}.  The most stringent bounds continue to be for the $\mathcal{Z}_2 \times \mathcal{Z}_9$ flavor symmetry even from the  observable $C_{\eps_K}$.   The bounds from the observable  $C_{\eps_K}$ on the flavor symmetries $\mathcal{Z}_2 \times \mathcal{Z}_{5}$ and $\mathcal{Z}_8 \times \mathcal{Z}_{22}$ are very close and are more stringent than that of the $\mathcal{Z}_2 \times \mathcal{Z}_{11}$ flavor symmetry, shown by the blue continuous curve.  In figure \ref{fig1c}, we show the bounds on the parameter space of the flavon of different $\mathcal{Z}_N \times \mathcal{Z}_M$ flavor symmetries  for the projected future sensitivities of the neutral meson mixing given in table \ref{hdsk} for  the soft symmetry-breaking and symmetry-conserving scenario again by the solid continuous curves.

In figures \ref{fig1a} and \ref{fig1c}, the bounds corresponding to the $\mathcal{Z}_2 \times \mathcal{Z}_{5}$ flavor symmetry are relatively loose from the observable $C_{\Delta m_K}$. The reason for this is a low value of the order parameter $\epsilon=0.1$, which dominates the behavior of the Wilson coefficients, in general, and allows a low value of the flavon VEV $f$. The symmetries $\mathcal{Z}_{2,8} \times \mathcal{Z}_{9,11,22}$ are showing a similar behavior since the order parameter $\epsilon$ is very close for them.  On the other side, in figure  \ref{fig1b} the contribution from the imaginary part of the product $y_{ij} y_{ij}$ is very large for the $\mathcal{Z}_2 \times \mathcal{Z}_{5}$ flavor symmetry, making it closer to the predictions of  $\mathcal{Z}_{2,8} \times \mathcal{Z}_{9,22}$ symmetries. The symmetry-conserving scenario is dominated by the values of the parameter $\epsilon$, leading to the strongest bounds for the $\mathcal{Z}_2 \times \mathcal{Z}_{9,11}$ flavor symmetries. This is a generic observation in the sense that it will continue to show up even for leptonic flavor observables except for the $D^0$ mixing.  Therefore, we shall discuss it only for the  $D^0$ mixing later.

For the symmetry-conserving scenario, the most stringent bounds from the observable  $C_{\Delta m_K}$ again appear for the $\mathcal{Z}_2 \times \mathcal{Z}_9$ flavor symmetry denoted by the orange dashed line  in figure \ref{fig1a}.  We must note that for the symmetry-conserving case, the dashed straight lines are the only allowed parameter space, and the rest of the region is excluded by the corresponding mixing observables.  The bounds for the projected future sensitivities of the neutral meson mixing given in table \ref{hdsk}  in the case of the symmetry-conserving scenario are shown by the dashed straight lines in figure \ref{fig1c} for different $\mathcal{Z}_N \times \mathcal{Z}_M$ flavor symmetries.  

\begin{figure}[H]
	\centering
	\begin{subfigure}[]{0.4\linewidth}
    \includegraphics[width=\linewidth]{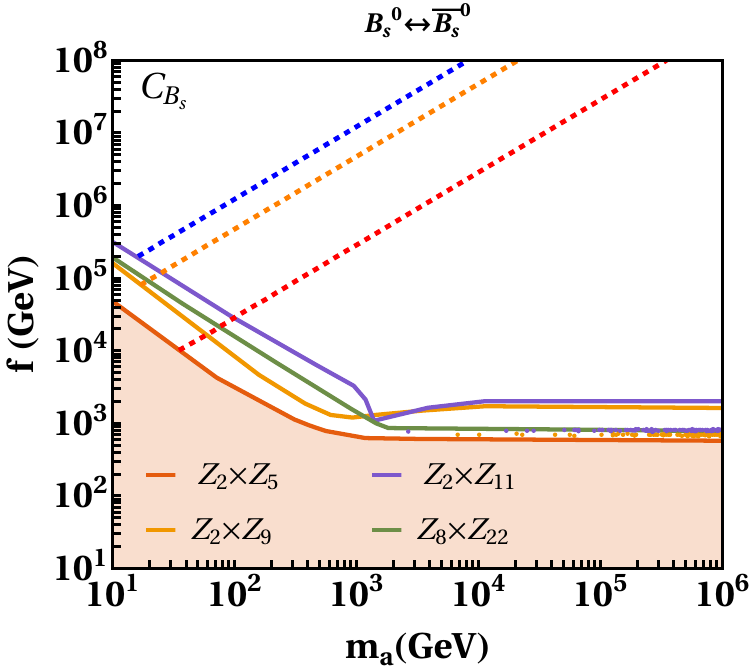}
    \caption{}
         \label{fig2a}	
\end{subfigure}
 \begin{subfigure}[]{0.4\linewidth}
 \includegraphics[width=\linewidth]{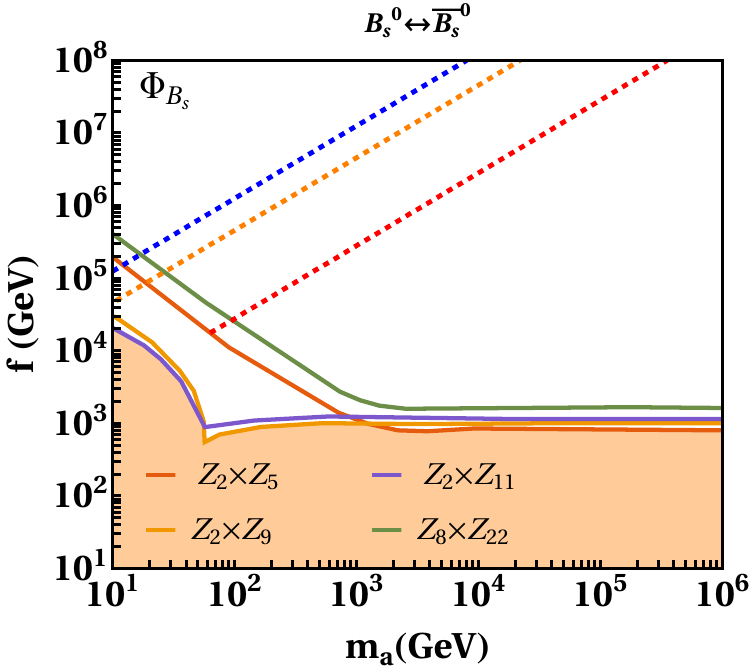}
 \caption{}
         \label{fig2b}
 \end{subfigure} 
\begin{subfigure}[]{0.4\linewidth}
 \includegraphics[width=\linewidth]{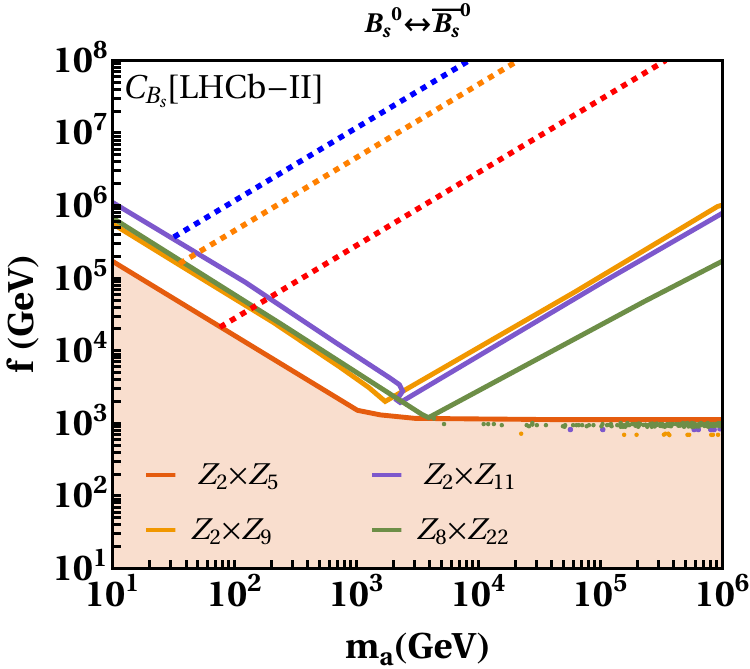}
 \caption{}
\label{fig2c}
 \end{subfigure} 
 \caption{Bounds on the parameter space of the flavon of different $\mathcal{Z}_N \times \mathcal{Z}_M$ flavor symmetries  in $f-m_a$ plane from the neutral $B_s$-meson mixing.  On the top left, we show the constraints arising from the observable $C_{B_{s}}$, and on the top right, the bounds from the observable $\Phi_{B_{s}}$ are shown  for the value  $\lambda_\chi= 2$. The figure \ref{fig2c} depicts the  allowed parameter space for projected sensitivities of $C_{B_{s}}$ in the LHCb  Phase-\rom{2}. Moreover, the region above the continuous curves represent the allowed $f-m_a$ parameter space in the soft symmetry-breaking framework, and dashed curves are for the symmetry-preserving scenario defined by equation \ref{VN}. The colored region below the continuous curves indicates the $f-m_a$ parameter space excluded by all four $\mathcal{Z}_N \times \mathcal{Z}_{M}$ models in the soft symmetry-breaking scenario.}
  \label{fig2}
	\end{figure}

The bounds on the parameter space of the flavon of the $\mathcal{Z}_N \times \mathcal{Z}_M$ flavor symmetries arising from the neutral $B_s$-meson mixing are represented in figure \ref{fig2}.  On the left in figure \ref{fig2a}, we show the constraints on the parameter space of the flavon derived from the observable $C_{B_{s}}$. For the soft symmetry-breaking case, the bounds are depicted by the solid continuous curves, and the region below these curves is excluded by the experimental measurement of the observable $C_{B_{s}}$. We notice that the observable $C_{B_{s}}$, unlike the neutral-kaon mixing, which was more constraining to the $\mathcal{Z}_2 \times \mathcal{Z}_9$ flavor symmetry,  place stronger bounds on the  $\mathcal{Z}_2 \times \mathcal{Z}_{11}$ flavor symmetry for the flavon mass approximately below 1 TeV.  Above the flavon mass 1 TeV, the bounds are similar for the   $\mathcal{Z}_2 \times \mathcal{Z}_{11}$ and $\mathcal{Z}_2 \times \mathcal{Z}_{9}$ flavor symmetries.  The observable $\Phi_{B_{s}}$ is more suitable to constrain the parameter space corresponding to the $\mathcal{Z}_8 \times \mathcal{Z}_{22}$ flavor symmetry, as observed in figure \ref{fig2b}.  The future projected sensitivity of the observable $C_{B_{s}}$ in the LHCb phases-\rom{1}, \rom{2} further constrains the parameter space corresponding to the $\mathcal{Z}_2 \times \mathcal{Z}_{11}$ flavor symmetry for the flavon mass approximately up to 1 TeV. The constraints for the future projected sensitivity of  the observable $C_{B_{s}}$ in  the LHCb phase-\rom{1} are almost similar to that of phase-\rom{2}.  Therefore, we show them only for phase-\rom{2} in this work. Above the flavon mass 1 TeV, the bounds are further stronger and again similar for the $\mathcal{Z}_2 \times \mathcal{Z}_{11}$ and $\mathcal{Z}_2 \times \mathcal{Z}_{9}$ flavor symmetries.

 \begin{figure}[H]
	\centering
	\begin{subfigure}[]{0.4\linewidth}
    \includegraphics[width=\linewidth]{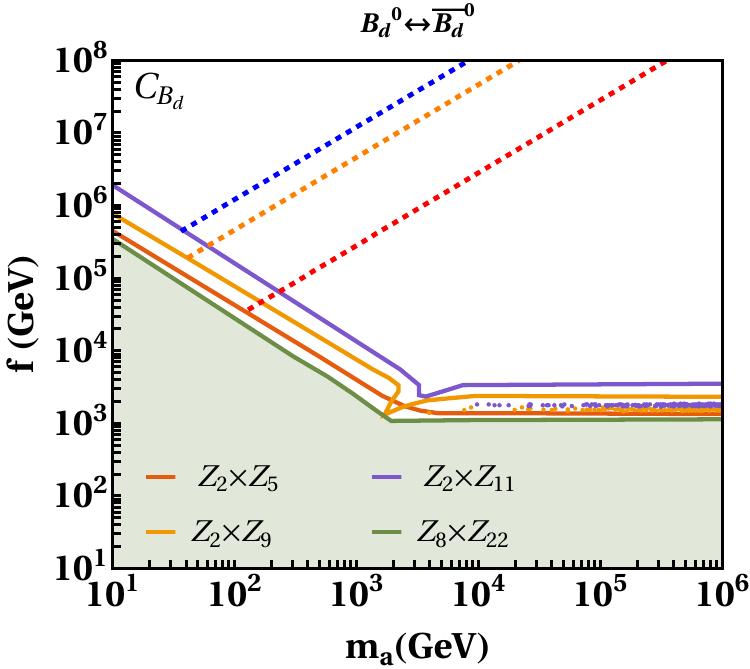}
    \caption{}
         \label{fig3a}	
\end{subfigure}
 \begin{subfigure}[]{0.4\linewidth}
 \includegraphics[width=\linewidth]{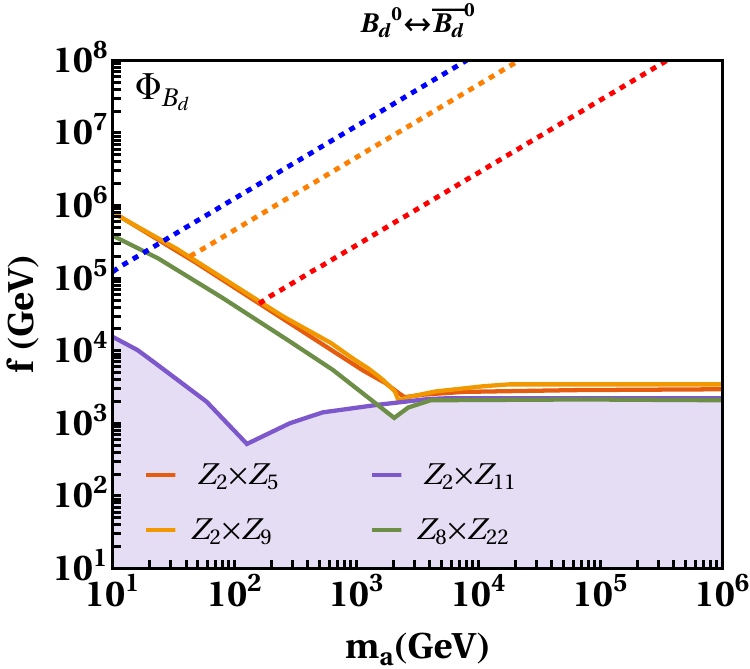}
 \caption{}
         \label{fig3b}
 \end{subfigure}
 \medskip
\begin{subfigure}[]{0.4\linewidth}
 \includegraphics[width=\linewidth]{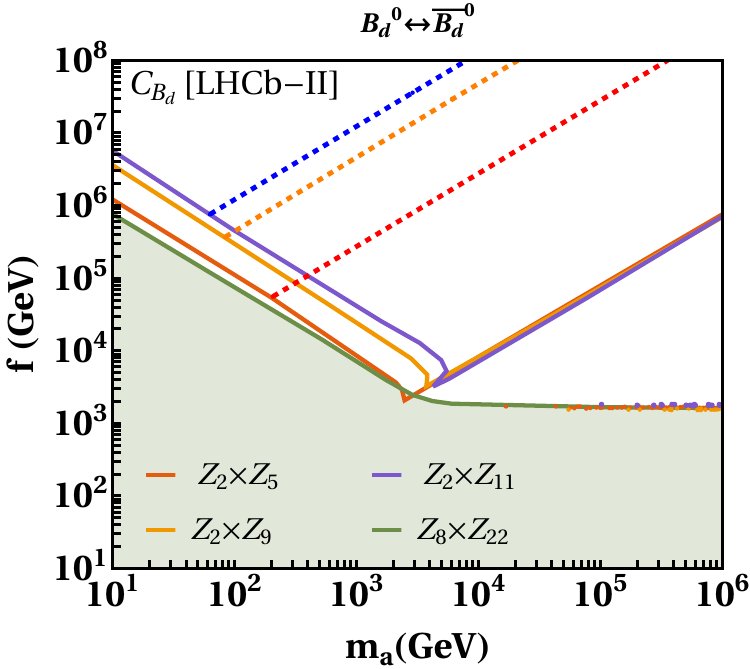}
 \caption{}
         \label{fig3c}
 \end{subfigure}
 \caption{Bounds on the parameter space of the flavon of different $\mathcal{Z}_N \times \mathcal{Z}_M$ flavor symmetries  in $f-m_a$ plane from the neutral $B_d$-meson mixing.  On the top left, we show the constraints arising from the observable $C_{B_{d}}$, and on the right, the bounds from the observable $\Phi_{B_{d}}$ are shown for the value  $\lambda_\chi= 2$.  The figure \ref{fig3c} depicts the  allowed parameter space for projected sensitivities of $C_{B_{d}}$ in the LHCb  Phase-\rom{2}. Moreover, the region above the continuous curves represent the allowed $f-m_a$ parameter space in the soft symmetry-breaking framework, and dashed curves are for the symmetry-preserving scenario defined by equation \ref{VN}.}
  \label{figbdmixing}
	\end{figure}

For the symmetry-conserving scenario, the most stringent constraints are derived for the symmetry $\mathcal{Z}_2 \times \mathcal{Z}_{11}$ from both the observables $C_{B_{s}}$ and  $\Phi_{B_{s}}$, as shown by the dashed blue straight lines in figures \ref{fig2a}, \ref{fig2b}, and in \ref{fig2c} for the future projected sensitivity of the observable $C_{B_{s}}$ in  the LHCb phase-\rom{2}.  After the $\mathcal{Z}_2 \times \mathcal{Z}_{11}$ flavor symmetry, the parameter space corresponding to the flavon of the $\mathcal{Z}_2 \times \mathcal{Z}_{9}$ flavor symmetry receives the stronger bounds denoted by the orange dashed straight lines in figure \ref{fig2}.

We notice that the bounds from the observable  $C_{B_{s}}$ depend on the value $|y_{ij} y_{ij}|$, which is small for the  $\mathcal{Z}_2 \times \mathcal{Z}_{5}$ flavor symmetry allowing a large parameter space for the order parameter $\epsilon=0.1$.  The parameter $\epsilon$ is largest for the $\mathcal{Z}_{2} \times \mathcal{Z}_{11}$ flavor symmetry resulting in the most stringent bounds.  The bounds for the rest of the symmetries are closer to that of the $\mathcal{Z}_{2} \times \mathcal{Z}_{11}$ flavor symmetry due to the closer value of the $\epsilon$ parameter.  The observable  $\Phi_{B_{s}}$  is sensitive to the  $\text{arg}(y_{ij} y_{ij})$, which is large for the  $\mathcal{Z}_2 \times \mathcal{Z}_{5}$ flavor symmetry leading to a more constrained parameter space, which is close to that of the  $\mathcal{Z}_8 \times \mathcal{Z}_{22}$ flavor symmetry.  The value of the   $\text{arg}(y_{ij} y_{ij})$ is small and close for the $\mathcal{Z}_{2} \times \mathcal{Z}_{9,11}$ flavor symmetries  producing similar bounds.

For the $B_d$-meson mixing, we can make observations similar to that of the  $B_s$-meson mixing discussed earlier.  The absolute values of the product  $|y_{ij} y_{ij}|$ together with the value of the parameter $\epsilon$ determines the bounds from the observable  $C_{B_{d}}$.  On the other side, the $\text{arg}(y_{ij} y_{ij})$ together with the   value of the parameter $\epsilon$ determines the bounds from the observable $\Phi_{B_{d}}$.

Thus, the observable $C_{B_{d}}$ places the most stringent bounds on the parameter space of the flavon of the $\mathcal{Z}_2 \times \mathcal{Z}_{11}$ flavor  symmetry for the soft symmetry-breaking scenario, as shown in figure \ref{fig3a}.  This observation is similar to that obtained from the neutral $B_s$-meson mixing in figure \ref{fig2}.  This feature continues to be seen even in the future projected sensitivity of the observable $C_{B_{d}}$ in the LHCb phase-\rom{2}, as shown in figure \ref{fig3c}.  The bounds for the future projected sensitivity of  the observable $C_{B_{d}}$ in  the LHCb phase-\rom{1} are very similar to that of phase-\rom{2}.  Therefore, we do not show them in this work.  On the other side, the observable $\Phi_{B_{d}}$  provides more stringent bounds on the parameter space of the flavon of the $\mathcal{Z}_2 \times \mathcal{Z}_{5,9}$ flavor  symmetries for the soft symmetry-breaking scenario as shown in figure \ref{fig3b}.

In the symmetry-conserving scenario, the $B_d$-meson mixing places bounds on the parameter space of the flavon of $\mathcal{Z}_N \times \mathcal{Z}_M$ flavor symmetries similar to that of the $B_s$-meson mixing, as can be observed from figure \ref{fig3a}-\ref{fig3c}, where the allowed parameter space is shown by dashed straight lines.  The strongest bounds in this case turn out to be for the flavon of the  $\mathcal{Z}_2 \times \mathcal{Z}_{11}$ flavor  symmetry, followed by the flavon of the  $\mathcal{Z}_2 \times \mathcal{Z}_{9}$ flavor symmetry.

Large hadronic uncertainties exist in the SM contribution of the  $D^0 - \bar D^0$ mixing.  For this reason, the flavon contribution to  the $D^0 - \bar D^0$ mixing is chosen to lie within the $2\sigma$  experimental bound \cite{Bona:2017kam}, that is,

\begin{align}
|M_{12}^D|= |\langle D^0|\mathcal{H}^{\Delta F=2}|\bar D^0\rangle | < 7.5 \times 10^{-3}  ps^{-1}
\end{align}

 \begin{figure}[H]
	\centering
	\includegraphics[width=0.4\linewidth]{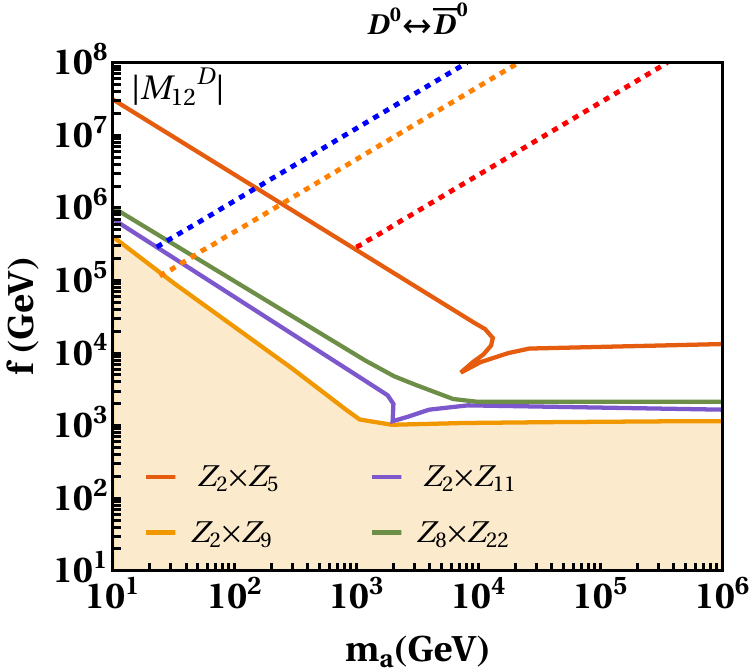}
\caption{Bounds on the parameter space of the flavon of different $\mathcal{Z}_N \times \mathcal{Z}_M$ flavor symmetries  in $f-m_a$ plane from the neutral $D^0 - \bar D^0$ meson mixing for the soft symmetry-breaking as well as for the symmetry-conserving scenarios for   $\lambda_\chi= 2$.}
 \label{fig4}
 \end{figure}

In figure \ref{fig4}, we show the bounds arising from the $D$-meson mixing.  In this case, the absolute values of the product  $|y_{ij} y_{ij}|$ dictate the behavior of bounds from the observable $|M_{12}^D|$.  For instance, the absolute values of the product  $|y_{ij} y_{ij}|$ is very large  for the $\mathcal{Z}_2 \times \mathcal{Z}_5$ flavor symmetry, leading to more stringent bounds for the soft symmetry-breaking, as well as for the symmetry-conserving scenarios.

 \subsection{Two body leptonic decays of pseudoscalar mesons}
The two-body leptonic decays of a pseudoscalar meson into charged leptons can be described by the following effective Hamiltonian,

\begin{align}
\mathcal{H}_\text{eff}=-\frac{G_F^2m_W^2}{\pi^2}\,\left(C_S^{ij}\, (\bar q_iP_L q_j)\bar \ell \ell +\tilde  C_S^{ij}\, (\bar q_iP_R q_j)\bar \ell \ell + C_P^{ij}(\bar q_iP_L q_j)\bar \ell \gamma_5 \ell+\tilde C_P^{ij}(\bar q_iP_R q_j)\bar \ell \gamma_5 \ell \right)+ \text{H.c.}.
\end{align}

The  branching ratio for such a leptonic decay of meson can be written as,
\begin{align}
\br(M\rightarrow \ell^+ \ell^- )=&
\frac{G_F^4m_W^4}{8\pi^5} \beta \,m_M f_M^2 m_\ell^2 \tau_M \notag\\
&\left(   \left|\frac{m_M^2\big(C_P^{ij}-\tilde C^{ij}_P\big)}{2m_\ell (m_i+m_j)}-C_A^\text{SM}\right|^2+
\left|\frac{m_M^2\big(C_S^{ij}-\tilde C^{ij}_S\big)}{2m_\ell (m_i+m_j)}\right|^2\beta^2
\right) \; ,
\end{align}
where $\beta(x)=\sqrt{1-4x^2}$ with $x = m_\ell/m_M$.  

The tree-level contribution of the flavon to the corresponding  Wilson coefficients reads as \cite{Buras:2013rqa,Crivellin:2013wna},

\begin{align}
C_S^{ij}& =\frac{\pi^2}{2G_F^2 m_W^2}\frac{2y_{\ell\ell}y_{ji}}{m_s^2}  \notag\\
\tilde C_S^{ij}&=\frac{\pi^2}{2G_F^2 m_W^2}\frac{2y_{\ell\ell}y_{ij}}{m_s^2} \notag\\
C_P^{ij}&=\frac{\pi^2}{2G_F^2 m_W^2}\frac{2y_{\ell\ell}y_{ji}}{m_a^2}  \notag\\
\tilde C_P^{ij}&=\frac{\pi^2}{2G_F^2 m_W^2}\frac{2y_{\ell\ell}y_{ij}}{m_a^2}\; .
\end{align}

In the SM, the two-body leptonic decays of a pseudoscalar meson into two charged leptons occur at one-loop, and their contribution is given by \cite{Crivellin:2013wna},

\begin{align}
\label{CA}
C_A^\text{SM}= -V_{tb}^*V_{ts}\,Y\left(\frac{m_t^2}{m_W^2}\right) -V_{cb}^*V_{cs}\,Y\left(\frac{m_c^2}{m_W^2}\right) \; ,
\end{align}
where $Y(x)$  is the Inami-Lim function, given by \cite{Inami:1980fz},
\begin{align}
Y(x)=\eta_\text{QCD}  \frac{x}{8}\left[\frac{4-x}{1-x}+\frac{3x}{(1-x)^2} \log x \right] ,
\end{align}
where $\eta_\text{QCD}=1.0113$  shows the  NLO QCD effects~\cite{Buras:2012ru}.    For $B_d$ meson,  the SM predictions can be  obtained by  replacing the indices  in equation \ref{CA}.

\subsubsection{The $B_{s,d}$ meson decays} 

The HFLAV group provides the branching fraction of    $B_{s}   \rightarrow \mu^+\mu^-$  to be \cite{HFLAV:2022pwe}, 

\begin{align}
\label{Hflav}
\br(B_s \rightarrow \mu^+\mu^-) & =(3.45 \pm 0.29 )  \times 10^{-9}.
\end{align}

The  branching ratio of the  $B_{d}   \rightarrow \mu^+\mu^-$  decay is \cite{LHCb:2021vsc,LHCb:2021awg},

\begin{align}
\br(B_d\rightarrow \mu^+\mu^-)& <  2.6 \times  10^{-10}\,.
\end{align}

To convert the theoretical branching ratio  of the $B_s$ meson to  the experimental branching ratio, we multiply by the factor $(1-y_s)^{-1}$ \cite{DeBruyn:2012wk},  where $y_s=0.088\pm0.014$~\cite{Fleischer:2012bu}. This is done due to the sizeable width difference  of the $B_s$ meson, and this correction can be ignored in the case of the $B_d$ meson.

\begin{table}[h!]
\centering
\begin{tabular}{l|ccccc}
\text{Observables} & \text{Current} & \text{LHCb-\rom{1}} & \text{LHCb-\rom{2}}  & \text{CMS}  & \text{ATLAS} \\
\hline
$\br(B_s\rightarrow \mu^+\mu^-)   (\times 10^9)$ & $\pm 0.38$    &   $\pm 0.30$    &  $\pm 0.16$   &  $-$      & $\pm 0.50$  \\
$ \mathcal{R}_{\mu\mu} $ & $\sim 70 \%$ &  $\sim 34 \%$& $\sim 10 \%$ &  $\sim 21 \%$   & $-$\\
$\tau_{\mu\mu}$ & $\sim 12 \%$  & $\pm 0.16$ ps  &   $\pm 0.04$ ps  &   $-$     & $-$\\
\hline
\end{tabular}
\caption{ The values of the rare  $B$ decays observables for the current and expected experimental precision.  The  \text{LHCb-\rom{1}} corresponds to $23 fb^{-1}$,   \text{LHCb-\rom{2}} corresponds to $300 fb^{-1}$,  and the  CMS and the ATLAS correspond to $3 ab^{-1}$\cite{CMS:2022dbz,Altmannshofer:2022hfs}.}
\label{Bsmmu}
\end{table}

We note that the  LHCb collaboration  has also  measured the ratio of the $\br(B_d\rightarrow \mu^+\mu^-) $ and $\br(B_s\rightarrow \mu^+\mu^-) $ branching fractions, $ \mathcal{R}_{\mu\mu} $\cite{LHCb:2021vsc,LHCb:2021awg}.  This observable can also be used to constrain the parameter space of flavon of different $\mathcal{Z}_N \times \mathcal{Z}_M$ flavor symmetries, and we shall show later that it provides the striking bounds of the parameter space of flavon of different $\mathcal{Z}_N \times \mathcal{Z}_M$ flavor symmetries.  It should be noted  that the ratio $ \mathcal{R}_{\mu\mu}$ turns out to be  an excellent observable to probe the minimal flavor violation\cite{Altmannshofer:2022hfs}.  Moreover, the effective lifetime,  $\tau_{\mu\mu}$,  of the $B_s\rightarrow \mu^+\mu^- $ decay is also measured by the CMS \cite{CMS:2022dbz}.  The effective lifetime,  $\tau_{\mu\mu}$, is capable of distinguishing between the contributions due to any possible new scalar and pseudoscalar mediators \cite{Altmannshofer:2022hfs}.

The  ratio of branching fractions, $ \mathcal{R}_{\mu\mu} $  is   \cite{LHCb:2021vsc,LHCb:2021awg},

\bea
 \mathcal{R}_{\mu\mu} &=& \frac{\br(B_d\rightarrow \mu^+\mu^-) }{\br(B_s\rightarrow \mu^+\mu^-)} = 0.039^{+0.030+ 0.006}_{-0.024-0.004}.
\eea

The CMS has measured the  effective lifetime $\tau_{\mu\mu}$  and the branching fraction of $B_s\rightarrow \mu^+\mu^-$, which are \cite{CMS:2022dbz},

\bea
 \tau_{\mu\mu} &=& 1.83^{+0.23+ 0.04}_{-0.20-0.04}  ~\rm{ps},
\eea

\begin{align}
\br(B_s\rightarrow \mu^+\mu^-) & =3.83^{+0.38+ 0.19+0.14}_{-0.36-0.16-0.13}   \times 10^{-9}.  
\end{align}

The  HFLAV measurement average of the  effective lifetime $\tau_{\mu\mu}$  is \cite{HFLAV:2022pwe}, 
\bea
\label{t_hflav}
 \tau_{\mu\mu} &=& 2.00^{+0.27}_{-0.26}  ~\rm{ps},
\eea
with $\br(B_s\rightarrow \mu^+\mu^-)$ given in equation \ref{Hflav}.

We summarize the current and future sensitivities of these observables in table \ref{Bsmmu}.  The effective lifetime can be expressed as \cite{DeBruyn:2012wj},

\be
 \tau_{\mu\mu} =  \tau_{B_s} \frac{(B_s\rightarrow \mu^+\mu^-)^{\rm experiment}}{(B_s\rightarrow \mu^+\mu^-)^{\rm theory}},
\ee
where the SM value of the  final state dependent observable is $\mathcal{A}^f_{\Delta \Gamma} =1$ \cite{Fleischer:2012bu}.

\subsubsection{The $K_{L}$  meson decays} 

The short-distance (SD) part of  $K_L \rightarrow \mu^+ \mu^-$ decay can be estimated in a reliable manner \cite{Buras:2013rqa}. Its SM prediction is given in reference  \cite{Buras:2013rqa}, and reads as,

\begin{align}
C_A^\text{SM}= -V_{ts}^*V_{td}\,   Y\left(\frac{m_t^2}{m_W^2}\right) -  V_{cs}^* V_{cd}         Y_{\rm NNL},
\end{align}
where at NNLO $Y_{\rm NNL} = \lambda^4 P_c(Y)$,  $\lambda = |V_{us}|$ and $P_c(Y) = 0.113 \pm 0.017$\cite{Gorbahn:2006bm}.   The short-distance contribution is  extracted from the experimental measurement,  and its  upper limit is \cite{Crivellin:2013wna},
\begin{align}
\br(K_L\rightarrow \mu^+\mu^-)_{\rm SD} & < 2.5   \times 10^{-9}.
\end{align}

\begin{figure}[H]
	\centering
	\begin{subfigure}[]{0.4\linewidth}
    \includegraphics[width=\linewidth]{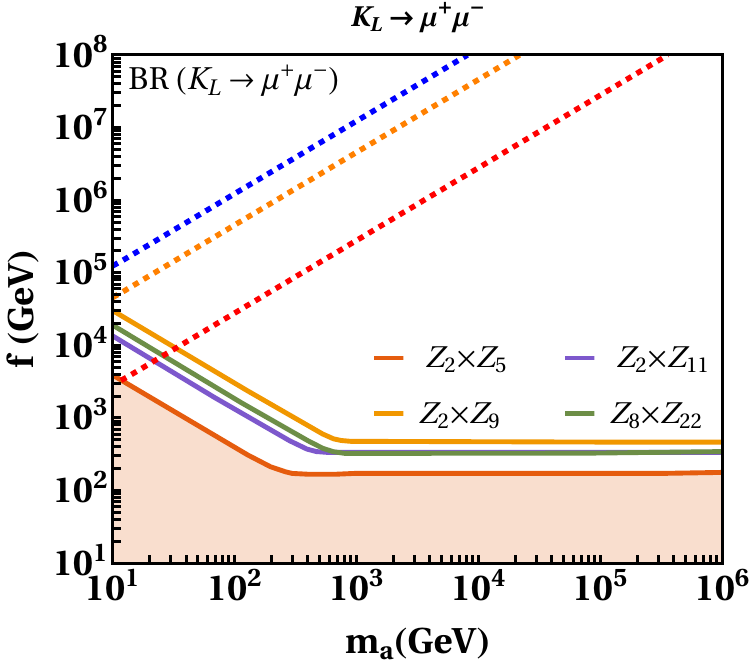}
    \caption{}
         \label{fig5a}	
\end{subfigure}
 \begin{subfigure}[]{0.4\linewidth}
 \includegraphics[width=\linewidth]{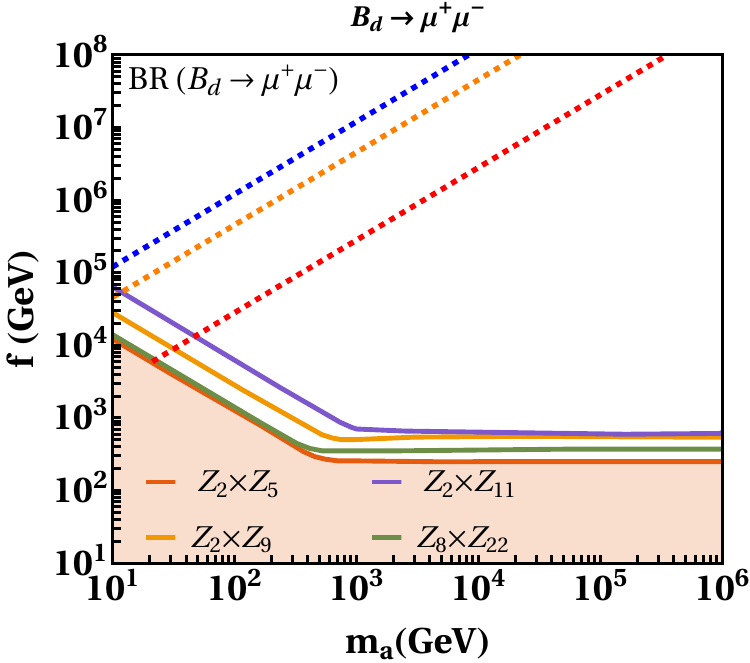}
 \caption{}
         \label{fig5b}
 \end{subfigure} 
 \begin{subfigure}[]{0.4\linewidth}
    \includegraphics[width=\linewidth]{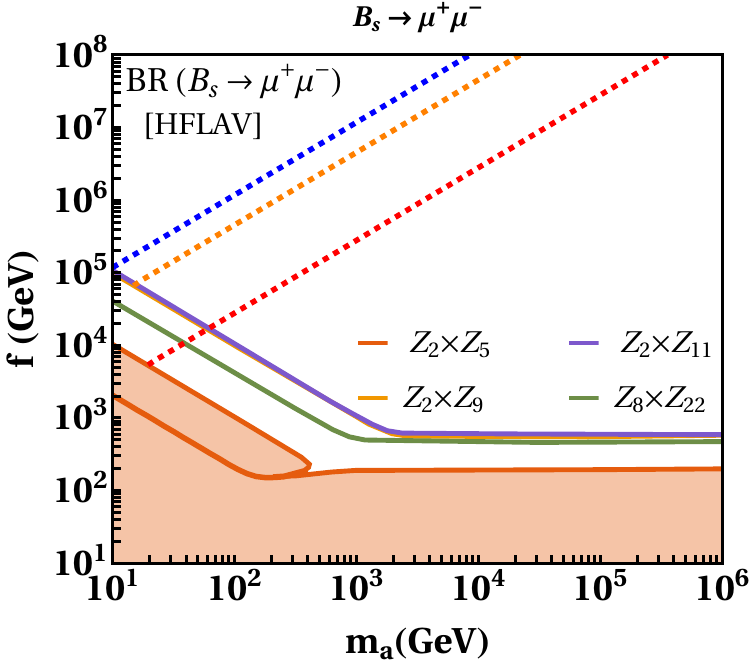}
    \caption{}
         \label{fig5c}	
\end{subfigure}
 \caption{ The coloured region represents the excluded parameter space for the flavon of all four $\mathcal{Z}_N \times \mathcal{Z}_M$ flavor symmetries  by the limits on branching ratios $\br(K_L\rightarrow \mu^+\mu^-)_{\rm SD}$,   $\br(B_d\rightarrow \mu^+\mu^-)$  and the   $\br(B_s\rightarrow \mu^+\mu^-)$ for the soft symmetry-breaking scenario.  The dashed line represents the allowed parameter space by the same observables for symmetry-conserving scenario, with  $\lambda_\chi= 2$ for both the scenarios.  }
  \label{fig5}
	\end{figure}

\subsubsection{The $D$ meson decays} 

The SM contribution of the $D \rightarrow \mu^+ \mu^-$ decay is marred by the large non-perturbative effects.  Therefore, we assume that the flavon contribution to this decay rate lies within the experimental upper bound on the branching ratio,  given at $90\%$ C.L.\cite{LHCb:2013jyo},

\begin{align}
\br(D \rightarrow \mu^+\mu^-) & < 6.2   \times 10^{-9}.
\end{align}

We note that the bounds from  the two-body leptonic decays of a pseudoscalar meson depend on the product $|y_{ij}^q y_{ij}^\ell| $.  This quantity is very small for the $\mathcal{Z}_2 \times \mathcal{Z}_5$ flavor symmetry, and closer for the rest of the flavor symmetries for the two-body leptonic decays of a pseudoscalar meson.

In figure \ref{fig5}, we show the bounds on the parameter space of flavon of different $\mathcal{Z}_N \times \mathcal{Z}_M$ flavor symmetries arising from the branching ratio of the $K_L \rightarrow \mu^+ \mu^-$ decay.  In the case of the soft symmetry-breaking scenario, the branching ratio of the $K_L \rightarrow \mu^+ \mu^-$ decay imposes stronger bounds on the  $\mathcal{Z}_2 \times \mathcal{Z}_9$ flavor symmetry.  After this, the constraints are stronger for the $\mathcal{Z}_2 \times \mathcal{Z}_{11}$ flavor symmetry.   The decay $B_{d}   \rightarrow \mu^+\mu^-$  constrains more parameter space of flavon of  the $\mathcal{Z}_2 \times \mathcal{Z}_{11}$ flavor symmetry, followed by the $\mathcal{Z}_2 \times \mathcal{Z}_9$ flavor symmetry.  On the other hand, the branching ratio of the decay $B_s\rightarrow \mu^+\mu^-$ bounds both the $\mathcal{Z}_2 \times \mathcal{Z}_9$ and $\mathcal{Z}_2 \times \mathcal{Z}_{11}$ flavor symmetries together with similar strength, while the bounds are comparatively weak in the case of $\mathcal{Z}_8 \times \mathcal{Z}_{22}$ flavor symmetry. In the symmetry-preserving scenario, all three decays place the stronger bounds on the $\mathcal{Z}_2 \times \mathcal{Z}_{11}$ flavor symmetry.  This is followed by the $\mathcal{Z}_2 \times \mathcal{Z}_9$ flavor symmetry, and $\mathcal{Z}_2 \times \mathcal{Z}_5$ flavor symmetry. We do not show the bounds from $BR(D \rightarrow \mu^+ \mu^-)$ as it imposes extremely weak constraint on the parameter space of all four $\mathcal{Z}_N \times \mathcal{Z}_M$ flavor symmetries.

\begin{figure}[H]
	\centering
 \includegraphics[width=7cm]{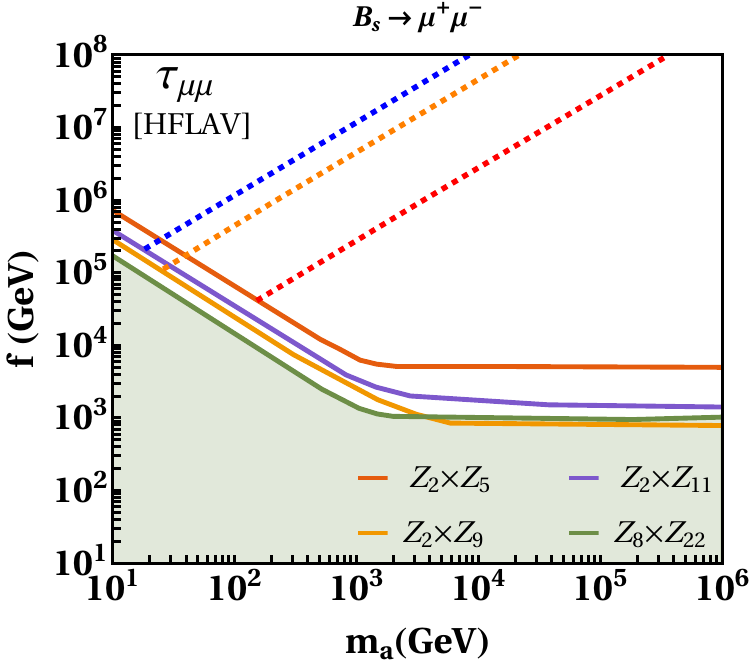}
 \caption{The allowed parameter space by  the observable $ \tau_{\mu\mu} $ for the flavon of different $\mathcal{Z}_N \times \mathcal{Z}_M$ flavor symmetries for the recent measurement.}
  \label{fig6}
	\end{figure}

The observable $ \tau_{\mu\mu} $  places the strongest bounds on the parameter space of flavon of  the $\mathcal{Z}_2 \times \mathcal{Z}_{5}$ flavor symmetry in the soft symmetry-breaking scenario, as observed in figure \ref{fig6}. This is expected since the lifetime is inverse of the decay-width, and the branching ratio of the decay $B_s\rightarrow \mu^+\mu^-$ places the most relaxed constraints on the flavon of the $\mathcal{Z}_2 \times \mathcal{Z}_{5}$ flavor symmetry.  This is followed by the flavon of the $\mathcal{Z}_2 \times \mathcal{Z}_{11}$ flavor symmetry.  For the symmetry-conserving scenario, the most stringent constraints arise for the flavon of the $\mathcal{Z}_2 \times \mathcal{Z}_{11}$ flavor symmetry, followed by the flavon of the $\mathcal{Z}_2 \times \mathcal{Z}_{9}$ flavor symmetry.  There is no substantial improvement over these bounds in the future projected sensitivities of the LHCb.  Therefore, we do not show them in this work.

\begin{figure}[H]
	\centering
	\begin{subfigure}[]{0.4\linewidth}
    \includegraphics[width=\linewidth]{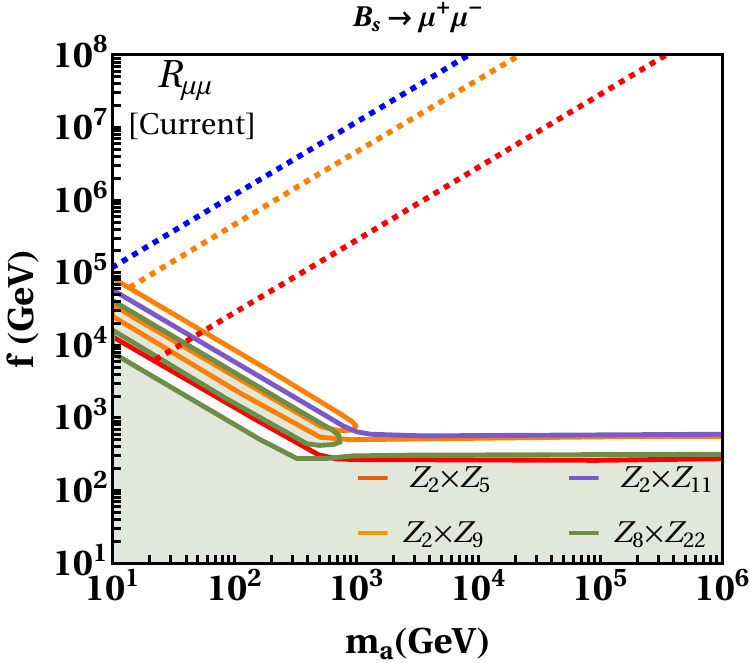}
    \caption{}
         \label{fig7a}	
\end{subfigure}
 \begin{subfigure}[]{0.4\linewidth}
 \includegraphics[width=\linewidth]{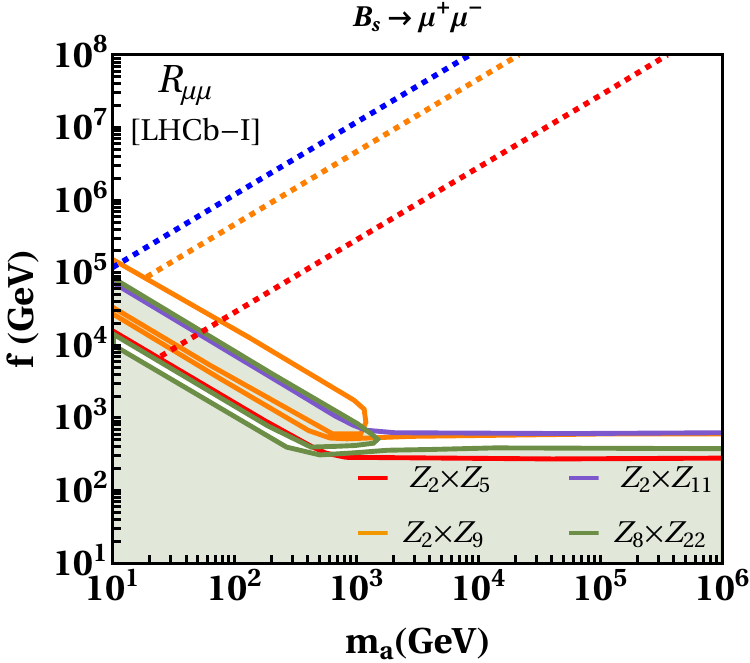}
 \caption{}
         \label{fig7b}
 \end{subfigure} 
 \begin{subfigure}[]{0.4\linewidth}
    \includegraphics[width=\linewidth]{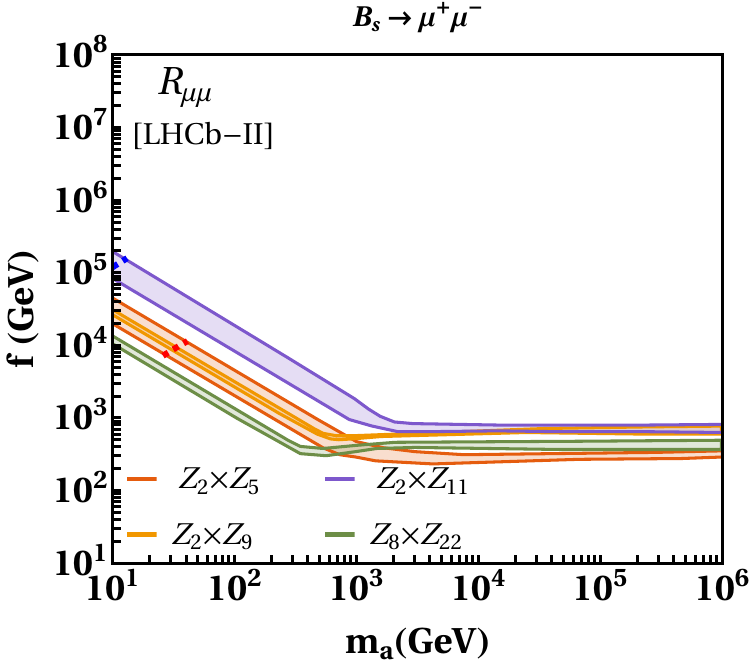}
    \caption{}
         \label{fig7c}	
\end{subfigure}
 \caption{The blank region above the continuous curves in figure \ref{fig7a} and \ref{fig7b} shows the allowed $f-m_a$ parameter space of the flavon of different $\mathcal{Z}_N \times \mathcal{Z}_{M}$ models by current measurement as well as the future projected sensitivities in the LHCb  Phase-\rom{1} for the observable $ \mathcal{R}_{\mu\mu} $, with $\lambda_\chi= 2$, in the soft symmetry-breaking scenario. The coloured region in figure \ref{fig7c} depicts the same allowed parameter space for projected sensitivities of $ \mathcal{R}_{\mu\mu} $ in the LHCb  Phase-\rom{2}, while the dashed lines represent the allowed parameter space in the symmetry-conserving scenario.}
  \label{fig7}
	\end{figure}

The ratio $ \mathcal{R}_{\mu\mu} $ is one of the  most important observables  to test the $\mathcal{Z}_N \times \mathcal{Z}_M$ flavor symmetries.  Its future projected sensitivity will play a decisive role in constraining the parameter space of flavon of $\mathcal{Z}_N \times \mathcal{Z}_M$ flavor symmetries.  The predictions from this observable are shown in figure \ref{fig7}.   The bounds from the ratio $ \mathcal{R}_{\mu\mu} $  are minimally sensitive to the product $y_{ij} \epsilon^{n_{ij}}  y_{ji} \epsilon^{n_{ji}} $ since it appears in the numerator as well as denominator of the ratio $ \mathcal{R}_{\mu\mu} $.  Therefore, all the  $\mathcal{Z}_N \times \mathcal{Z}_M$ flavor symmetries exhibit a close behavior providing relatively relaxed bounds from the current measurements.

In the case of soft symmetry-breaking scenario, the strongest bounds turn out to be for the $\mathcal{Z}_2 \times \mathcal{Z}_{9}$ and $\mathcal{Z}_2 \times \mathcal{Z}_{11}$ flavor symmetries for the current measurement of the observable $ \mathcal{R}_{\mu\mu} $, as observed in figure \ref{fig7a}.  This feature continues to grow stronger further in the future high luminosity  phase-\rom{1} of the LHCb, as can be seen in figure \ref{fig7b}.  In the  symmetry-conserving scenario, the parameter space of the flavon of the $\mathcal{Z}_2 \times \mathcal{Z}_{11}$ is more constrained than any other flavor symmetry for the current measurement as well as for the  high luminosity  phase-\rom{1} of the LHCb, as can be seen in figure \ref{fig7a} and \ref{fig7b}.  This is followed by the $\mathcal{Z}_2 \times \mathcal{Z}_{9}$ flavor symmetry.

 We obtain very stringent  bounds on the parameter space of the flavon of  different $\mathcal{Z}_N \times \mathcal{Z}_M$ flavor symmetries from the future high luminosity phase-\rom{2} of the LHCb, as can be seen in figure \ref{fig7c}.  We emphasize to note that the colored stripes in this figure are allowed parameter space for the soft symmetry-breaking scenario, and the blank region is the excluded parameter space.  We observe from \ref{fig7c} that the allowed  parameter space is very narrow and different for  $\mathcal{Z}_N \times \mathcal{Z}_M$ flavor symmetries for $m_a < 1$ TeV.  The stripe corresponding to the $\mathcal{Z}_2 \times \mathcal{Z}_{9}$ flavor symmetry resides inside the stripe corresponding to the $\mathcal{Z}_2 \times \mathcal{Z}_{5}$ flavor symmetry in this case. Above $m_a > 1$ TeV, the stripes corresponding to the $\mathcal{Z}_2 \times \mathcal{Z}_{9}$ and $\mathcal{Z}_2 \times \mathcal{Z}_{11}$ flavor symmetries  overlap around $10^4$ TeV.

 The  remarkable predictions are obtained in the case of the symmetry-conserving scenario.  As evident from figure \ref{fig7c}, there is a very small allowed parameter space for symmetries $\mathcal{Z}_2 \times \mathcal{Z}_{5}$ and $\mathcal{Z}_2 \times \mathcal{Z}_{11}$, shown by the dashed-red and dashed-blue lines within the stripes, respectively. There is no allowed parameter space for other symmetries for the symmetry-conserving scenario.  Thus, the future high luminosity  phase-\rom{2} of the LHCb will be crucial in ruling out the symmetry-conserving scenario discussed in this work.

\subsection{Constraints from $ b \to s \mu^+ \mu^-$ transition} 
In this section, we explore constraints arising from a plethora of observables in diverse decay channels triggered by the quark-level transition $b \to s \mu^+ \mu^-$. This also encompasses  the branching ratio of $B_s \to \mu^+ \mu^-$, which was addressed in the preceding sub-section. The effective Hamiltonian for   $b \to s \mu^+ \mu^-$ transition can  be written as follows:
\begin{equation}
    \mathcal{H}_{\rm eff} = - \frac{ 4 G_F}{\sqrt{2}}\frac{ e^2}{16 \pi^2} m_b\,V_{ts}^* V_{tb}  \left[\mathcal{C}_{\rm S}^{\mu \mu}(\overline{s}_L b_R)(\overline{\mu} \mu)+ \mathcal{C}_{\rm S}^{'\mu \mu}(\overline{s}_R b_L)(\overline{\mu} \mu) + \mathcal{C}_{\rm P}^{\mu \mu} (\overline{s}_L b_R)(\overline{\mu} \gamma_5 \mu) + \mathcal{C}_{\rm P}^{'\mu \mu} (\overline{s}_R b_L)(\overline{\mu} \gamma_5 \mu)\right]\,.
\end{equation}
Here, $\mathcal{C}_{\rm S}^{\mu \mu}$, $\mathcal{C}_{\rm P}^{\mu \mu}$, $\mathcal{C}_{\rm P}^{'\mu \mu}$, and $\mathcal{C}_{\rm P}^{'\mu \mu}$ represent the new physics scalar/pseudoscalar Wilson coefficients (WCs). In the model under consideration, these WCs are generated at the tree level and are expressed in terms of model parameters. For various flavor symmetries, these WCs are given in Table \ref{wcs}.

\begin{table}[h!]
\centering
\begin{tabular}{l|cccc}
\text{WCs} & $\mathcal{Z}_2 \times \mathcal{Z}_5$ & $\mathcal{Z}_2 \times \mathcal{Z}_9$ & $\mathcal{Z}_2 \times \mathcal{Z}_{11}$  & $\mathcal{Z}_8 \times \mathcal{Z}_{22}$\\
\hline
$\mathcal{C}_{\rm S}^{\mu \mu}$ &  $\mathcal{N} (9\epsilon^6/m_s^2) y_{23}^d y_{22}^{\ell *}$ &  $\mathcal{N} (25\epsilon^{10}/m_s^2) y_{23}^d y_{22}^{\ell *}$ & $\mathcal{N} (49\epsilon^{14}/m_s^2) y_{23}^d y_{22}^{\ell *}$ & $\mathcal{N} (25\epsilon^{10}/m_s^2) y_{23}^d y_{22}^{\ell *}$\\
$\mathcal{C}_{\rm S}^{'\mu \mu}$ & $\mathcal{N} (3\epsilon^4/m_s^2) y_{32}^{d*} y_{22}^{\ell *}$ & $\mathcal{N} (15\epsilon^8/m_s^2) y_{32}^{d*} y_{22}^{\ell *}$ & $\mathcal{N} (21\epsilon^{10}/m_s^2) y_{32}^{d*} y_{22}^{\ell *}$  & $\mathcal{N} (15\epsilon^{8}/m_s^2) y_{32}^{d*} y_{22}^{\ell *}$\\ 
$\mathcal{C}_{\rm P}^{\mu \mu}$ & $\mathcal{N} (9\epsilon^6/m_a^2) y_{23}^d y_{22}^{\ell *}$   & $\mathcal{N} (25\epsilon^{10}/m_a^2) y_{23}^d y_{22}^{\ell *}$ & $\mathcal{N} (49\epsilon^{14}/m_a^2) y_{23}^d y_{22}^{\ell *}$ & $\mathcal{N} (25\epsilon^{10}/m_a^2) y_{23}^d y_{22}^{\ell *}$\\
$\mathcal{C}_{\rm P}^{'\mu \mu}$ & $\mathcal{N} (3\epsilon^4/m_a^2) y_{32}^{d*} y_{22}^{\ell *}$ & $\mathcal{N} (15\epsilon^8/m_a^2) y_{32}^{d*} y_{22}^{\ell *}$ & $\mathcal{N} (21\epsilon^{10}/m_a^2) y_{32}^{d*} y_{22}^{\ell *}$ & $\mathcal{N} (15\epsilon^{8}/m_a^2) y_{32}^{d*} y_{22}^{\ell *}$ \\
\hline
\end{tabular}
\caption{Scalar and pseudoscalar WCs for $b \to s \mu^+ \mu^-$ decay in terms of model parameters for various flavor symmetries. Here $\mathcal{N}=-\frac{2\sqrt{2}\pi^2 v^2}{G_F m_b V_{tb}V^*_{ts}e^2 f^2}$ whereas the benchmark values of $y_{23}^d$ ($y_{32}^d$), and $y_{22}^\ell$ couplings are given in the appendix for all four $\mathcal{Z}_N \times \mathcal{Z}_M$ flavor symmetries.}
\label{wcs}
\end{table}

These new physics WCs, which are complex, are subject to constraints imposed by a multitude of $b \to s \mu^+ \mu^-$ observables (158 observables), which include:
\begin{enumerate}

\item The branching ratio of $B_s \to \mu^+ \mu^-$ \cite{HFLAV:2022pwe,Ciuchini:2022wbq} and $B \to X_{s}\mu^{+}\mu^{-}$ in both low and high-$q^2$ bins \cite{Lees:2013nxa}. Here, we consider the modified world average of the branching ratio of $B_s \to \mu^+ \mu^-$ \cite{Ciuchini:2022wbq} obtained after the recent CMS update, which is based on the full Run 2 dataset \cite{CMS:2022dbz}. The updated world average now aligns with the predictions of the SM.

\item  The differential branching ratios of $B^0 \to K^{*0} \mu^+ \mu^- $ \cite{LHCb:2016ykl,CDFupdate,Khachatryan:2015isa}, $B^{+} \to K^{*+}\mu^{+}\mu^{-}$, $B^{0}\to K^{0} \mu^{+}\mu^{-}$, $B^{+}\rightarrow K^{+}\mu^{+}\mu^{-}$ \cite{Aaij:2014pli,CDFupdate} and $B_s \to \phi \mu^+ \mu^-$ \cite{bsphilhc3} in various  $q^2$ intervals.  It should be noted that for all of these observables, as well as for other $b \to s \mu^+ \mu^-$ observables utilized in our analysis, we have  excluded measurements within the $6\, {\rm GeV}^2 \le q^2 \le 14.0 \, {\rm GeV}^2$ bin.

\item In addition to branching fractions, our analysis incorporates various angular observables as constraints within the fit. These observables encompass the longitudinal polarization fraction $f_L$, forward-backward asymmetry $A_{FB}$, and $S_{3,4,5,7,8,9}$ observables measured in the $B^0 \to K^{*0} \mu^+\mu^-$ decay across multiple $q^2$ bins, as reported by the LHCb collaboration \cite{LHCb:2020lmf}. Furthermore, we include the observables $f_L$, $P_1$, $P'_4$, $P'_5$, $P'_6$, and $P'_8$ measured in $B^0 \to K^{*0} \mu^+\mu^-$ decay by ATLAS \cite{ATLAS:2018gqc}, along with $P_1$ and $P'_5$ measurements from CMS \cite{CMS:2017rzx}. Additionally, measurements of $f_L$ and $A_{FB}$ by the CDF and CMS collaborations \cite{CDFupdate,Khachatryan:2015isa} are also integrated into our fit.
 
 \item  The $f_L$  and $P_1 - P_8 '$  angular observables 
in $B^+ \to K^{*+} \mu^+\mu^-$ decay mode are also included \cite{LHCb:2020gog} in the fit.

\item  The fit also incorporates $f_L$, $S_3$, $S_4$, and $S_7$ observables obtained from the measurements of the $B_s \to \phi \mu^+\mu^-$ decay by the LHCb collaboration, as reported in \cite{LHCb:2021xxq}.

\end{enumerate}

The constraints on new physics couplings are obtained by performing a $\chi^2$ fit using the  CERN minimization code {\tt MINUIT} \cite{James:1975dr}. The available experimental correlations  \cite{LHCb:2020lmf, LHCb:2020gog,LHCb:2021xxq}, as well as theoretical correlations, are included in the analysis. The theoretical correlations, along with the expressions for various observables, are computed using {\tt flavio} \cite{Straub:2018kue}. The observables in {\tt flavio} are pre-implemented on the basis of refs. \cite{Bharucha:2015bzk,Gubernari:2018wyi}. For the fit, we closely follow the methodology adopted in \cite{Alok:2022pjb,SinghChundawat:2022zdf,SinghChundawat:2022ldm}.   

The value of $\chi^2_{SM}$ is 164.80, whereas the $\chi^2$ at the best-fit point is 164.26, indicating that the addition of new physics in the form of scalar and pseudoscalar interactions only provides an extremely marginal improvement over the SM (Pull = 0.54). The fit results are (1$\sigma$ allowed range of real and complex components of the WCs)
\begin{eqnarray}
\mathcal{C}_{\rm S}^{\mu \mu} &=& [(-0.13, 0.13),(-0.13,0.13)]\nonumber\\
\mathcal{C}_{\rm P}^{\mu \mu} &=& [(-0.07, 0.13),(-0.12,0.12)]\nonumber\\
\mathcal{C}_{\rm S}^{'\mu \mu}  &=& [(-0.11, 0.10),(-0.12,0.12)]\nonumber\\
\mathcal{C}_{\rm P}^{'\mu \mu}  &=& [(-0.07, 0.13),(-0.11,0.11)].
\label{mi-sps}
\end{eqnarray}

\begin{figure}[h!]
	\centering
	\includegraphics[width=0.4\linewidth]{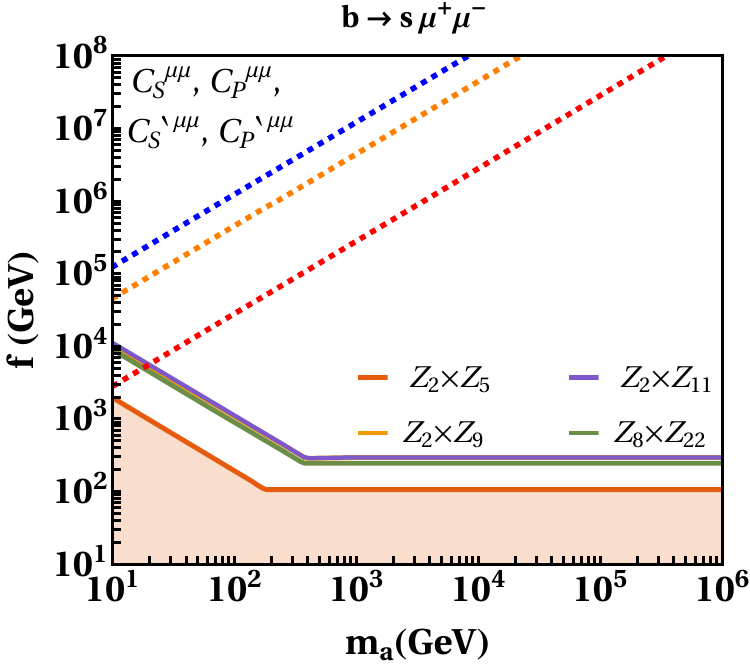}
\caption{Bounds on the $f- m_a$ parameter space of the flavon of different $\mathcal{Z}_N \times \mathcal{Z}_M$ models by obtained limits on $\mathcal{C}_{\rm S}^{\mu \mu}$ , $\mathcal{C}_{\rm P}^{\mu \mu}$, $\mathcal{C}_{\rm S}^{'\mu \mu}$ and $\mathcal{C}_{\rm P}^{'\mu \mu}$ as in equation  \ref{mi-sps} for the soft symmetry-breaking as well as symmetry-conserving scenario.  }
 \label{figwc}
 \end{figure}
 
 The model-independent fit results for the Wilson coefficients of $b \to s \mu^+ \mu^-$ obtained in equation \ref{mi-sps} can be utilized to constrain free parameters for the different $\mathcal{Z}_N \times \mathcal{Z}_M$ symmetries, as shown in figure \ref{figwc}. In the case of soft symmetry-breaking, the region above the continuous curves represents the allowed parameter space, and it is evident that the bounds are weakest for the minimal $\mathcal{Z}_2 \times \mathcal{Z}_5$ flavor symmetry, while the $\mathcal{Z}_2 \times \mathcal{Z}_9$, $\mathcal{Z}_2 \times \mathcal{Z}_{11}$, and $\mathcal{Z}_8 \times \mathcal{Z}_{22}$ impose almost similar constraints on the free parameters. In the symmetry-conserving scenario, the corresponding allowed parameter space for $\mathcal{Z}_2 \times \mathcal{Z}_{5, 9, 11}$ models are represented by the dashed straight lines, among which, the parameter space for the flavon of $\mathcal{Z}_2 \times \mathcal{Z}_{11}$ is most constrained.

\subsection{Muon forward-backward asymmetry in $B \to K \mu^+ \mu^-$}

Within the framework of the SM, the muon forward-backward asymmetry in the context of $B \to K \mu^+ \mu^-$ decay is inherently predicted to have a value of approximately zero. This prediction holds even when considering the effects of new physics scenarios involving vector-axial vector interactions. However, it's noteworthy that if new physics manifests in the form of scalar or pseudoscalar interactions, it has the potential to introduce a non-zero value to $A_{FB}$ \cite{Bobeth:2007dw,Alok:2008wp}. Given this intriguing possibility, it is prudent to investigate this observable within the framework of flavons. Such an exploration could yield valuable insights into whether the new physics parameter space allowed in the context of this model has the potential to enhance $A_{FB}$ to a value that can be  measured experimentally.

The muon forward-backward asymmetry in $B \to K \mu^+ \mu^-$ is given by \cite{Bobeth:2007dw}
\begin{equation}
A_{FB} = \frac{\int_{q^2_{\rm min}}^{{q^2_{max}}}  b_{\mu}(q^2) \, dq^2}{
\int_{q^2_{\rm min}}^{{q^2_{max}}} \left(2a_{\mu}(q^2) +\frac{2}{3} c_{\mu}(q^2)\right)dq^2}
\,,
\end{equation}
where
\begin{eqnarray}
a_{\mu}(q^2) &=& N(q^2) \Bigg[ q^2 |F_P|^2 + \frac{\lambda}{4} \left( |F_V|^2 + |F_A|^2\right)+ 2m_{\mu} \left( m_B^2 -m_K^2 +q^2\right){\rm Re} (F_P F_A^{*}) \nonumber\\
&& + 4 m_{\mu}^2 m_B^2 |F_A|^2  \Bigg]\,,\\
b_{\mu}(q^2) &=& 2 N(q^2) \Bigg[ m_{\mu} \beta_{\mu} \sqrt{\lambda_K} {\rm Re} (F_S F_V^{*})  \Bigg]\,,\\
c_{\mu}(q^2) &=& -\frac{\lambda}{4} \beta_{\mu}^2 N(q^2) \left( |F_V|^2 + |F_A|^2\right)\,,
\end{eqnarray}
with
\begin{equation}
N(q^2) = \frac{G_F^2\alpha |V_{tb} V_{ts}^*|^2}{512 \pi^5 m_B^3}\beta_{\mu}\lambda_K f_+(q^2).
\end{equation}
Here $\lambda = m^4_B+m^4_{K}+q^4-2(m^2_B m^2_{K} +m^2_{B}q^2+m^{2}_{K}q^2)$ and $\beta_{\mu}= \sqrt{1-4m^2_{\mu}/q^2}$. The $F$ functions are written in terms of WCs and form-factors as
\begin{eqnarray}
F_A &=& C_{10} \,,\\
F_V &=& C_9 + \frac{2m_B}{M_B} \frac{T_P(q^2)}{f_+(q^2)}\,, \\
F_S &=& \frac{M_B^2 - M_K^2}{2(m_b - m_s)} \frac{f_0(q^2}{f_+(q^2)} m_b \left(\mathcal{C}_{\rm S}^{\mu \mu} + \mathcal{C}_{\rm S}^{'\mu \mu}\right)\,, \\
F_P &=& \frac{M_B^2 - M_K^2}{2(m_b - m_s)} \frac{f_0(q^2}{f_+(q^2)} m_b \left(\mathcal{C}_{\rm P}^{\mu \mu} + \mathcal{C}_{\rm P}^{'\mu \mu}\right) 
+ m_{\mu} C_{10} \left[ \frac{M_B^2 - M_K^2}{q^2} \left(\frac{f_0(q^2}{f_+(q^2)} - 1\right) -1\right]
\end{eqnarray}

Here $f_0(q^2)$ and $f_+(q^2)$ are $B \to K$ form-factors whereas $C_{9, 10}$ are SM WCs. 
The function $T_P(q^2)$ in the expression of $F_V$ incorporates contributions from virtual one-photon exchange between the hadrons and the lepton pair, in addition to the hard scattering contributions \cite{Bobeth:2007dw}.  At low-$q^2$ values, all form factors simplify to a single soft form factor \cite{Charles:1998dr,Beneke:2000wa}.
It is evident that in the absence of new physics scalar and pseudoscalar couplings, the muon-forward backward asymmetry remains zero. Consequently, in the current scenario, we anticipate a non-zero value for $A_{FB}$. Nonetheless, it would be intriguing to determine whether the constraints imposed by the new physics scalar and pseudoscalar couplings, as derived from the existing $b \to s \mu^+ \mu^-$ data, permit any meaningful enhancements or not.

\begin{figure}[h!]
	\centering
	\begin{subfigure}[]{0.41\linewidth}
    \includegraphics[width=\linewidth]{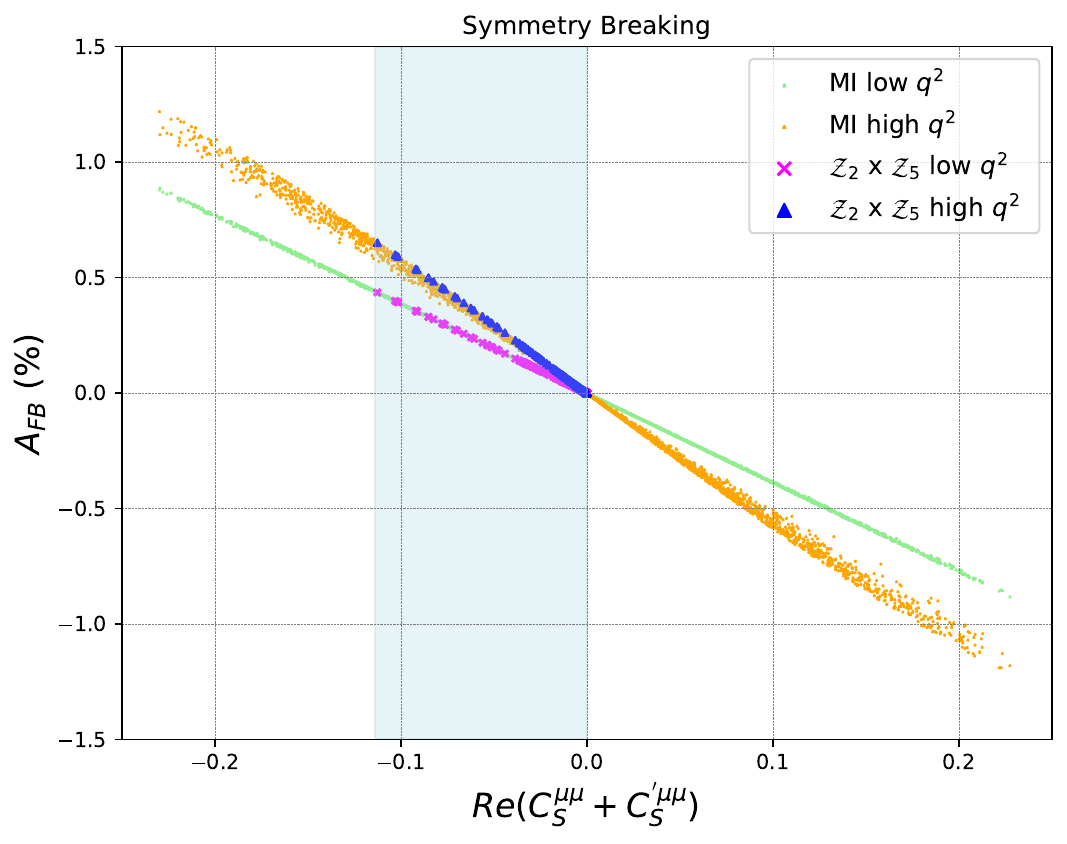}
    \caption{}
         \label{fig_afb1}	
\end{subfigure}
 \begin{subfigure}[]{0.41\linewidth}
 \includegraphics[width=\linewidth]{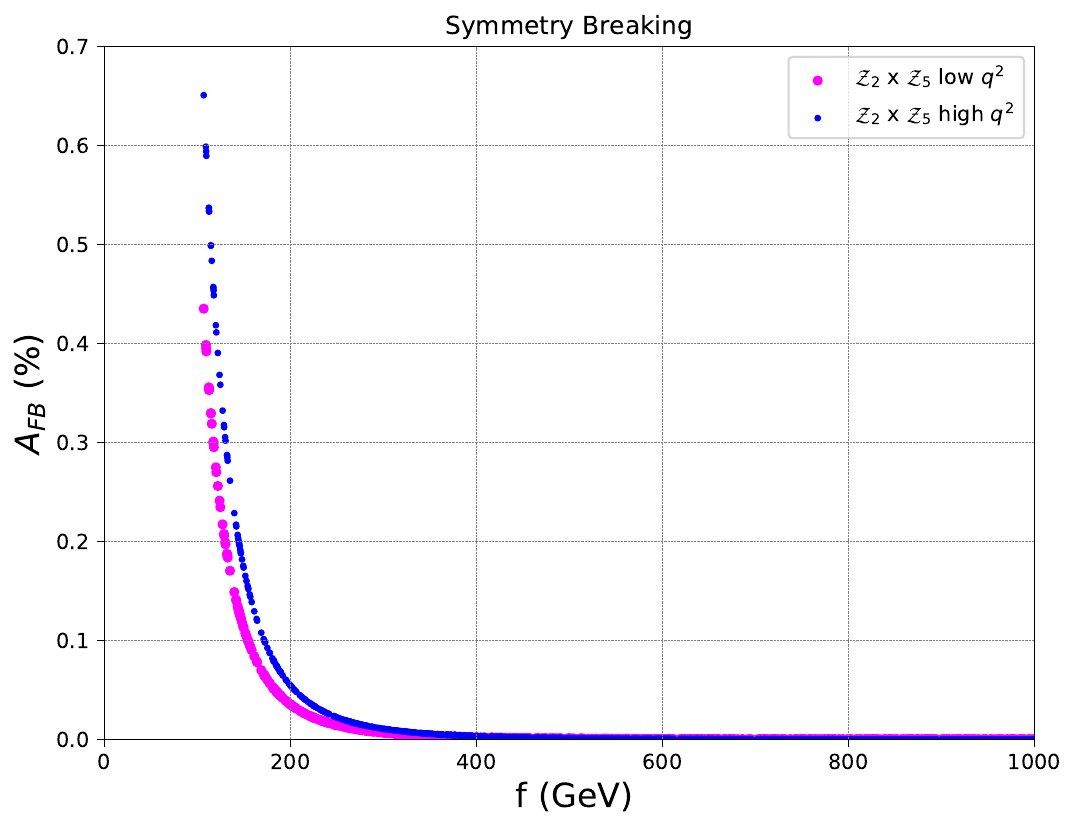}
 \caption{}
         \label{fig_afb2}
 \end{subfigure} 
 \caption{Predictions for the muon FB asymmetry in the $B^+ \to K^+ \mu^+ \mu^-$ decay.}
   \label{afb}
	\end{figure}

Utilizing the model-independent constraints on the new physics scalar and pseudoscalar couplings as derived in the previous sub-section, the FB asymmetry in both the low and high-$q^2$ regions is illustrated in Fig.~\ref{afb}. The figure clearly demonstrates that the existing $b \to s \mu^+ \mu^-$ data permits an enhancement in $A_{FB}$, reaching up to a percent level.

We now explore the prediction of $A_{FB}$ within the flavon model, considering both symmetry-conserving and symmetry-breaking scenarios. In the symmetry-conserving scenario, the flavon masses, as discussed in equation \ref{mphi1}, are dependent on its VEV ($f$). Therefore, for all these symmetries, $f$ serves as the sole free parameter. It is observed that, in order to satisfy the model-independent constraints on scalar as well as pseudoscalar couplings derived in eq. \ref{mi-sps}, the values of $f$ need to be large. This holds true for all four flavor symmetries.  However, such large values of $f$ result in exceedingly small values of $\mathcal{C}_{\rm S}^{\mu \mu}$ and $\mathcal{C}_{\rm S}^{'\mu \mu}$ couplings, leading to predicted values of $A_{FB}$ that are too negligible to be observed.

In the scenario of symmetry breaking, both $f$ and $m_a$ serve as free parameters. The parameter space of $(m_a - f)$ is constrained by eq. \ref{mi-sps}. For all flavor symmetries except $\mathcal{Z}_2 \times \mathcal{Z}_5$, the constraints derived from eq. \ref{mi-sps} limits the $(m_a - f)$ parameter space to such an extent that any significant enhancement in $A_{FB}$ is precluded.
The prediction of $A_{FB}$ for the $\mathcal{Z}_2 \times \mathcal{Z}_5$ flavor symmetry is illustrated in Fig.~\ref{afb}. As depicted in the left panel of the figure, $A_{FB}$ can be enhanced by up to half a percent level, both in the low and high-$q^2$ regions. This enhancement is smaller as compared to the model independent results because the allowed $(\mathcal{C}_{\rm S}^{\mu \mu} + \mathcal{C}_{\rm S}^{'\mu \mu})$ values (indicated by the light blue band), which are pertinent to $A_{FB}$, are more restrictive compared to the model-independent bounds. The right panel displays $A_{FB}$ as a function of the parameter $f$. It is apparent from the plot that a meaningful enhancement of $A_{FB}$ is achievable only for smaller values of $f$, specifically $f<200$ GeV.

\section{Leptonic flavor physics of  the flavon of   the $\mathcal{Z}_N \times \mathcal{Z}_M$ flavor symmetries}
\label{lepton_flavor}
The current and  future experiments on the charged lepton flavor violation (CLFV) processes have a great potential to place stringent bounds on the parameter space of the flavon of the $\mathcal{Z}_N \times \mathcal{Z}_M$ flavor symmetries, and even may provide an improvement over the constraints arising from the quark flavor physics. The sensitivities of the CLFV processes for the current as well as future projected experiments are shown in table \ref{lfv_exp}.

The bounds from the leptonic flavor physics are sensitive to the  $|y_{ij}^\ell  \epsilon^{n_{ij}}  y_{ji}^\ell \epsilon^{n_{ji}}|$, which is minimum for the  $\mathcal{Z}_2 \times \mathcal{Z}_5$ flavor symmetry, providing the most relaxed constraints on the flavon corresponding to this symmetry. For other flavor symmetries, the value of the $|y_{ij}^\ell  \epsilon^{n_{ij}}  y_{ji}^\ell \epsilon^{n_{ji}}|$ is close, resulting in bounds, that are close in magnitude.  This is a generic observation about the leptonic flavor physics.

\begin{table}[h!]\centering
\begin{tabular}{l|cccc}
\text{Observables} & \text{ Current sensitivity}  & \text{Ref.}   & \text{Future projection} & \text{Ref.}  \\
\hline
BR($\meg $ )& $ < 4.2 \times 10^{-13}$ & MEG~\cite{MEG:2016leq}
 &$ 6 \times  10^{-14}$   &     MEG\rom{2}~\cite{MEGII:2018kmf} \\
 BR($\tau\to e \gamma$) & $<3.3\times 10^{-8}$&  Babar~\cite{BaBar:2009hkt} & $ \sim 10^{-9} $& Belle \rom{2}~\cite{Belle-II:2018jsg} \\
  BR( $\tau\to \mu \gamma$) & $< 4.4 \times 10^{-8}$&  Babar~\cite{BaBar:2009hkt} & $ \sim 10^{-9} $& Belle \rom{2}~\cite{Belle-II:2018jsg} \\
BR $\text{(}\mu  \to e \text{)}^{\rm Au} $& 
$< 7 \times 10^{-13}$ & SINDRUM \rom{2}~\cite{SINDRUMII:2006dvw}  &
 $ -$  &   $-$    \\
  BR $\text{(} \mu  \to e \text{)}^{\rm Al} $& 
$ -$ & $-$ &
 $ 3 \times 10^{-15}$  &   COMET Phase-\rom{1}~\cite{Wong:2015fzj,Mu2e-2}     \\
  BR $\text{(}\mu  \to e  \text{)}^{\rm Al} $& 
$ -$ & $-$ &
 $ 6 \times 10^{-17}$  &   COMET Phase-\rom{2}~\cite{Wong:2015fzj}     \\
 BR $\text{(}\mu  \to e  \text{)}^{\rm Al} $& 
$ -$ & $-$ &
 $ 6 \times 10^{-17}$  &  Mu2e~\cite{Mu2e:2008sio}     \\
 BR $\text{(}\mu  \to e  \text{)}^{\rm Al} $& 
$ -$ & $-$ &
 $ 3 \times 10^{-18}$  &  Mu2e \rom{2}~\cite{Mu2e-2}     \\
 BR $\text{(}\mu  \to e  \text{)}^{\rm Si} $& 
$ - $ & $-$  &
 $ 2 \times 10^{-14}$   &  DeeMe  \cite{Teshima:2018ugo}     \\
  BR $  \text{(} \mu  \to e  \text{)}^{\rm Ti} $ & & &
  $ \sim 10^{- 20} -  10^{-18}$   &  PRISM/PRIME~\cite{Davidson:2022nnl,Kuno:2012pt} \\
BR(  $\meee $)& $ < 1.0 \times 10^{-12}$
 &  SINDRUM~\cite{SINDRUM:1987nra}   &   $ \sim 10^{-16}$ &   Mu3e~\cite{Blondel:2013ia}  \\
BR($\tau\to 3\mu$ )& $<$ $ 2.1 \times 10^{-8}$  & Belle~\cite{Hayasaka:2010np} & $ \sim 10^{-9} $    &   Belle \rom{2}~\cite{Belle-II:2018jsg}\\
BR($\tau\to 3e$ )& $<$ $ 2.7 \times 10^{-8}$  & Belle~\cite{Hayasaka:2010np}
& $ \sim 10^{-9} $    &   Belle \rom{2}~\cite{Belle-II:2018jsg} \\
\hline
\end{tabular}
\caption{Experimental upper limits on various Leptonic flavor violation (LFV) processes.}
\label{lfv_exp}
\end{table}

\subsection{Radiative leptonic decays}
We write the following effective Lagrangian  for the radiative leptonic decays,
\begin{align}
\lag_{\text{eff}}&=m_{\ell'}\, C_T^L\,\bar \ell \sigma^{\rho\lambda}P_L\,\ell' \,F_{\rho\lambda}+m_{\ell'}\, C_T^R\,\bar \ell \sigma^{\rho\lambda}P_R\,\ell'\,F_{\rho\lambda}.
\label{meg}
\end{align}

The branching ratio of the radiative leptonic decays is given by,

\begin{align}
\br(\ell'\rightarrow  \ell\gamma)=\frac{m_{\ell'}^5}{4\pi \Gamma_{\ell'}}\left(|C_T^L|^2+|C_T^R|^2\right) \; .
\end{align}

\begin{figure}[H]
	\centering
    \includegraphics[width=0.35\linewidth]{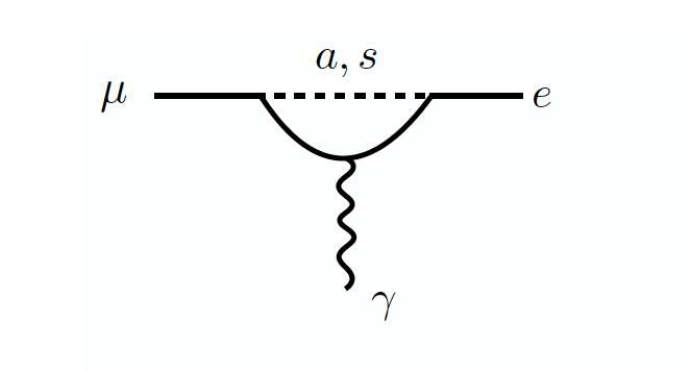}	
 \caption{Feynman diagram representing  $\mu   \rightarrow  e  \gamma$ decay.}
 \label{llg}
	\end{figure}

\begin{figure}[H]
	\centering
	\begin{subfigure}[]{0.4\linewidth}
    \includegraphics[width=\linewidth]{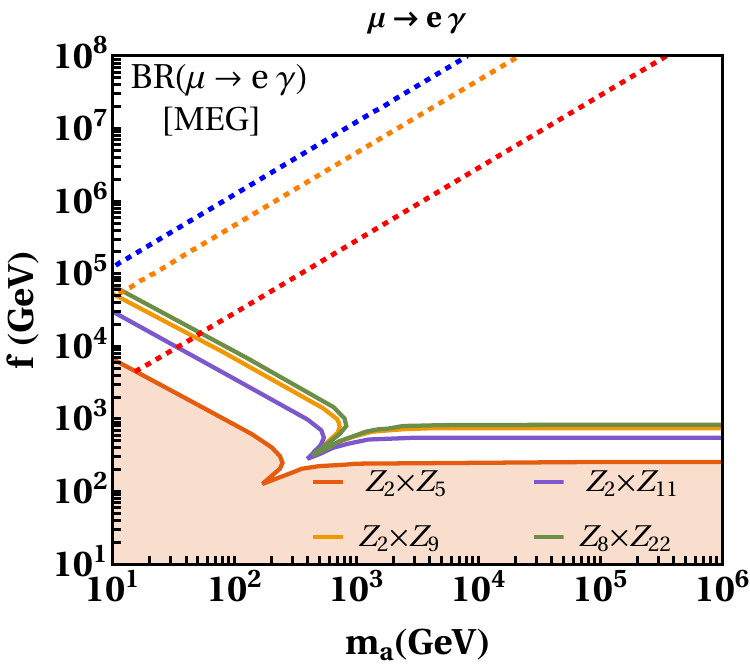}
    \caption{}
         \label{figmega}	
\end{subfigure}
 \begin{subfigure}[]{0.4\linewidth}
 \includegraphics[width=\linewidth]{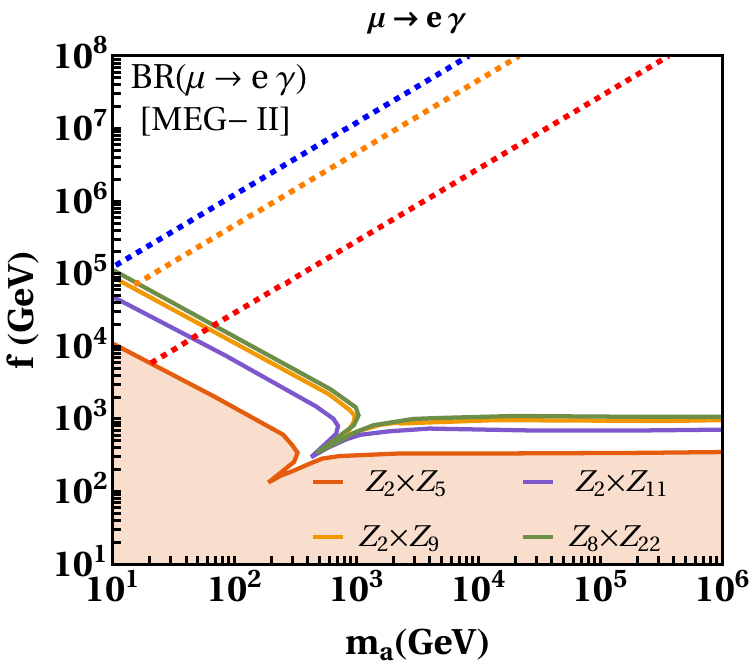}
 \caption{}
         \label{figmegb}
 \end{subfigure} 
 \caption{Bounds on the $f-m_a$ parameter space of the flavon of different $\mathcal{Z}_N \times \mathcal{Z}_{M}$ models  by BR($\mu\rightarrow e \gamma$) for current as well as projected measurements from MEG and MEG-\rom{2} experiments. The region above continuous curves represents the allowed parameter space for the corresponding $\mathcal{Z}_N \times \mathcal{Z}_{M}$ flavor symmetries in the soft symmetry-breaking scenario. The dashed lines show the allowed parameter space in the symmetry-conserving scenario. For both the scenario,  $\lambda_{\chi}=2$.}
  \label{figmeg}
	\end{figure}
 
The radiative leptonic decays receive the one-loop contribution, as shown in figure \ref{llg}.  The Wilson coefficients corresponding to the one-loop contribution read as \cite{Bauer:2015kzy},
\begin{align}
C_T^L = (C_T^R)^*
=\frac{e}{32\pi^2}\sum_{k=e,\mu,\tau} &\bigg\{ \frac{1}{6}\left( \,y^*_{\ell k}y_{\ell' k}+\frac{m_\ell}{m_k}y^*_{k \ell }y_{k\ell' }\right)\left(\frac{1}{m_s^2}-
\frac{1}{m_a^2}\right)\notag\\
&-y_{\ell k}y_{k\ell'}\frac{m_k}{m_{\ell'}}
\left[ \frac{1}{m_s^2}
\left(\frac{3}{2}+\log\frac{m_{\ell'}^2}{m_s^2}\right)-\frac{1}{m_a^2}
\left(\frac{3}{2}+
\log\frac{m_{\ell'}^2}{m_a^2}
\right)\right]
\bigg\} \; .
\label{megw}
\end{align}

The $\mu\rightarrow e \gamma$ decay impose the most stringent constraints on the  parameter space of flavon of  the $\mathcal{Z}_8 \times \mathcal{Z}_{22}$ and  $\mathcal{Z}_2 \times \mathcal{Z}_{9}$  flavor symmetries in the soft symmetry-breaking scenario, as observed in figure \ref{figmega}  and \ref{figmegb} for the MEG and MEG-\rom{2} experiments.  This is followed by the $\mathcal{Z}_2 \times \mathcal{Z}_{11}$ flavor symmetry.  For the symmetry-conserving case, the strongest bounds arise for the flavon of the $\mathcal{Z}_2 \times \mathcal{Z}_{11}$ flavor symmetry, followed by the flavon of the $\mathcal{Z}_2 \times \mathcal{Z}_{9}$ flavor symmetry.

\subsection{$A~\mu\rightarrow A~e$ conversion}
We write the following effective Lagrangian describing  $A~ \mu\rightarrow A ~e$ conversion,
\begin{align}
\lag_{\text{eff}}=
C_{qq}^{VL}\,\bar e \gamma^\nu P_L \mu\, \bar q \gamma_\nu q
+m_\mu m_q\,C_{qq}^{SL}\bar e P_R \mu \,  \bar q q
+m_\mu \alpha_s C_{gg}^L\,\bar e P_R \mu \,G_{\rho\nu}G^{\rho\nu}\,+ (R\leftrightarrow L) \; ,
\label{eq:ConvLag}
\end{align}

\begin{figure}[h!]
	\centering
    \includegraphics[width=0.4\linewidth]{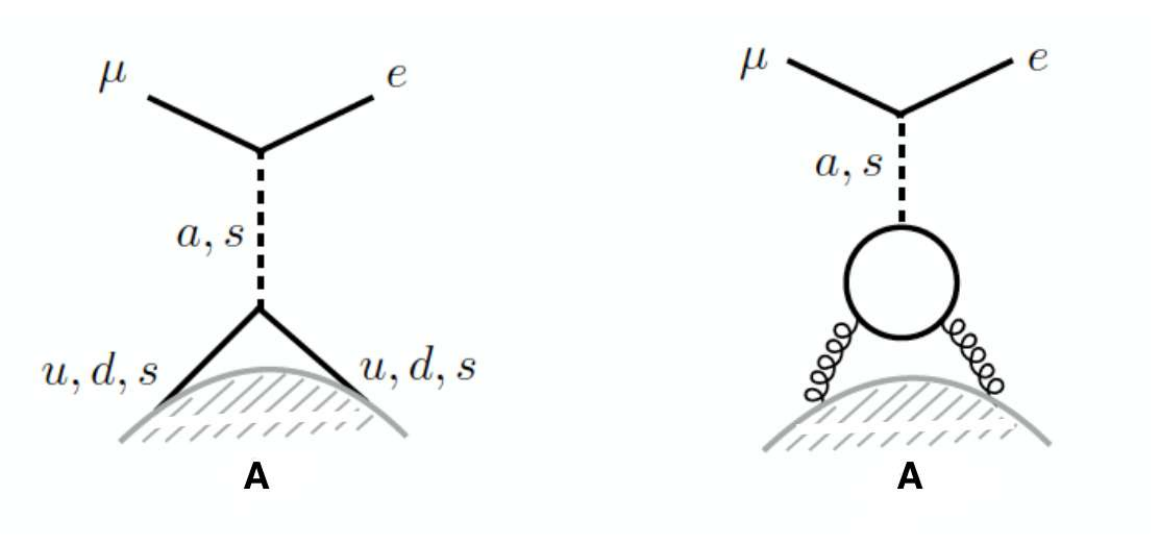}	
 \caption{Feynman diagram showing   $A~ \mu\rightarrow A~ e$ conversion.}
 \label{mec}
	\end{figure}

Moreover, the dipole operators given in equation \eqref{meg} provide additional contribution to $A~ \mu\rightarrow A~ e$ conversion.  We show the Feynman diagram for  $A~ \mu\rightarrow A~ e$ conversion  in figure \ref{mec}.  The  Wilson coefficients for the diagram on the left in figure \ref{mec} are given by~\cite{Bauer:2015kzy},
\begin{align}
C^{SL}_{qq}&=\left(\frac{1}{m_s^2}+\frac{1}{m_a^2}\right)y_{\mu e}^*\text{Re}(y_{qq})\,,\notag\\
C^{SR}_{qq}&=\left(\frac{1}{m_s^2}-\frac{1}{m_a^2}\right)y_{ e\mu} \text{Re}(y_{qq})\,.
\end{align}

The nucleon-level Wilson coefficients which include the  nuclear effects of quarks inside the nucleons as well as the contribution of the Feynman diagram on the right side of figure \ref{mec} can be written as,
\begin{align}
\tilde C_p^{VL} &=\sum_{q=u,d} C^{VL}_{qq}\, f_{V_q}^p, \\ \nonumber 
\tilde C_p^{SL} &=\sum_{q=u,d,s} C^{SL}_{qq}\, f_{q}^p-\sum_{Q=c,b,t} C^{SL}_{QQ} \,f_\text{heavy}^p\,,
\end{align}
where  the  proton quark content is parameterized by the vector and scalar couplings $f_{V_q}^{p}, f_q^p$, and $f_\text{heavy}^p=2/27\big(1-f_u^p-f_d^p-f_s^p\big)$ \cite{Shifman:1978zn}.   For right-handed operators, analogous expressions can be obtained by replacing $L$ with $R$,  and for the neutron $p$ is replaced by $n$. The contribution of the  vector operators is much less  than that of the scalar operators. Therefore, it can be ignored \cite{Bauer:2015kzy}.  We use the numerical values of vector and scalar couplings given in references \cite{Crivellin:2013ipa, Crivellin:2014cta},  which are obtained after using the lattice average given in  reference~\cite{Junnarkar:2013ac},
\begin{align}
f_u^p&=0.0191,  \quad f_u^n=0.0171,
 \qquad f_d^p=0.0363, \quad f_d^n=0.0404, 
\qquad f_s^p=f_s^n=0.043\,.
\end{align}

\begin{figure}[H]
	\centering
	\begin{subfigure}[]{0.4\linewidth}
    \includegraphics[width=\linewidth]{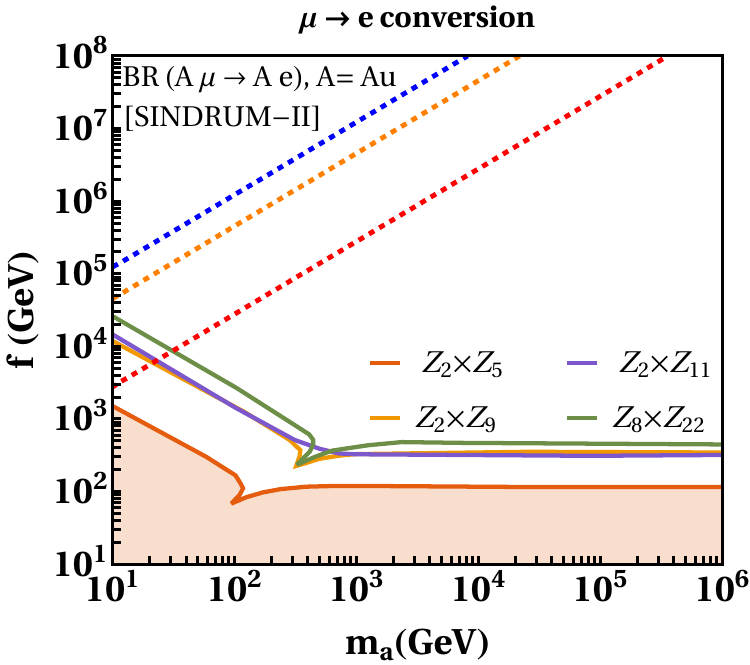}
    \caption{}
         \label{fig11a}	
\end{subfigure}
 \begin{subfigure}[]{0.4\linewidth}
 \includegraphics[width=\linewidth]{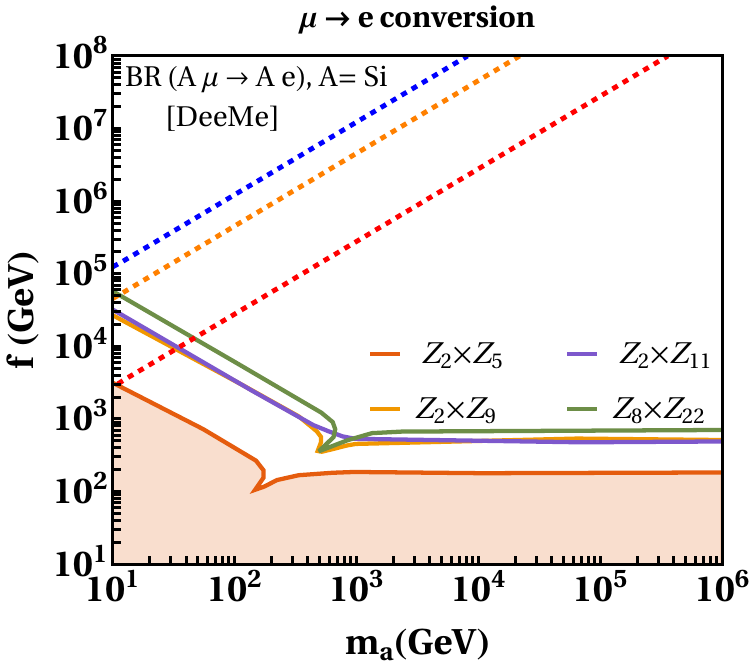}
 \caption{}
         \label{fig11b}
 \end{subfigure} \\
 \begin{subfigure}[]{0.4\linewidth}
    \includegraphics[width=\linewidth]{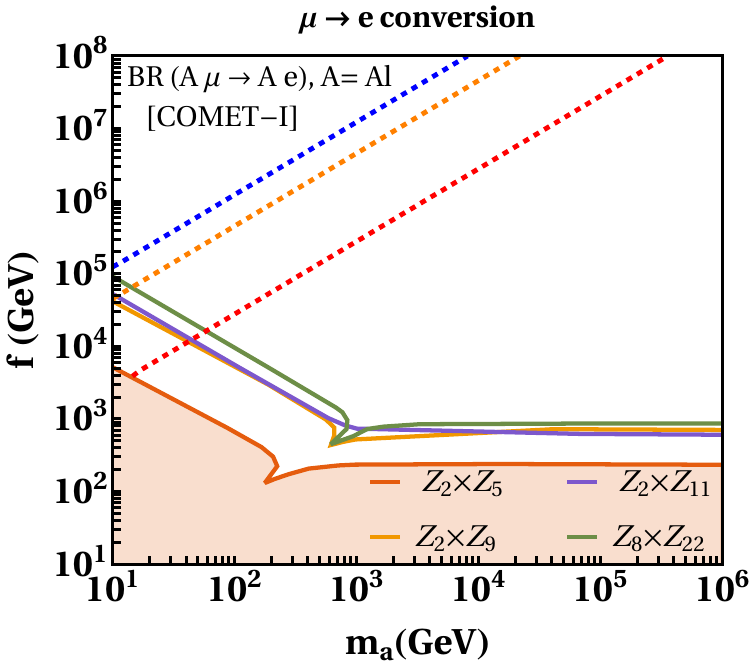}
    \caption{}
         \label{fig11c}	
\end{subfigure}
 \begin{subfigure}[]{0.4\linewidth}
 \includegraphics[width=\linewidth]{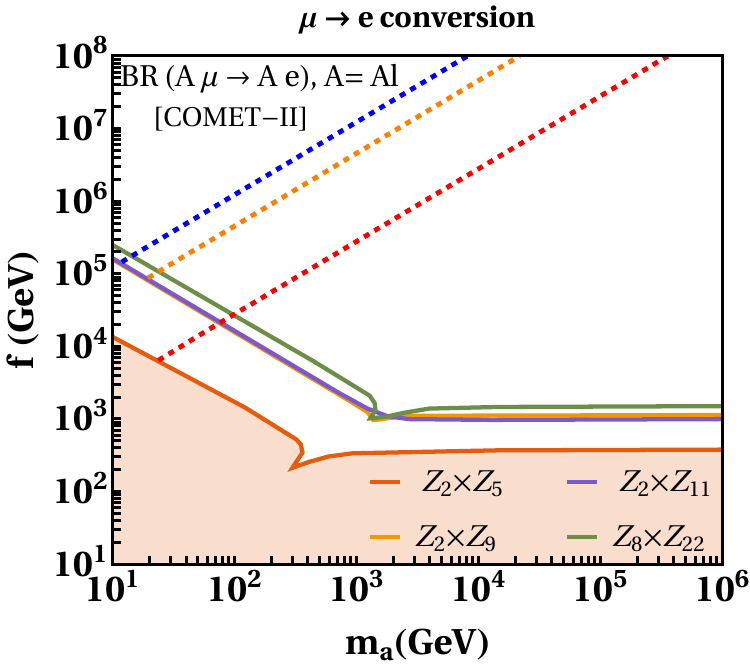}
 \caption{}
         \label{fig11d}
 \end{subfigure} 
 \begin{subfigure}[]{0.4\linewidth}
    \includegraphics[width=\linewidth]{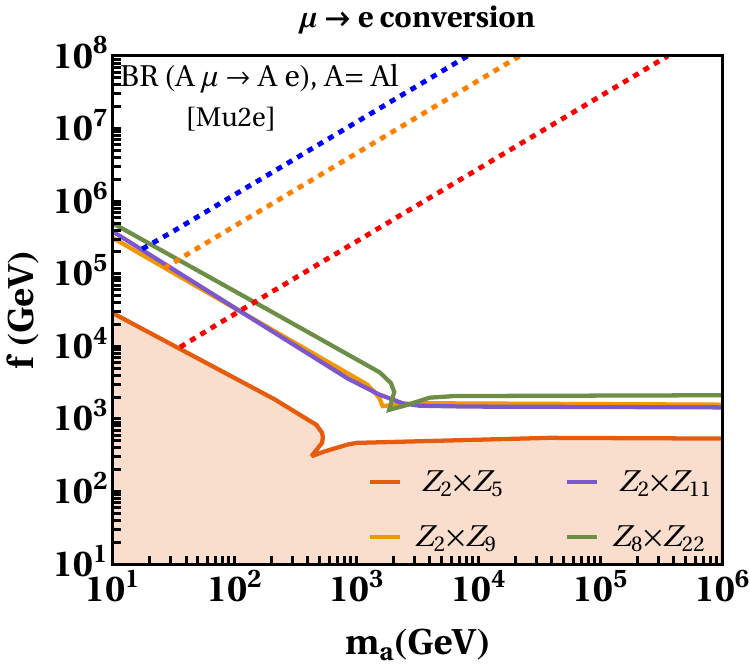}
    \caption{}
         \label{fig11e}	
\end{subfigure}
 \begin{subfigure}[]{0.4\linewidth}
    \includegraphics[width=\linewidth]{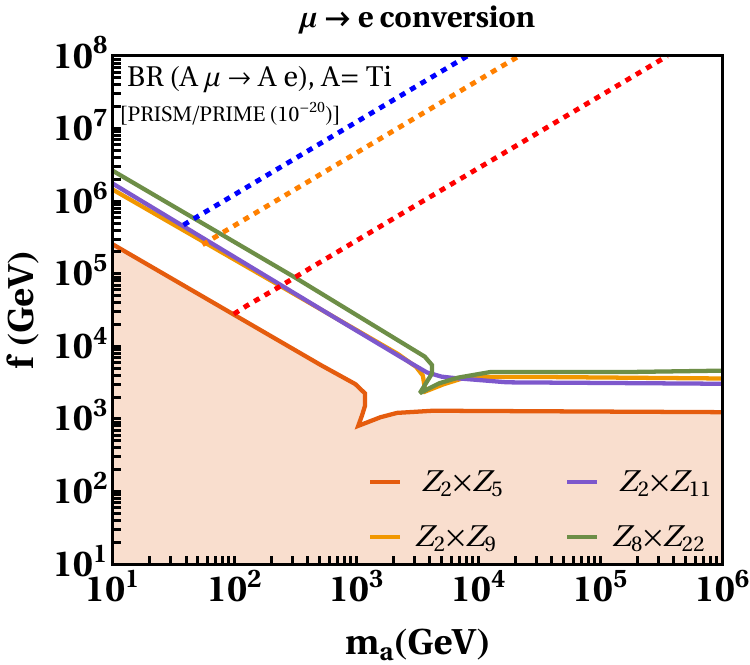}
    \caption{}
         \label{fig11g}	
\end{subfigure}
 \caption{Bounds on the $f-m_a$ parameter space of the flavon of different $\mathcal{Z}_N \times \mathcal{Z}_{M}$ symmetries by BR$(A \mu\rightarrow A e)$ for the current as well as projected sensitivities of the experiments given in table \ref{lfv_exp}. The coloured region represents the excluded parameter space by all four $\mathcal{Z}_N \times \mathcal{Z}_{M}$ symmetries in the soft symmetry-breaking scenario, while the dashed lines represent the allowed parameter space in the symmetry-conserving scenario.}
  \label{figmutoe}
	\end{figure}

After including nuclear effects, the $A~ \mu\rightarrow A~ e$ conversion rate reads as \cite{Bauer:2015kzy},
\begin{align}
\Gamma_{A ~\mu\rightarrow A~ e}=\frac{m_\mu^5}{4}\left|C_T^L D +4\left[m_\mu m_p\tilde C_p^{SL}+\tilde C_p^{VL}V^p+ (p\rightarrow n) \right]\right|^2\, + L \rightarrow R,
\label{eq:Convrate}
\end{align}
where the  dimensionless coefficients $D, S^{p,n}$, and $V^{p,n}$ are functions of the  overlap integrals of the initial state muon and the final-state electron wave functions with the target nucleus.  We use  their numerical values  given in table \ref{nucl}~\cite{Kitano:2002mt}.

\begin{table}[H]
\centering
\begin{tabular}{p{2.0cm}|p{1.5cm} p{1.5cm} p{1.5cm} p{1.5cm} p{1.5cm} p{2cm}}
\hline
\text{Target}& $D$& $S^p$ & $S^n$ & $V^p$& $V^n$&$\Gamma_\text{capt} [10^{6} \text{s}^{-1}]$\\\hline
\text{Au}&  0.189& 0.0614&0.0918&0.0974&0.146&13.06\\
\text{Al}&  0.0362& 0.0155&0.0167&0.0161&0.0173&0.705\\
\text{Si}&  0.0419& 0.0179&0.0179&0.0187&0.0187&0.871 \\ 
\text{Ti}&  0.0864& 0.0368&0.0435&0.0396&0.0468&2.59 \\ \hline
\end{tabular}
\caption{Numerical values of the  dimensionless coefficients $D, S^{p,n}$,  $V^{p,n}$ and  the muon capture rate for different nuclei where  $\Gamma_\text{capt}$ denotes  the muon capture rate.}
\label{nucl}
\end{table}

The most stringent constraints from the  $A~ \mu\rightarrow A~ e$ conversion rate on the  parameter space of flavon of $\mathcal{Z}_N \times \mathcal{Z}_M$ flavor symmetries  arise for the $\mathcal{Z}_8 \times \mathcal{Z}_{22}$ and  $\mathcal{Z}_2 \times \mathcal{Z}_{9, 11}$  flavor symmetries.  The bounds from the future experiments are remarkably stronger than the existing limits.    This can be seen in figure \ref{figmutoe} for different ongoing and future experiments.    This is followed by the $\mathcal{Z}_2 \times \mathcal{Z}_{5}$ flavor symmetries.  In the symmetry-conserving scenario, the most stringent bounds are for the $\mathcal{Z}_2 \times \mathcal{Z}_{11}$ flavor symmetry, followed by the $\mathcal{Z}_2 \times \mathcal{Z}_{9,5}$ flavor symmetries.

 \subsection{ $\mu\rightarrow 3e$ and $\tau \rightarrow 3 \mu$ decays}
 \begin{figure}[H]
	\centering
	\begin{subfigure}[]{0.39\linewidth}
    \includegraphics[width=\linewidth]{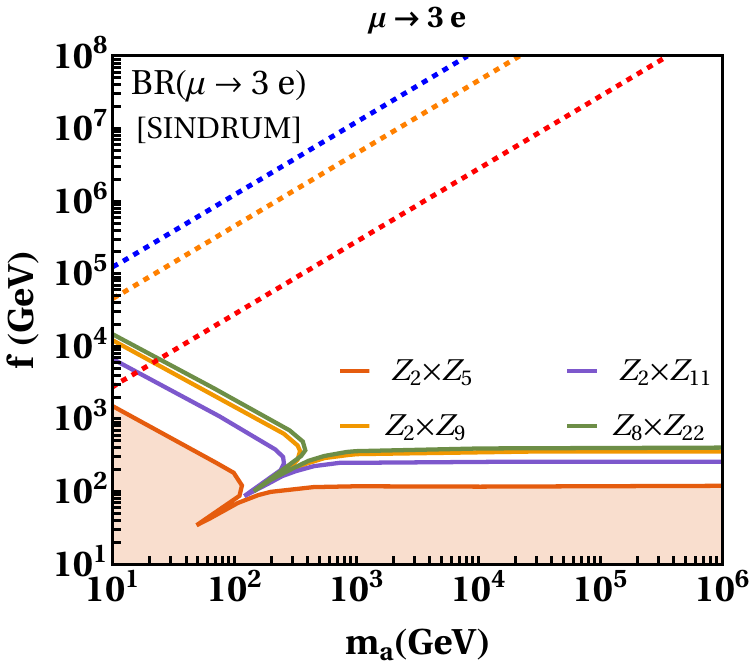}
    \caption{}
         \label{figm3ec}	
\end{subfigure}
 \begin{subfigure}[]{0.39\linewidth}
 \includegraphics[width=\linewidth]{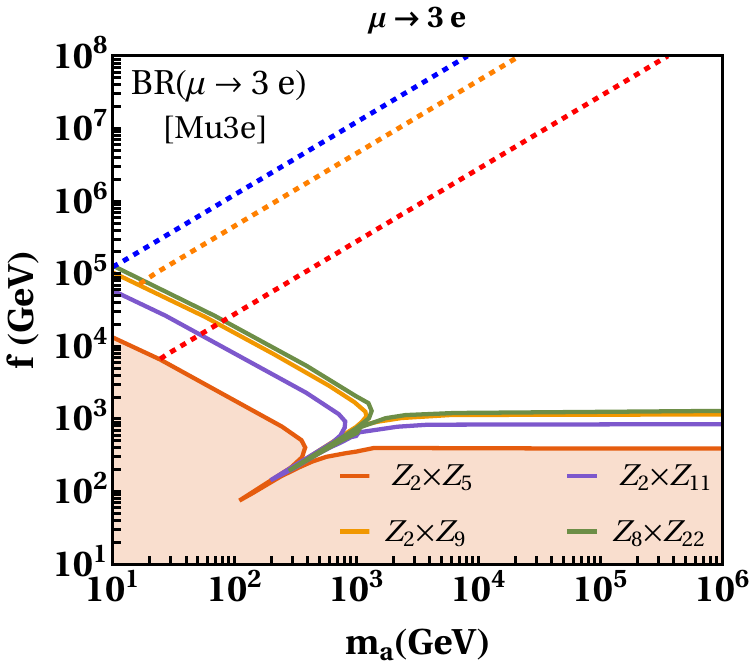}
 \caption{}
         \label{figm3epro}
 \end{subfigure} 
 \caption{Bounds on the $f-m_a$ parameter space of the flavon of different $\mathcal{Z}_N \times \mathcal{Z}_{M}$ symmetries by BR$(\mu\rightarrow 3e)$  for sensitivities of the SINDRUM and Mu3e experiments. The coloured region represents the excluded parameter space by all four $\mathcal{Z}_N \times \mathcal{Z}_{M}$ symmetries in the soft symmetry-breaking scenario, while the dashed lines represent the allowed parameter space in the symmetry-conserving scenario. }
  \label{figm3e}
	\end{figure}
The  additional probes of the dipole operators given in equation \eqref{meg} can emerge from the three body flavor violating  leptonic decays  $\mu \to 3 e $ and   $\tau \to 3 \ell $ where $\ell= e, \mu$.   The decay width of these decays  is given by \cite{Bauer:2015kzy},
\begin{align}
\Gamma(\ell'\rightarrow 3 \ell)=\frac{ \alpha m_\ell'^5}{12 \pi^2}\left|log \frac{{m_\ell'}^2}{{m_\ell}^2} - \frac{11}{4}\right|\left(|C_T^L|^2 + |C_T^R|^2\right).
\end{align}
where the tree-level contribution is negligible due to the strong chiral-suppression caused  by the logarithmic enhancement of the dipole operators\cite{Bauer:2015kzy}. We ignore other contributions, such as $Z$-mediated penguin, due to their strong suppression \cite{Goto:2015iha}.

The $\mu\rightarrow 3e$  decays  places the most stringent bounds on the parameter space of flavon of the $\mathcal{Z}_8 \times \mathcal{Z}_{22}$ flavor symmetry for the soft symmetry-breaking case as shown in figure \ref{figm3e} for the SINDRUM and Mu3e experiments.  The bounds on the flavor symmetry $\mathcal{Z}_2 \times \mathcal{Z}_{9}$ are almost similar to that of the $\mathcal{Z}_8 \times \mathcal{Z}_{22}$ flavor symmetry.  This is followed by the $\mathcal{Z}_2 \times \mathcal{Z}_{11}$  flavor symmetry.  For the symmetry-conserving scenario, the strongest bounds are for the $\mathcal{Z}_2 \times \mathcal{Z}_{11}$ flavor symmetry, followed by the $\mathcal{Z}_2 \times \mathcal{Z}_{9,5}$ flavor symmetries.

\subsection{Summary of the flavor bounds}
\begin{figure}[H]
	\centering
	\begin{subfigure}[]{0.4\linewidth}
    \includegraphics[width=\linewidth]{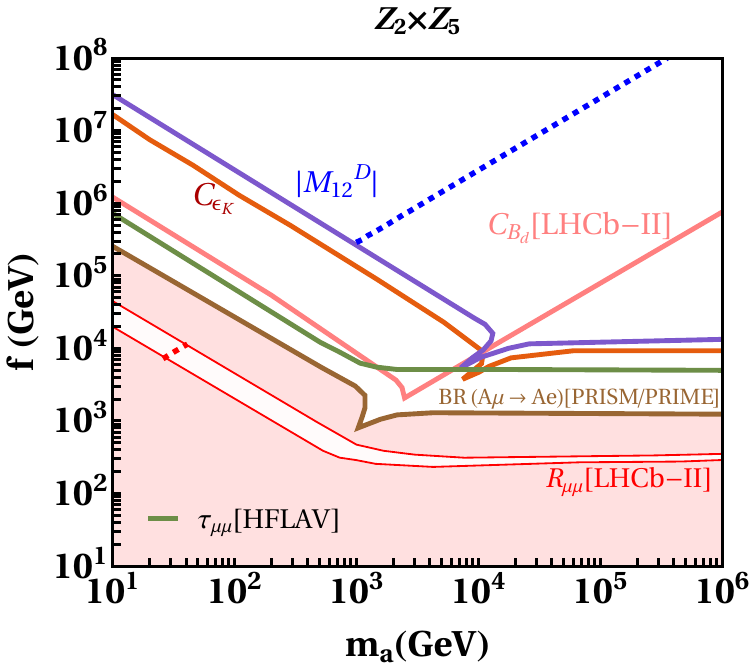}
    \caption{}
         \label{summ_z5}	
\end{subfigure}
 \begin{subfigure}[]{0.4\linewidth}
 \includegraphics[width=\linewidth]{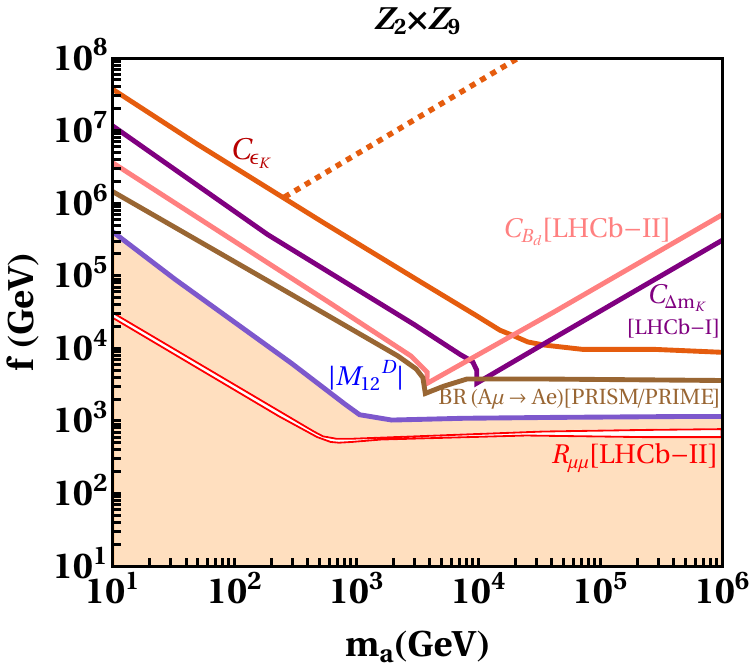}
 \caption{}
         \label{summ_z9}
 \end{subfigure} 
 \begin{subfigure}[]{0.4\linewidth}
 \includegraphics[width=\linewidth]{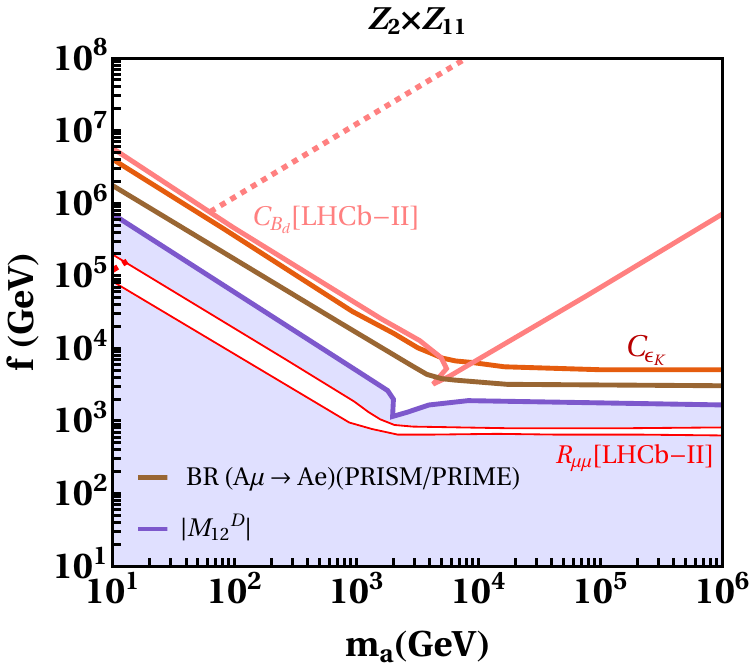}
 \caption{}
         \label{summ_z11}
 \end{subfigure} 
 \begin{subfigure}[]{0.4\linewidth}
 \includegraphics[width=\linewidth]{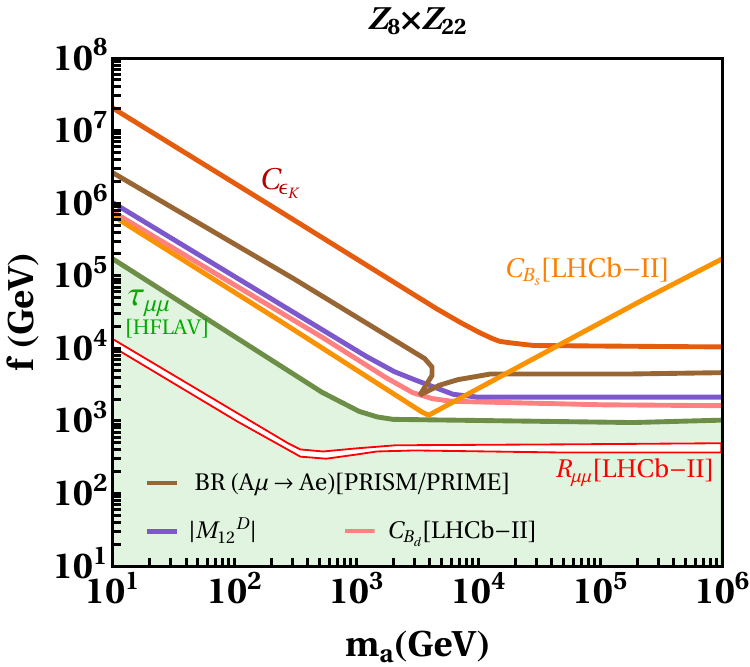}
 \caption{}
         \label{summ_z8}
 \end{subfigure} 
 \caption{Summary of the excluded parameter space constrained significantly by the quark and leptonic flavor observables for different $\mathcal{Z}_N \times \mathcal{Z}_M$ flavor symmetries are shown with the coloured region in the soft symmetry-breaking scenario. The dashed line represents the allowed parameter space, which is most stringently constrained by the corresponding flavor observable in the symmetry-conserving scenario, except for the $\mathcal{Z}_8 \times \mathcal{Z}_{22}$ flavor symmetry.}
  \label{figsumm}
	\end{figure}

In figure \ref{figsumm}, we present the summary of the most significant bounds on the parameter space of the flavon of different  $\mathcal{Z}_{\rm N} \times \mathcal{Z}_{\rm M}$ flavor symmetries. We notice that in the soft symmetry-breaking scenario the most stringent bound for the $\mathcal{Z}_{\rm 2} \times \mathcal{Z}_{\rm 5}$ flavor symmetry arise from the observable $|M_{12}^D|$, while for the $\mathcal{Z}_{\rm 2} \times \mathcal{Z}_{\rm 9}$ and $\mathcal{Z}_{\rm 8} \times \mathcal{Z}_{\rm 22}$ flavor symmetries, bounds from the observable $C_{\eps_K}$ are most stringent, and in the case of the $\mathcal{Z}_{\rm 2} \times \mathcal{Z}_{\rm 11}$ the observable $C_{B_d}$ in the phase-\rom{2} of the LHCb places the strongest constraint. For the symmetry-conserving scenario, the most significant bounds for every  $\mathcal{Z}_{\rm N} \times \mathcal{Z}_{\rm M}$ flavor symmetry are shown with the dashed straight-lines in figure \ref{figsumm}.

\section{Hadron collider physics of the flavon of the  $\mathcal{Z}_N \times \mathcal{Z}_M$ flavor symmetries} 
\label{collider}
We now discuss  the collider signatures of the flavon of the  $\mathcal{Z}_N \times \mathcal{Z}_M$ flavor symmetries through its decays and production processes.   We first discuss the decays of flavon to a pair of fermions, as well as decays of top quark involving flavon.  After this, we investigate the production mechanisms of the flavon of the  $\mathcal{Z}_N \times \mathcal{Z}_M$ flavor symmetries.  Our aim is to identify different  $\mathcal{Z}_N \times \mathcal{Z}_M$ flavor symmetries as much as possible through the collider signatures. Therefore, we  shall finally discuss the sensitivity of signatures of different  $\mathcal{Z}_N \times \mathcal{Z}_M$ flavor symmetries to the HL-LHC, the HE-LHC,  and a 100 TeV collider.  We use  the square root of the luminosity to scale the sensitivity of the HL-LHC, HE-LHC, and 100 TeV collider, which is a conservative estimate.

For collider investigation, we employ the parameter space of the flavon of the  $\mathcal{Z}_N \times \mathcal{Z}_M$ flavor symmetries provided by the observable $R_{\mu \mu }$ in figure \ref{fig7c}.  The observable $R_{\mu \mu }$ is sufficiently capable of differentiating the   $\mathcal{Z}_N \times \mathcal{Z}_M$ flavor symmetries investigated in this work.  For instance, we use the lower boundary of the allowed region of the parameter space corresponding to the $\mathcal{Z}_2 \times \mathcal{Z}_5$ flavor symmetry, which does not overlap with the parameter space corresponding to the $\mathcal{Z}_2 \times \mathcal{Z}_9$ flavor symmetry.

We notice that in our framework, there is no Higgs-flavon mixing. Therefore, the flavon does not directly  couple to the gauge bosons. This fact leads to no constraints on low masses of flavon from the direct LHC searches  \cite{CMS:2018amk}. Moreover, the WW and ZZ decyas of flavon occur at one-loop level, and are much suppressed. The WW and ZZ final states searches set limits on masses above 300 GeV from ATLAS \cite{ATLAS:2022eap} and 200 GeV from CMS  \cite{CMS:2019bnu}. The searches in the di-photon channel place bounds on the masses above 200 GeV from the ATLAS \cite{ATLAS:2021uiz}, and above 500 GeV from the CMS \cite{CMS:2018dqv}. 

\subsection{Flavon decays to a pair of fermions}

The flavon of the  $\mathcal{Z}_N \times \mathcal{Z}_M$ flavor symmetries decays to a pair of fermions  at tree-level, while the decays $a \to gg$ and $a \to \gamma \gamma$ are loop-induced.  This is due to the non-zero off-diagonal couplings of the flavon to fermions.  The partial decay width of the decay $a \to f_i \bar{f}_j$  at the tree-level can be written as, 
\begin{align}
    \Gamma(a \to f_i \bar{f}_j) = \, & 
    \frac{N_c m_a}{16 \pi} 
    \left[ \frac{(m_a^2 - (m_i + m_j)^2)
           (m_a^2 - (m_i - m_j)^2)}{m_a^4}
    \right]^{1/2}\\ \notag
   &\left[ 
           \left( |y_{ij}|^2+|y_{ji}|^2 \right)
           \left(1 - \frac{m_i^2+m_j^2}{m_a^2} \right) 
       - 2 \left( y_{ij} y_{ji} + y^*_{ij} y^*_{ji}\right)
           \frac{m_i m_j}{m_a^2}
    \right]  ,
\end{align}
 where $N_c$ is the number of colors. 
 
\begin{figure}[H]
	\centering
	\begin{subfigure}[]{0.4\linewidth}
    \includegraphics[width=\linewidth]{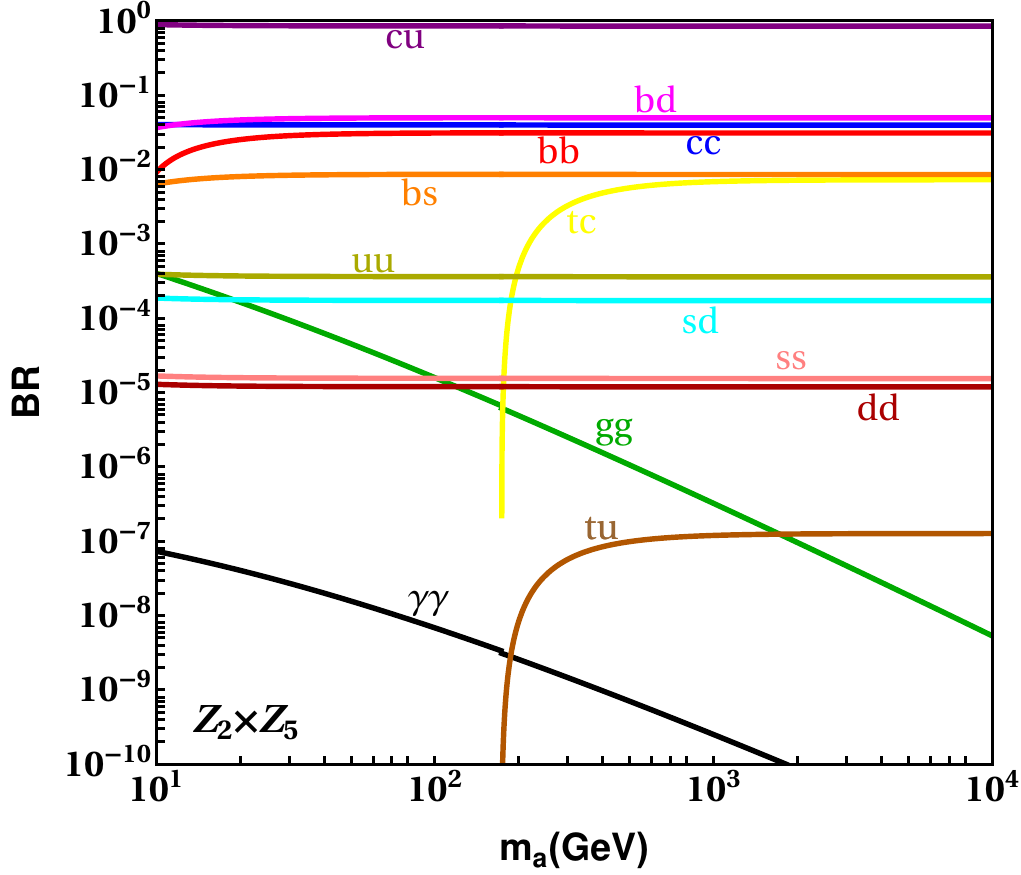}
    \caption{}
         \label{fdtoq_z2z5}	
\end{subfigure}
 \begin{subfigure}[]{0.4\linewidth}
 \includegraphics[width=\linewidth]{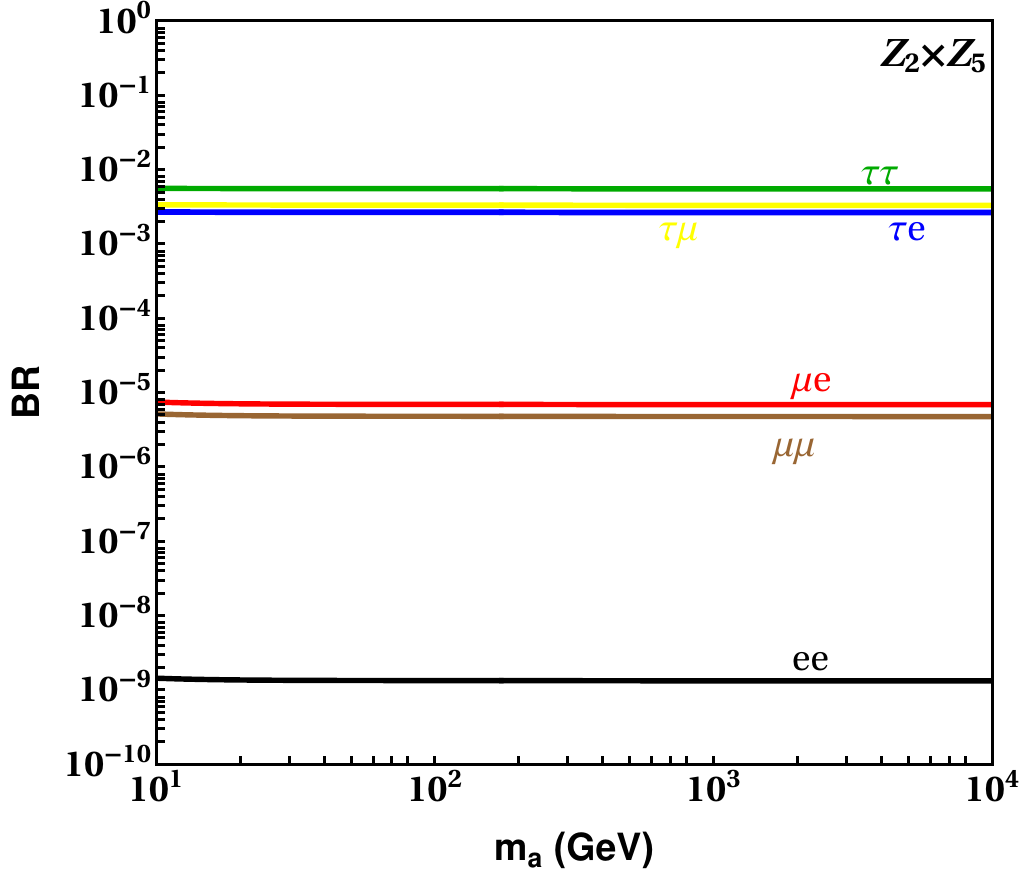}
 \caption{}
         \label{fdtol_z2z5}
 \end{subfigure}
 \begin{subfigure}[]{0.4\linewidth}
 \includegraphics[width=\linewidth]{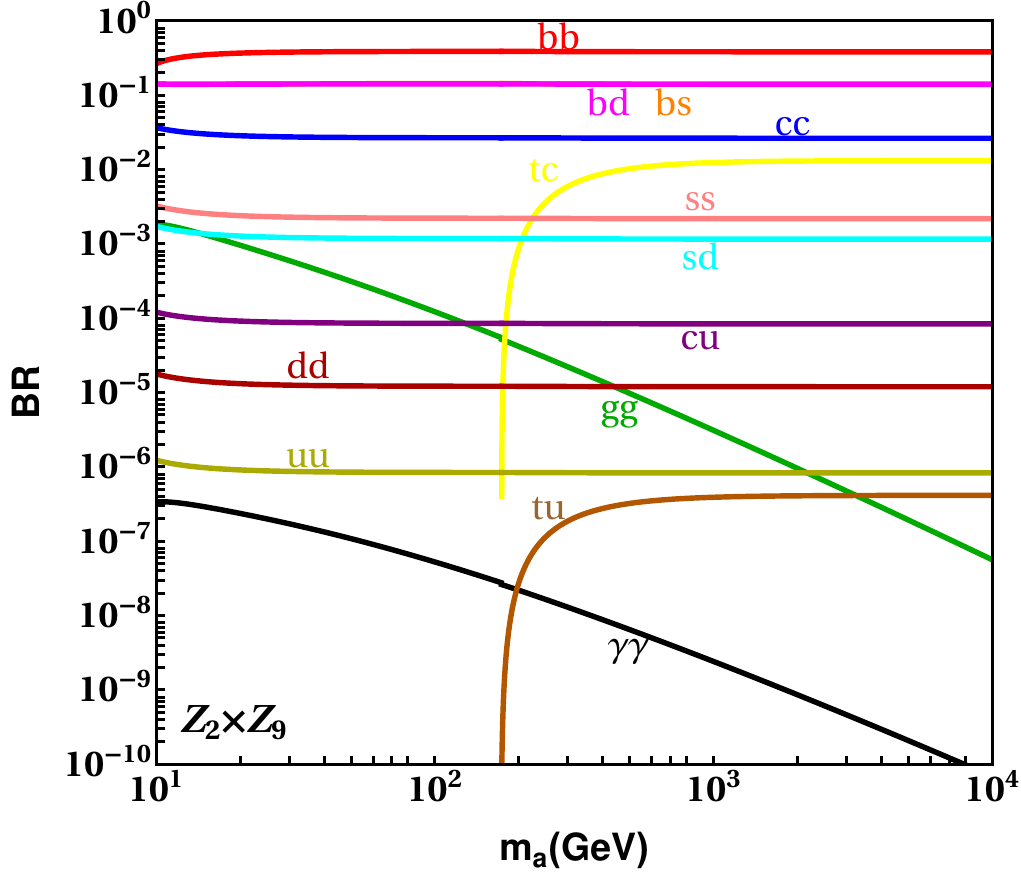}
 \caption{}
         \label{fdtoq_z2z9}
 \end{subfigure} 
 \begin{subfigure}[]{0.4\linewidth}
 \includegraphics[width=\linewidth]{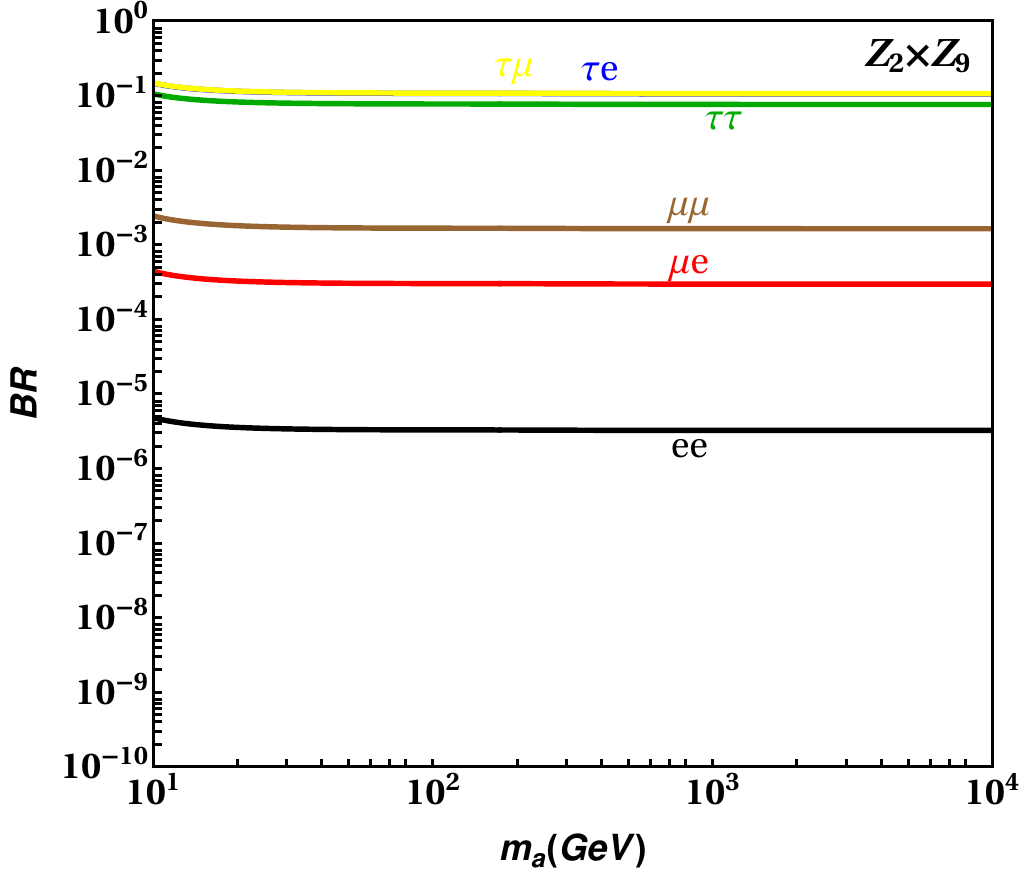}
 \caption{}
         \label{fdtol_z2z9}
 \end{subfigure} 
 \begin{subfigure}[]{0.4\linewidth}
 \includegraphics[width=\linewidth]{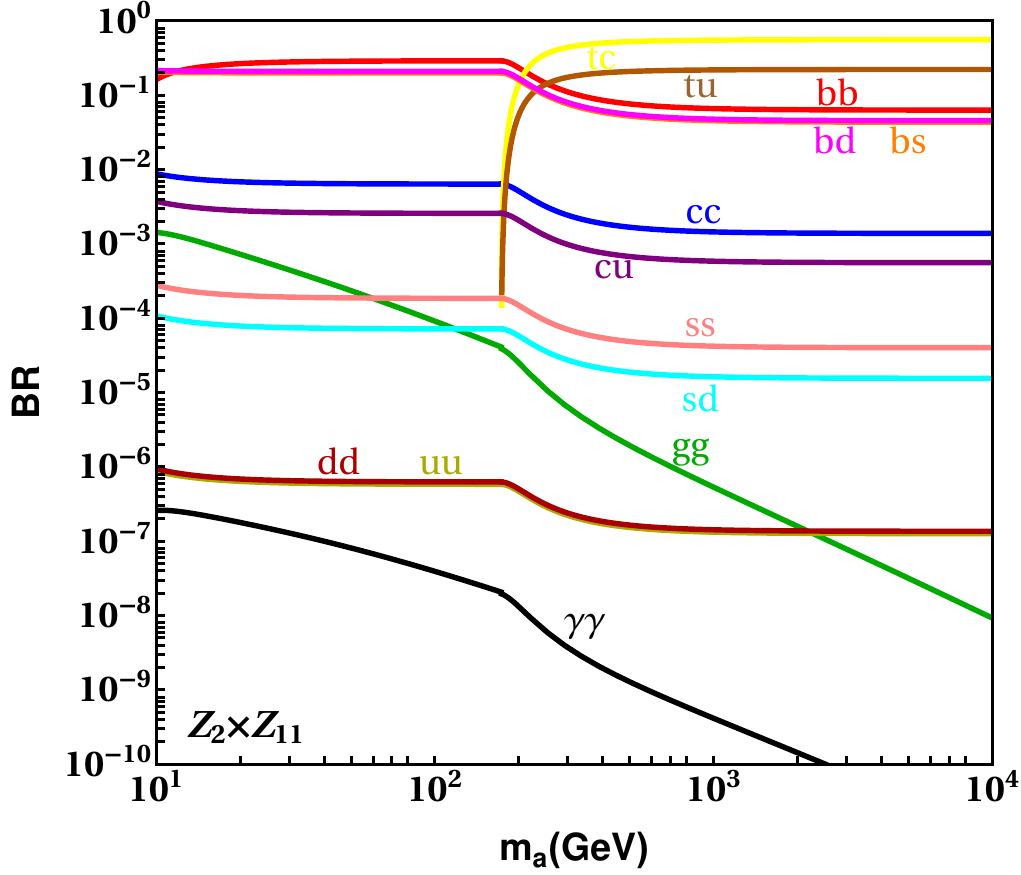}
 \caption{}
         \label{fdtoq_z2z11}
 \end{subfigure} 
 \begin{subfigure}[]{0.4\linewidth}
 \includegraphics[width=\linewidth]{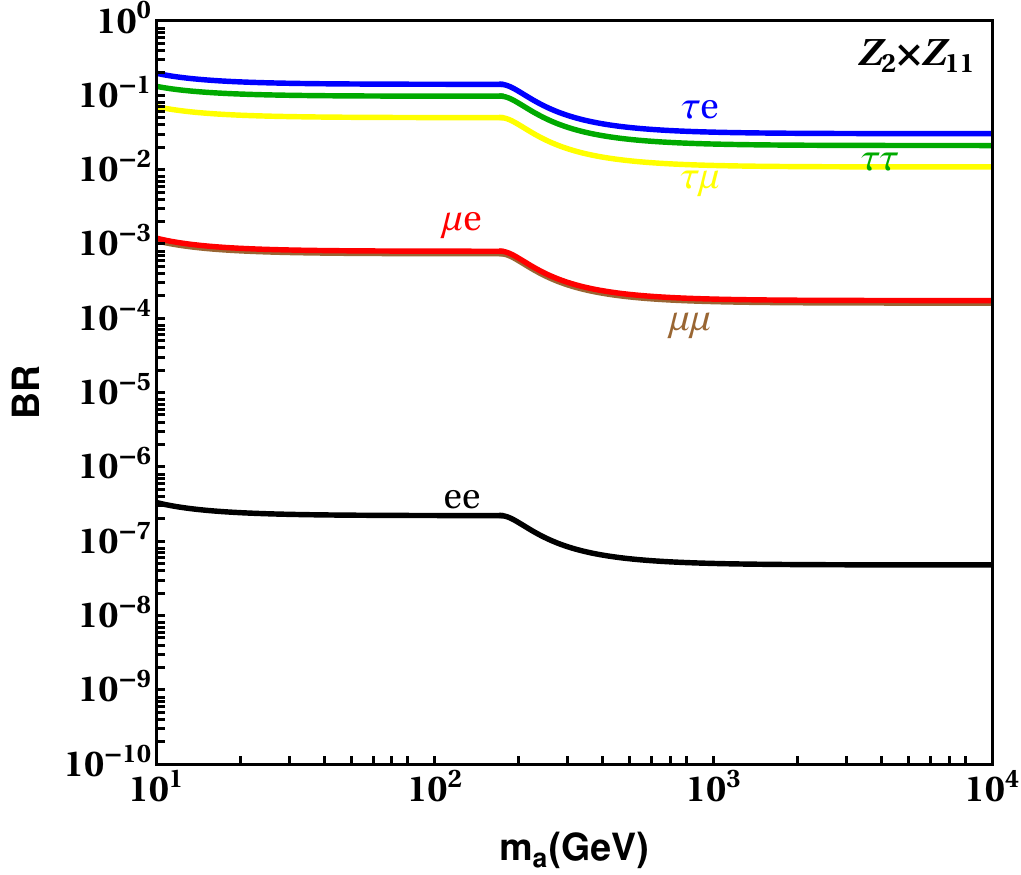}
 \caption{}\textit{}
         \label{fdtol_z2z11}
 \end{subfigure} 
\caption{ Branching ratios of the various possible decay modes of the flavon into quark and lepton pairs for $\mathcal{Z}_2 \times \mathcal{Z}_{5}$, $\mathcal{Z}_2 \times \mathcal{Z}_{9}$, and $\mathcal{Z}_2 \times \mathcal{Z}_{11}$ flavor symmetries. }
  \label{fig_fdecaysz2z5}
	\end{figure}

\begin{figure}[h!]
	\centering
 \begin{subfigure}[]{0.4\linewidth}
    \includegraphics[width=\linewidth]{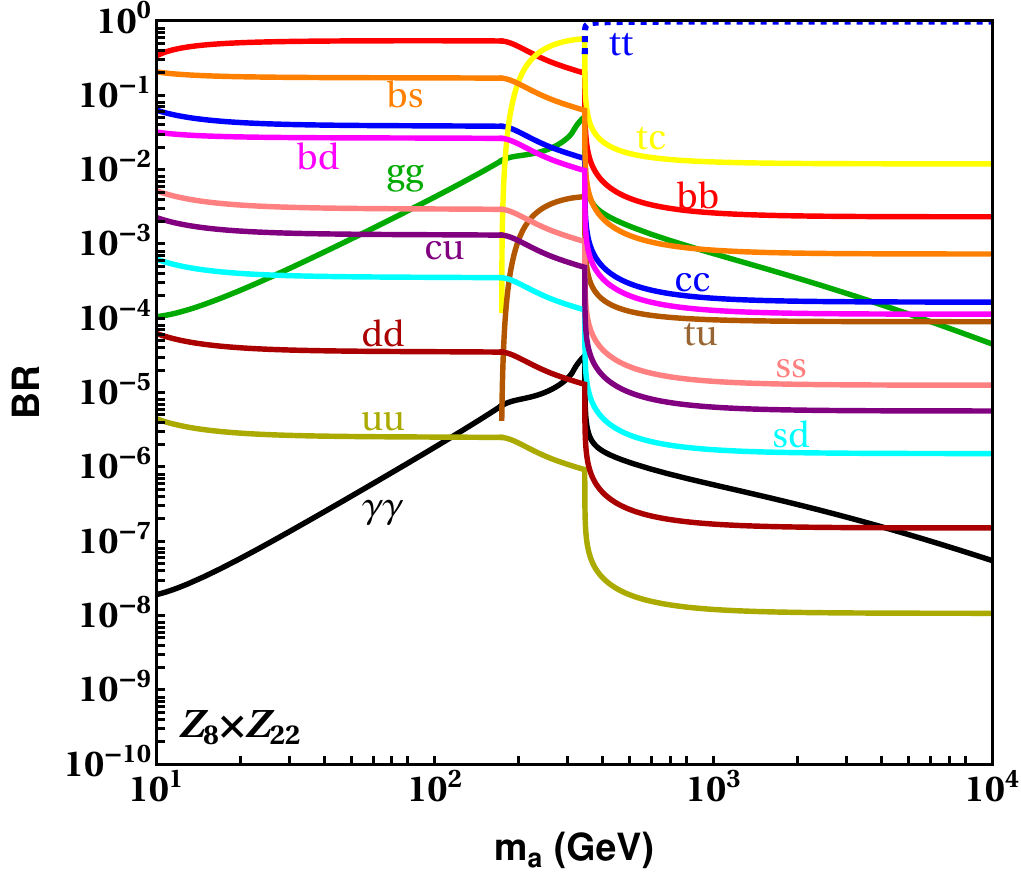}
    \caption{}
         \label{fdtoq_z8z22}	
\end{subfigure}
	\begin{subfigure}[]{0.4\linewidth}
    \includegraphics[width=\linewidth]{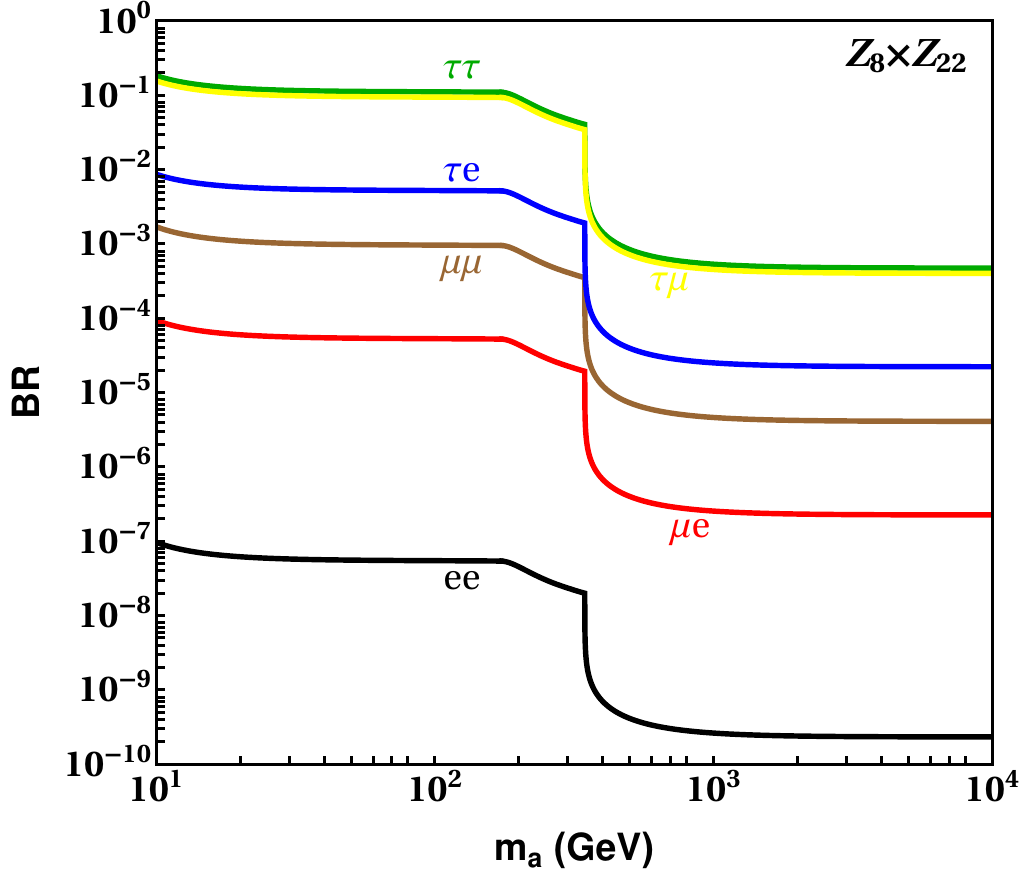}
    \caption{}
         \label{fdtol_z8z22}	
\end{subfigure}
\caption{ Branching ratios of the various possible decay modes of the flavon into quark and lepton pairs for $\mathcal{Z}_8 \times \mathcal{Z}_{22}$ flavor symmetry. }
  \label{fig_flavondecays}
	\end{figure}

We present the branching ratios of $a \to f_i \bar{f}_j$,  $a \to gg$ and $a \to \gamma \gamma$ in figure \ref{fig_fdecaysz2z5} for $\mathcal{Z}_2 \times \mathcal{Z}_{5,9,11}$  flavor symmetries , and in figure  \ref{fig_flavondecays} for the $\mathcal{Z}_8 \times \mathcal{Z}_{22}$  flavor symmetry. In the quark sector,  the dominant decay mode is $a \rightarrow b \bar{b} $ below the top quark threshold  except for the minimal $\mathcal{Z}_2 \times \mathcal{Z}_5$ symmetry.   For the  minimal $\mathcal{Z}_2 \times \mathcal{Z}_5$ symmetry the mode $cu  (c \bar{u}, \bar{c}u)$ prevails over other modes.  This is due to the specific flavor structure of models  based on the  $\mathcal{Z}_N \times \mathcal{Z}_M$ flavor symmetries and the corresponding  Yukawa couplings  given in the appendix.  As $m_a$ approaches $m_t$, decay modes involving top quark, particularly, $tc (t \overline{c}, \overline{t} c) $, become increasingly significant for both the $\mathcal{Z}_2 \times \mathcal{Z}_{11}$ and $\mathcal{Z}_8 \times \mathcal{Z}_{22}$ models. These top quark related decay modes, however, remain comparatively less prominent than the decay modes of the flavon involving bottom and charm quark, such as, $bd, bs, cc, cu$, and  $bb$ for $\mathcal{Z}_2 \times \mathcal{Z}_{5}$ as well as $\mathcal{Z}_2 \times \mathcal{Z}_{9}$ symmetries, even beyond the top-threshold.  Among the leptonic decay modes,  the decay channels involving at least one $\tau $  dominate for all $\mathcal{Z}_N \times \mathcal{Z}_{M}$ flavor symmetries.

At $m_a > m_t$, in contrast to the other three $\mathcal{Z}_N \times \mathcal{Z}_{M}$  flavor symmetries, the $\mathcal{Z}_8 \times \mathcal{Z}_{22}$ model exhibits a dominant decay of the flavon into top quark pairs $(t \overline{t})$, as shown by the dashed-blue line in figure \ref{fdtoq_z8z22}. This striking feature arises from the construction of the $\mathcal{Z}_8 \times \mathcal{Z}_{22}$ model, where the top quark couples to the flavon with a suppression factor of order $\epsilon$. Notably, this feature is absent in the remaining three $\mathcal{Z}_N \times \mathcal{Z}_{M}$ symmetry-based models under investigation.  This feature keeps apart the  $\mathcal{Z}_8 \times \mathcal{Z}_{22}$ based model from the other models in terms of collider signatures, which are discussed in the later part of this work.

  \subsection{Top quark decays to flavon}
\begin{figure}[h!]
	\centering
	\begin{subfigure}[]{0.4\linewidth}
    \includegraphics[width=\linewidth]{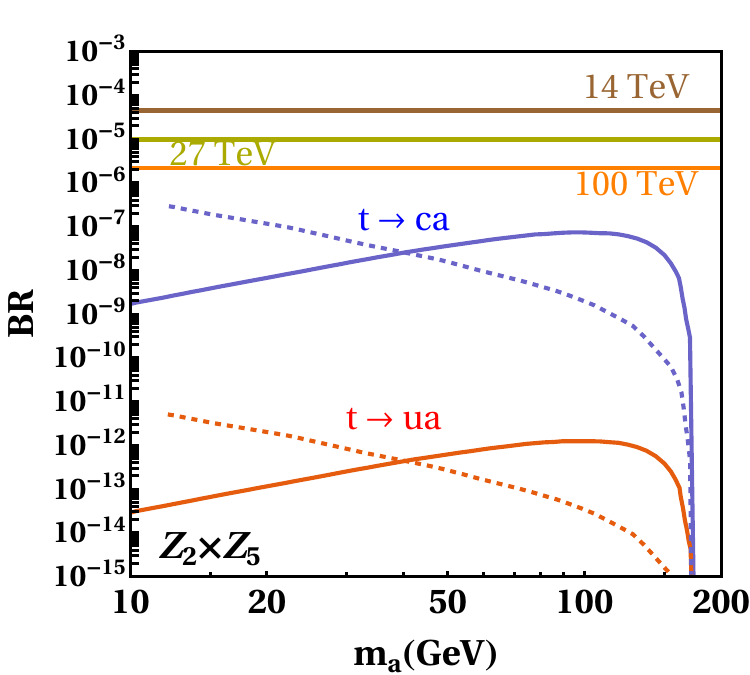}
    \caption{}
         \label{t_fa_z5}	
\end{subfigure}
 \begin{subfigure}[]{0.4\linewidth}
 \includegraphics[width=\linewidth]{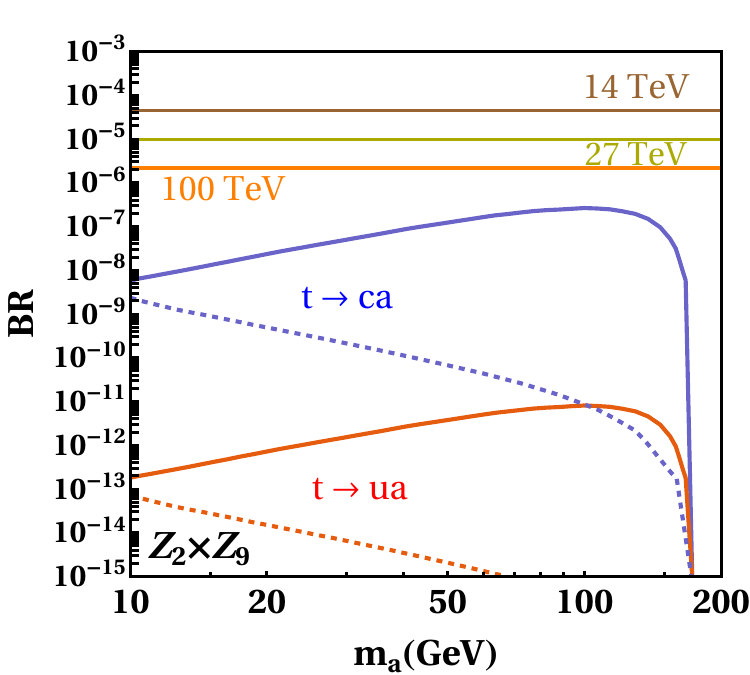}
 \caption{}
         \label{t_fa_z9}
 \end{subfigure} 
 \begin{subfigure}[]{0.4\linewidth}
 \includegraphics[width=\linewidth]{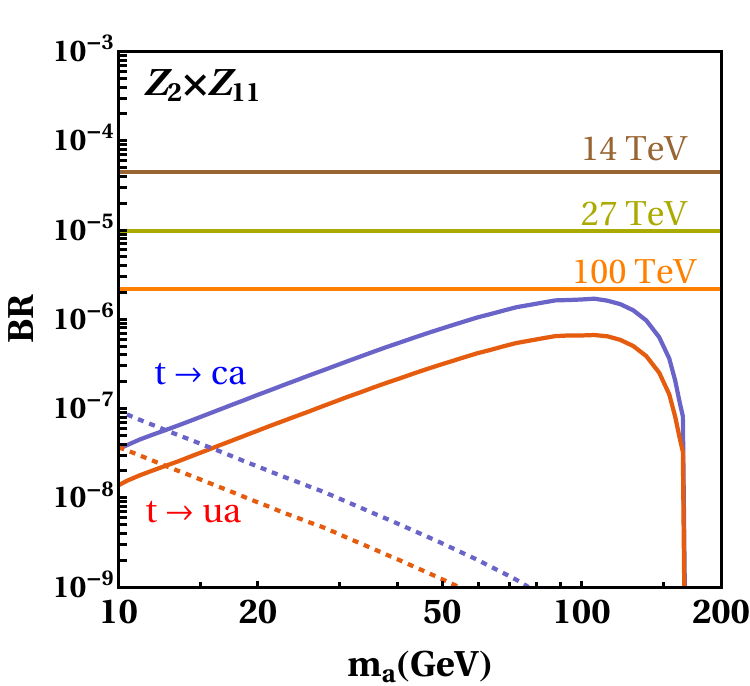}
 \caption{}
         \label{t_fa_z11}
 \end{subfigure} 
 \begin{subfigure}[]{0.4\linewidth}
 \includegraphics[width=\linewidth]{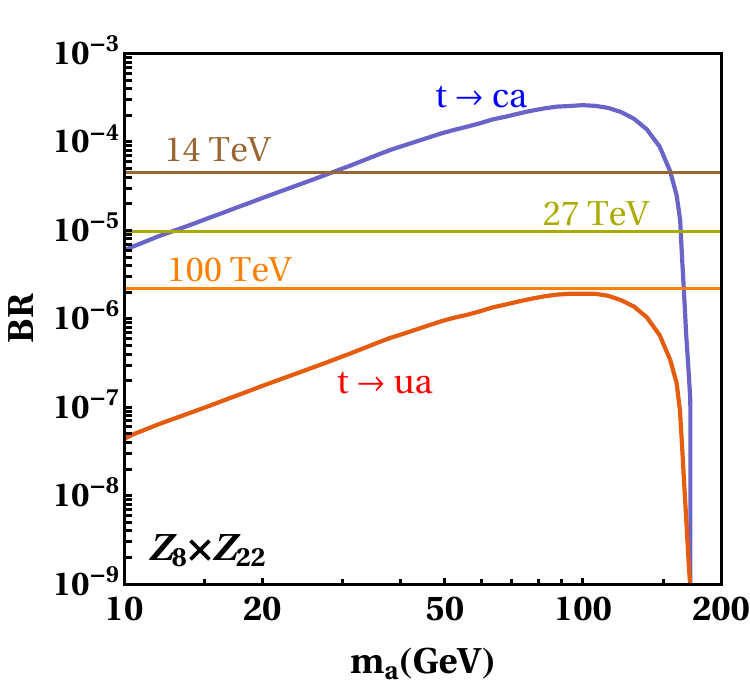}
 \caption{}
         \label{t_fa_z8}
 \end{subfigure} 
 \caption{The branching ratios of top quark decays into flavon of $\mathcal{Z}_N \times \mathcal{Z}_{M}$ flavor symmetries as a function of the flavon mass  $(m_a)$ along the parameter space allowed by the observable $R_{\mu \mu}$ in the phase-\rom{2} of the LHCb for soft-symmetry breaking scenario, and $BR(K_L \rightarrow \mu^+ \mu^-)$ for symmetry-conserving scenario.}
  \label{fig_topdecays}
	\end{figure} 

The top quark decays to flavon play an important role in constraining a low-mass flavon. For the flavon mass, which is less than the mass of the top quark,  anomalous top decays $ t \rightarrow a c (a u) $ provide distinguished signatures of flavon at hadron colliders. For Higgs production via such anomalous top-quark decays, the limits for the expected reach of the current $14$ TeV LHC, HE-LHC,  as well as a future high luminosity $100$ TeV hadron collider are known, which can also be utilized for the flavon production. The current and expected limits on such processes at the LHC and a 100 TeV hadron collider are~\cite{limit_top_decay,Bauer:2016rxs},
\begin{align}
BR_{8~ TeV}(t\rightarrow Hc)&< 5.6 \e{-3}, \nonumber  \\
BR_{14~ TeV, 3~ ab^{-1}}(t \rightarrow Hc)&< 4.5 \e{-5}, \nonumber  \\
BR_{27~ TeV, 15~ ab^{-1}}(t \rightarrow Hc)&< 9.7 \e{-6} , \nonumber  \\
BR_{100~ TeV, 30~ ab^{-1}}(t \rightarrow Hc)&< 2.2 \e{-6}.
\label{tch}
\end{align}

In figure  \ref{fig_topdecays}, we show the predictions of branching ratios for decays $t \rightarrow (c,u) a$  in the framework of different $\mathcal{Z}_N \times \mathcal{Z}_{M}$  flavor symmetries.  We note that in the case of the soft symmetry-breaking scenario, shown by the solid continuous curves, the top decays to the flavon are  investigated along the boundaries of parameter space allowed by the observable $R_{\mu \mu}$ in the phase-\rom{2} of the LHCb, shown in figure \ref{fig7c}.  With mass range $m_a$ lying between 10 up to 200 GeV, the corresponding ranges for the VEV of the flavon $f$ varies approximately as $ \sim (2 \times 10^3, 4\times 10^4)$ GeV for $\mathcal{Z}_2 \times \mathcal{Z}_{5}$, $ \sim  (10^3, 2 \times 10^4)$ GeV for $\mathcal{Z}_2 \times \mathcal{Z}_{9}$, $ \sim (9 \times 10^3 - 1.4 \times 10^5)$ GeV for $\mathcal{Z}_2 \times \mathcal{Z}_{11}$, and  $ \sim (4 \times 10^2 - 10^4)$ GeV for $\mathcal{Z}_8 \times \mathcal{Z}_{22}$ flavor symmetry. We observe that only the flavon of the $\mathcal{Z}_8 \times \mathcal{Z}_{22}$ has the potential to reach the sensitivity of the 14 TeV LHC, the HE-LHC,  and a 100 TeV future hadron collider.   This prediction is very specific to test the  $\mathcal{Z}_8 \times \mathcal{Z}_{22}$ flavor symmetry.

 For the symmetry-conserving scenario, represented by the dashed curves in figure \ref{fig_topdecays}, we analyze  $t \rightarrow (c,u) a$   decays along the parameter space allowed by the observable $BR(K_L \rightarrow \mu^+ \mu^-)$, where VEV $f$ ranges approximately as $ \sim (3 \times 10^3, 6\times 10^4)$ GeV for $\mathcal{Z}_2 \times \mathcal{Z}_{5}$, $ \sim (4\times 10^4, 9\times 10^5)$ GeV for $\mathcal{Z}_2 \times \mathcal{Z}_{9}$, and $ \sim (1.2 \times 10^5, 2\times 10^6)$ GeV for $\mathcal{Z}_2 \times \mathcal{Z}_{11}$ flavor symmetry. It turns out  that the decays $ t \rightarrow a c (a u) $ in the case of symmetry-conserving scenario for a light flavon are beyond the reach of the LHC, the HE-LHC, and a future 100 TeV  hadron collider.

\begin{figure}[h!]
	\centering
	\begin{subfigure}[]{0.4\linewidth}
    \includegraphics[width=\linewidth]{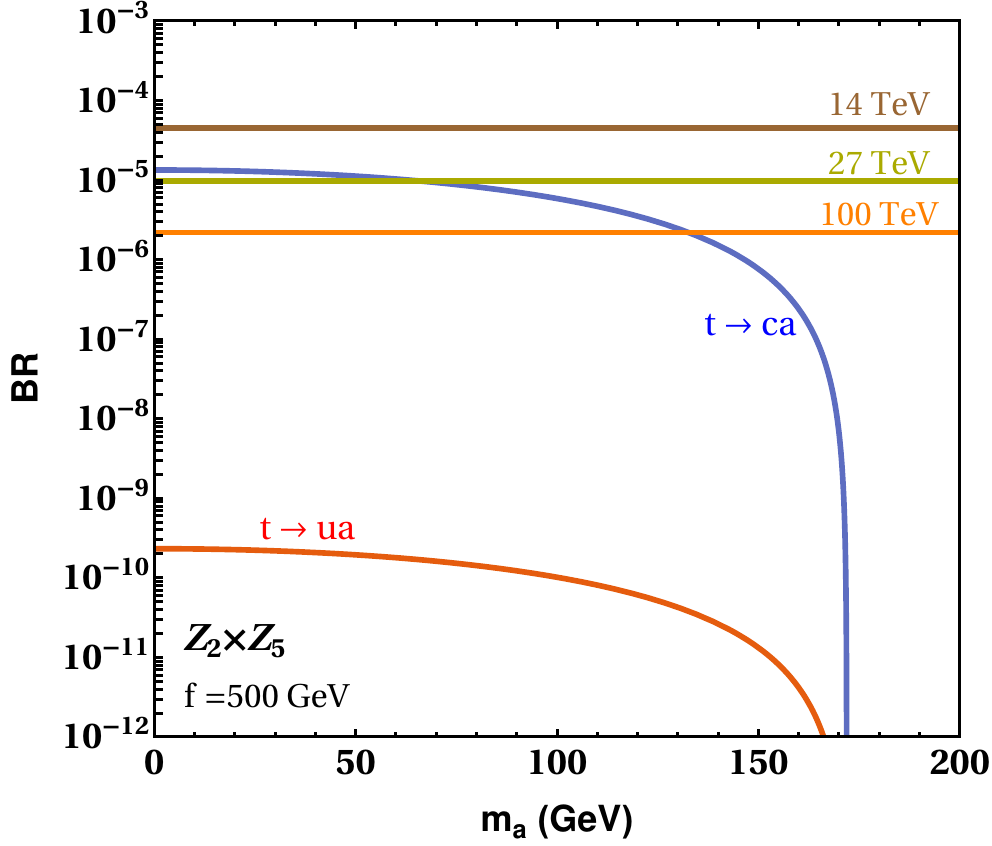}
    \caption{}
         \label{t_fa_z5}	
\end{subfigure}
 \begin{subfigure}[]{0.4\linewidth}
 \includegraphics[width=\linewidth]{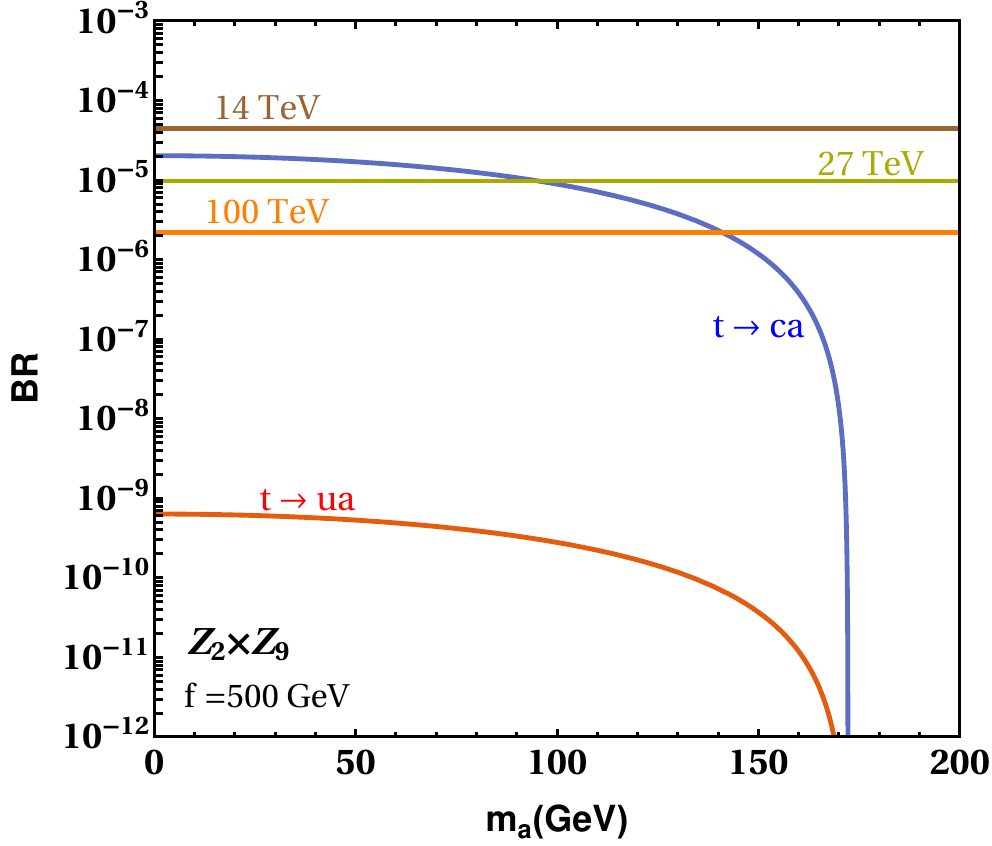}
 \caption{}
         \label{t_fa_z9}
 \end{subfigure} 
 \begin{subfigure}[]{0.4\linewidth}
 \includegraphics[width=\linewidth]{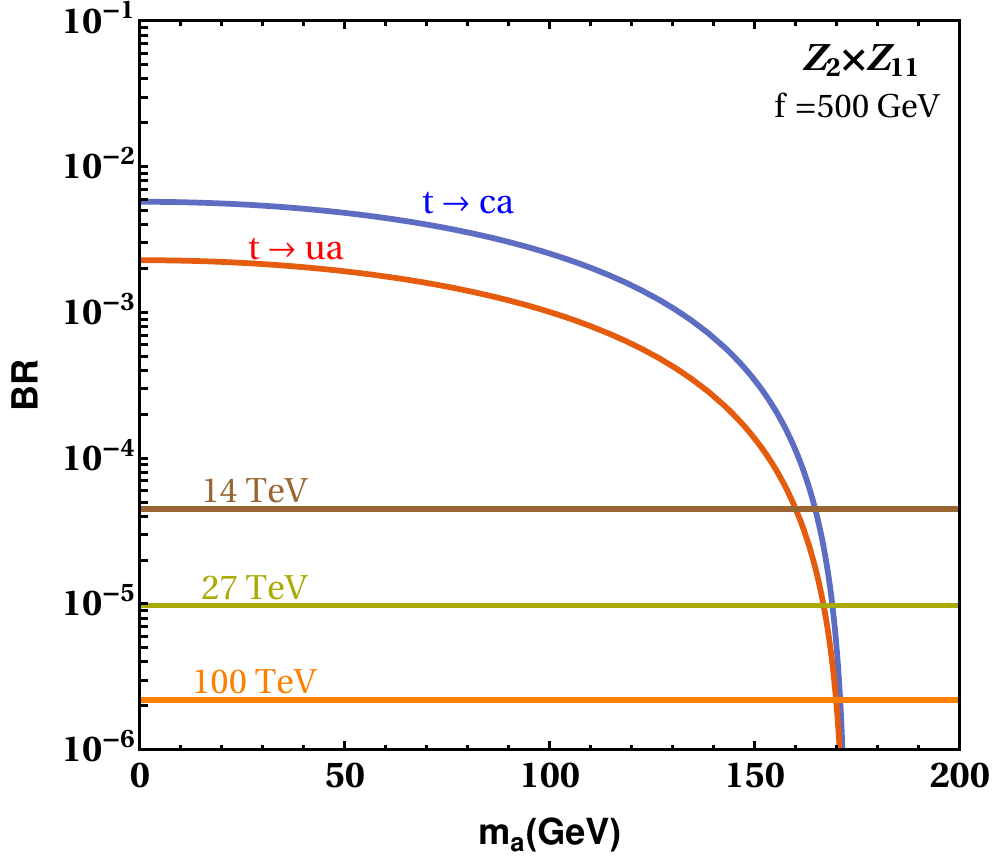}
 \caption{}
         \label{t_fa_z11}
 \end{subfigure} 
 \begin{subfigure}[]{0.4\linewidth}
 \includegraphics[width=\linewidth]{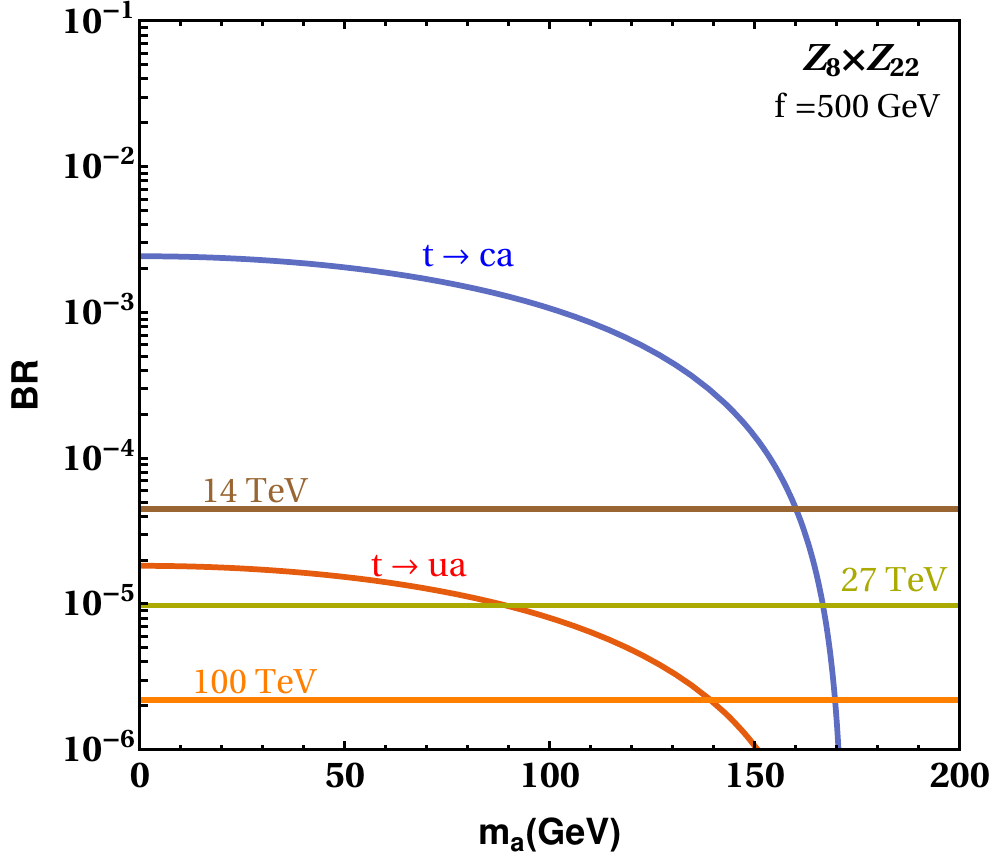}
 \caption{}
         \label{t_fa_z8}
 \end{subfigure} 
 \caption{Branching ratios of top quark decays into flavon of $\mathcal{Z}_2 \times \mathcal{Z}_{5}$, $\mathcal{Z}_2 \times \mathcal{Z}_{9}$, $\mathcal{Z}_2 \times \mathcal{Z}_{11}$ and $\mathcal{Z}_8 \times \mathcal{Z}_{22}$ flavor symmetries as a function of $m_a$, assuming the flavon VEV  $f= 500$ GeV. }
  \label{fig_topdecays_2}
	\end{figure}

In figure  \ref{fig_topdecays_2}, we show the prediction of branching ratios of decays $t \rightarrow (c,u) a$  for  different $\mathcal{Z}_N \times \mathcal{Z}_{M}$  flavor symmetries for  the flavon VEV $f=500$ GeV in the soft symmetry-breaking case.  This is a conventional choice in literature \cite{Bauer:2016rxs}.  In this scenario, the branching ratio of the decay $t \rightarrow c a$ is accessible to a future 100 TeV hadron collider in a certain mass range for all   $\mathcal{Z}_N \times \mathcal{Z}_{M}$  flavor symmetries discussed in this work. For the symmetries $\mathcal{Z}_2 \times \mathcal{Z}_{5,9}$, the branching ratio of the decay $t \rightarrow c a$ is of the order $10^{-6}-10^{-5}$ for $m_a \leq 150$ GeV.  The branching ratio of the decay $t \rightarrow c a$ is larger,  and is maximally of the order of $10^{-3}$ for the $\mathcal{Z}_2 \times \mathcal{Z}_{11}$ and $\mathcal{Z}_8 \times \mathcal{Z}_{22}$ flavor symmetries.  

The decay  $t \rightarrow c a$  can be within the limits of the LHC, HE-LHC and   a 100 TeV fulture hadron collider for the $\mathcal{Z}_2 \times \mathcal{Z}_{11}$ and $\mathcal{Z}_8 \times \mathcal{Z}_{22}$ flavor symmetries. This is a remarkable prediction of this work.   The decay  $t \rightarrow c a$ is within the reach of the limits of the HE-LHC  for the $\mathcal{Z}_2 \times \mathcal{Z}_{5,9}$ flavor symmetries.  We notice another important prediction of this work for the branching ratio of the decay $t \rightarrow u a$, which is in the reach of LHC, HE-LHC and a 100 TeV collider only for the $\mathcal{Z}_2 \times \mathcal{Z}_{11}$ flavor symmetry.  This decay is accessible to the HE-LHC only for the $\mathcal{Z}_8 \times \mathcal{Z}_{22}$ flavor symmetry.  Thus, we conclude that the decays $t \rightarrow (c,u) a$ can be a good test ground for different $\mathcal{Z}_N \times \mathcal{Z}_{M}$  flavor symmetries for a low mass flavon.

\begin{figure}[H]
	\centering
 \begin{subfigure}[]{0.329\linewidth}
 \includegraphics[width=\linewidth]{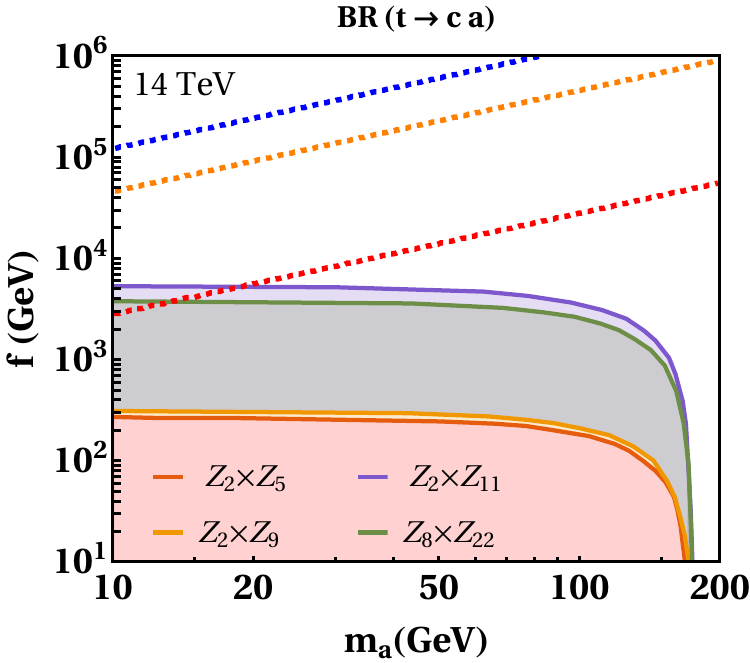}
 \caption{}
         \label{ttoac14}
 \end{subfigure} 
 \begin{subfigure}[]{0.329\linewidth}
 \includegraphics[width=\linewidth]{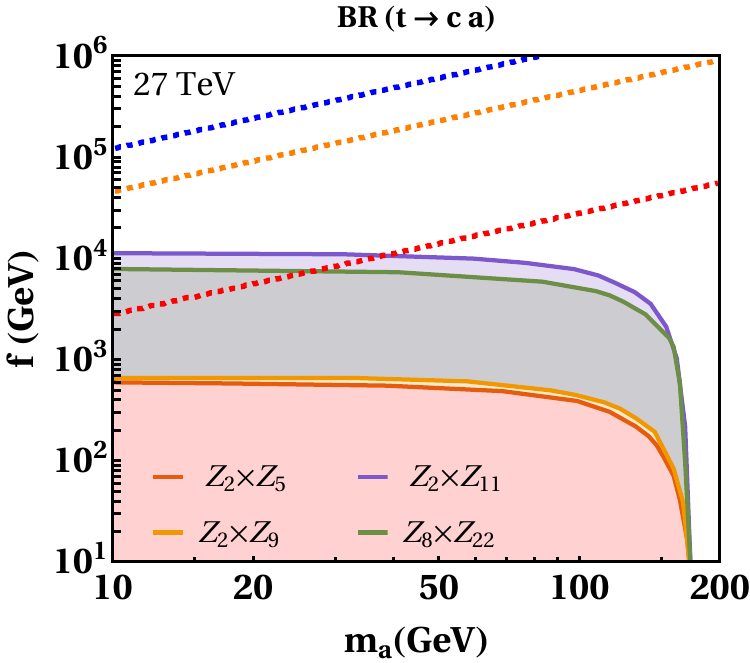}
 \caption{}
         \label{ttoac100}
 \end{subfigure} 
 \begin{subfigure}[]{0.329\linewidth}
 \includegraphics[width=\linewidth]{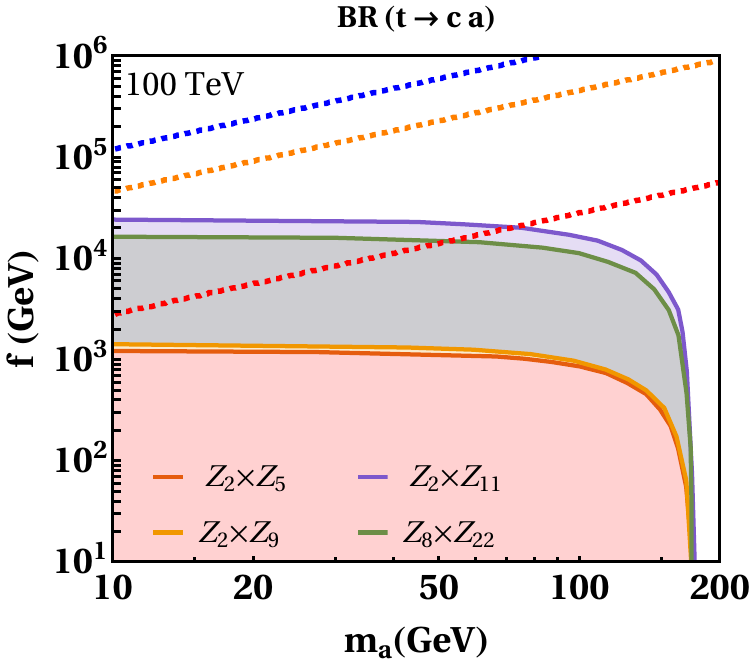}
 \caption{}
         \label{ttoac100}
 \end{subfigure} 
 \caption{ The parameter space excluded by the limits on BR ($t \to H c$) from  14 TeV LHC, 27 TeV HE-LHC, and a 100 TeV collider  for $\mathcal{Z}_2 \times \mathcal{Z}_{5}$, $\mathcal{Z}_2 \times \mathcal{Z}_{9}$, $\mathcal{Z}_2 \times \mathcal{Z}_{11}$ and $\mathcal{Z}_8 \times \mathcal{Z}_{22}$ flavor symmetries in the soft symmetry-breaking scenario is shown with the coloured regions. The dashed lines represents the allowed parameter space in the symmetry-conserving scenario. }
  \label{fig_ttoac_fma}
	\end{figure}

The  limits in equation \ref{tch} can be used to place bounds on the parameter space of the flavon of the $\mathcal{Z}_N \times \mathcal{Z}_{M}$  flavor symmetries, as shown in figure \ref{fig_ttoac_fma}.  We must note that dashed straight lines denote  our predictions for the symmetry-conserving scenarios, which are beyond the reach of the sensitivities given in equation \ref{tch}.

 \subsection{Flavon production at the HL-LHC, HE-LHC, and a 100 TeV hadron collider}
We now discuss the production of the  flavon of the $\mathcal{Z}_N \times \mathcal{Z}_{M}$  flavor symmetries at the LHC, HE-LHC,  and at a 100 TeV future hadron collider. 
The analytical results for the production cross-sections of the flavon in various channels have been derived using the {\tt MSTW2008} PDF \cite{Martin:2009iq}.  Additionally,  {\tt FeynRules} \cite{Alloul:2013bka} has been utilized to produce the model {\tt UFO} files for all four $\mathcal{Z}_N \times \mathcal{Z}_{M}$  flavor symmetries discussed in this work. We used these files to perform production cross-section calculations with {\tt MadGraph} \cite{Alwall:2014hca}, and verify our analytical results.

The inclusive production cross-sections of flavon of  the $\mathcal{Z}_N \times \mathcal{Z}_{M}$  flavor symmetries at the LHC, HE-LHC,  and at a 100 TeV future hadron collider are shown in figures \ref{fig_fprod1}  and  \ref{fig_fprod2} for the $\mathcal{Z}_2 \times \mathcal{Z}_{5,9,11}$  and the $\mathcal{Z}_8 \times \mathcal{Z}_{22}$ flavor symmetries.  For the soft symmetry-breaking scenario, the boundaries of parameter space allowed by the observable $R_{\mu \mu}$ in the phase-\rom{2} of the LHCb, shown in figure \ref{fig7c}, are used.  The dashed curves are the cross-sections for the symmetry-conserving scenarios using the parameter space allowed by the observable $BR(K_L \rightarrow \mu^+ \mu^-)$.

 We use the parameter space allowed by the projected sensitivities of $ \mathcal{R}_{\mu\mu} $ in the LHCb  Phase-\rom{2} in figure \ref{fig7c} to show the production cross-sections of a heavy flavon  in  the different $\mathcal{Z}_N \times \mathcal{Z}_{M}$ models.  Furthermore, the most optimistic scenario is obtained  using the flavon VEV $f=500$ GeV, which is  approximately allowed for a heavy flavon by the  bounds obtained from the current measurement of the observable $R_{\mu \mu}$ in figure \ref{fig7a}.  This choice is also used in previous works, such as references \cite{Bauer:2016rxs, Tsumura:2009yf}. We continue to use the flavon VEV $f=500$ GeV even for the production cross-sections of a low mass flavon.   

In the soft symmetry-breaking scenario,  the  projected sensitivities of $ \mathcal{R}_{\mu\mu} $ in the LHCb  Phase-\rom{2}  allows the  mass range $m_a$ lying between $10$ up to $10^4$ GeV, the corresponding ranges for the VEV of the flavon $f$ vary approximately as $ \sim (3 \times 10^2, 4\times 10^4)$ GeV for $\mathcal{Z}_2 \times \mathcal{Z}_{5}$, $ \sim (6 \times 10^2, 2.6 \times 10^4)$ GeV for $\mathcal{Z}_2 \times \mathcal{Z}_{9}$, $\sim (7.8 \times 10^2, 1.4 \times 10^5)$ GeV for $\mathcal{Z}_2 \times \mathcal{Z}_{11}$, and  $\sim (3.6 \times 10^2, 10^4)$ GeV for $\mathcal{Z}_8 \times \mathcal{Z}_{22}$ flavor symmetry.  

For the symmetry-conserving scenario, we analyze  flavon production channels with the parameter space allowed the observable BR$(K_L \rightarrow \mu^+ \mu^-)$, where VEV $f$ ranges approximately as $ \sim (3 \times 10^3, 3\times 10^6)$ GeV for $\mathcal{Z}_2 \times \mathcal{Z}_{5}$, $ \sim (4\times 10^4, 4\times 10^7)$ GeV for $\mathcal{Z}_2 \times \mathcal{Z}_{9}$, and $ \sim ( 10^5, 9\times 10^7)$ GeV for $\mathcal{Z}_2 \times \mathcal{Z}_{11}$ flavor symmetry.

\subsubsection{Inclusive production}

\begin{figure}[h!]
	\centering
	\begin{subfigure}[]{0.315\linewidth}
    \includegraphics[width=\linewidth]{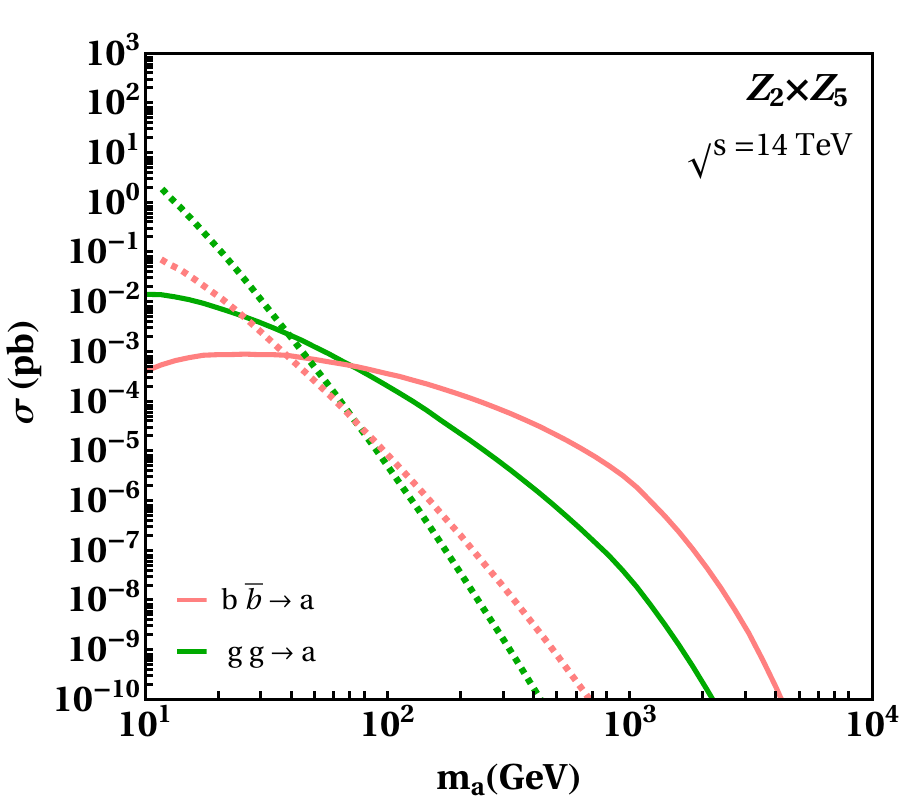}
    \caption{}
         \label{fprod14_z2z5}	
\end{subfigure}
 \begin{subfigure}[]{0.315\linewidth}
 \includegraphics[width=\linewidth]{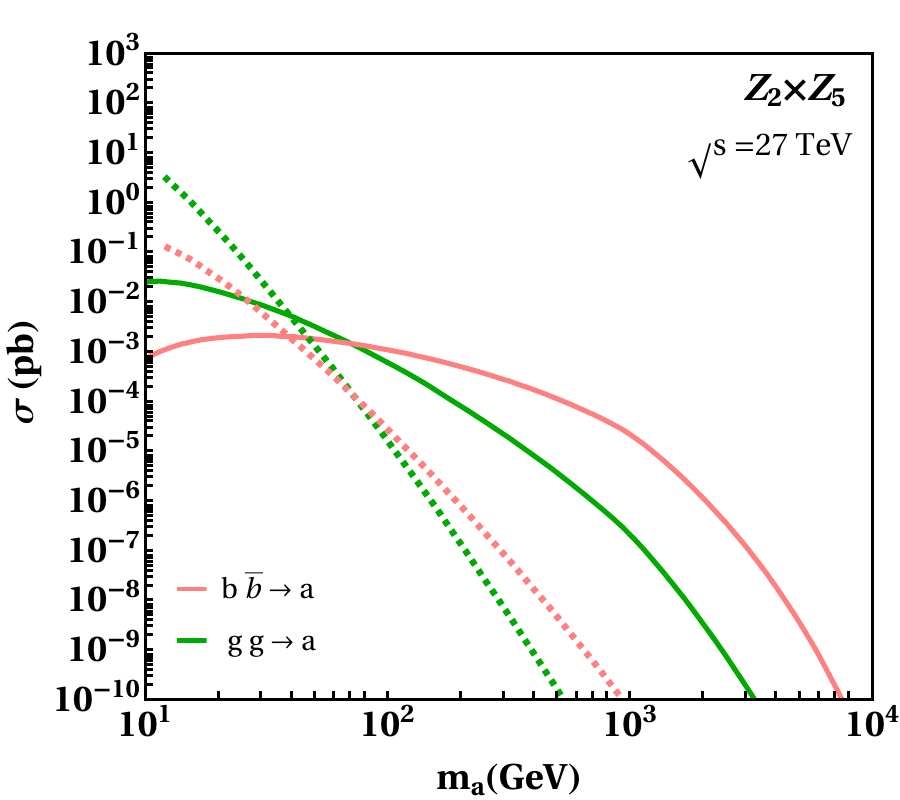}
 \caption{}
         \label{fprod27_z2z5}
 \end{subfigure} 
 \begin{subfigure}[]{0.315\linewidth}
 \includegraphics[width=\linewidth]{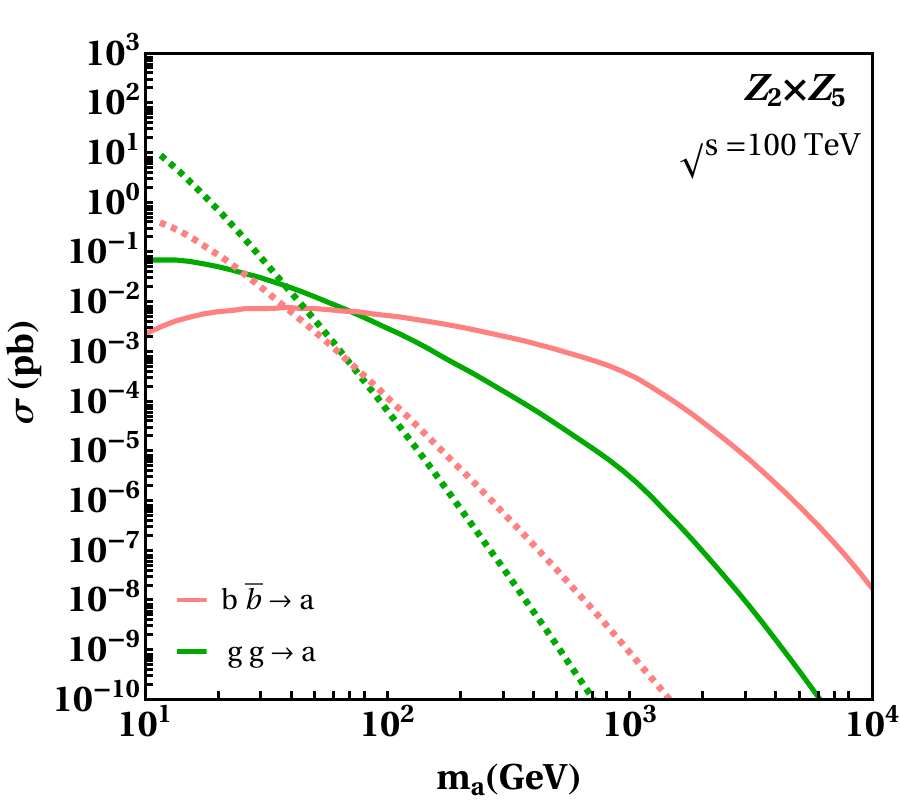}
 \caption{}
         \label{fprod100_z2z5}
 \end{subfigure} 
 \begin{subfigure}[]{0.315\linewidth}
    \includegraphics[width=\linewidth]{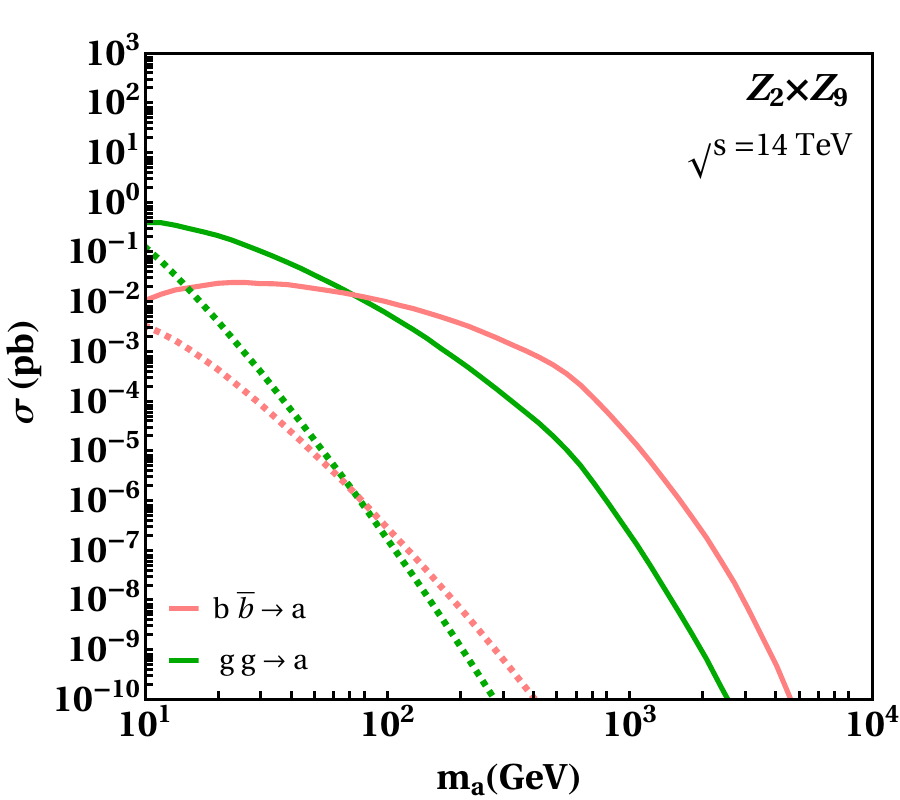}
    \caption{}
         \label{fprod14_z2z9}	
\end{subfigure}
 \begin{subfigure}[]{0.315\linewidth}
 \includegraphics[width=\linewidth]{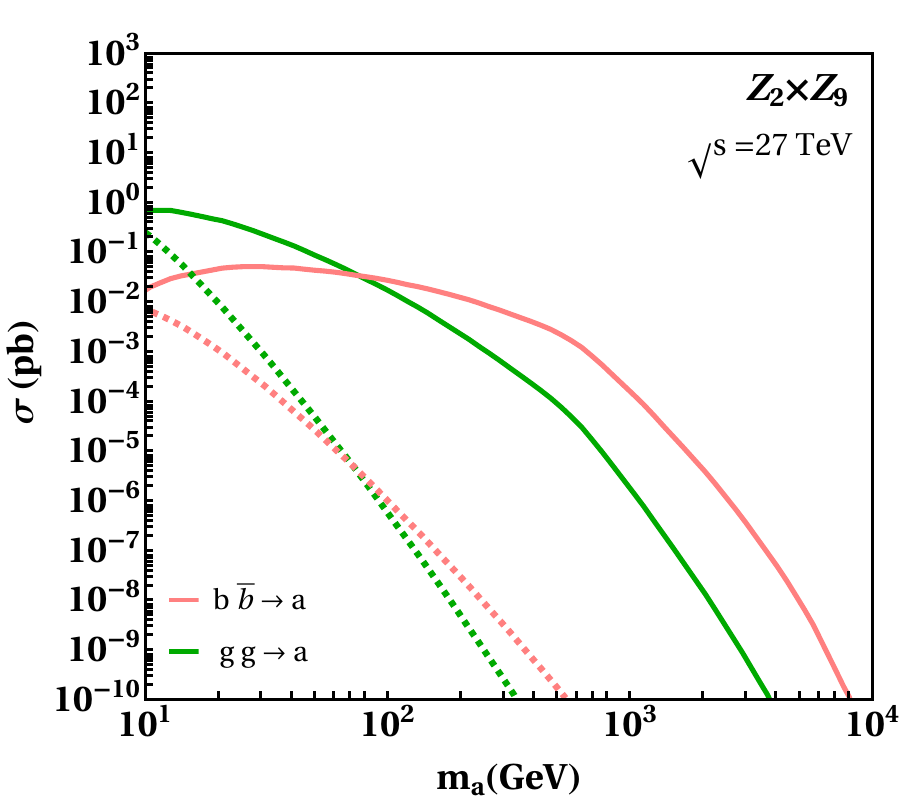}
 \caption{}
         \label{fprod27_z2z9}
 \end{subfigure} 
 \begin{subfigure}[]{0.315\linewidth}
 \includegraphics[width=\linewidth]{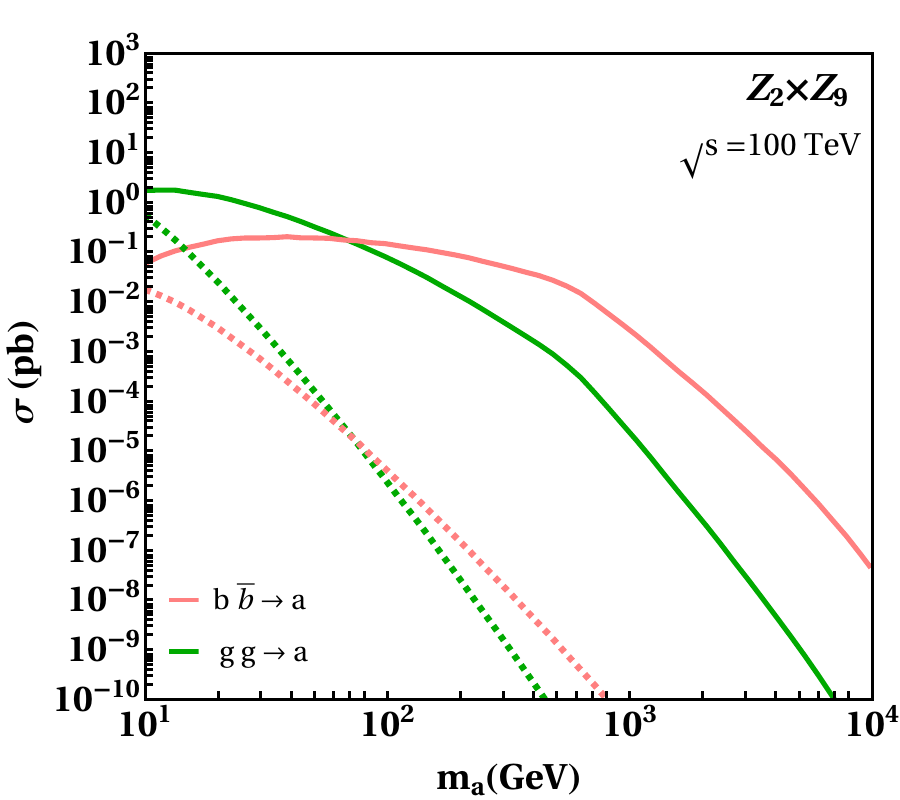}
 \caption{}
         \label{fprod100_z2z9}
 \end{subfigure} 
 \caption{Production cross-sections of the flavon of $\mathcal{Z}_2 \times \mathcal{Z}_5$, $\mathcal{Z}_2 \times \mathcal{Z}_9$ flavor symmetries with respect to its mass through different channels for the 14 TeV HL-LHC, 27 TeV HE-LHC, and future 100 TeV hadron collider. The solid lines represents the production cross section along the boundaries of parameter space allowed by the observable $R_{\mu \mu}$ for soft symmetry-breaking scenario, while the dashed lines correspond to that in the symmetry-conserving scenario along the allowed parameter space by the observable $BR(K_L\rightarrow \mu \mu)_{SD}$.}
  \label{fig_fprod1}
	\end{figure}

\begin{figure}[h!]
	\centering
	\begin{subfigure}[]{0.327\linewidth}
    \includegraphics[width=\linewidth]{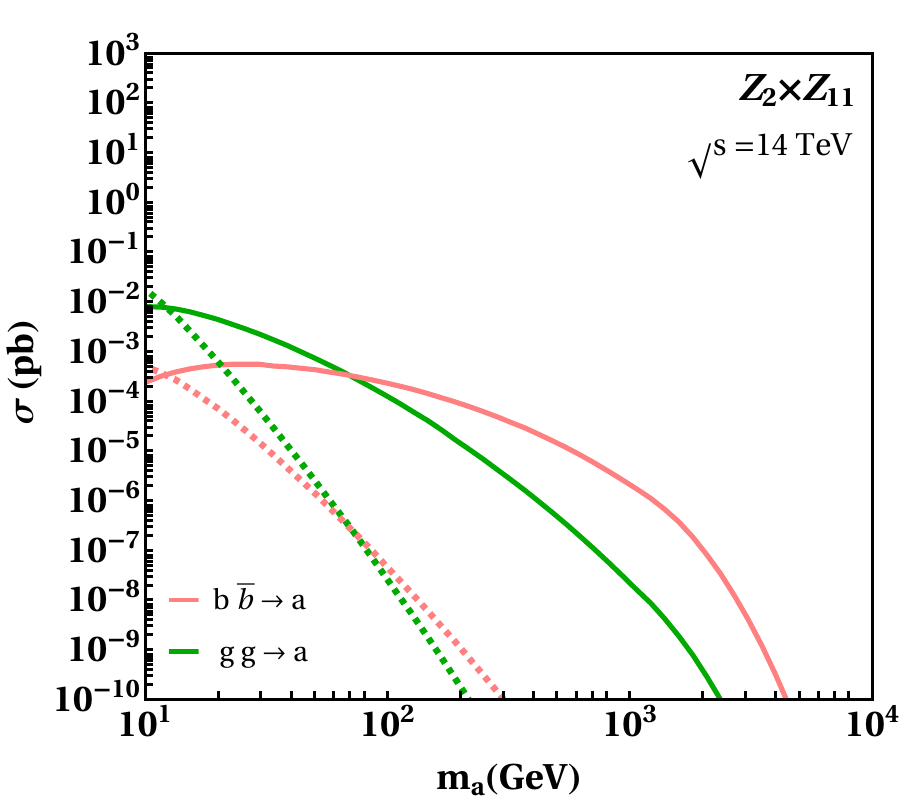}
    \caption{}
         \label{fprod14_z2z5}	
\end{subfigure}
 \begin{subfigure}[]{0.327\linewidth}
 \includegraphics[width=\linewidth]{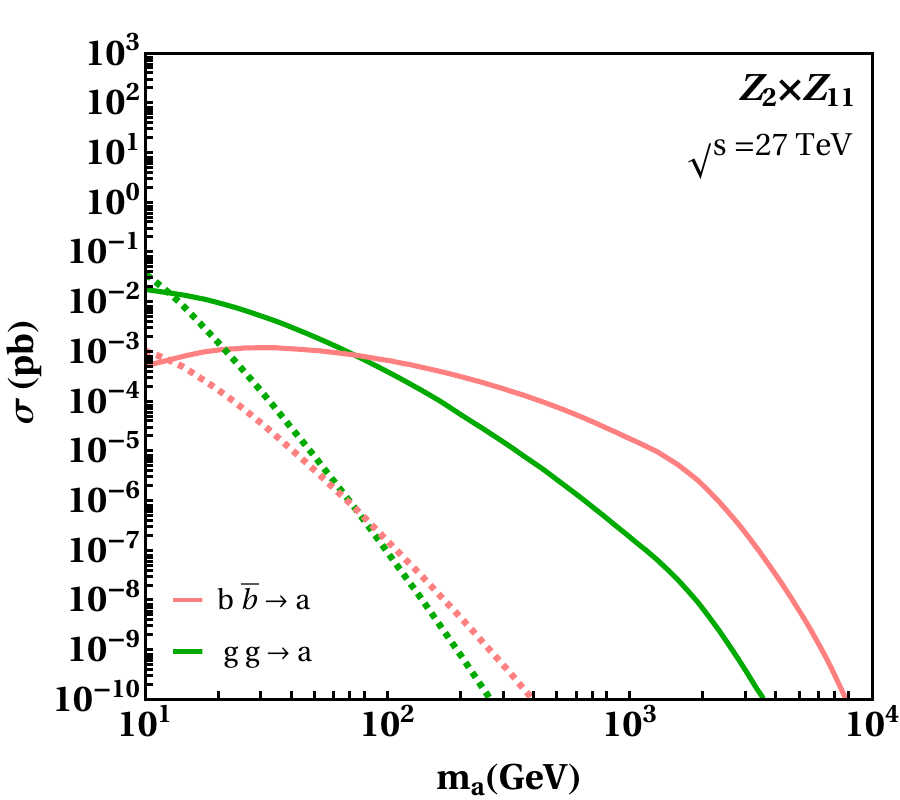}
 \caption{}
         \label{fprod27_z2z5}
 \end{subfigure} 
 \begin{subfigure}[]{0.327\linewidth}
 \includegraphics[width=\linewidth]{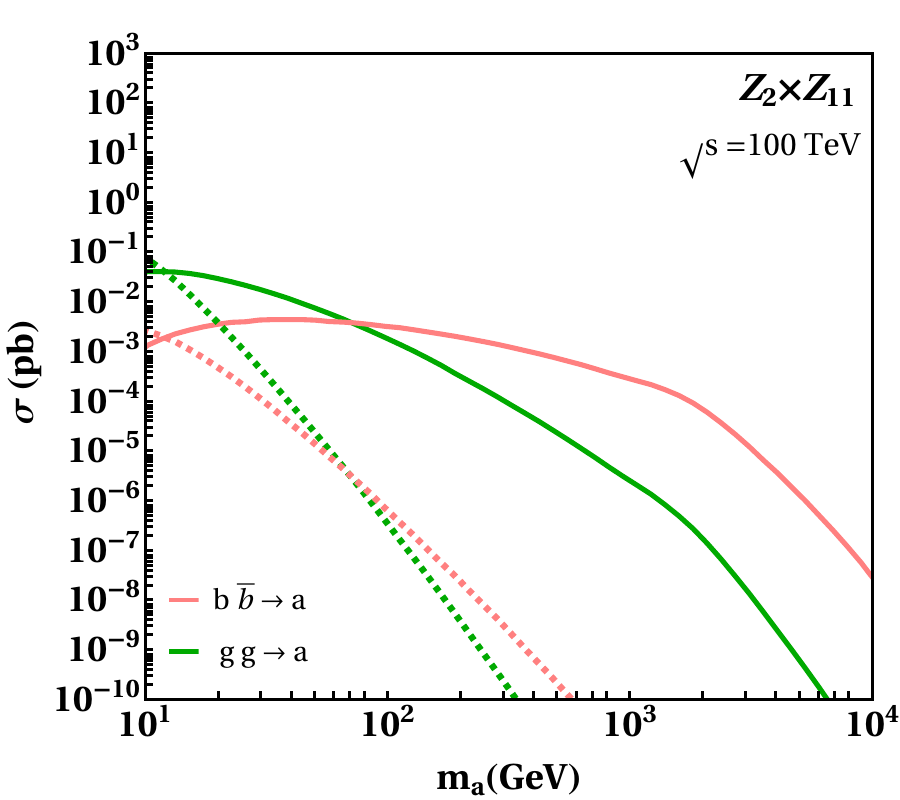}
 \caption{}
         \label{fprod100_z2z5}
 \end{subfigure} 
 \begin{subfigure}[]{0.327\linewidth}
    \includegraphics[width=\linewidth]{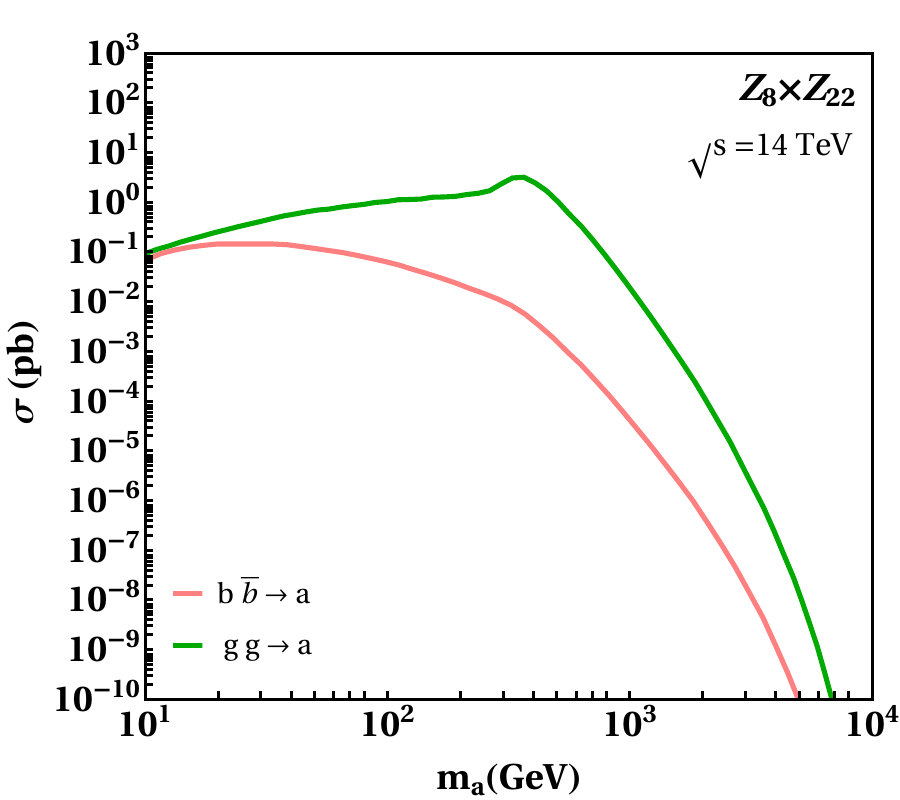}
    \caption{}
         \label{z8a}	
\end{subfigure}
 \begin{subfigure}[]{0.327\linewidth}
 \includegraphics[width=\linewidth]{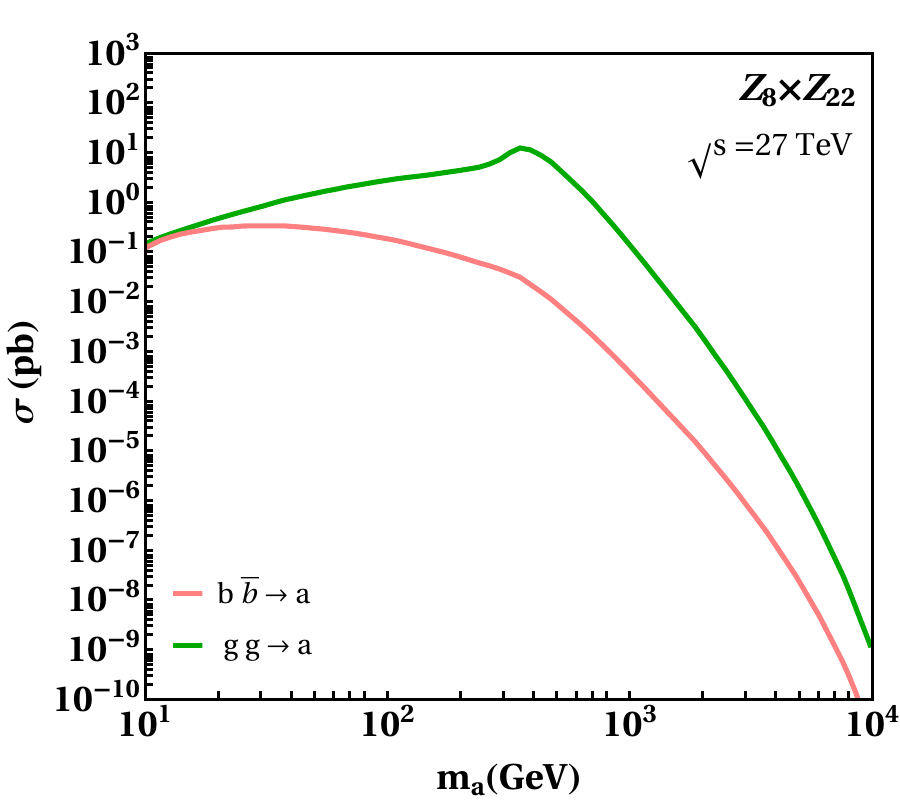}
 \caption{}
         \label{z8b}
 \end{subfigure} 
 \begin{subfigure}[]{0.327\linewidth}
 \includegraphics[width=\linewidth]{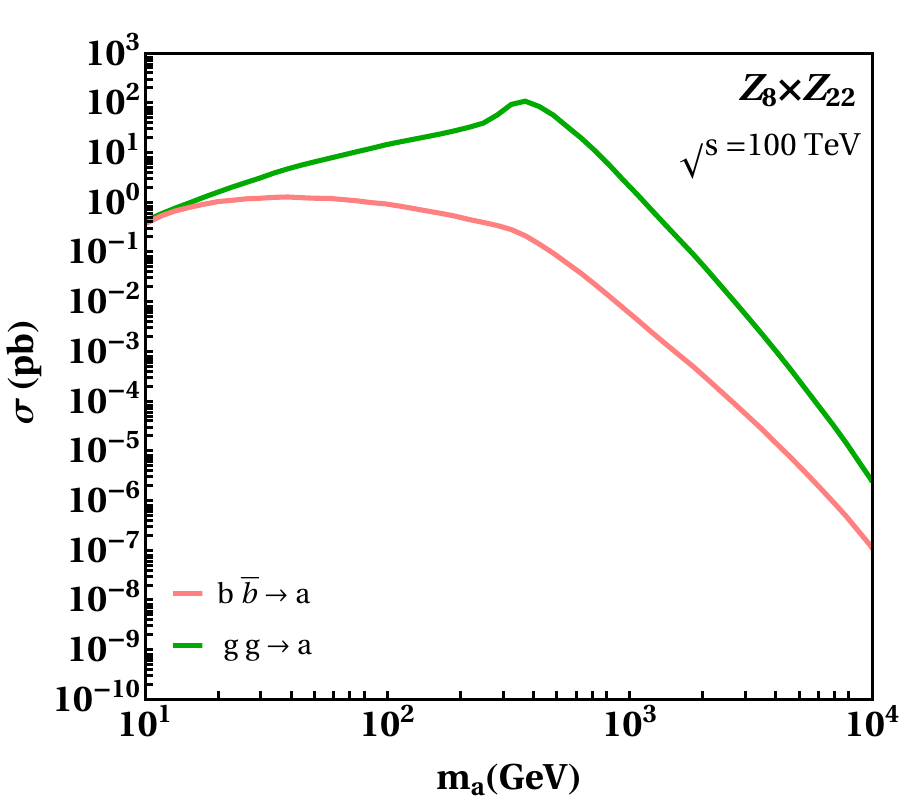}
 \caption{}
         \label{z8c}
 \end{subfigure} 
 \caption{Production cross-sections of the flavon of  $\mathcal{Z}_2 \times \mathcal{Z}_{11}$, and $\mathcal{Z}_8 \times \mathcal{Z}_{22}$ flavor symmetries with respect to its mass through different channels for the 14 TeV HL-LHC, 27 TeV HE-LHC, and future 100 TeV hadron collider. The solid lines represents the production cross section along the boundaries of parameter space allowed by the observable $R_{\mu \mu}$ for soft symmetry-breaking scenario, while the dashed lines correspond to that in the symmetry-conserving scenario along the allowed parameter space by the observable $BR(K_L\rightarrow \mu \mu)_{SD}$.}
  \label{fig_fprod2}
	\end{figure}

A flavon can be produced through the inclusive channels,
\be
gg, bb \rightarrow a.
\ee
As it can be noted from figure \ref{fig_fprod1}  and  \ref{fig_fprod2}, the gluon-fusion production cross-sections for the soft symmetry-breaking scenario,  is very small for the $\mathcal{Z}_2 \times \mathcal{Z}_{5,9,11}$  flavor symmetries due to the absence of the flavor-diagonal coupling of the flavon to top quarks.  However, this observation dramatically changes   for the $\mathcal{Z}_8 \times \mathcal{Z}_{22}$ flavor symmetry, where  the flavor-diagonal coupling of the flavon to top quarks is allowed.  This can be seen in figures \ref{z8a}-\ref{z8c}.   In the case of symmetry-conserving scenario, the gluon-fusion production cross-section turns out to be very small and, as will be shown later, is beyond the reach of the LHC, HE-LHC, and a 100 TeV future collider.

\begin{figure}[h!]
	\centering
	\begin{subfigure}[]{0.327\linewidth}
    \includegraphics[width=\linewidth]{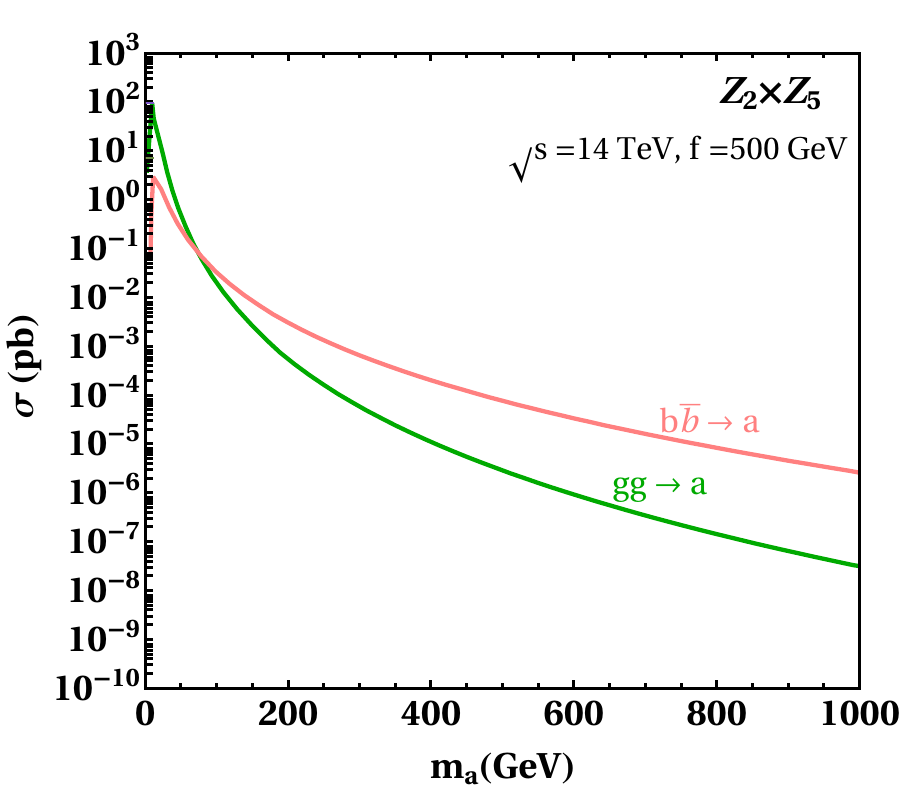}
    \caption{}
         \label{fprod14_z2z5}	
\end{subfigure}
 \begin{subfigure}[]{0.327\linewidth}
 \includegraphics[width=\linewidth]{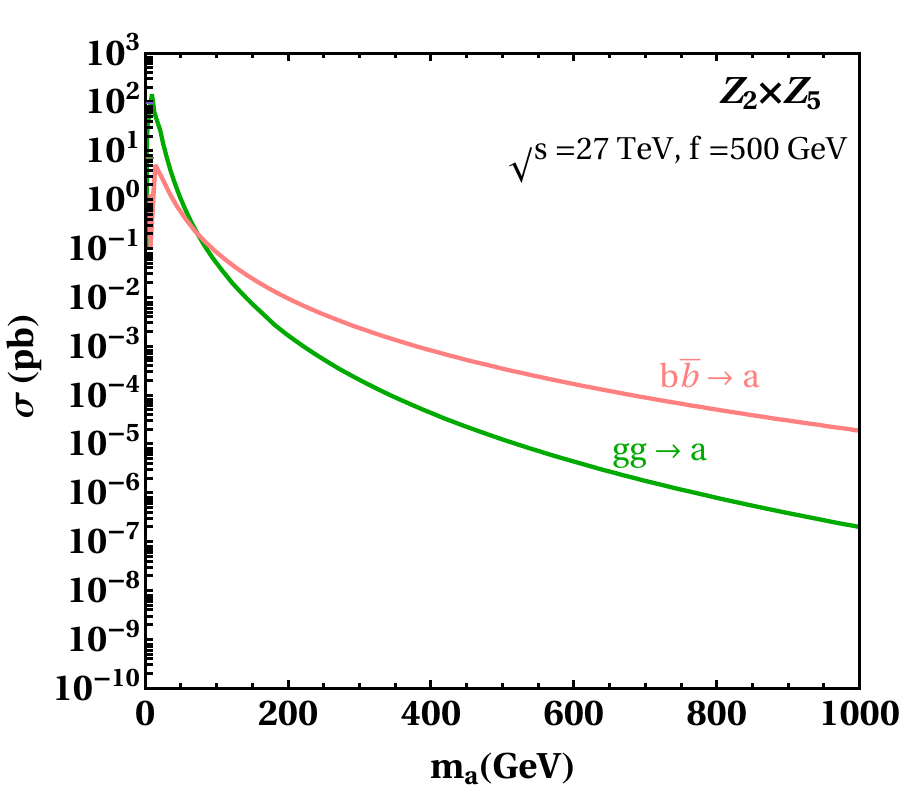}
 \caption{}
         \label{fprod27_z2z5}
 \end{subfigure} 
 \begin{subfigure}[]{0.327\linewidth}
 \includegraphics[width=\linewidth]{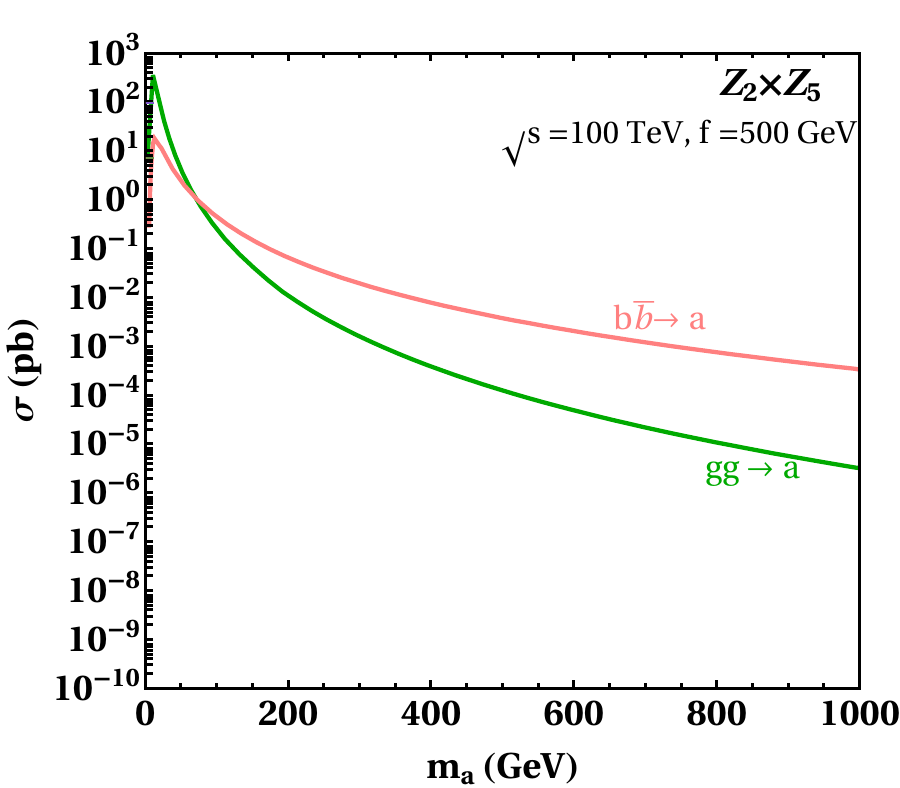}
 \caption{}
         \label{fprod100_z2z5}
 \end{subfigure} 
 \begin{subfigure}[]{0.327\linewidth}
    \includegraphics[width=\linewidth]{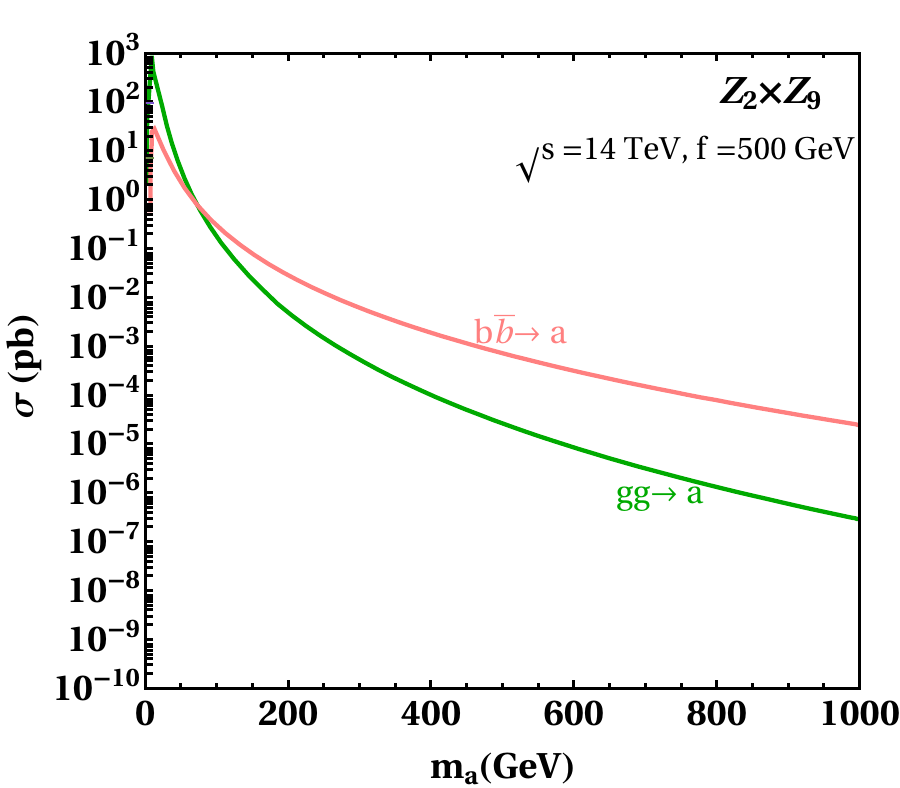}
    \caption{}
         \label{fprod14_z2z5}	
\end{subfigure}
 \begin{subfigure}[]{0.327\linewidth}
 \includegraphics[width=\linewidth]{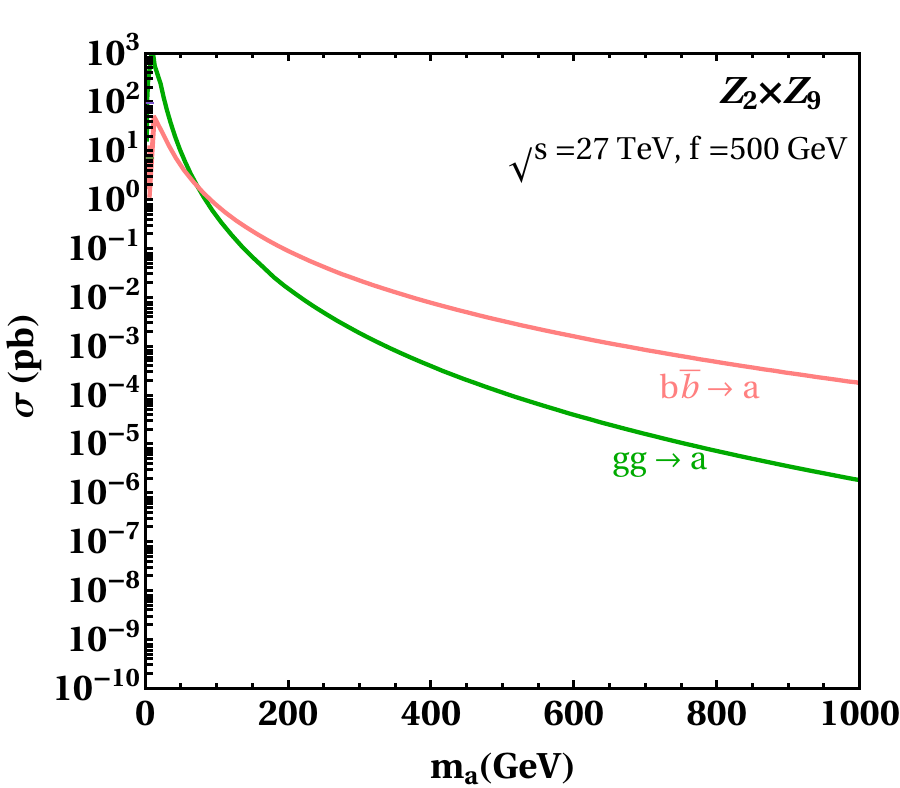}
 \caption{}
         \label{fprod27_z2z5}
 \end{subfigure} 
 \begin{subfigure}[]{0.327\linewidth}
 \includegraphics[width=\linewidth]{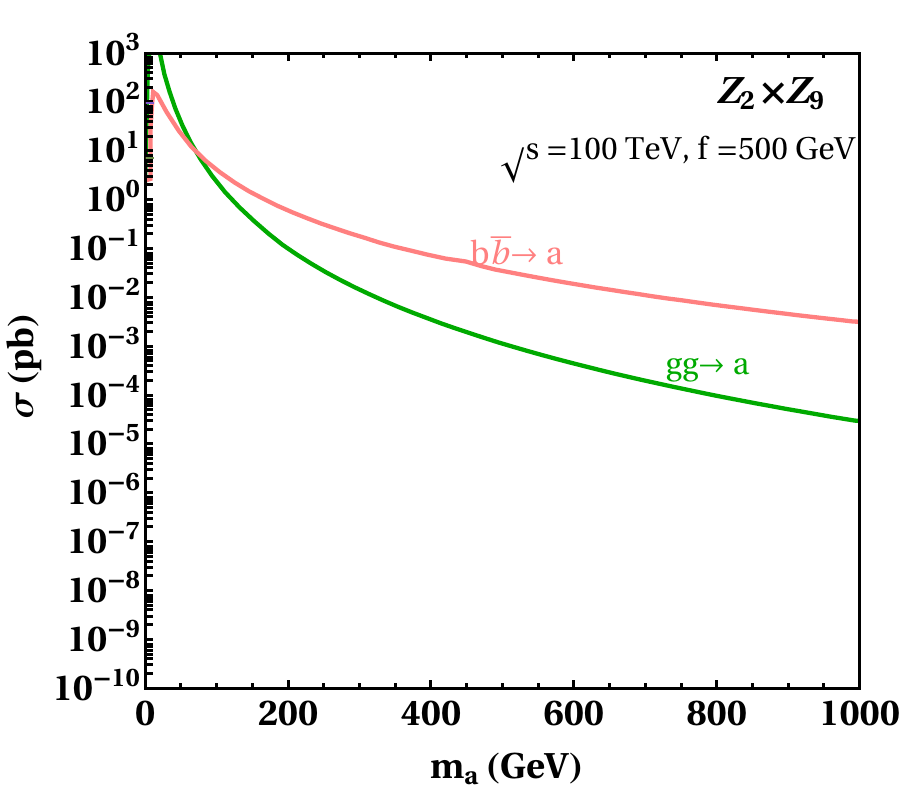}
 \caption{}
         \label{fprod100_z2z5}
 \end{subfigure} 
 \caption{Production cross-sections of the flavon of  $\mathcal{Z}_2 \times \mathcal{Z}_{5}$, and $\mathcal{Z}_2 \times \mathcal{Z}_{9}$ flavor symmetries with respect to its mass through different channels for 14 TeV HL-LHC, 27 TeV HE-LHC, and future 100 TeV hadron collider, where the flavon VEV $f=500$ GeV.}
  \label{fig_fprod500a}
	\end{figure}

\begin{figure}[h!]
	\centering
	\begin{subfigure}[]{0.327\linewidth}
    \includegraphics[width=\linewidth]{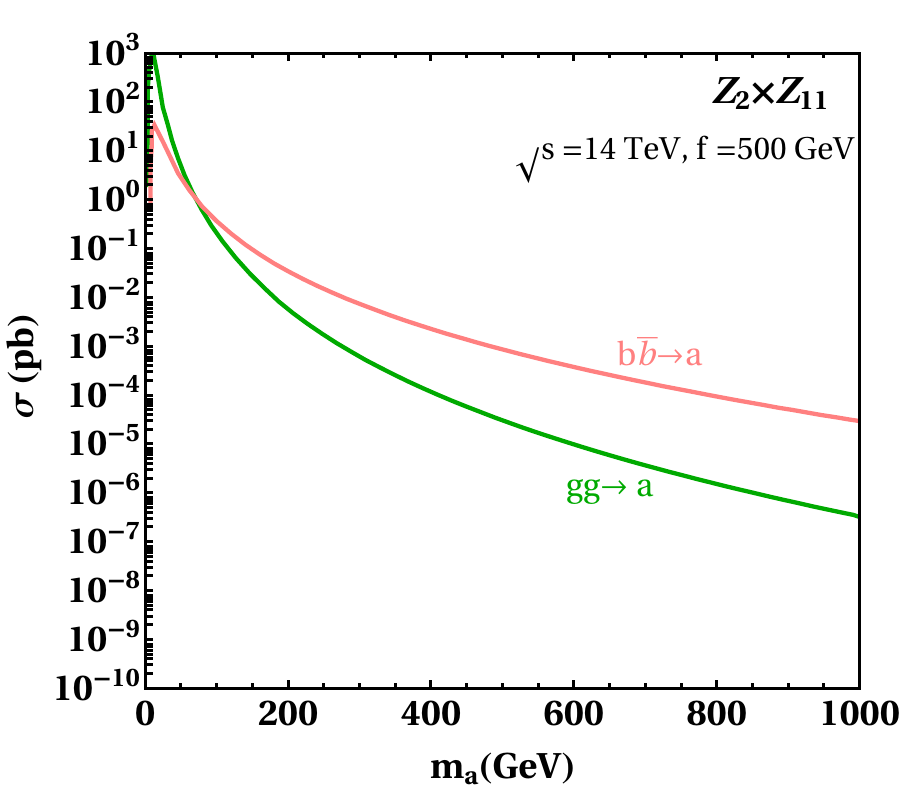}
    \caption{}
         \label{fprod14_z2z5}	
\end{subfigure}
 \begin{subfigure}[]{0.327\linewidth}
 \includegraphics[width=\linewidth]{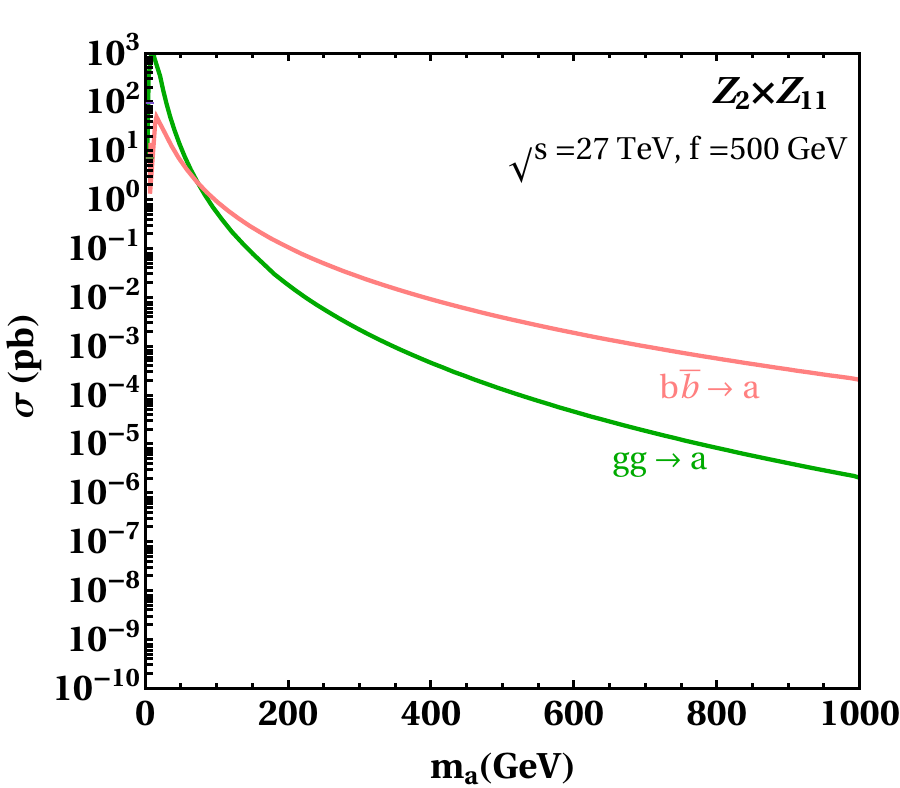}
 \caption{}
         \label{fprod27_z2z5}
 \end{subfigure} 
 \begin{subfigure}[]{0.327\linewidth}
 \includegraphics[width=\linewidth]{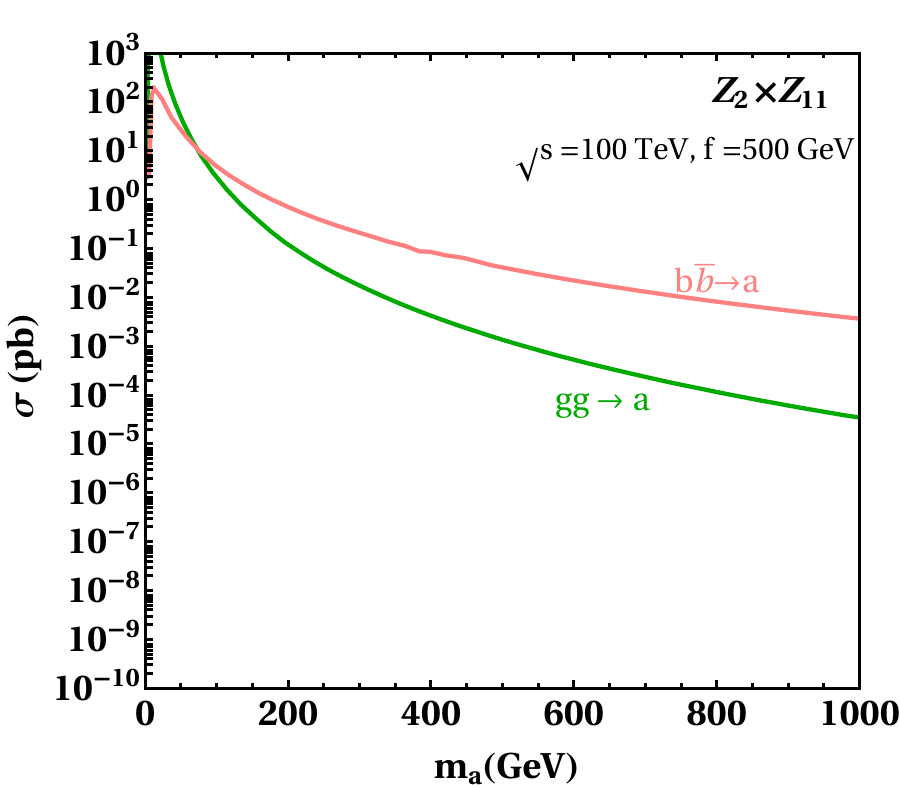}
 \caption{}
         \label{fprod100_z2z5}
 \end{subfigure} 
 \begin{subfigure}[]{0.327\linewidth}
    \includegraphics[width=\linewidth]{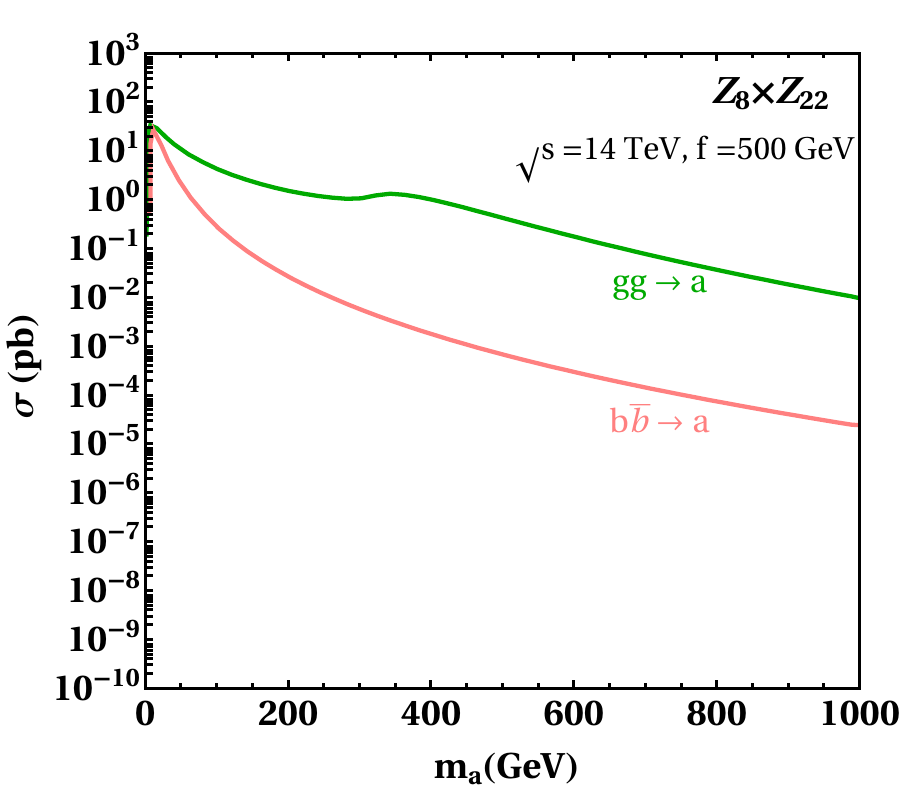}
    \caption{}
         \label{fprod14_z2z5}	
\end{subfigure}
 \begin{subfigure}[]{0.327\linewidth}
 \includegraphics[width=\linewidth]{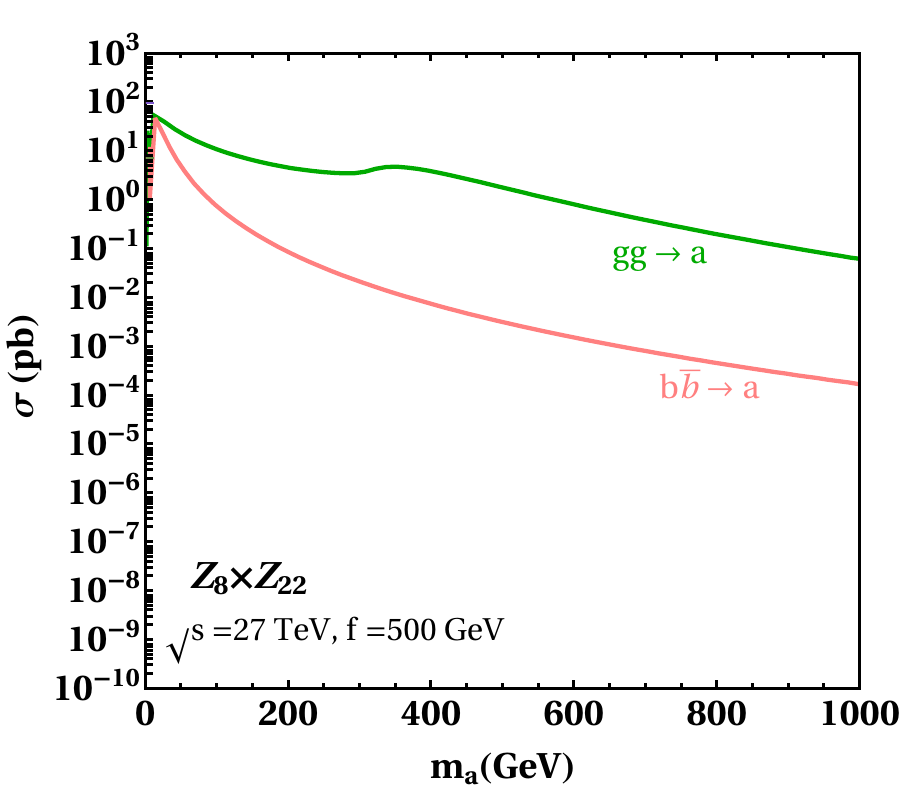}
 \caption{}
         \label{fprod27_z2z5}
 \end{subfigure} 
 \begin{subfigure}[]{0.327\linewidth}
 \includegraphics[width=\linewidth]{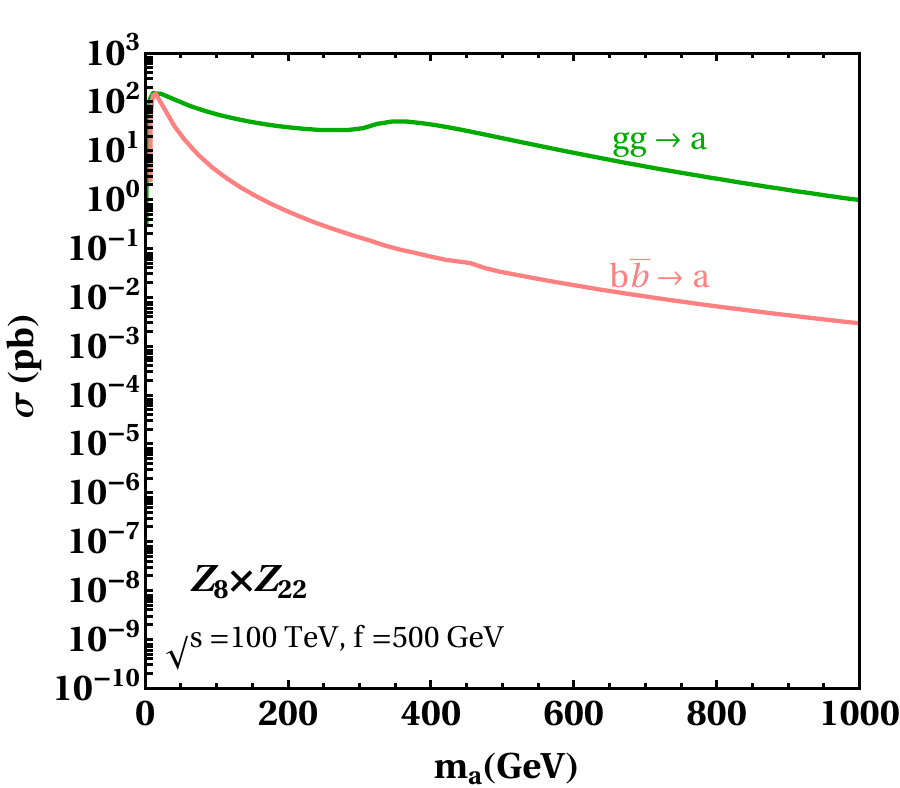}
 \caption{}
         \label{fprod100_z2z5}
 \end{subfigure} 
 \caption{Production cross-sections of the flavon of  $\mathcal{Z}_2 \times \mathcal{Z}_{11}$, and $\mathcal{Z}_8 \times \mathcal{Z}_{22}$ flavor symmetries with respect to its mass through different channels for 14 TeV HL-LHC, 27 TeV HE-LHC, and future 100 TeV high-luminosity hadron collider, where the flavon VEV $f=500$ GeV.}
  \label{fig_fprod500b}
	\end{figure}

In figures \ref{fig_fprod500a} and \ref{fig_fprod500b},  we show the production cross-sections of the flavon of different $\mathcal{Z}_{\rm N}\times \mathcal{Z}_{\rm M}$ flavor symmetries for the VEV $f =500$ GeV for the soft symmetry-breaking scenario.  In this scenario, the production cross-sections  can be sufficiently large for the $\mathcal{Z}_{\rm 2}\times \mathcal{Z}_{\rm 5, 9,11}$  and $\mathcal{Z}_{8}\times \mathcal{Z}_{22}$ flavor symmetries.  Therefore, our  investigation of the inclusive modes will be performed only for this choice from now on for the heavy mass flavon.   

The above results can be used to look for resonance searches of the flavon.  For example, a generic search is of the type,
\begin{align}
p p \to a \to
b \bar{b} /  \ell_i \ell_i / \gamma \gamma\ .
\end{align}

On the other side, due to the off-diagonal flavon couplings of flavon to a fermionic pair, searches with a specific  signature of the $\mathcal{Z}_{\rm N}\times \mathcal{Z}_{\rm M}$ flavor symmetries are,
\begin{align}
p p \to a \to
t \bar{c} / t \bar{u} .
\end{align}

\begin{table}[H]
\setlength{\tabcolsep}{6pt} 
\renewcommand{\arraystretch}{1} 
\centering
\begin{tabular}{l|cc|cc|cc}
\toprule
& \multicolumn{2}{c|}{$\mathcal{L}[fb^{-1}]$ [References]} & \multicolumn{2}{c|}{ATLAS 13 TeV} & \multicolumn{2}{c}{CMS 13 TeV} \\
$m_{a}$~[GeV] &  \myalign{c}{ATLAS} & \myalign{c|}{CMS} & \myalign{c}{500} & \myalign{c|}{1000} & \myalign{c}{500} & \myalign{c}{1000}   \\
\midrule
jet-jet ~[pb]          & $139$ \cite{ATLAS:2019fgd} &  $137$ \cite{CMS:2019gwf}  &  &0.1   &            &  0.2                 \\
$\tau \tau$~[pb]  & $36.1$ \cite{ATLAS:2017eiz} & 
 $35.9$ \cite{CMS:2018rmh}  & $8\e{-2}$ & $10^{-2}$  & $6\e{-2}$ & $10^{-2}$   \\
$e e$, $\mu \mu$~[pb]   & $139$ \cite{ATLAS:2019erb}&  
 $140$ \cite{CMS:2021ctt}  & $8\e{-4}$ &\phantom{xx}  $2\e{-4}$  & $2\e{-3}$ &\phantom{xx}  $4\e{-4}$  \\
$\mu e$~[pb]   & $138$ \cite{ATLAS:2020tre}&  $139$ \cite{CMS:2022fsw}  &  &\phantom{xx}  $3\e{-4}$  & $4\e{-3}$ &\phantom{xx}  $3\e{-4}$   \\
$ \mu \tau$~[pb]   & $138$ \cite{ATLAS:2020tre}&  $139$ \cite{CMS:2022fsw}  &  &\phantom{xx}  $1\e{-3}$  & $7\e{-3}$ &\phantom{xx}  $1\e{-3}$  \\
$e \tau$~[pb]   & $138$ \cite{ATLAS:2020tre}&  $139$ \cite{CMS:2022fsw}  &  &\phantom{xx}  $1\e{-3}$  & $5\e{-3}$ &\phantom{xx}  $1\e{-3}$  \\
$b  \bar{b}$~[pb]   & $139$ \cite{ATLAS:2019fgd}&  $138$ \cite{CMS:2022eud}  &  &\phantom{xx} $1\e{-2}$   &  &\phantom{xx} $4\e{-2}$   \\
$\gamma \gamma$~[pb]& $139$ \cite{ATLAS:2021uiz}  & $35.9$ \cite{CMS:2018dqv}  & $5\e{-4}$ & $1\e{-4}$  & $4\e{-3}$ &   $8\e{-4}$        \\
$t  \bar{t}$~[pb]& $36.1$ \cite{ATLAS:2019npw,ATLAS:2020lks}  & $35.9$ \cite{CMS:2018rkg} & $2\e{2}$ & $2$  & $30 $ &     $0.4$       \\
\bottomrule
\end{tabular}
\caption{Current limits for production cross section times branching ratio ($\sigma \times BR$) at 13 TeV LHC by ATLAS and CMS in high mass resonance searches for inclusive flavon production channels.}
\label{tab:limits_LHC}
\end{table}

In table \ref{tab:limits_LHC}, we show the present sensitivities of different inclusive flavon production channels at the 14 TeV LHC.  These results  will be used to estimate the sensitivities of the HL-LHC, HE-LHC, and  a 100 TeV collider for a heavy flavon production and decay. We show these sensitivities in table \ref{tab:futurelimits1}. These sensitivities are estimated using simple square root scaling of the luminosity of the LHC by,
\begin{equation}
 \mathcal{S} \simeq \frac{S}{\sqrt{B}} \simeq    \sqrt{\mathcal{L}} \frac{\sigma_s}{\sqrt{\sigma_B}},
\end{equation} 
where $S$ is the number of signal events, $B$ denotes the background events, $\sigma_s$ is the signal cross-section, and 
$\sigma_B$ stands for the background cross-section.

We adopt a conservative  approach for estimating the sensitivities of the HL-LHC, HE-LHC, and  a 100 TeV collider.  We make the following assumptions for this purpose. 
\begin{enumerate}
    \item   The significance $\mathcal{S} \simeq \frac{S}{\sqrt{B}}$ remains constant among the colliders.
    \item We assume that there is no appreciable change in the reconstruction efficiencies and  background rejection.
\end{enumerate}
We  note that similar assumptions are also used in the ``Collider Reach" tool, which remarkably provides   an estimate of the  mass of a BSM physics that can be searched at the LHC and a future collider \cite{CR}.

Thus, the required sensitive cross-section of any flavon signal at a future collider (FC) is given by 
\begin{align}
  \sigma_s^{\rm FC}  = \sqrt{\dfrac{\mathcal{L}_{\rm LHC}}{\mathcal{L}_{\rm FC}}} \sqrt{\dfrac{\sigma_{ B}^{ FC}}{\sigma_{ B}^{ LHC}}} \sigma_{s}^{\rm LHC},
\end{align}
where FC= HL-LHC, HE-LHC, and  a 100 TeV collider.  We have  computed $\sigma_{ B}^{ FC}$ and $\sigma_{ B}^{ LHC}$ through MadGraph, and it turns out that  $\sigma_{ B}^{ HE-LHC} \leq 2 \sigma_{ B}^{ LHC}$ and $\sigma_{ B}^{ 100 TeV} \leq 10 \sigma_{ B}^{ LHC}$,  and  $ \sigma_{ s}^{ LHC}$ is given in table \ref{tab:limits_LHC}.

\begin{table}[H]
\setlength{\tabcolsep}{6pt} 
\renewcommand{\arraystretch}{1} 
\centering
\begin{tabular}{l|cc|cc|cc}
\toprule
 & \multicolumn{2}{c|}{HL-LHC [14 TeV, $3~\iab$] } & \multicolumn{2}{c|}{HE-LHC [27 TeV, $15~\iab$]} & \multicolumn{2}{c}{100 TeV, $30~\iab$} \\
$m_{a}$~[GeV] &  500 &  1000 &  500 &  1000 &  500 &  1000 \\
\midrule
jet-jet [pb]        &           & $4\e{-2}$   &            &   $3\e{-2}$         &            &   $4\e{-2}$       \\
$\tau \tau$ [pb]  & $7\e{-3}$ & $1\e{-3}$ & $4\e{-3}$ & $7\e{-4}$ & $5\e{-3}$ & $8\e{-4}$   \\
$e e$, $\mu \mu$ [pb] & $2\e{-4}$ & $4\e{-5}$  & $1\e{-4}$ & $3\e{-5}$  & $1\e{-4}$ & $3\e{-5}$  \\
$\mu e$ [pb]     &  $9\e{-4}$     & $7\e{-5}$ & $7\e{-4}$ & $5\e{-5}$  & $1\e{-3}$ & $1\e{-4}$  \\
$\mu \tau$ [pb]  &  $2\e{-3}$        &  $2\e{-4}$ & $1\e{-3}$ & $2\e{-4}$   &  $2\e{-3}$ & $3\e{-4}$  \\
$e \tau$ [pb]    & $1\e{-3}$           & $2\e{-4}$ & $8\e{-4}$ & $2\e{-4}$  &  $1\e{-3}$ & $3\e{-4}$  \\
$b  \bar{b}$ [pb]    & & $9\e{-3}$    &  & $5\e{-3}$    &  & $7\e{-3}$  \\
$\gamma \gamma$ [pb] & $1\e{-4}$  & $2\e{-5}$  & $6\e{-5}$ & $1\e{-5}$  & $7\e{-5}$ & $1\e{-5}$  \\
$t \bar{t}$ [pb]     & 4   & $5\e{-2}$      & 3      & $4\e{-2}$    & $8$ & $0.1$  \\
\bottomrule
\end{tabular}
\caption{Estimated reach ($\sigma \times BR$) of the HL-LHC, HE-LHC and a 100 TeV hadron collider for high flavon mass ($m_a$) in inclusive flavon production channels.}
\label{tab:futurelimits1}
\end{table}

We show our benchmark predictions  for the processes given in table \ref{tab:limits_LHC}, for different $\mathcal{Z}_{\rm N} \times \mathcal{Z}_{\rm M}$ flavor symmetries in tables  \ref{tab:limits_bench14a}- \ref{tab:limits_bench100a} for $f=500$ GeV and $m_a= 500,1000$ for the soft symmetry-breaking scenarios.  The modes that are accessible to the HL-LHC, HE-LHC, and a 100 TeV collider are marked with a box \fbox{\begin{minipage}{0.2cm}  \end{minipage}}.  

From table  \ref{tab:limits_bench14a}, we observe that only the $\mathcal{Z}_2 \times \mathcal{Z}_{5}$ and the $\mathcal{Z}_8 \times \mathcal{Z}_{22}$ flavor symmetries are accessible to the HL-LHC, and both the symmetries can be differentiated due to the distinct accessible processes.  This situation changes sufficiently for the HE-LHC as shown in table  \ref{tab:limits_bench27a}, as more processes are under the reach of the HE-LHC.  However, even now, mainly the $\mathcal{Z}_2 \times \mathcal{Z}_{5}$ and the $\mathcal{Z}_8 \times \mathcal{Z}_{22}$ flavor symmetries are under the reach of the HE-LHC.  The $\mathcal{Z}_2 \times \mathcal{Z}_{9}$ flavor symmetry can be probed through the $e\tau$ channel.




 \begin{table}[h!]
\setlength{\tabcolsep}{4.8pt} 
\renewcommand{\arraystretch}{1.12} 
\centering
\begin{tabular}{@{}l|rr|rr|rr|rr@{}}
\toprule
 & \multicolumn{2}{c|}{Benchmark} & \multicolumn{2}{c|}{Benchmark} & \multicolumn{2}{c|}{Benchmark} & \multicolumn{2}{c}{Benchmark}\\
 & \multicolumn{2}{c|}{$\mathcal{Z}_2 \times \mathcal{Z}_{5}$} & \multicolumn{2}{c|}{$\mathcal{Z}_2 \times \mathcal{Z}_{9}$} & \multicolumn{2}{c|}{$\mathcal{Z}_2 \times \mathcal{Z}_{11}$} & \multicolumn{2}{c}{$\mathcal{Z}_8 \times \mathcal{Z}_{22}$}\\
 $m_{a}$~[GeV] & \myalign{c}{500} & \myalign{c|}{1000} & \myalign{c}{500} & \myalign{c|}{1000} & \myalign{c}{500} & \myalign{c|}{1000} &\myalign{c}{500} & \myalign{c}{1000}\\
\midrule
jet-jet~[pb]              &  & \fbox{$3.6\e{-2}$}  &           &  $1.5\e{-6}$         &      &   $2.3\e{-7}$    &    &  $1.4\e{-3}$ \\
$\tau \tau$~[pb]      & $1.2 \e{-3}$ & $9.2\e{-5}$  & $8.0\e{-5}$ & $3.4\e{-6}$ & $2.9\e{-5}$ & $1.6\e{-6}$ & $3.4\e{-3}$ & $6.1\e{-5}$  \\
$\mu \tau$~[pb]      & $1.4\e{-3}$ & $1.1\e{-4}$  & $2.3\e{-4}$ & $9.5\e{-6}$ & $3\e{-5}$ & $1.7\e{-6}$ &  \fbox{$5.8\e{-3}$}  & $1\e{-4}$  \\
$e \tau$~[pb]      & \fbox{$1.1\e{-3}$} & $8.9\e{-5}$  & $2.2\e{-4}$ & $9.4\e{-6}$ & $8.5\e{-5}$ & $4.7\e{-6}$ & $3.2\e{-4}$ & $5.8\e{-6}$  \\
$\mu \mu$~[pb]        & $1.1\e{-6}$ & $8.3\e{-8}$  & $1.7\e{-6}$ &  $7.3\e{-8}$ & $2.2\e{-7}$ & $1.2\e{-8}$ & $2.9\e{-5}$  & $5.3\e{-7}$ \\
$\mu e$~[pb]        & $2.9\e{-6}$ &  $2.3\e{-7}$  & $6.3\e{-7}$ &  $2.6\e{-8}$ & $4.8\e{-7}$ & $2.7\e{-8}$ & $3.3\e{-6}$  &  $5.8\e{-8}$ \\
$e e$~[pb]        & $2.5\e{-10}$ &  $2\e{-11}$  & $3.4\e{-9}$ &  $1.4\e{-10}$ & $6.7\e{-11}$ &  $3.7\e{-12}$ & $1.7\e{-9}$ & $3\e{-11}$  \\
$\gamma \gamma$~[pb]      & $1.3\e{-7}$ & $3.6\e{-9}$  & $8.2\e{-10}$ &   $1.2\e{-11}$      & $1.5\e{-10}$  & $3\e{-12}$ & \fbox{$6.6\e{-4}$} & $1\e{-5}$ \\
$b  \bar{b}$~[pb]      & $9.8\e{-3}$ & $6.3\e{-4}$  & $4.7\e{-4}$ & $1.9\e{-5}$ & $1.2\e{-4}$ & $5.7\e{-6}$ & $1.9\e{-2}$ & $3.2\e{-4}$  \\
$t  c$~[pb]      & $3.6\e{-3}$  & $2.8\e{-4}$ & $2.5\e{-5}$ & $1.2\e{-6}$ &$ 1.5\e{-3}$ & $ 8.5\e{-5}$   & $0.152$ & $3.1\e{-3}$   \\
$t  u$~[pb]      & $6.5\e{-8}$  & $5.1\e{-9}$  & $7.8\e{-10}$ & $3.8\e{-11}$ &$ 6.6\e{-4}$ &  $ 3.8\e{-5}$ & $ 1.1\e{-3}$ & $2.3\e{-5}$   \\
$t  \bar{t}$~[pb]      &   &  &  & & &   & \fbox{$4.42$} & $0.12$   \\
\bottomrule
\end{tabular}
\caption{Benchmark points for different  $\mathcal{Z}_N \times \mathcal{Z}_{M}$  flavor symmetries  for inclusive flavon production channels with high flavon mass ($m_a$) in case of the soft symmetry-breaking scenario at the 14 TeV HL-LHC, assuming VEV ($f$) $= 500$ GeV. Accessible channels are represented within the box.}
\label{tab:limits_bench14a}
\end{table}

\begin{table}[H]
\setlength{\tabcolsep}{4.8pt} 
\renewcommand{\arraystretch}{1.12} 
\centering
\begin{tabular}{@{}l|rr|rr|rr|rr@{}}
\toprule
 & \multicolumn{2}{c|}{Benchmark} & \multicolumn{2}{c|}{Benchmark} & \multicolumn{2}{c|}{Benchmark} & \multicolumn{2}{c}{Benchmark}\\
 & \multicolumn{2}{c|}{$\mathcal{Z}_2 \times \mathcal{Z}_{5}$} & \multicolumn{2}{c|}{$\mathcal{Z}_2 \times \mathcal{Z}_{9}$} & \multicolumn{2}{c|}{$\mathcal{Z}_2 \times \mathcal{Z}_{11}$} & \multicolumn{2}{c}{$\mathcal{Z}_8 \times \mathcal{Z}_{22}$}\\
 $m_{a}$~[GeV] & \myalign{c}{500} & \myalign{c|}{1000} & \myalign{c}{500} & \myalign{c|}{1000} & \myalign{c}{500} & \myalign{c|}{1000} &\myalign{c}{500} & \myalign{c}{1000}\\
\midrule
jet-jet~[pb]              &  & \fbox{$0.133$} &    &  $8.2\e{-6}$         &     &   $9.4\e{-7}$         & & $9.8\e{-3}$ \\
$\tau \tau$~[pb]      & $2.6 \e{-3}$ & $2.8\e{-4}$  & $2.9\e{-4}$ & $1.7\e{-5}$ & $8\e{-5}$ & $5.6\e{-6}$ & $1.5\e{-2}$  & $4\e{-4}$  \\
$\mu \tau$~[pb]      & \fbox{$3.2\e{-3}$} & \fbox{$3.5\e{-4}$}  & $8.3\e{-4}$ & $4.8\e{-5}$ & $8.4\e{-5}$ & $5.8\e{-6}$ & \fbox{$2.5\e{-2}$}  & \fbox{$6.8\e{-4}$} \\
$e \tau$~[pb]      & \fbox{$2.6\e{-3}$} & \fbox{$2.8\e{-4}$}  & \fbox{$8.2\e{-4}$} & $4.8\e{-5}$ & $2.3\e{-4}$ & $1.6\e{-5}$ & \fbox{$1.4\e{-3}$} & $3.8\e{-5}$  \\
$\mu \mu$~[pb]        & $2.4\e{-6}$ &  $2.6\e{-7}$  & $6.4\e{-6}$ &  $3.7\e{-7}$ & $6.2\e{-7}$ &  $4.3\e{-8}$ & \fbox{$1.3\e{-4}$}  & $3.5\e{-6}$ \\
$\mu e$~[pb]        & $6.6\e{-6}$ &  $7.1\e{-7}$  & $2.3\e{-6}$ &  $1.4\e{-7}$ & $1.3\e{-6}$ & $9.3\e{-8}$ & $1.4\e{-5}$  & $3.8\e{-7}$ \\
$e e$~[pb]        & $5.6\e{-10}$ & $6.1\e{-11}$  & $1.3\e{-8}$ &  $7.4\e{-10}$ & $1.8\e{-10}$ &  $1.3\e{-11}$ & $7.2\e{-9}$ & $1.9\e{-10}$  \\
$\gamma \gamma$~[pb]      & $2.9\e{-7}$ & $1.1\e{-8}$  & $3\e{-9}$ &   $6.3\e{-11}$      & $4.2\e{-10}$  & $1.1\e{-11}$ & \fbox{$2.8\e{-3}$} & \fbox{$6.7\e{-5}$}\\
$b  \bar{b}$~[pb]      & $2.7\e{-2}$ & $2.3\e{-3}$  & $1.9\e{-3}$ & $1\e{-4}$ & $3.8\e{-4}$ & $2.3\e{-5}$ & $8.8\e{-2}$ & $2.2\e{-3}$  \\
$t  c$~[pb]      & $1\e{-2}$  & $1\e{-3}$ & $1\e{-4}$ & $6.6\e{-6}$ &$ 4.6\e{-3}$ & $ 3.4\e{-4}$   & $0.702$ & $2.1\e{-2}$  \\
$t  u$~[pb]      & $1.8\e{-7}$  & $1.9\e{-8}$  & $3.1\e{-9}$ & $2\e{-10}$ &$ 2\e{-3}$ &  $ 1.5\e{-4}$ & $ 5.3\e{-3}$ & $1.6\e{-4}$   \\
$t  \bar{t}$~[pb]      &   &  &  & & &   & \fbox{$ 20.46$} & \fbox{$0.83$}   \\
\bottomrule
\end{tabular}
\caption{Benchmark points for different  $\mathcal{Z}_N \times \mathcal{Z}_{M}$  flavor symmetries for inclusive flavon production channels with high flavon mass ($m_a$) in case of the soft symmetry-breaking scenario at the 27 TeV HE-LHC , assuming $f = 500$ GeV. }
\label{tab:limits_bench27a}
\end{table}

A 100 TeV collider provides a more dramatic scenario for the $\mathcal{Z}_{\rm N} \times \mathcal{Z}_{\rm M}$ flavor symmetries. As shown in table \ref{tab:limits_bench100a}, all symmetries except the $\mathcal{Z}_2 \times \mathcal{Z}_{11}$  become sensitive in most of the inclusive processes.   The $\mathcal{Z}_2 \times \mathcal{Z}_{11}$ flavor symmetry is accessible only  through the $e\tau$ channel.  
\begin{table}[h!]
\setlength{\tabcolsep}{2.1pt} 
\renewcommand{\arraystretch}{1.15} 
\centering
\begin{tabular}{@{}l|rr|rr|rr|rr@{}}
\toprule
 & \multicolumn{2}{c|}{Benchmark} & \multicolumn{2}{c|}{Benchmark} & \multicolumn{2}{c|}{Benchmark} & \multicolumn{2}{c}{Benchmark}\\
 & \multicolumn{2}{c|}{$\mathcal{Z}_2 \times \mathcal{Z}_{5}$} & \multicolumn{2}{c|}{$\mathcal{Z}_2 \times \mathcal{Z}_{9}$} & \multicolumn{2}{c|}{$\mathcal{Z}_2 \times \mathcal{Z}_{11}$} & \multicolumn{2}{c}{$\mathcal{Z}_8 \times \mathcal{Z}_{22}$}\\
 $m_{a}$~[GeV] & \myalign{c}{500} & \myalign{c|}{1000} & \myalign{c}{500} & \myalign{c|}{1000} & \myalign{c}{500} & \myalign{c|}{1000} &\myalign{c}{500} & \myalign{c}{1000}\\
\midrule
jet-jet~[pb]              &  & \fbox{$0.95$}  &            &  $1.1\e{-4}$         &      &   $8.1\e{-6}$         &  & \fbox{$0.18$}\\
$\tau \tau$~[pb]      & \fbox{$1.1 \e{-2}$} & \fbox{$1.4\e{-3}$}  & $2.2\e{-3}$ & $1.9\e{-4}$ & $4.8\e{-4}$ & $3.8\e{-5}$ & \fbox{$0.14$} & \fbox{$6.2\e{-3}$}  \\
$\mu \tau$~[pb]      & \fbox{$1.3\e{-2}$} & \fbox{$1.7\e{-3}$}  & \fbox{$6.3\e{-3}$} & \fbox{$5.5\e{-4}$} & $5\e{-4}$ & $3.9\e{-5}$ & \fbox{$0.23$} & \fbox{$1.0\e{-2}$} \\
$e \tau$~[pb]      & \fbox{$1.1\e{-2}$} & \fbox{$1.4\e{-3}$}  & \fbox{$6.2\e{-3}$} & \fbox{$5.4\e{-4}$} & \fbox{$1.3\e{-3}$} & $1.1\e{-4}$ & \fbox{$1.3\e{-2}$} & \fbox{$5.9\e{-4}$}  \\
$\mu \mu$~[pb]        & $9.9\e{-5}$ &\phantom{xx}  $1.3\e{-6}$  & $4.8\e{-5}$ &\phantom{xx}  $4.2\e{-6}$ & $3.7\e{-6}$ &\phantom{xx}  $2.9\e{-7}$ & \fbox{$1.2\e{-3}$}  &\phantom{xx}  \fbox{$5.4\e{-5}$} \\
$\mu e$~[pb]        & $2.8\e{-5}$ &\phantom{xx}  $3.5\e{-6}$  & $1.7\e{-5}$ &\phantom{xx}  $1.5\e{-6}$ & $8\e{-6}$ &\phantom{xx}  $6.3\e{-7}$ & $1.3\e{-4}$  &\phantom{xx}  $5.9\e{-6}$ \\
$e e$~[pb]        & $2.4\e{-9}$ &\phantom{xx}  $3.0\e{-10}$  & $9.6\e{-8}$ &\phantom{xx}  $8.4\e{-9}$ & $1.1\e{-9}$ &\phantom{xx}  $8.8\e{-11}$ & $6.9\e{-8}$ & $3.0\e{-9}$  \\
$\gamma \gamma$~[pb]      & $1.2\e{-6}$ & $5.6\e{-8}$  & $2.3\e{-8}$ &   $7.2\e{-10}$      & $2.5\e{-9}$  & $7.2\e{-11}$ & \fbox{$2.7\e{-2}$} & \fbox{$1\e{-3}$}\\
$b  \bar{b}$~[pb]      & $0.15$ & \fbox{$1.7\e{-2}$}  & $1.7\e{-2}$ & $1.3\e{-3}$ & $2.7\e{-3}$ & $2\e{-4}$ & $1.03$ & \fbox{$4.1\e{-2}$}  \\
$t  c$~[pb]      & $5.4\e{-2}$  & $7.6\e{-3}$ & $9.2\e{-4}$ & $8.9\e{-5}$ &$ 3.2\e{-2}$ & $ 3\e{-3}$   & $8.26$ & $0.39$   \\
$t  u$~[pb]      & $9.8\e{-7}$  & $1.4\e{-7}$  & $2.8\e{-8}$ & $2.7\e{-9}$ &$ 1.4\e{-2}$ &  $ 1.3\e{-3}$ & $ 6.2\e{-2}$ & $2.9\e{-3}$   \\
$t  \bar{t}$~[pb]      &   &  &  & & &   & \fbox{$ 241.4$} & \fbox{$15.4$ }  \\
\bottomrule
\end{tabular}
\caption{Benchmark points for different  $\mathcal{Z}_N \times \mathcal{Z}_{M}$  flavor symmetries for inclusive flavon production channels with high flavon mass ($m_a$) in case of the soft symmetry-breaking scenario at a 100 TeV hadron collider, assuming $f = 500$ GeV.}
\label{tab:limits_bench100a}
\end{table}

The LHC is actively searching for  light pseudoscalar particles.  For a light flavon, present sensitivities  of the LHC  are given in table  \ref{tab:limits_lighta}. Using these sensitivities given in table  \ref{tab:limits_lighta}, the estimated sensitivities for the HL-LHC, HE-LHC, and a 100 TeV collider are given in table \ref{tab:futurelimits_lighta}.  The benchmark predictions for the processes given in table \ref{tab:limits_lighta}, for different $\mathcal{Z}_{\rm N} \times \mathcal{Z}_{\rm M}$ flavor symmetries, are given in tables  \ref{tab:limits_bench_light14a}- \ref{tab:limits_bench_light100a} for  $m_a= 20,60$ GeV for the soft symmetry-breaking scenario.

\begin{table}[H]
\centering
\begin{tabular}{l|cc|cc|cc}
\toprule
& \multicolumn{2}{c|}{$\mathcal{L}[fb^{-1}]$ [References]} & \multicolumn{2}{c|}{ATLAS 13 TeV} & \multicolumn{2}{c}{CMS 13 TeV}  \\
$m_{a}$~[GeV] &  \myalign{c}{ATLAS} & \myalign{c|}{CMS} & \myalign{c}{20} & \myalign{c|}{60} & \myalign{c}{20} & \myalign{c}{60} \\
\midrule
$\tau \tau$~[pb]  &  & 
 $138$ \cite{CMS:2022goy}  &  &   &  & $4$    \\
$\gamma \gamma$~[pb]& $138$ \cite{ATLAS:2022abz,ATLAS:95}  & \cite{CMS:2023yay}  & $6$ & $7$  &  &    \\
\bottomrule
\end{tabular}
\caption{Current limits of $\sigma \times BR$ at 13 TeV LHC by ATLAS and CMS in low mass resonance searches for inclusive flavon production channels. }
\label{tab:limits_lighta}
\end{table}

 \begin{table}[H]
\setlength{\tabcolsep}{5pt} 
\renewcommand{\arraystretch}{0.7} 
\centering
\begin{tabular}{l|cc|cc|cc}
\toprule
 & \multicolumn{2}{c|}{HL-LHC [14 TeV, $3~\iab$] } & \multicolumn{2}{c|}{HE-LHC [27 TeV, $15~\iab$]} & \multicolumn{2}{c}{100 TeV, $30~\iab$} \\
$m_{a}$~[GeV] &  20 &  60 &  20 &  60 &  20 &  60 \\
\midrule
$\tau \tau$ [pb]  & & $0.9$ & & $0.5$ &  & $0.6$  \\
$\gamma \gamma$ [pb] & $1.3$  & $1.5$  & $0.7$ & $0.8$  & $0.8$ & $0.9$  \\
\bottomrule
\end{tabular}
\caption{Estimated reach ($\sigma \times BR$) of the HL-LHC, HE-LHC and a 100 TeV collider for low flavon mass ($m_a$) in inclusive flavon production channels.}
\label{tab:futurelimits_lighta}
\end{table}


We observe from table \ref{tab:limits_bench_light14a} that the HL-LHC provides access only to probe the $\mathcal{Z}_8 \times \mathcal{Z}_{22}$ flavor symmetry through the $\tau \tau$ channel for the mass $m_a= 60$ GeV.  The situation changes greatly as we approach the HE-LHC and a 100 TeV hadron collider, as shown in tables \ref{tab:limits_bench_light27a} and \ref{tab:limits_bench_light100a}.  Now, the  $\tau \tau$ channel is accessible for all $\mathcal{Z}_{\rm N} \times \mathcal{Z}_{\rm M}$ flavor symmetries.  The $\gamma \gamma$ channel and the flavon mass $m_a= 20$ GeV remain beyond the reach of the HL-LHC, HE-LHC and a 100 TeV collider for all flavor symmetries.

\begin{table}[H]
\setlength{\tabcolsep}{6pt} 
\renewcommand{\arraystretch}{1.1} 
\centering
\begin{tabular}{@{}l|rr|rr|rr|rr@{}}
\toprule
 & \multicolumn{2}{c|}{Benchmark} & \multicolumn{2}{c|}{Benchmark} & \multicolumn{2}{c|}{Benchmark} & \multicolumn{2}{c}{Benchmark}\\
 & \multicolumn{2}{c|}{$\mathcal{Z}_2 \times \mathcal{Z}_{5}$} & \multicolumn{2}{c|}{$\mathcal{Z}_2 \times \mathcal{Z}_{9}$} & \multicolumn{2}{c|}{$\mathcal{Z}_2 \times \mathcal{Z}_{11}$} & \multicolumn{2}{c}{$\mathcal{Z}_8 \times \mathcal{Z}_{22}$}\\
 $m_{a}$~[GeV] & \myalign{c}{20} & \myalign{c|}{60} & \myalign{c}{20} & \myalign{c|}{60} & \myalign{c}{20} & \myalign{c|}{60} &\myalign{c}{20} & \myalign{c}{60}\\
\midrule
$\tau \tau$~[pb]      & $4.7\e{-3}$ & $0.58$  & $1.1\e{-2}$ & $0.62$ & $2.2\e{-2}$ & $0.69$  & $7.7\e{-3}$ & \fbox{$9.62$}  \\
$\gamma \gamma$~[pb]      & $5.6\e{-4}$ & $1\e{-3}$  & $1.5\e{-4}$ & $9.9\e{-5}$ & $1.1\e{-4}$ & $5.8\e{-5}$ & $9\e{-5}$ & $1.7\e{-2}$  \\
\bottomrule
\end{tabular}
\caption{Benchmark points for different $\mathcal{Z}_N \times \mathcal{Z}_{M}$ flavor symmetries for inclusive flavon production channels with low flavon mass ($m_a$) in case of the soft symmetry-breaking scenario at the 14 TeV HL-LHC  , assuming $f= 500$ GeV.}
\label{tab:limits_bench_light14a}
\end{table}

\begin{table}[H]
\setlength{\tabcolsep}{6pt} 
\renewcommand{\arraystretch}{1.1} 
\centering
\begin{tabular}{@{}l|rr|rr|rr|rr@{}}
\toprule
 & \multicolumn{2}{c|}{Benchmark} & \multicolumn{2}{c|}{Benchmark} & \multicolumn{2}{c|}{Benchmark} & \multicolumn{2}{c}{Benchmark}\\
 & \multicolumn{2}{c|}{$\mathcal{Z}_2 \times \mathcal{Z}_{5}$} & \multicolumn{2}{c|}{$\mathcal{Z}_2 \times \mathcal{Z}_{9}$} & \multicolumn{2}{c|}{$\mathcal{Z}_2 \times \mathcal{Z}_{11}$} & \multicolumn{2}{c}{$\mathcal{Z}_8 \times \mathcal{Z}_{22}$}\\
 $m_{a}$~[GeV] & \myalign{c}{20} & \myalign{c|}{60} & \myalign{c}{20} & \myalign{c|}{60} & \myalign{c}{20} & \myalign{c|}{60} &\myalign{c}{20} & \myalign{c}{60}\\
\midrule
$\tau \tau$~[pb]      & $8\e{-3}$ & \fbox{$1.06$}  & $2.2\e{-2}$ & \fbox{$1.38$} & $4.2\e{-2}$ & \fbox{$1.55$}  & $1.5\e{-2}$ & \fbox{$21.78$}  \\
$\gamma \gamma$~[pb]      & $9.4\e{-4}$ & $1.8\e{-3}$  & $2.9\e{-4}$ & $2.2\e{-4}$ & $2.2\e{-4}$ & $1.3\e{-4}$ & $1.7\e{-4}$ & $3.8\e{-2}$ \\
\bottomrule
\end{tabular}
\caption{Benchmark points for different $\mathcal{Z}_N \times \mathcal{Z}_{M}$ flavor symmetries for inclusive flavon production channels with low flavon mass ($m_a$) in case of the soft symmetry-breaking scenario at the 27 TeV HE-LHC, assuming $f= 500$ GeV.}
\label{tab:limits_bench_light27a}
\end{table}

\begin{table}[H]
\setlength{\tabcolsep}{6pt} 
\renewcommand{\arraystretch}{1.1} 
\centering
\begin{tabular}{@{}l|rr|rr|rr|rr@{}}
\toprule
 & \multicolumn{2}{c|}{Benchmark} & \multicolumn{2}{c|}{Benchmark} & \multicolumn{2}{c|}{Benchmark} & \multicolumn{2}{c}{Benchmark}\\
 & \multicolumn{2}{c|}{$\mathcal{Z}_2 \times \mathcal{Z}_{5}$} & \multicolumn{2}{c|}{$\mathcal{Z}_2 \times \mathcal{Z}_{9}$} & \multicolumn{2}{c|}{$\mathcal{Z}_2 \times \mathcal{Z}_{11}$} & \multicolumn{2}{c}{$\mathcal{Z}_8 \times \mathcal{Z}_{22}$}\\
 $m_{a}$~[GeV] & \myalign{c}{20} & \myalign{c|}{60} & \myalign{c}{20} & \myalign{c|}{60} & \myalign{c}{20} & \myalign{c|}{60} &\myalign{c}{20} & \myalign{c}{60}\\
\midrule
$\tau \tau$~[pb]      & $2.2\e{-2}$ & \fbox{$3.29$}  & $7.5\e{-2}$ & \fbox{$5.72$} & $0.15$ & \fbox{$6.43$}  & $5.1\e{-2}$ & \fbox{$92.23$}  \\
$\gamma \gamma$~[pb]      & $2.6\e{-3}$ & $5.7\e{-3}$  & $9.5\e{-4}$ & $9.1\e{-4}$ & $7.5\e{-4}$ & $5.3\e{-4}$ & $5.5\e{-4}$ & $0.16$  \\
\bottomrule
\end{tabular}
\caption{Benchmark points for different $\mathcal{Z}_N \times \mathcal{Z}_{M}$ flavor symmetries for inclusive flavon production channels with low flavon mass ($m_a$) in case of the soft symmetry-breaking scenario at a 100 TeV collider, assuming $f= 500$ GeV.}
\label{tab:limits_bench_light100a}
\end{table}

\begin{table}[H]
\setlength{\tabcolsep}{6pt} 
\renewcommand{\arraystretch}{1.1} 
\centering
\begin{tabular}{@{}l|rr|rr|rr@{}}
\toprule
 & \multicolumn{2}{c|}{Benchmark} & \multicolumn{2}{c|}{Benchmark} & \multicolumn{2}{c}{Benchmark} \\
 & \multicolumn{2}{c|}{$\mathcal{Z}_2 \times \mathcal{Z}_{5}$} & \multicolumn{2}{c|}{$\mathcal{Z}_2 \times \mathcal{Z}_{9}$} & \multicolumn{2}{c}{$\mathcal{Z}_2 \times \mathcal{Z}_{11}$}\\
 $m_{a}$~[GeV] & \myalign{c}{20} & \myalign{c|}{60} & \myalign{c}{20} & \myalign{c|}{60} & \myalign{c}{20} & \myalign{c}{60}\\
\midrule
$\tau \tau$~[pb]      & $1.9\e{-7}$ & $5.1\e{-4}$  & $1.1\e{-11}$ & $2\e{-6}$ & $2.4\e{-13}$ & $3.3\e{-7}$   \\
$\gamma \gamma$~[pb]      & $3.8\e{-7}$ & $8.8\e{-7}$  & $4.5\e{-10}$ & $3.2\e{-10}$ & $3.8\e{-11}$ & $2.7\e{-11}$   \\
\bottomrule
\end{tabular}
\caption{Benchmark points for different $\mathcal{Z}_N \times \mathcal{Z}_{M}$ flavor symmetries for inclusive flavon production channels with low flavon mass ($m_a$) in case of the symmetry-conserving scenario at the 14 TeV HL-LHC.}
\label{tab:limits_bench_sc14a}
\end{table}

In the symmetry-conserving case, we observe from figures  \ref{fig_fprod1}  and  \ref{fig_fprod2} that the cross-section can be relatively significant only for a low flavon mass.  Therefore, we investigate the symmetry-conserving scenario only for $m_a= 20, 60$ GeV.  We show benchmark predictions of symmetry-conserving scenario for different $\mathcal{Z}_{\rm N} \times \mathcal{Z}_{\rm M}$ flavor symmetries in tables  \ref{tab:limits_bench_sc14a}- \ref{tab:limits_bench_sc100a} for  $m_a= 20,60$ GeV. It turns out that the inclusive processes in this scenario are beyond the reach of the HL-LHC, HE-LHC, and even a 100 TeV collider.

\begin{table}[H]
\setlength{\tabcolsep}{6pt} 
\renewcommand{\arraystretch}{1.1} 
\centering
\begin{tabular}{@{}l|rr|rr|rr@{}}
\toprule
 & \multicolumn{2}{c|}{Benchmark} & \multicolumn{2}{c|}{Benchmark} & \multicolumn{2}{c}{Benchmark} \\
 & \multicolumn{2}{c|}{$\mathcal{Z}_2 \times \mathcal{Z}_{5}$} & \multicolumn{2}{c|}{$\mathcal{Z}_2 \times \mathcal{Z}_{9}$} & \multicolumn{2}{c}{$\mathcal{Z}_2 \times \mathcal{Z}_{11}$}\\
 $m_{a}$~[GeV] & \myalign{c}{20} & \myalign{c|}{60} & \myalign{c}{20} & \myalign{c|}{60} & \myalign{c}{20} & \myalign{c}{60}\\
\midrule
$\tau \tau$~[pb]      & $2\e{-7}$ & $9.2\e{-4}$  & $1.8\e{-11}$ & $4.5\e{-6}$ & $4\e{-13}$ & $7.3\e{-7}$   \\
$\gamma \gamma$~[pb]      & $6.3\e{-7}$ & $1.6\e{-6}$  & $8.8\e{-10}$ & $7.2\e{-10}$ & $7.4\e{-11}$ & $6\e{-11}$   \\
\bottomrule
\end{tabular}
\caption{Benchmark points for different $\mathcal{Z}_N \times \mathcal{Z}_{M}$ flavor symmetries for inclusive flavon production channels with low flavon mass ($m_a$) in case of the symmetry-conserving scenario at the 27 TeV HE-LHC.}
\label{tab:limits_bench_sc27a}
\end{table}

\begin{table}[H]
\setlength{\tabcolsep}{6pt} 
\renewcommand{\arraystretch}{1.1} 
\centering
\begin{tabular}{@{}l|rr|rr|rr@{}}
\toprule
 & \multicolumn{2}{c|}{Benchmark} & \multicolumn{2}{c|}{Benchmark} & \multicolumn{2}{c}{Benchmark} \\
 & \multicolumn{2}{c|}{$\mathcal{Z}_2 \times \mathcal{Z}_{5}$} & \multicolumn{2}{c|}{$\mathcal{Z}_2 \times \mathcal{Z}_{9}$} & \multicolumn{2}{c}{$\mathcal{Z}_2 \times \mathcal{Z}_{11}$}\\
 $m_{a}$~[GeV] & \myalign{c}{20} & \myalign{c|}{60} & \myalign{c}{20} & \myalign{c|}{60} & \myalign{c}{20} & \myalign{c}{60}\\
\midrule
$\tau \tau$~[pb]      & $2.7\e{-10}$ & $2.8\e{-3}$  & $4.9\e{-11}$ & $1.9\e{-5}$ & $1.1\e{-12}$ & $3\e{-6}$   \\
$\gamma \gamma$~[pb]      & $1.7\e{-6}$ & $4.9\e{-6}$  & $2.8\e{-9}$ & $2.9\e{-9}$ & $2.4\e{-10}$ & $2.5\e{-10}$   \\
\bottomrule
\end{tabular}
\caption{Benchmark points for different $\mathcal{Z}_N \times \mathcal{Z}_{M}$ flavor symmetries for inclusive flavon production channels with low flavon mass ($m_a$) in case of the symmetry-conserving scenario at a 100 TeV collider.}
\label{tab:limits_bench_sc100a}
\end{table}


\subsubsection{Associative production}
\begin{figure}[h!]
\centering
 \begin{subfigure}[]{0.32\linewidth}
 \includegraphics[width=\linewidth]{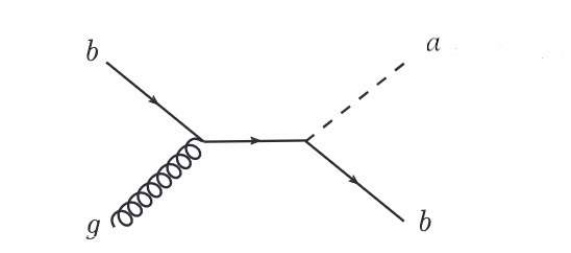}
 \caption{}
         \label{fprod14_z2z11}
 \end{subfigure} 
 \begin{subfigure}[]{0.32\linewidth}
 \includegraphics[width=\linewidth]{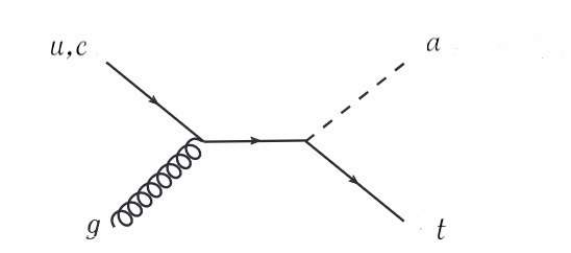}
 \caption{}
         \label{fprod14_z2z11}
 \end{subfigure} \\
 \begin{subfigure}[]{0.32\linewidth}
 \includegraphics[width=\linewidth]{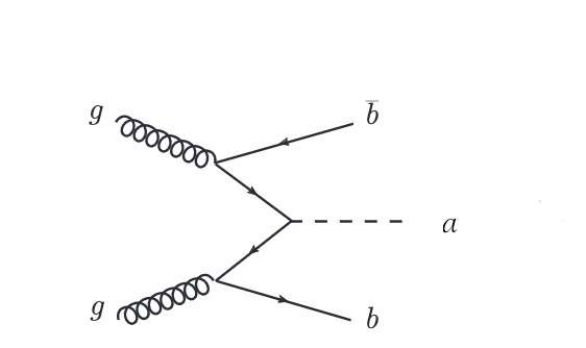}
 \caption{}
         \label{fprod100_z2z11}
 \end{subfigure} 
\begin{subfigure}[]{0.32\linewidth}
 \includegraphics[width=\linewidth]{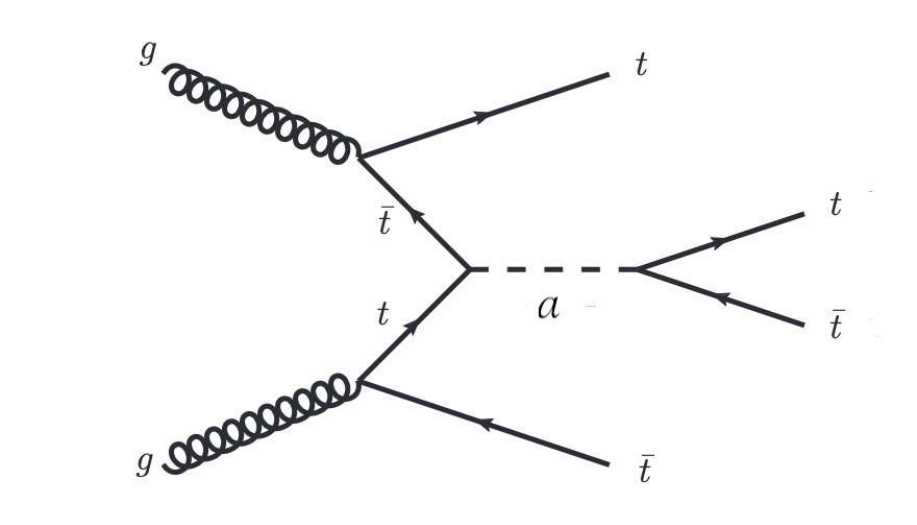}
 \caption{}
         \label{fprod14_z8z22}
 \end{subfigure} 
 \caption{Feynman diagrams for the production of the flavon through different associative channels.}
 \label{assoc_prod}
	\end{figure}

 We notice that there are other channels through which flavon can be produced, as shown in figure \ref{assoc_prod}.  For instance, the following associated production processes can be used to produce the flavon,
\be
bg \rightarrow b a,~ ug, c g  \rightarrow t a,  b\bar{b} a \rightarrow  b\bar{b} \tau^+ \tau^-,  t\bar{t} a \rightarrow  t\bar{t}  t\bar{t},
\ee
where the first process undergoes through the  flavor-diagonal coupling of the flavon to the $b\bar{b}$, and the second process occurs through the flavor-violating flavon-quark coupling.

\begin{figure}[h!]
	\centering
	\begin{subfigure}[]{0.32\linewidth}
    \includegraphics[width=\linewidth]{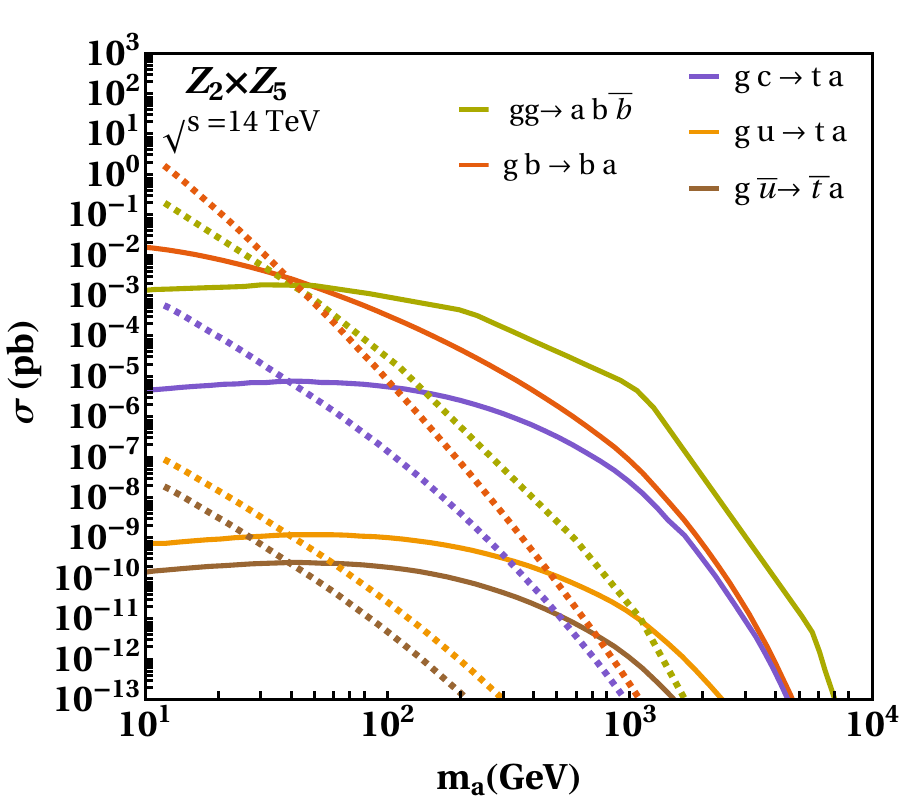}
    \caption{}
         \label{fprod14_z2z5}	
\end{subfigure}
 \begin{subfigure}[]{0.32\linewidth}
 \includegraphics[width=\linewidth]{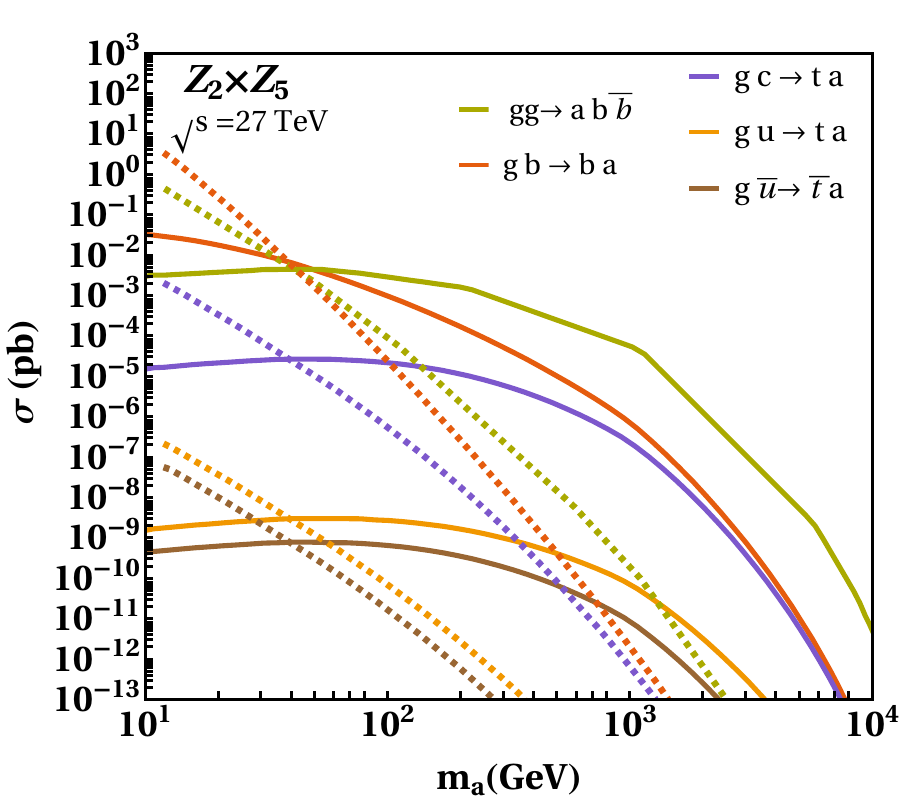}
 \caption{}
         \label{fprod27_z2z5}
 \end{subfigure} 
 \begin{subfigure}[]{0.32\linewidth}
 \includegraphics[width=\linewidth]{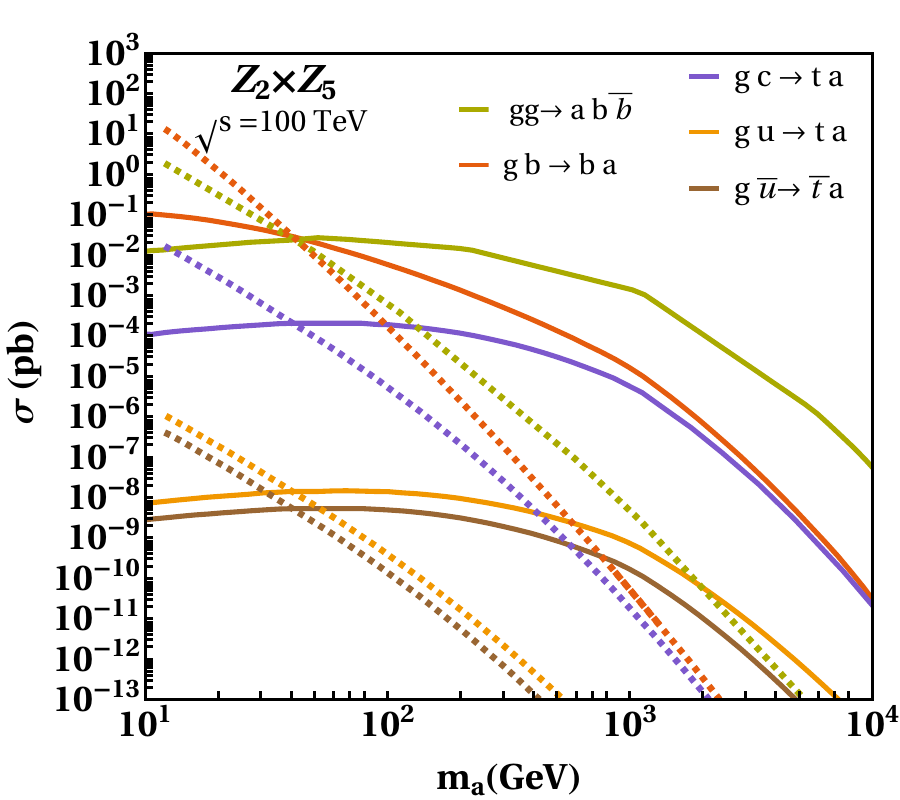}
 \caption{}
         \label{fprod100_z2z5}
 \end{subfigure} 
 \begin{subfigure}[]{0.32\linewidth}
    \includegraphics[width=\linewidth]{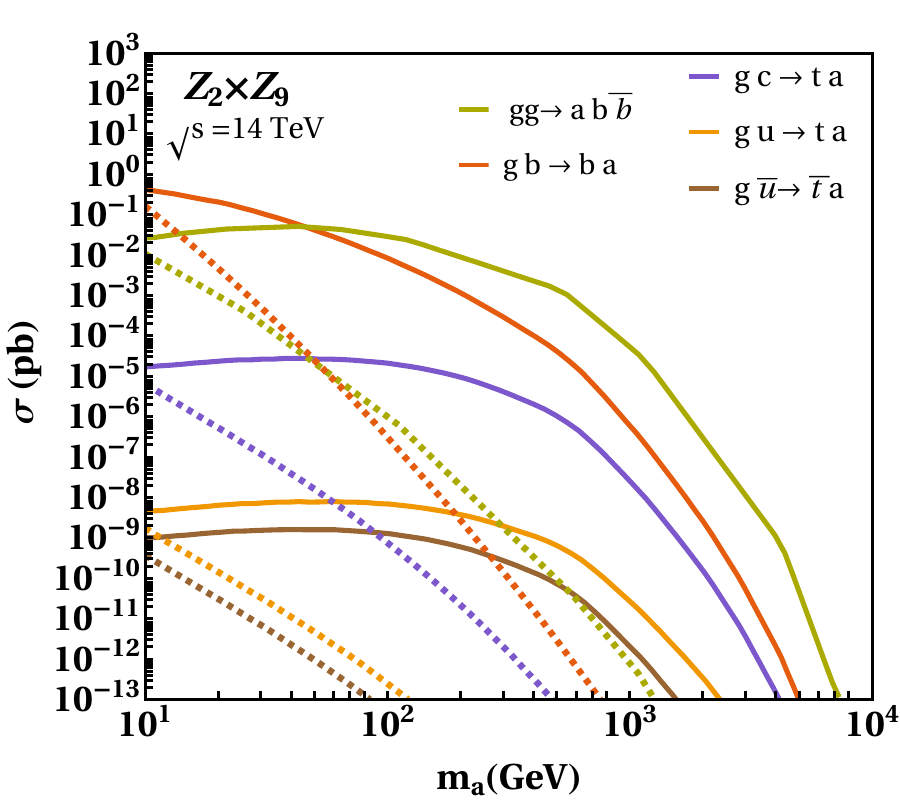}
    \caption{}
         \label{fprod14_z2z5}	
\end{subfigure}
 \begin{subfigure}[]{0.32\linewidth}
 \includegraphics[width=\linewidth]{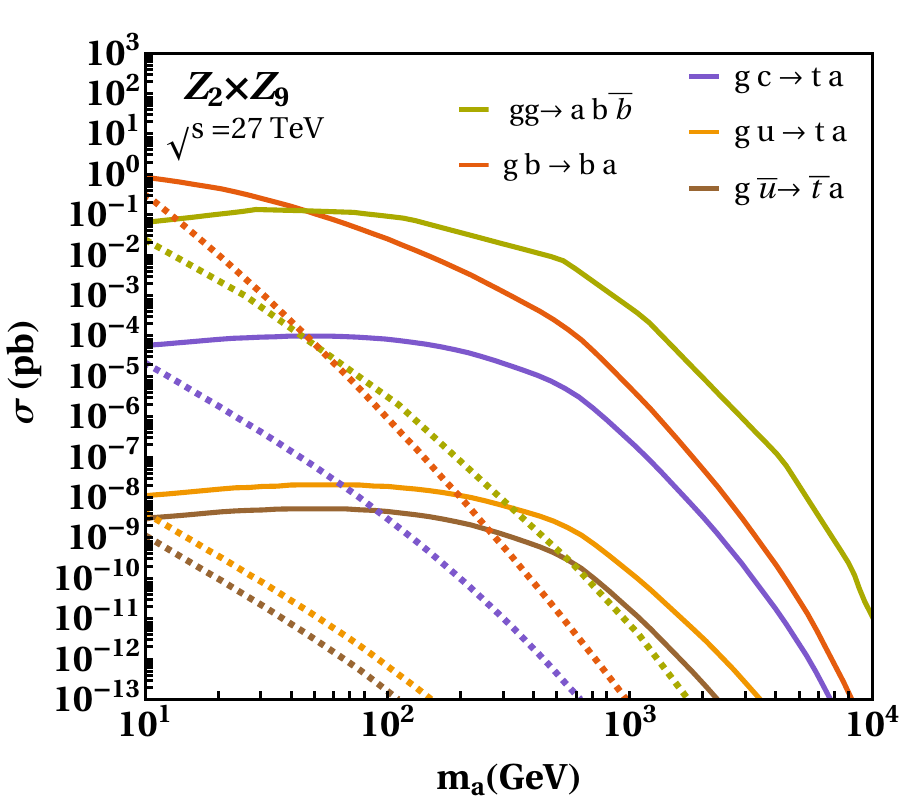}
 \caption{}
         \label{fprod27_z2z5}
 \end{subfigure} 
 \begin{subfigure}[]{0.32\linewidth}
 \includegraphics[width=\linewidth]{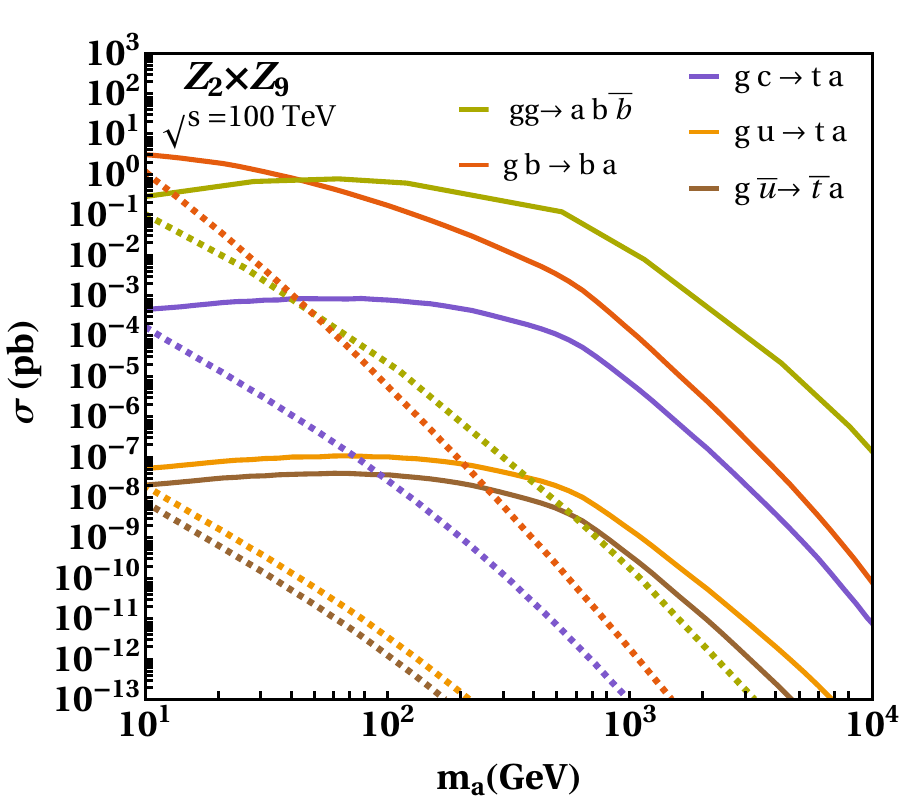}
 \caption{}
         \label{fprod100_z2z5}
 \end{subfigure} 
 \caption{Production cross-sections of the flavon of  $\mathcal{Z}_2 \times \mathcal{Z}_{5}$, and $\mathcal{Z}_2 \times \mathcal{Z}_{9}$ flavor symmetries with respect to its mass through different channels for 14 TeV HL-LHC, 27 TeV HE-LHC and future 100 TeV collider. The solid lines represents the production cross section along the boundaries of parameter space allowed by the observable $R_{\mu \mu}$ for soft symmetry-breaking scenario, while the dashed lines correspond to that in the symmetry-conserving scenario along the allowed parameter space by the observable $BR(K_L\rightarrow \mu \mu)_{SD}$. }
  \label{assoc_prod1}
	\end{figure}

\begin{figure}[h!]
	\centering
	\begin{subfigure}[]{0.322\linewidth}
    \includegraphics[width=\linewidth]{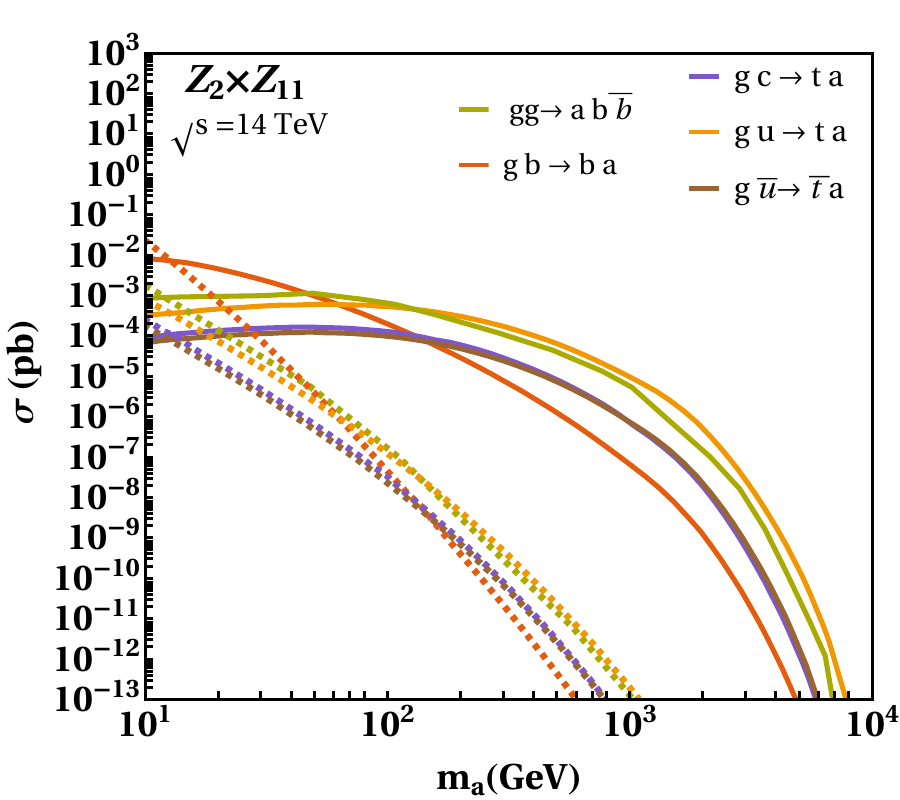}
    \caption{}
         \label{fprod14_z2z5}	
\end{subfigure}
 \begin{subfigure}[]{0.322\linewidth}
 \includegraphics[width=\linewidth]{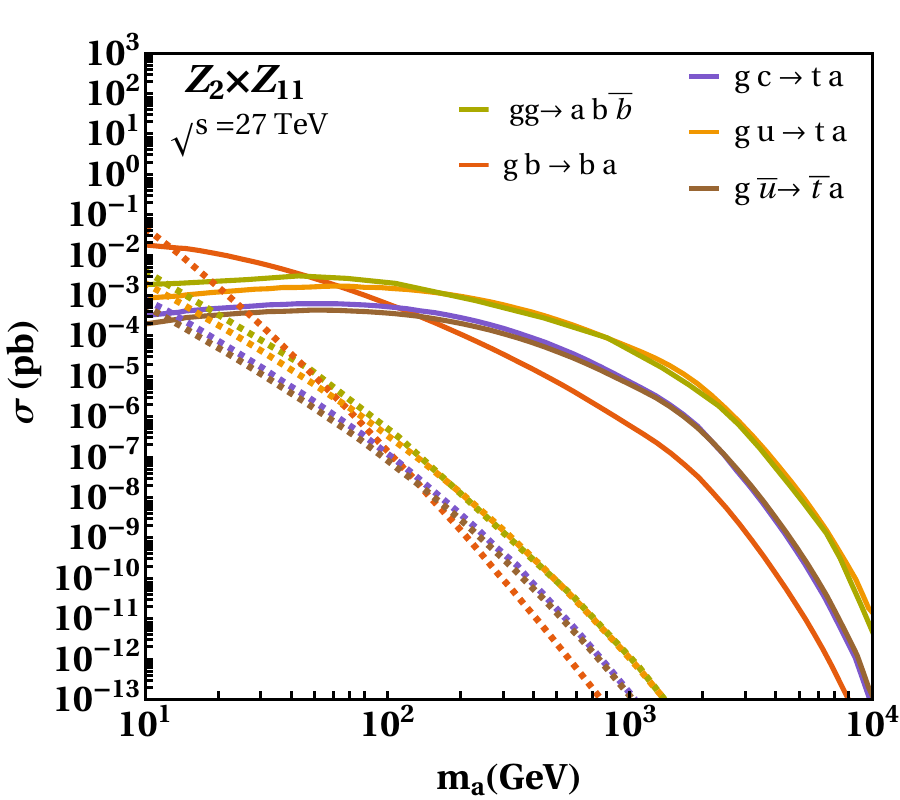}
 \caption{}
         \label{fprod27_z2z5}
 \end{subfigure} 
 \begin{subfigure}[]{0.322\linewidth}
 \includegraphics[width=\linewidth]{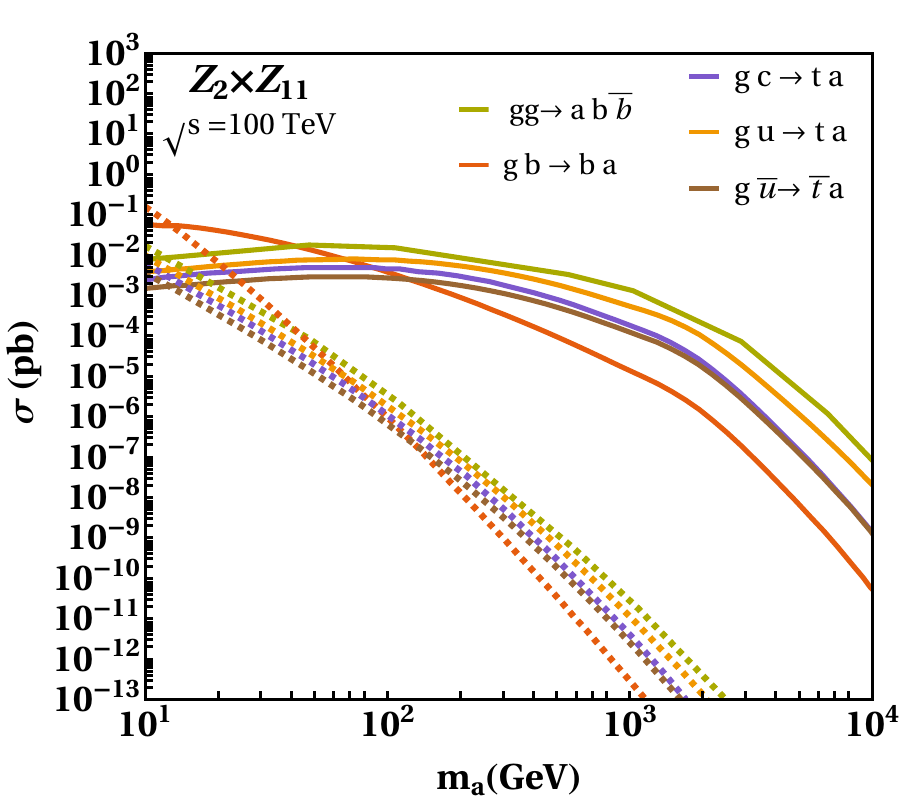}
 \caption{}
         \label{fprod100_z2z5}
 \end{subfigure} 
 \begin{subfigure}[]{0.322\linewidth}
    \includegraphics[width=\linewidth]{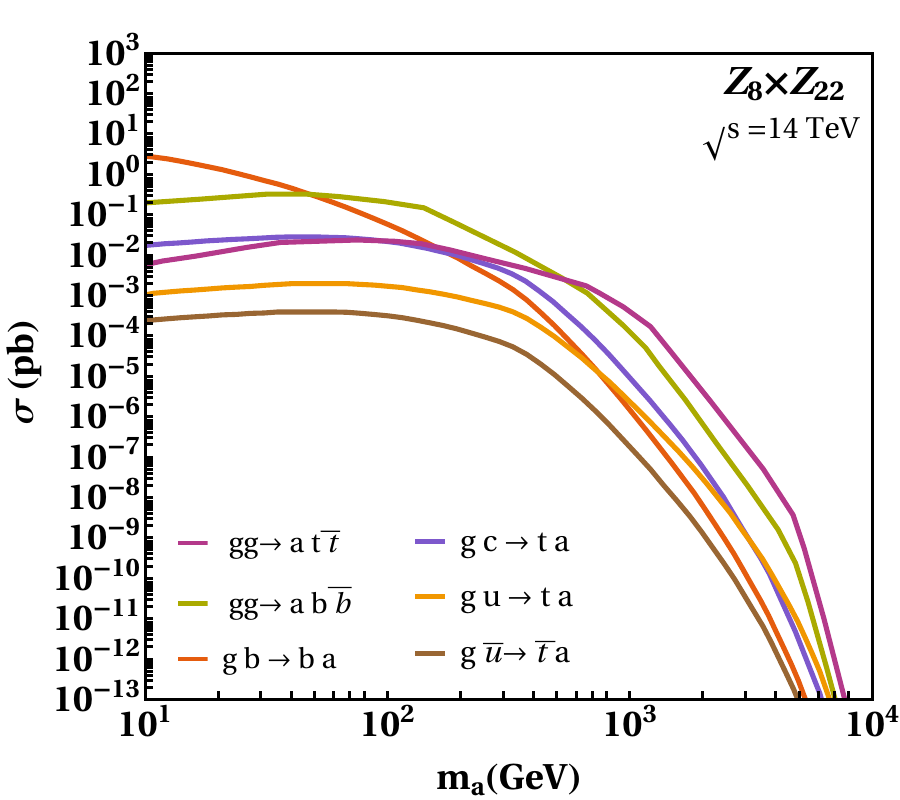}
    \caption{}
         \label{fprod14_z2z5}	
\end{subfigure}
 \begin{subfigure}[]{0.322\linewidth}
 \includegraphics[width=\linewidth]{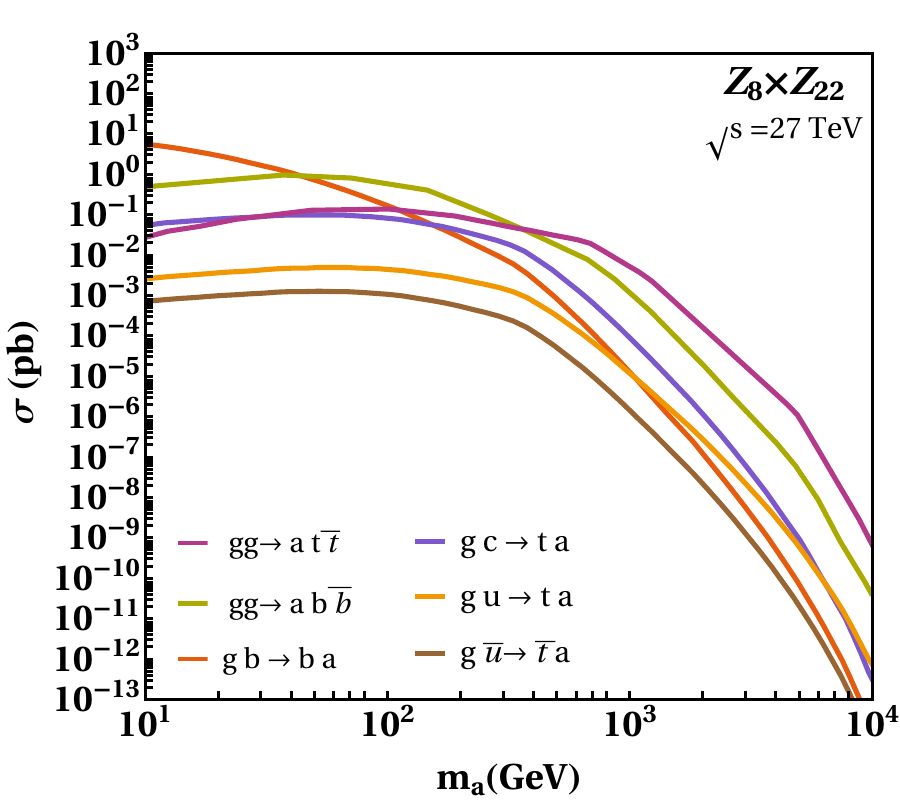}
 \caption{}
         \label{fprod27_z2z5}
 \end{subfigure} 
 \begin{subfigure}[]{0.327\linewidth}
 \includegraphics[width=\linewidth]{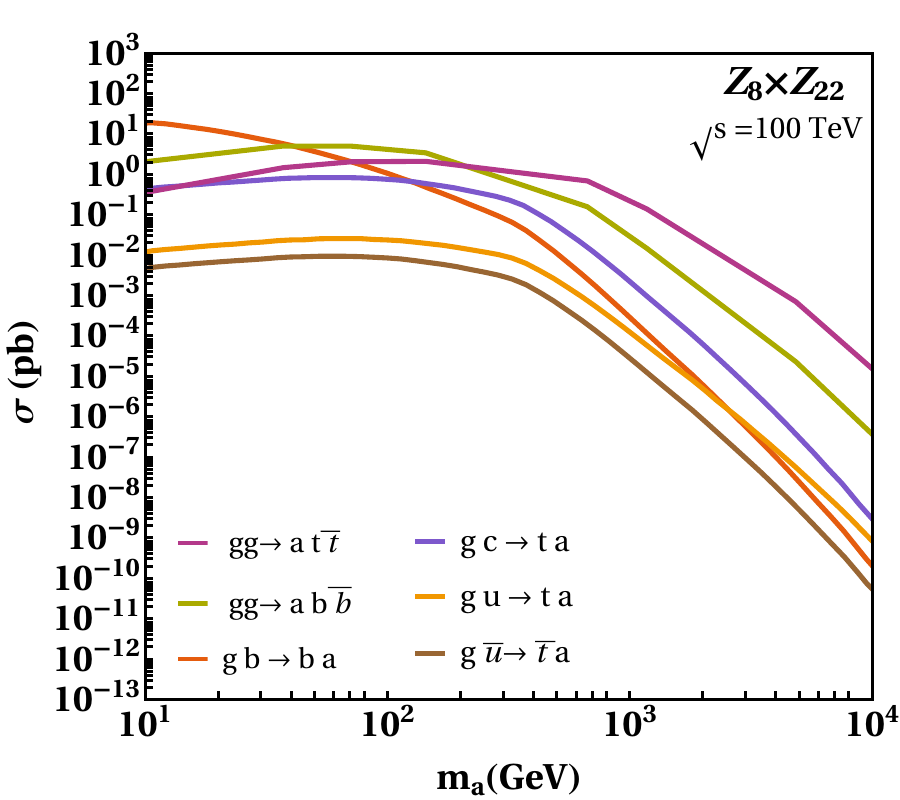}
 \caption{}
         \label{fprod100_z2z5}
 \end{subfigure} 
 \caption{Production cross-sections of the flavon of  $\mathcal{Z}_2 \times \mathcal{Z}_{11}$, and $\mathcal{Z}_8 \times \mathcal{Z}_{22}$ flavor symmetries with respect to its mass through different channels for 14 TeV HL-LHC, 27 TeV HE-LHC and future 100 TeV collider. The solid lines represents the production cross section along the boundaries of parameter space allowed by the observable $R_{\mu \mu}$ for soft symmetry-breaking scenario, while the dashed lines correspond to that in the symmetry-conserving scenario along the allowed parameter space by the observable $BR(K_L\rightarrow \mu \mu)_{SD}$. }
  \label{assoc_prod2}
	\end{figure}

The cross-sections of different associative production channels are also shown in figures \ref{assoc_prod1} and \ref{assoc_prod2}.  We observe that the production cross sections are, in general, smaller for the soft symmetry-breaking as well as symmetry-conserving scenarios, and can be relatively large only for low flavon mass.  

The process $t  \bar{t} a \rightarrow t  \bar{t} t  \bar{t} $ can occur only for the  $\mathcal{Z}_8 \times \mathcal{Z}_{22}$ flavor symmetry where  the flavor-diagonal coupling of the flavon to top quarks is allowed.  This results in a relatively large cross-section involving  $ t  \bar{t} $ channels. Thus, the $\mathcal{Z}_8 \times \mathcal{Z}_{22}$ flavor symmetry is very special in this sense.

\begin{figure}[h!]
	\centering
	\begin{subfigure}[]{0.32\linewidth}
    \includegraphics[width=\linewidth]{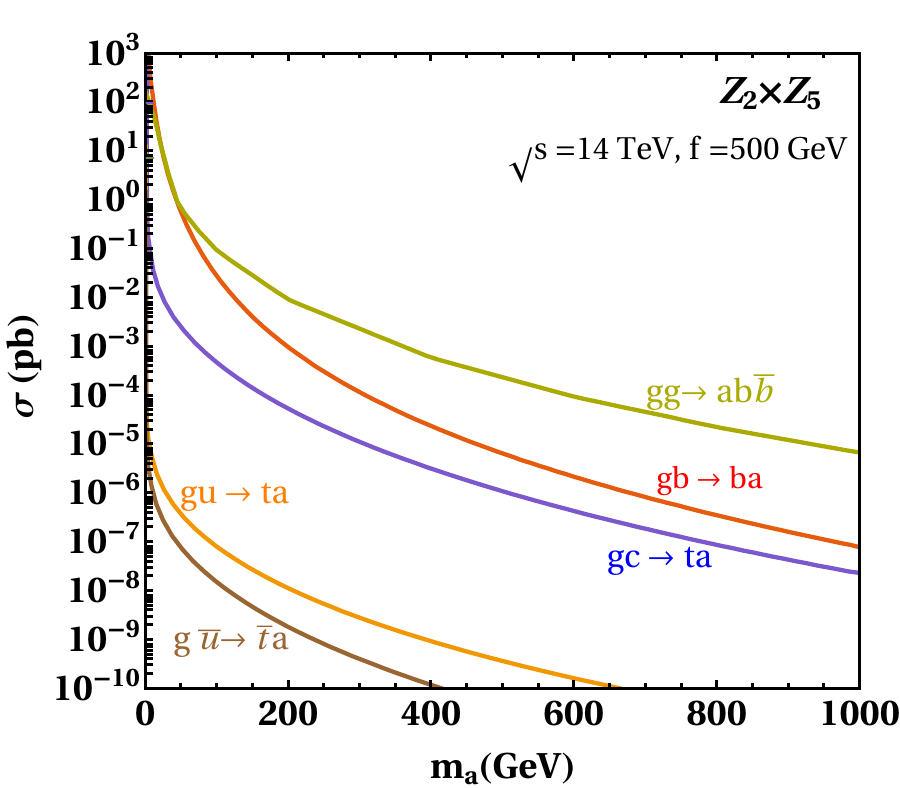}
    \caption{}
         \label{fprod14_z2z5}	
\end{subfigure}
 \begin{subfigure}[]{0.32\linewidth}
 \includegraphics[width=\linewidth]{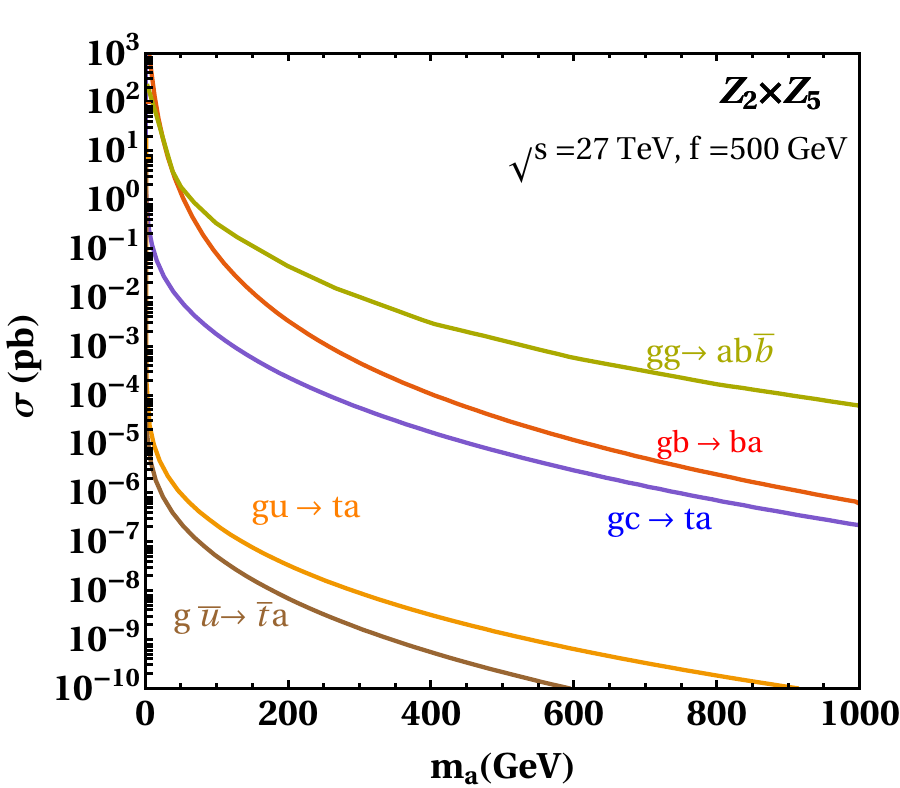}
 \caption{}
         \label{fprod27_z2z5}
 \end{subfigure} 
 \begin{subfigure}[]{0.32\linewidth}
 \includegraphics[width=\linewidth]{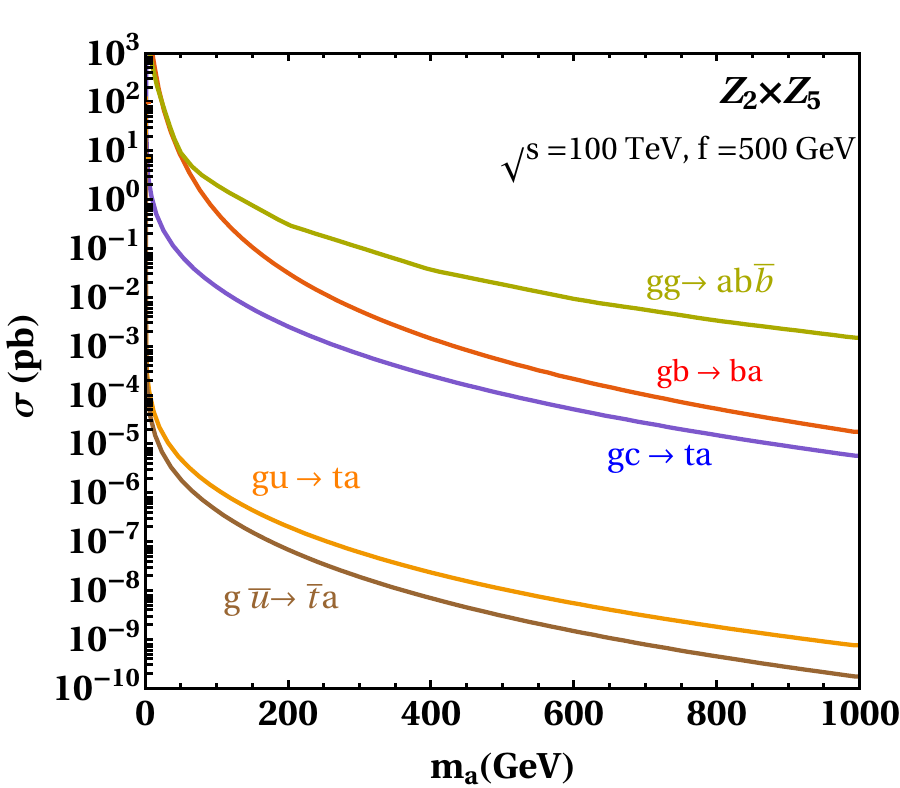}
 \caption{}
         \label{fprod100_z2z5}
 \end{subfigure} 
 \begin{subfigure}[]{0.32\linewidth}
    \includegraphics[width=\linewidth]{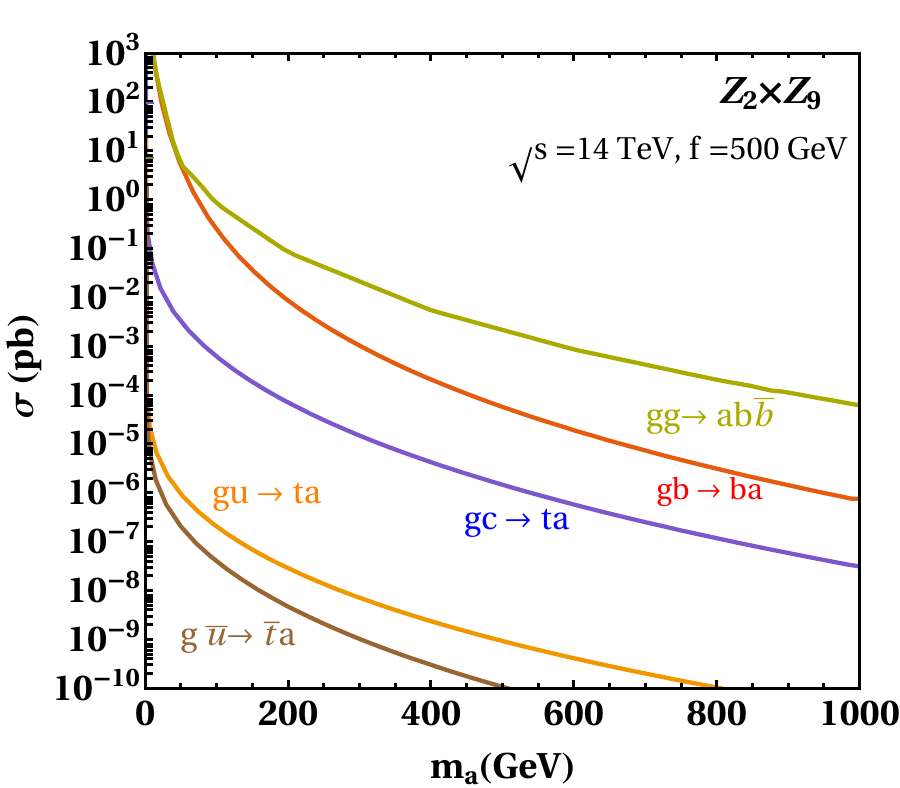}
    \caption{}
         \label{fprod14_z2z5}	
\end{subfigure}
 \begin{subfigure}[]{0.32\linewidth}
 \includegraphics[width=\linewidth]{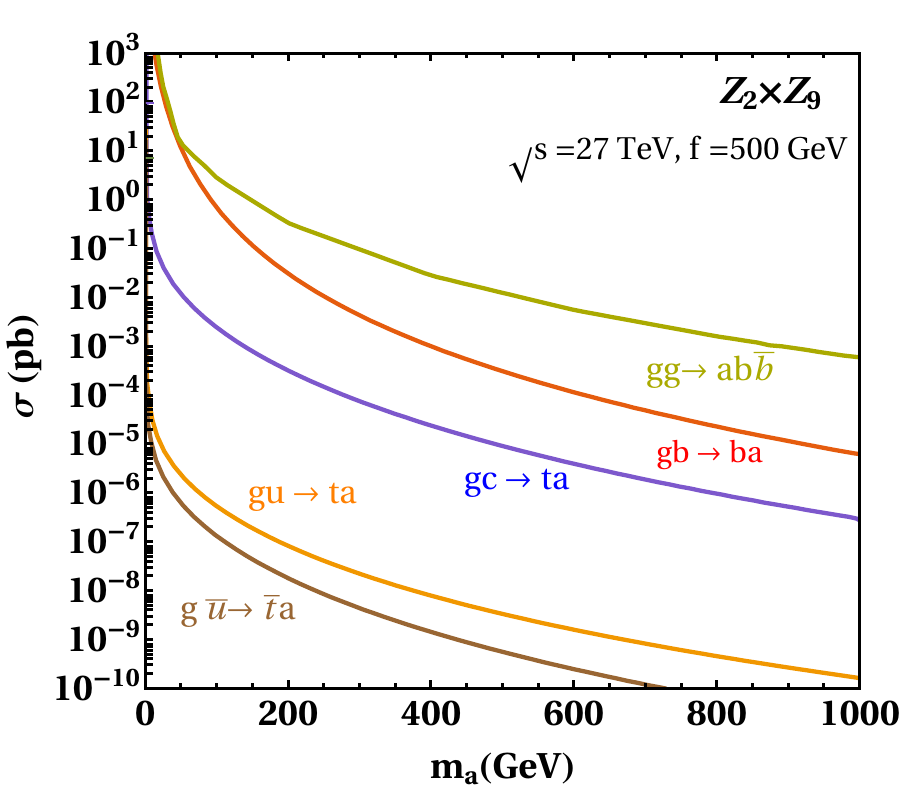}
 \caption{}
         \label{fprod27_z2z5}
 \end{subfigure} 
 \begin{subfigure}[]{0.32\linewidth}
 \includegraphics[width=\linewidth]{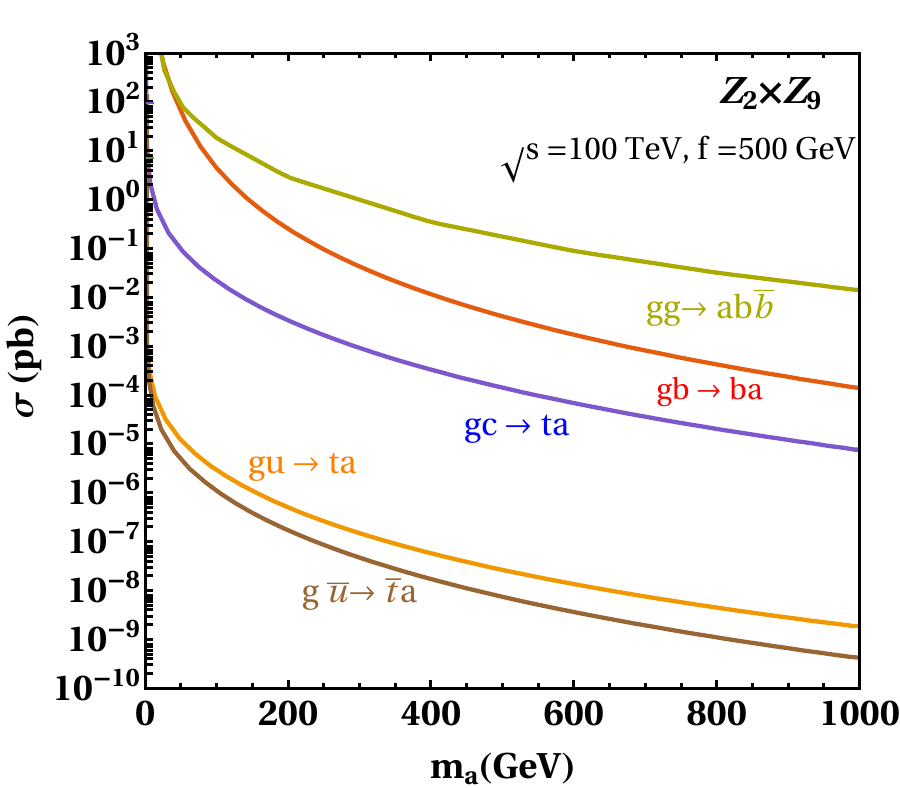}
 \caption{}
         \label{fprod100_z2z5}
 \end{subfigure} 
 \caption{Production cross-sections of the flavon of  $\mathcal{Z}_2 \times \mathcal{Z}_{5}$, and $\mathcal{Z}_2 \times \mathcal{Z}_{9}$ flavor symmetries with respect to its mass through different channels for 14 TeV HL-LHC, 27 TeV HE-LHC, and future 100 TeV hadron collider, where the flavon VEV $f=500$ GeV.}
  \label{assoc_prod1a}
	\end{figure}

\begin{figure}[h!]
	\centering
	\begin{subfigure}[]{0.32\linewidth}
    \includegraphics[width=\linewidth]{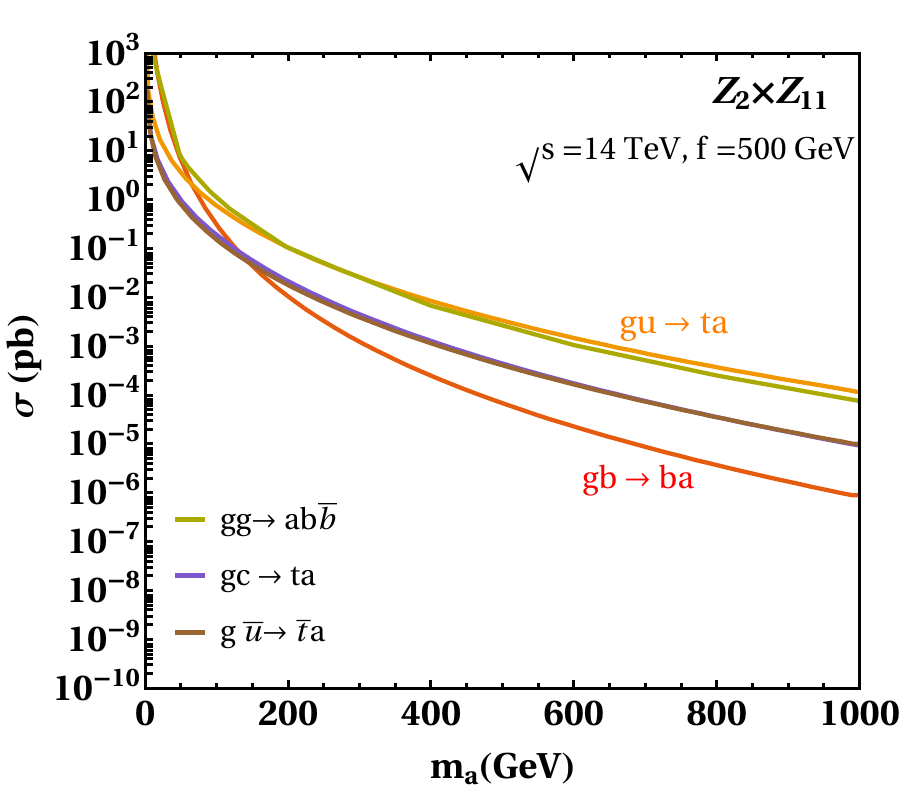}
    \caption{}
         \label{fprod14_z2z5}	
\end{subfigure}
 \begin{subfigure}[]{0.32\linewidth}
 \includegraphics[width=\linewidth]{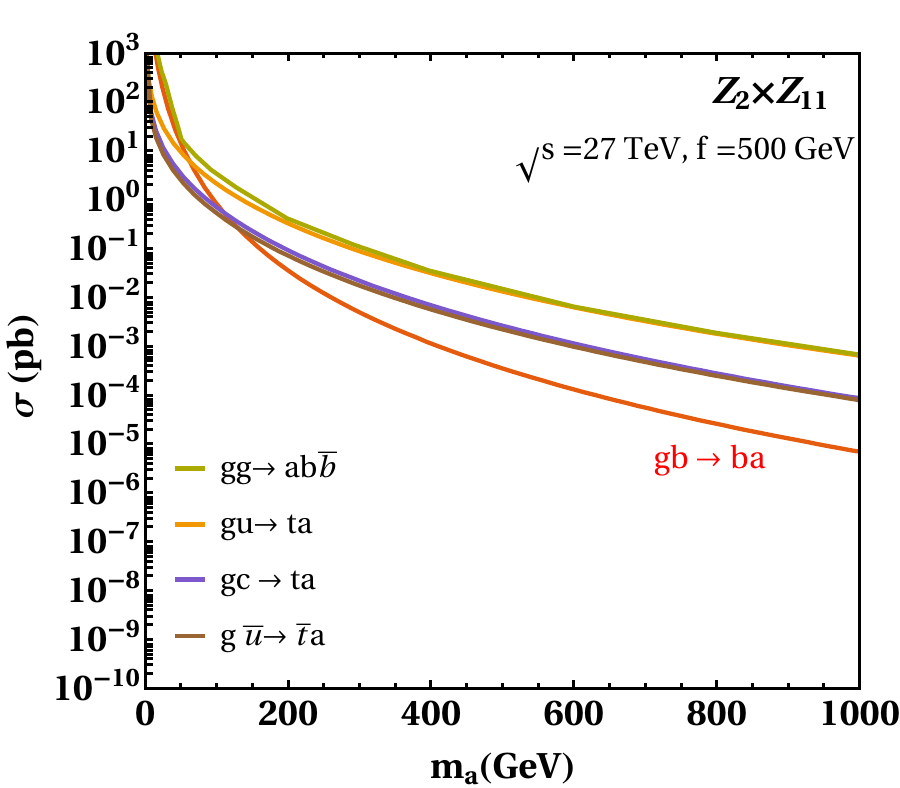}
 \caption{}
         \label{fprod27_z2z5}
 \end{subfigure} 
 \begin{subfigure}[]{0.32\linewidth}
 \includegraphics[width=\linewidth]{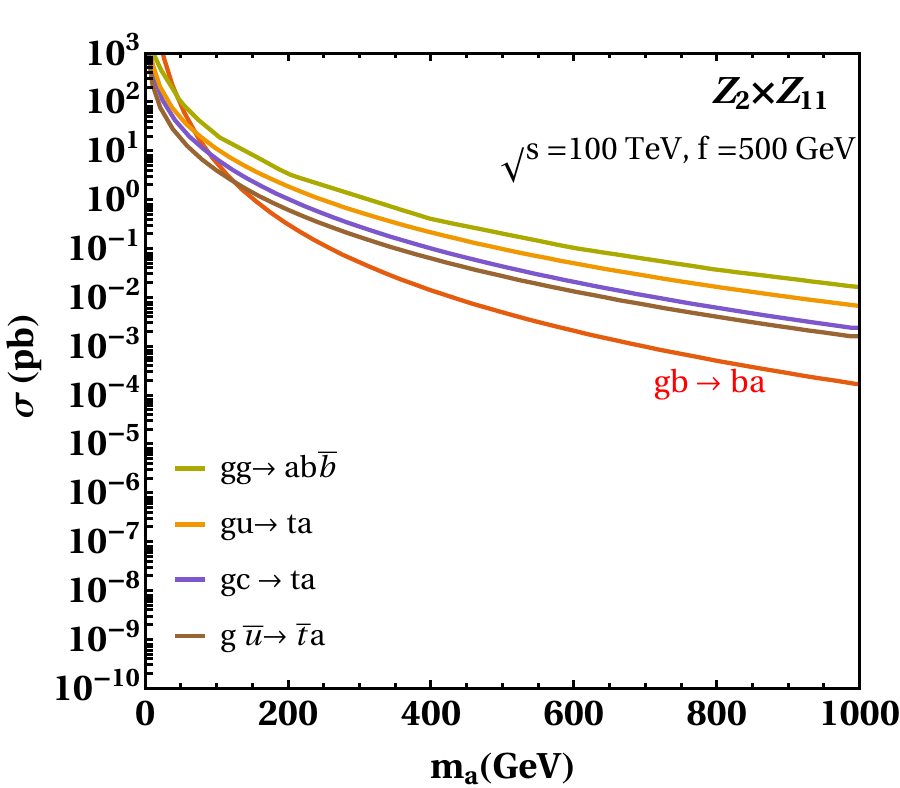}
 \caption{}
         \label{fprod100_z2z5}
 \end{subfigure} 
 \begin{subfigure}[]{0.32\linewidth}
    \includegraphics[width=\linewidth]{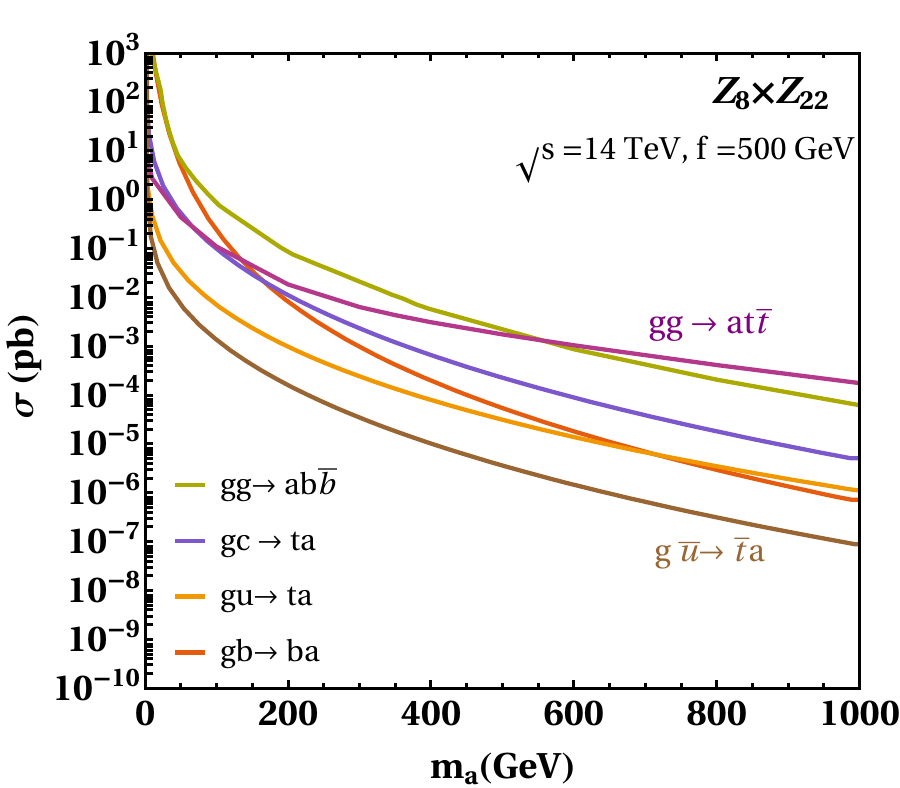}
    \caption{}
         \label{fprod14_z2z5}	
\end{subfigure}
 \begin{subfigure}[]{0.32\linewidth}
 \includegraphics[width=\linewidth]{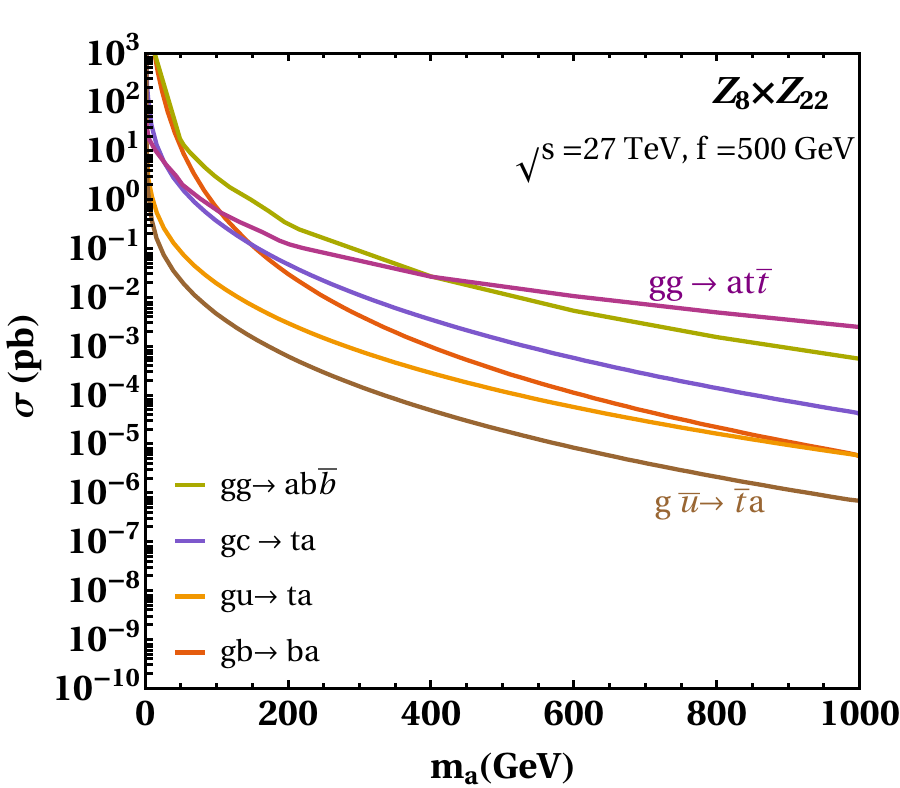}
 \caption{}
         \label{fprod27_z2z5}
 \end{subfigure} 
 \begin{subfigure}[]{0.32\linewidth}
 \includegraphics[width=\linewidth]{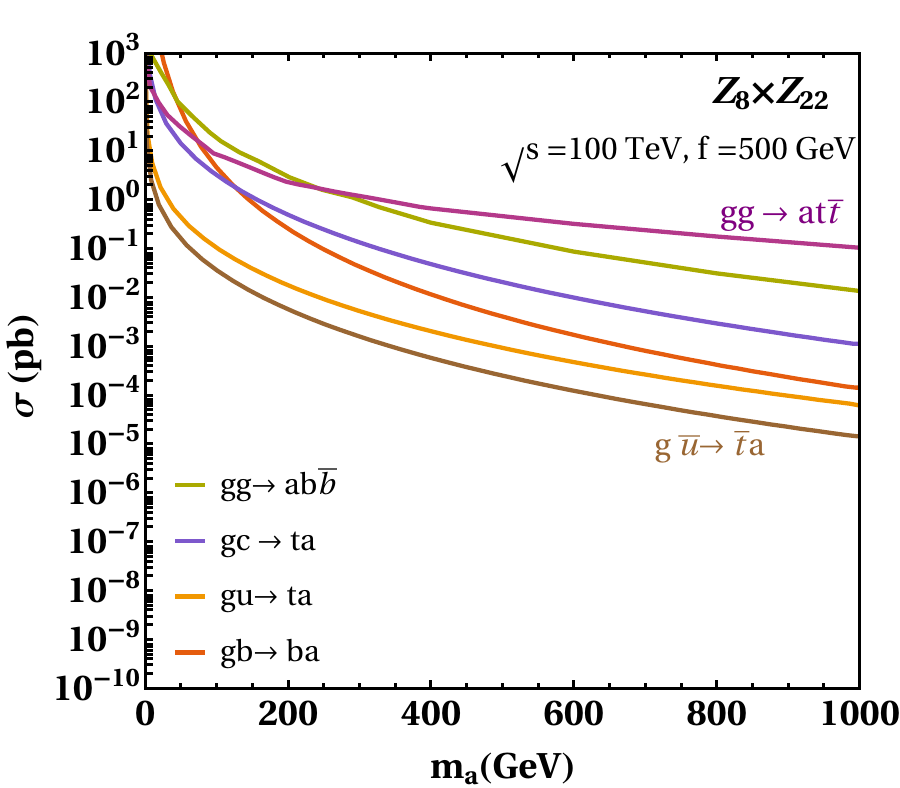}
 \caption{}
         \label{fprod100_z2z5}
 \end{subfigure} 
 \caption{Production cross-sections of the flavon of  $\mathcal{Z}_2 \times \mathcal{Z}_{11}$, and $\mathcal{Z}_8 \times \mathcal{Z}_{22}$ flavor symmetries with respect to its mass through different channels for 14 TeV HL-LHC, 27 TeV HE-LHC, and future 100 TeV hadron collider, where the flavon VEV $f=500$ GeV.}
  \label{assoc_prod2a}
	\end{figure}

As noticed in the case of inclusive production of the heavy flavon mass, we observe from figures \ref{assoc_prod1a} and \ref{assoc_prod2a} that the associative production cross-sections of the flavon of different $\mathcal{Z}_{\rm N}\times \mathcal{Z}_{\rm M}$ flavor symmetries for the VEV $f =500$ GeV for the soft symmetry-breaking scenario can be sufficiently large.   Therefore, we continue to use this choice even for the associative production  of the heavy flavon of different $\mathcal{Z}_{\rm N}\times \mathcal{Z}_{\rm M}$ flavor symmetries.

\begin{table}[H]
\setlength{\tabcolsep}{6pt} 
\renewcommand{\arraystretch}{1} 
\centering
\begin{tabular}{l|cc|cc|cc}
\toprule
& \multicolumn{2}{c|}{$\mathcal{L}[fb^{-1}]$ [References]} & \multicolumn{2}{c|}{ATLAS 13 TeV} & \multicolumn{2}{c}{CMS 13 TeV}  \\
$m_{a}$~[GeV] &  \myalign{c}{ATLAS} & \myalign{c|}{CMS} & \myalign{c}{500} & \myalign{c|}{1000} & \myalign{c}{500} & \myalign{c}{1000}  \\
\midrule
$t  \bar{t} a \rightarrow t  \bar{t} t  \bar{t} $~[pb]  & $139$ \cite{ATLAS:2022rws}  & $137$ \cite{CMS:2019rvj}  &  $1\e{-2}$ &\phantom{xx}  $6\e{-3}$   &  $2\e{-2}$ &\phantom{xx}    \\
$gg \rightarrow a b \bar{b} \rightarrow \tau \tau b \bar{b} $~[pb]    & 
 $36.1$ \cite{ATLAS:2017eiz} & $35.9$ \cite{CMS:2018rmh}  & $8\e{-2}$  & $9\e{-3}$   & $6\e{-2}$  & $1\e{-2}$  \\
$gb \rightarrow a b  \rightarrow \tau \tau b  $~[pb] )   & $36.1$ \cite{ATLAS:2017eiz} & $35.9$ \cite{CMS:2018rmh}  & $3\e{-2}$ & $4\e{-3}$  & $3\e{-2}$ & $4\e{-3}$     \\
\bottomrule
\end{tabular}
\caption{Current limits of $\sigma \times BR$ at 13 TeV LHC by ATLAS and CMS in resonance searches for associative flavon production channels.}
\label{tab:limits_associtive}
\end{table}

The present reach of the LHC for the associative production channels is shown in table  \ref{tab:limits_associtive}.   We show the sensitivities of these modes at the HL-LHC, HE-LHC,  and a 100 TeV collider in table  \ref{tab:futurelimits_associative}.  Our benchmark predictions for different $\mathcal{Z}_{\rm N} \times \mathcal{Z}_{\rm M}$ flavor symmetries, are given in tables  \ref{tab:limits_bench_asc14}- \ref{tab:limits_bench_asc100} for  the soft symmetry-breaking scenario.

 \begin{table}[H]
\setlength{\tabcolsep}{6pt} 
\renewcommand{\arraystretch}{1} 
\centering
\begin{tabular}{l|cc|cc|cc}
\toprule
 & \multicolumn{2}{c|}{HL-LHC [14 TeV, $3~\iab$] } & \multicolumn{2}{c|}{HE-LHC [27 TeV, $15~\iab$]} & \multicolumn{2}{c}{100 TeV, $30~\iab$} \\
$m_{a}$~[GeV] &  500 &  1000 &  500 &  1000 &  500 &  1000 \\
\midrule
$t  \bar{t} a \rightarrow t  \bar{t} t  \bar{t} $~[pb] & $2\e{-3}$ & $1\e{-3}$ & $3\e{-3}$ & $2\e{-3}$ & $1\e{-2}$ & $9\e{-3}$   \\
$gg \rightarrow a b \bar{b} \rightarrow \tau \tau b \bar{b} $~[pb] & $7\e{-3}$  & $1\e{-3}$  & $5\e{-3}$ & $8\e{-4}$  & $8\e{-3}$ & $1\e{-3}$  \\
$gb \rightarrow a b  \rightarrow \tau \tau b  $~[pb]  &   $3\e{-3}$ & $5\e{-4}$  & $2\e{-3}$  & $3\e{-4}$   & $4\e{-3}$  & $5\e{-4}$  \\
\bottomrule
\end{tabular}
\caption{Estimated reach ($\sigma \times BR$) of HL-LHC, HE-LHC and the 100 TeV collider for high flavon mass ($m_a$) in associative flavon production channels.}
\label{tab:futurelimits_associative}
\end{table}

We observe from table \ref{tab:limits_bench_asc14} that only the mode $t  \bar{t} a \rightarrow t  \bar{t} t  \bar{t} $ is within the reach of the HL-LHC for the flavon mass $m_a = 500$ GeV for the $\mathcal{Z}_8 \times \mathcal{Z}_{22}$ flavor symmetry.  This scenario improves slightly at the HE-LHC and at a 100 TeV collider for  the $\mathcal{Z}_8 \times \mathcal{Z}_{22}$ flavor symmetry such that the flavon mass $m_a =1000$ GeV becomes also accessible. The symmetry-conserving scenario remains beyond the reach of the  HL-LHC, HE-LHC and a 100 TeV collider  for the  associative production modes. Therefore, we do not discuss it anymore in this case.


\begin{table}[H]
\setlength{\tabcolsep}{3.6pt} 
\renewcommand{\arraystretch}{1} 
\centering
\begin{tabular}{@{}l|rr|rr|rr|rr@{}}
\toprule
 & \multicolumn{2}{c|}{Benchmark} & \multicolumn{2}{c|}{Benchmark} & \multicolumn{2}{c|}{Benchmark} & \multicolumn{2}{c}{Benchmark}\\
 & \multicolumn{2}{c|}{$\mathcal{Z}_2 \times \mathcal{Z}_{5}$} & \multicolumn{2}{c|}{$\mathcal{Z}_2 \times \mathcal{Z}_{9}$} & \multicolumn{2}{c|}{$\mathcal{Z}_2 \times \mathcal{Z}_{11}$} & \multicolumn{2}{c}{$\mathcal{Z}_8 \times \mathcal{Z}_{22}$}\\
 $m_{a}$~[GeV] & \myalign{c}{500} & \myalign{c|}{1000} & \myalign{c}{500} & \myalign{c|}{1000} & \myalign{c}{500} & \myalign{c|}{1000} &\myalign{c}{500} & \myalign{c}{1000}\\
\midrule
$t  \bar{t} a \rightarrow t  \bar{t} t  \bar{t} $~[pb]      & &   &  &  &  & & \fbox{$1.6\e{-3}$} & $1.6\e{-4}$  \\
$gg \rightarrow a b \bar{b} \rightarrow \tau \tau b \bar{b} $~[pb]      & $6\e{-7}$ & $1.8\e{-8}$  & $1\e{-4}$ & $3.2\e{-6}$ & $3.4\e{-5}$ & $8.9\e{-7}$ & $1.6\e{-6}$ & $3.2\e{-8}$  \\
$gb \rightarrow a b  \rightarrow \tau \tau b  $~[pb]      & $1.7\e{-7}$ & $5.6\e{-9}$  & $2.9\e{-5}$ & $9.7\e{-7}$ & $9.4\e{-6}$ & $2.7\e{-7}$ & $4.5\e{-7}$ & $9.7\e{-9}$  \\
\bottomrule
\end{tabular}
\caption{Benchmark points for different  $\mathcal{Z}_N \times \mathcal{Z}_{M}$  flavor symmetries for associative flavon production channels with high flavon mass ($m_a$) in case of the soft symmetry-breaking at the 14 TeV HL-LHC, assuming $f = 500$ GeV.}
\label{tab:limits_bench_asc14}
\end{table}

\begin{table}[H]
\setlength{\tabcolsep}{3.6pt} 
\renewcommand{\arraystretch}{1} 
\centering
\begin{tabular}{@{}l|rr|rr|rr|rr@{}}
\toprule
 & \multicolumn{2}{c|}{Benchmark} & \multicolumn{2}{c|}{Benchmark} & \multicolumn{2}{c|}{Benchmark} & \multicolumn{2}{c}{Benchmark}\\
 & \multicolumn{2}{c|}{$\mathcal{Z}_2 \times \mathcal{Z}_{5}$} & \multicolumn{2}{c|}{$\mathcal{Z}_2 \times \mathcal{Z}_{9}$} & \multicolumn{2}{c|}{$\mathcal{Z}_2 \times \mathcal{Z}_{11}$} & \multicolumn{2}{c}{$\mathcal{Z}_8 \times \mathcal{Z}_{22}$}\\
 $m_{a}$~[GeV] & \myalign{c}{500} & \myalign{c|}{1000} & \myalign{c}{500} & \myalign{c|}{1000} & \myalign{c}{500} & \myalign{c|}{1000} &\myalign{c}{500} & \myalign{c}{1000}\\
\midrule
$t  \bar{t} a \rightarrow t  \bar{t} t  \bar{t} $~[pb]      & &   &  &  &  & & \fbox{$1.5\e{-2}$} & \fbox{$2.3\e{-3}$}  \\
$gg \rightarrow a b \bar{b} \rightarrow \tau \tau b \bar{b} $~[pb]      & $3.3\e{-6}$ & $1.6\e{-7}$  & $5.8\e{-4}$ & $2.8\e{-5}$ & $1.9\e{-4}$ & $7.9\e{-6}$ & $9\e{-6}$ & $2.8\e{-7}$  \\
$gb \rightarrow a b  \rightarrow \tau \tau b  $~[pb]      & $9.3\e{-7}$ & $5.1\e{-8}$  & $1.6\e{-4}$ & $8.8\e{-6}$ & $5.3\e{-5}$ & $2.5\e{-6}$ & $2.5\e{-6}$ & $8.8\e{-8}$  \\
\bottomrule
\end{tabular}
\caption{Benchmark points for different  $\mathcal{Z}_N \times \mathcal{Z}_{M}$  flavor symmetries for associative flavon production channels with high flavon mass ($m_a$) in case of the soft symmetry-breaking at the 27 TeV HE-LHC, assuming $f = 500$ GeV.}
\label{tab:limits_bench_asc27}
\end{table}

\begin{table}[H]
\setlength{\tabcolsep}{3.6pt} 
\renewcommand{\arraystretch}{1.1} 
\centering
\begin{tabular}{@{}l|rr|rr|rr|rr@{}}
\toprule
 & \multicolumn{2}{c|}{Benchmark} & \multicolumn{2}{c|}{Benchmark} & \multicolumn{2}{c|}{Benchmark} & \multicolumn{2}{c}{Benchmark}\\
 & \multicolumn{2}{c|}{$\mathcal{Z}_2 \times \mathcal{Z}_{5}$} & \multicolumn{2}{c|}{$\mathcal{Z}_2 \times \mathcal{Z}_{9}$} & \multicolumn{2}{c|}{$\mathcal{Z}_2 \times \mathcal{Z}_{11}$} & \multicolumn{2}{c}{$\mathcal{Z}_8 \times \mathcal{Z}_{22}$}\\
 $m_{a}$~[GeV] & \myalign{c}{500} & \myalign{c|}{1000} & \myalign{c}{500} & \myalign{c|}{1000} & \myalign{c}{500} & \myalign{c|}{1000} &\myalign{c}{500} & \myalign{c}{1000}\\
\midrule
$t  \bar{t} a \rightarrow t  \bar{t} t  \bar{t} $~[pb]      & &   &  &  &  & & \fbox{$0.41$} & \fbox{$9.6\e{-2}$}  \\
$gg \rightarrow a b \bar{b} \rightarrow \tau \tau b \bar{b} $~[pb]      & $4.7\e{-5}$ & $4.0\e{-6}$  & $8.2\e{-3}$ & $6.9\e{-4}$ & $2.7\e{-3}$ & $1.9\e{-4}$ & $1.3\e{-4}$ & $6.8\e{-6}$  \\
$gb \rightarrow a b  \rightarrow \tau \tau b  $~[pb]      & $1.34\e{-5}$ & $1.23\e{-6}$  & $2.3\e{-3}$ & $2.1\e{-4}$ & $7.5\e{-4}$ & $6.0\e{-5}$ & $3.6\e{-5}$ & $2.1\e{-6}$  \\
\bottomrule
\end{tabular}
\caption{Benchmark points for different  $\mathcal{Z}_N \times \mathcal{Z}_{M}$  flavor symmetries for associative flavon production channels with high flavon mass ($m_a$) in case of the soft symmetry-breaking scenario at a 100 TeV collider, assuming $f = 500$ GeV.}
\label{tab:limits_bench_asc100}
\end{table}

\subsubsection{Di-flavon production}
\begin{figure}[H]
	\centering
	\begin{subfigure}[]{0.38\linewidth}
    \includegraphics[width=\linewidth]{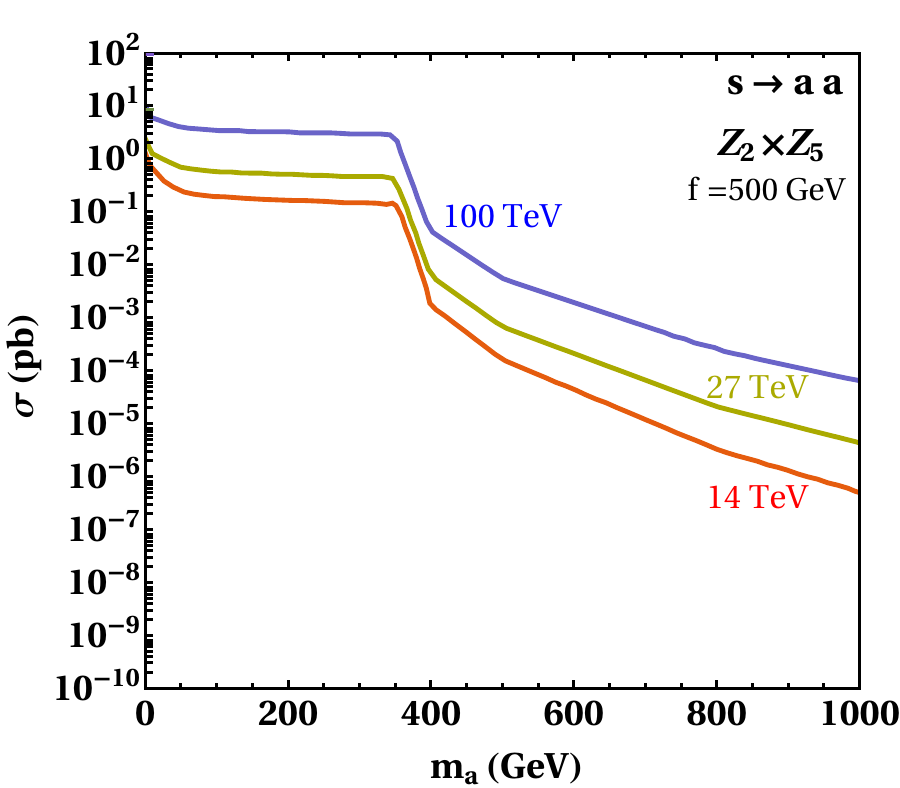}
    \caption{}
         \label{fprod14_z2z5}	
\end{subfigure}
 \begin{subfigure}[]{0.38\linewidth}
 \includegraphics[width=\linewidth]{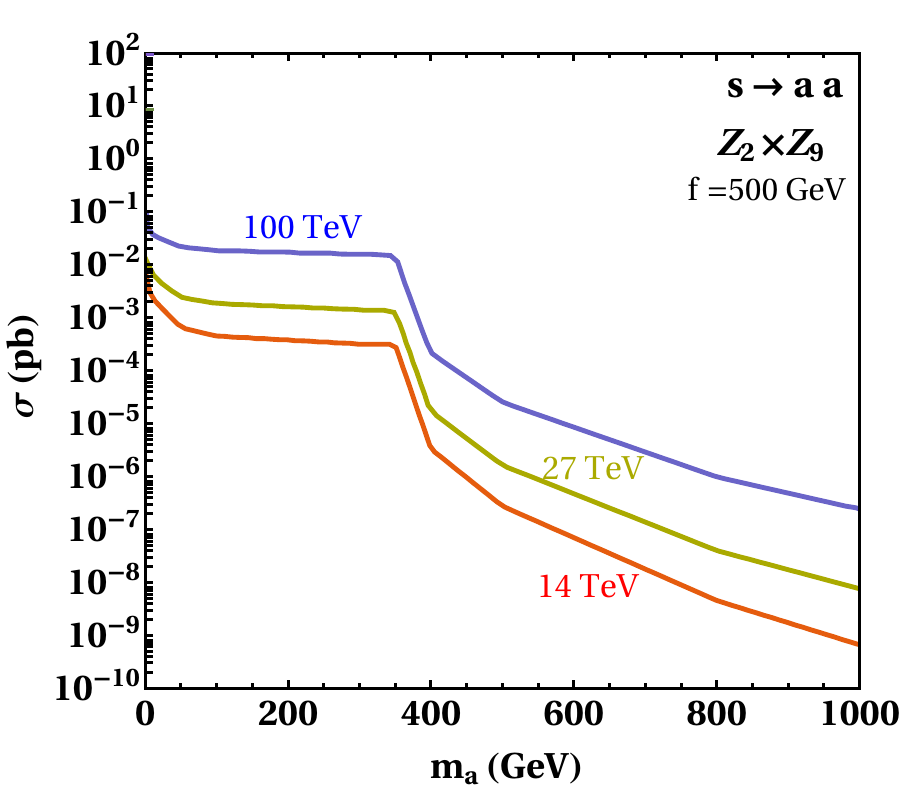}
 \caption{}
         \label{fprod27_z2z5}
 \end{subfigure} 
 \begin{subfigure}[]{0.38\linewidth}
 \includegraphics[width=\linewidth]{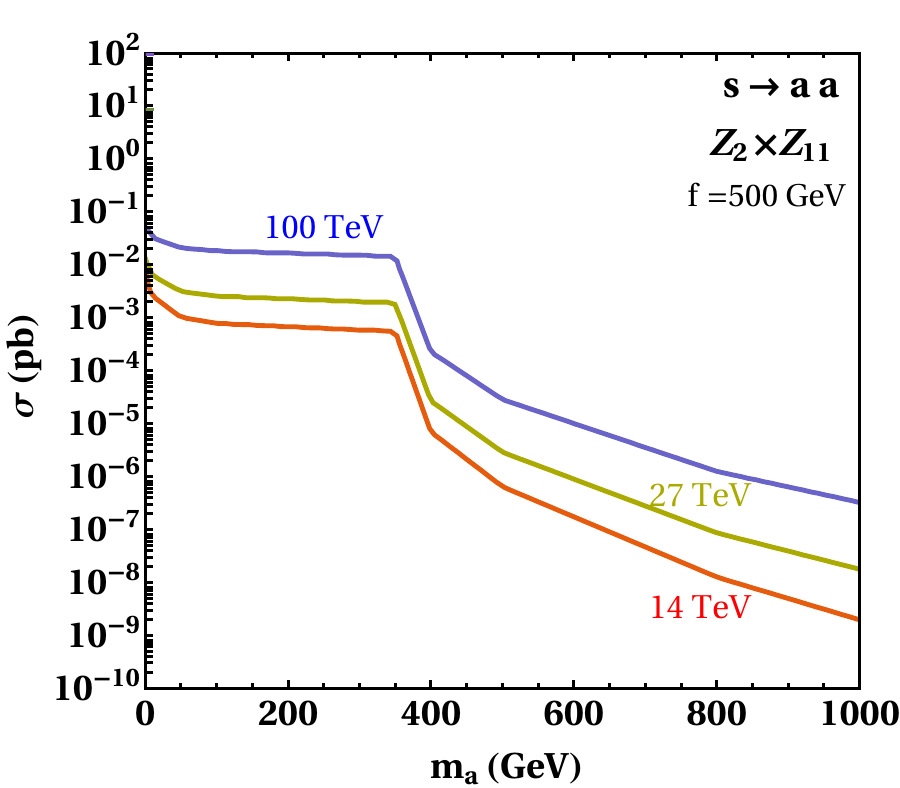}
 \caption{}
         \label{fprod100_z2z5}
 \end{subfigure} 
 \begin{subfigure}[]{0.38\linewidth}
    \includegraphics[width=\linewidth]{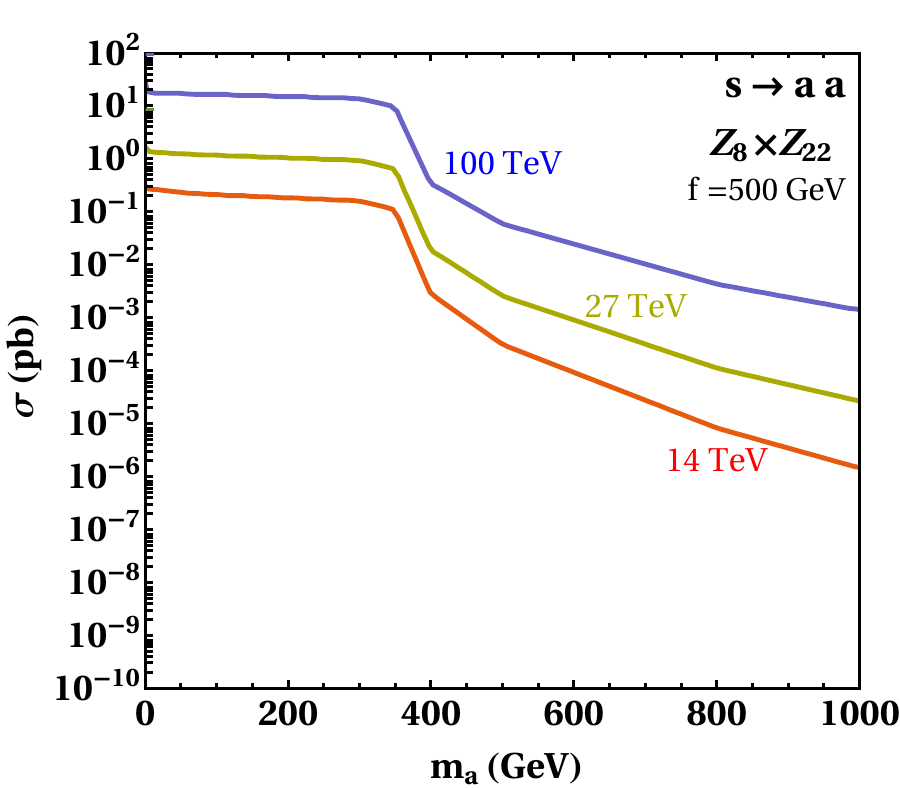}
    \caption{}
         \label{fprod14_z2z5}	
\end{subfigure}
  \caption{Production cross-sections of the di-flavon in  $\mathcal{Z}_2 \times \mathcal{Z}_{5}$, $\mathcal{Z}_2 \times \mathcal{Z}_{9}$, $\mathcal{Z}_2 \times \mathcal{Z}_{11}$, and $\mathcal{Z}_8 \times \mathcal{Z}_{22}$ flavor symmetries with respect to its mass for 14 TeV HL-LHC, 27 TeV HE-LHC, and 100 TeV hadron collider, where the flavon VEV $f=500$ GeV.}
  \label{diflav}
	\end{figure}

In this section, we discuss the production of di-flavon, which is an important mode to explore the flavon physics of different  $\mathcal{Z}_{\rm N}\times \mathcal{Z}_{\rm M}$ flavor symmetries.  We show the present LHC sensitivities of di-flavon channels in  table \ref{tab:limits_heavy_dif1}.  The reach of the HL-LHC, HE-LHC and a 100 TeV collider is given in table \ref{tab:futurelimits_dif1a}.  The di-flavon production cross-sections for different $\mathcal{Z}_{\rm N} \times \mathcal{Z}_{\rm M} $ flavor symmetries are shown in figure \ref{diflav} in the case of soft symmetry-breaking scenario.

The benchmark predictions for different $\mathcal{Z}_{\rm N} \times \mathcal{Z}_{\rm M} $ flavor symmetries, are given in tables \ref{tab:limits_benchdif14}- \ref{tab:limits_benchdif100} for  the soft symmetry-breaking scenario and heavy flavon masses.  We observe from tables \ref{tab:limits_benchdif14}- \ref{tab:limits_benchdif100} that the di-flavon production for heavy flavon mass is insensitive to the HL-LHC, HE-LHC and even to a 100 TeV collider for all $\mathcal{Z}_{\rm N} \times \mathcal{Z}_{\rm M} $ flavor symmetries. 

\begin{table}[H]
\setlength{\tabcolsep}{13pt} 
\renewcommand{\arraystretch}{1.1} 
\centering
\begin{minipage}{\textwidth}
\begin{tabular}{@{}l|rr| rr|rr@{}}
\toprule
& \multicolumn{2}{c|}{$\mathcal{L}[fb^{-1}]$ [References]} & \multicolumn{2}{c|}{ATLAS 13 TeV} & \multicolumn{2}{c}{CMS 13 TeV} \\
$m_{a}$~[GeV] &  \myalign{c}{ATLAS} & \myalign{c|}{CMS} & \myalign{c}{500} & \myalign{c|}{1000} & \myalign{c}{500} & \myalign{c}{1000}  \\
\midrule
$s \rightarrow aa \rightarrow b \bar{b} \ell \ell $~[pb]& $36.1$ \cite{ATLAS:2018uni}  & $19.7$ \cite{CMS:2017yfv}\footnote{The reference \cite{CMS:2017yfv} is at sqrt(s) = 8 TeV.} & $2\e{-2}$ & $1\e{-2}$  & $2\e{-1}$  &   $2\e{-2}$  \\
$s \rightarrow aa \rightarrow b \bar{b}  b \bar{b} $~[pb]& $36.1$ \cite{ATLAS:2018rnh}  & $35.9$ \cite{CMS:2018qmt}  & $4\e{-2}$ & $9\e{-3}$  & $7\e{-2}$ &   $1\e{-2}$\\
\bottomrule
\end{tabular}
\end{minipage}
\caption{Current limits of $\sigma \times BR$ at 13 TeV LHC by ATLAS and CMS in high mass resonance searches for di-flavon production channels.}
\label{tab:limits_heavy_dif1}
\end{table}

 \begin{table}[H]
\setlength{\tabcolsep}{6pt} 
\renewcommand{\arraystretch}{1.1} 
\centering
\begin{tabular}{l|cc|cc|cc}
\toprule
 & \multicolumn{2}{c|}{HL-LHC [14 TeV, $3~\iab$] } & \multicolumn{2}{c|}{HE-LHC [27 TeV, $15~\iab$]} & \multicolumn{2}{c}{100 TeV, $30~\iab$} \\
$m_{a}$~[GeV] &  500 &  1000 &  500 &  1000 &  500 &  1000 \\
\midrule
$s \rightarrow aa \rightarrow b \bar{b} \tau \tau $~[pb]    & $2\e{-3}$ & $1\e{-3}$ & $2\e{-3}$ & $8\e{-4}$ & $3\e{-3}$ & $1\e{-3}$   \\
$s \rightarrow aa \rightarrow b \bar{b}  b \bar{b} $~[pb]  & $5\e{-3}$  & $1\e{-3}$  & $3\e{-3}$ & $7\e{-4}$   & $5\e{-3}$  & $1\e{-3}$ \\

\bottomrule
\end{tabular}
\caption{Estimated reach ($\sigma \times BR$) of HL-LHC, HE-LHC and the 100 TeV collider for high flavon mass ($m_a$) in di-flavon production channels.}
\label{tab:futurelimits_dif1a}
\end{table}


\begin{table}[H]
 \setlength{\tabcolsep}{3.5pt} 
\renewcommand{\arraystretch}{1.1} 
\centering
\begin{tabular}{@{}l|rr|rr|rr|rr@{}}
\toprule
 & \multicolumn{2}{c|}{Benchmark} & \multicolumn{2}{c|}{Benchmark} & \multicolumn{2}{c|}{Benchmark} & \multicolumn{2}{c}{Benchmark}\\
 & \multicolumn{2}{c|}{$\mathcal{Z}_2 \times \mathcal{Z}_{5}$} & \multicolumn{2}{c|}{$\mathcal{Z}_2 \times \mathcal{Z}_{9}$} & \multicolumn{2}{c|}{$\mathcal{Z}_2 \times \mathcal{Z}_{11}$} & \multicolumn{2}{c}{$\mathcal{Z}_8 \times \mathcal{Z}_{22}$}\\
 $m_{a}$~[GeV] & \myalign{c}{500} & \myalign{c|}{1000} & \myalign{c}{500} & \myalign{c|}{1000} & \myalign{c}{500} & \myalign{c|}{1000} &\myalign{c}{500} & \myalign{c}{1000}\\
\midrule
$s \rightarrow aa \rightarrow b \bar{b} \tau \tau $~[pb]      & $1.4\e{-8}$ & $4.4\e{-11}$  & $7\e{-9}$ & $1.7\e{-11}$ & $7.5\e{-10}$ & $1.7\e{-12}$ & $2\e{-9}$ & $3.7\e{-12}$  \\
$s \rightarrow aa \rightarrow b \bar{b} \mu \mu $~[pb]      & $1.3\e{-11}$ & $4\e{-14}$  & $1.5\e{-10}$ & $3.6\e{-13}$ & $5.8\e{-12}$ & $1.3\e{-14}$ & $1.8\e{-11}$ & $3.2\e{-14}$  \\
$s \rightarrow aa \rightarrow b \bar{b}  b \bar{b} $~[pb]     & $4\e{-8}$  & $1.3\e{-10}$ & $1.8\e{-8}$ & $4.3\e{-11}$ & $1.1\e{-9}$ & $2.6\e{-12}$  & $4.9\e{-9}$ & $9.1\e{-12}$  \\
\bottomrule
\end{tabular}
\caption{Benchmark points for different  $\mathcal{Z}_N \times \mathcal{Z}_{M}$  flavor symmetries for di-flavon production channels with high flavon mass ($m_a$) in case of the soft symmetry-breaking scenario at the 14 TeV HL-LHC, assuming $f = 500$ GeV.}
\label{tab:limits_benchdif14}
\end{table}

\begin{table}[H]
 \setlength{\tabcolsep}{3.5pt} 
\renewcommand{\arraystretch}{1.1} 
\centering
\begin{tabular}{@{}l|rr|rr|rr|rr@{}}
\toprule
 & \multicolumn{2}{c|}{Benchmark} & \multicolumn{2}{c|}{Benchmark} & \multicolumn{2}{c|}{Benchmark} & \multicolumn{2}{c}{Benchmark}\\
 & \multicolumn{2}{c|}{$\mathcal{Z}_2 \times \mathcal{Z}_{5}$} & \multicolumn{2}{c|}{$\mathcal{Z}_2 \times \mathcal{Z}_{9}$} & \multicolumn{2}{c|}{$\mathcal{Z}_2 \times \mathcal{Z}_{11}$} & \multicolumn{2}{c}{$\mathcal{Z}_8 \times \mathcal{Z}_{22}$}\\
 $m_{a}$~[GeV] & \myalign{c}{500} & \myalign{c|}{1000} & \myalign{c}{500} & \myalign{c|}{1000} & \myalign{c}{500} & \myalign{c|}{1000} &\myalign{c}{500} & \myalign{c}{1000}\\
\midrule
$s \rightarrow aa \rightarrow b \bar{b} \tau \tau $~[pb]      & $5.9\e{-8}$ & $3.7\e{-10}$  & $4.2\e{-8}$ & $1.9\e{-10}$ & $3.4\e{-9}$ & $1.5\e{-11}$ & $1.6\e{-8}$ & $6.9\e{-11}$  \\
$s \rightarrow aa \rightarrow b \bar{b} \mu \mu $~[pb]      & $5.3\e{-11}$ & $3.4\e{-13}$  & $9.1\e{-10}$ & $4.2\e{-12}$ & $2.6\e{-11}$ & $1.2\e{-13}$ & $1.4\e{-10}$ & $5.9\e{-13}$  \\
$s \rightarrow aa \rightarrow b \bar{b}  b \bar{b} $~[pb]     & $1.7\e{-7}$  & $1.1\e{-9}$ & $1.1\e{-7}$ & $4.9\e{-10}$ & $5.1\e{-9}$ & $2.3\e{-11}$  & $4\e{-8}$ & $1.7\e{-10}$  \\
\bottomrule
\end{tabular}
\caption{Benchmark points for different  $\mathcal{Z}_N \times \mathcal{Z}_{M}$  flavor symmetries for di-flavon production channels with high flavon mass ($m_a$) in case of the soft symmetry-breaking scenario at the 27 TeV HE-LHC, assuming $f = 500$ GeV.}
\label{tab:limits_benchdif27}
\end{table}

 \begin{table}[H]
 \setlength{\tabcolsep}{3.5pt} 
\renewcommand{\arraystretch}{1.1} 
\centering
\begin{tabular}{@{}l|rr|rr|rr|rr@{}}
\toprule
 & \multicolumn{2}{c|}{Benchmark} & \multicolumn{2}{c|}{Benchmark} & \multicolumn{2}{c|}{Benchmark} & \multicolumn{2}{c}{Benchmark}\\
 & \multicolumn{2}{c|}{$\mathcal{Z}_2 \times \mathcal{Z}_{5}$} & \multicolumn{2}{c|}{$\mathcal{Z}_2 \times \mathcal{Z}_{9}$} & \multicolumn{2}{c|}{$\mathcal{Z}_2 \times \mathcal{Z}_{11}$} & \multicolumn{2}{c}{$\mathcal{Z}_8 \times \mathcal{Z}_{22}$}\\
 $m_{a}$~[GeV] & \myalign{c}{500} & \myalign{c|}{1000} & \myalign{c}{500} & \myalign{c|}{1000} & \myalign{c}{500} & \myalign{c|}{1000} &\myalign{c}{500} & \myalign{c}{1000}\\
\midrule
$s \rightarrow aa \rightarrow b \bar{b} \tau \tau $~[pb]      & $4.8\e{-7}$ & $5.5\e{-9}$  & $6.5\e{-7}$ & $6.2\e{-9}$ & $3.3\e{-8}$ & $2.8\e{-10}$ & $3.7\e{-7}$ & $3.5\e{-9}$  \\
$s \rightarrow aa \rightarrow b \bar{b} \mu \mu $~[pb]      & $4.2\e{-10}$ & $4.9\e{-12}$  & $1.4\e{-8}$ & $1.3\e{-10}$ & $2.5\e{-10}$ & $2.2\e{-12}$ & $3.2\e{-9}$ & $3.1\e{-11}$  \\
$s \rightarrow aa \rightarrow b \bar{b}  b \bar{b} $~[pb]     & $1.4\e{-6}$  & $1.6\e{-8}$ & $1.7\e{-6}$ & $1.6\e{-8}$ & $5\e{-8}$ & $4.3\e{-10}$  & $9.1\e{-7}$ & $8.6\e{-9}$  \\
\bottomrule
\end{tabular}
\caption{Benchmark points for different  $\mathcal{Z}_N \times \mathcal{Z}_{M}$  flavor symmetries for di-flavon production channels with high flavon mass ($m_a$) in case of the soft symmetry-breaking scenario at a 100 TeV collider, assuming $f = 500$ GeV.}
\label{tab:limits_benchdif100}
\end{table}


We observed earlier that the di-flavon production of a heavy mass flavon is beyond the reach of HL-LHC, HE-LHC, and a 100 TeV collider for different $\mathcal{Z}_{\rm N} \times \mathcal{Z}_{\rm M} $ flavor symmetries.  However, this scenario changes in the case of a light flavon.  We show the present sensitivities of the light di-flavon searches of the LHC in table \ref{tab:diflav_light_limits}.  The sensitivities of the HL-LHC, HE-LHC, and a 100 TeV collider for a light di-flavon productions are shown in table \ref{tab:diflav_light_sens}.

\begin{table}[H]
\setlength{\tabcolsep}{6pt} 
\renewcommand{\arraystretch}{1} 
\centering
\begin{tabular}{@{}l|rr| rr|rr@{}}
\toprule
& \multicolumn{2}{c|}{$\mathcal{L}[fb^{-1}]$ [References]} & \multicolumn{2}{c|}{ATLAS 13 TeV} & \multicolumn{2}{c}{CMS 13 TeV} \\
$m_{a}$~[GeV] &  \myalign{c}{ATLAS} & \myalign{c|}{CMS} & \myalign{c}{20} & \myalign{c|}{60} & \myalign{c}{20} & \myalign{c}{60}   \\
\midrule
$s \rightarrow aa \rightarrow b \bar{b} \ell \ell $~[pb]& $139$ \cite{ATLAS:2021hbr}  & $138$ \cite{CMS:2024uru}  & $1\e{-4}$ & $1\e{-4}$  & $9\e{-5}$ &   $8\e{-5}$ \\
$s \rightarrow aa \rightarrow b \bar{b}  b \bar{b} $~[pb]& $36.1$ \cite{ATLAS:2018pvw}  & $138$ \cite{CMS:2024zfv}  & $3$ & $1$  & $1$ &   $3\e{-1}$  \\
\bottomrule
\end{tabular}
\caption{Current limits of $\sigma \times BR$ at 13 TeV LHC by ATLAS and CMS in low mass resonance searches for di-flavon channels.}
\label{tab:diflav_light_limits}
\end{table}

 \begin{table}[H]
\setlength{\tabcolsep}{6pt} 
\renewcommand{\arraystretch}{1} 
\centering
\begin{tabular}{l|cc|cc|cc}
\toprule
 & \multicolumn{2}{c|}{HL-LHC [14 TeV, $3~\iab$] } & \multicolumn{2}{c|}{HE-LHC [27 TeV, $15~\iab$]} & \multicolumn{2}{c}{100 TeV, $30~\iab$} \\
$m_{a}$~[GeV] &  20 &  60 &  20 &  60 &  20 &  60 \\
\midrule
$s \rightarrow aa \rightarrow b \bar{b} \ell \ell $~[pb]    & $2\e{-5}$ & $2\e{-5}$ & $1\e{-5}$ & $1\e{-5}$ & $2\e{-5}$ & $2\e{-5}$   \\
$s \rightarrow aa \rightarrow b \bar{b}  b \bar{b} $~[pb]  & 0.2  & $7\e{-2}$ & 0.16 & $5\e{-2}$   & 0.2 & $7\e{-2}$ \\

\bottomrule
\end{tabular}
\caption{Estimated reach ($\sigma \times BR$) of HL-LHC, HE-LHC and the 100 TeV collider for low flavon mass ($m_a$) in di-flavon production channels.}
\label{tab:diflav_light_sens}
\end{table}

We show our benchmark predictions of  di-flavon production cross-sections  in tables \ref{tab:diflav_light14}-\ref{tab:diflav_light100} for low flavon masses in the soft symmetry-breaking scenario.  We observe  that $ aa \rightarrow b \bar{b} \tau \tau $ mode is accessible for all $\mathcal{Z}_{\rm N} \times \mathcal{Z}_{\rm M} $ flavor symmetries at the HL-LHC, HE-LHC, and a 100 TeV collider.  All three channels are sensitive to the HL-LHC, HE-LHC, and a 100 TeV collider only for the $\mathcal{Z}_8 \times \mathcal{Z}_{22}$ flavor symmetry.  The production cross-sections  in the symmetry-conserving scenario are too small to be under the reach of any collider, as observed in tables \ref{tab:limits_bench_sc14}-\ref{tab:limits_bench_sc100}.


\begin{table}[H]
\setlength{\tabcolsep}{2.3pt} 
\renewcommand{\arraystretch}{1} 
\centering
\begin{tabular}{@{}l|rr|rr|rr|rr@{}}
\toprule
 & \multicolumn{2}{c|}{Benchmark} & \multicolumn{2}{c|}{Benchmark} & \multicolumn{2}{c|}{Benchmark} & \multicolumn{2}{c}{Benchmark}\\
 & \multicolumn{2}{c|}{$\mathcal{Z}_2 \times \mathcal{Z}_{5}$} & \multicolumn{2}{c|}{$\mathcal{Z}_2 \times \mathcal{Z}_{9}$} & \multicolumn{2}{c|}{$\mathcal{Z}_2 \times \mathcal{Z}_{11}$} & \multicolumn{2}{c}{$\mathcal{Z}_8 \times \mathcal{Z}_{22}$}\\
 $m_{a}$~[GeV] & \myalign{c}{20} & \myalign{c|}{60} & \myalign{c}{20} & \myalign{c|}{60} & \myalign{c}{20} & \myalign{c|}{60} &\myalign{c}{20} & \myalign{c}{60}\\
\midrule
$s \rightarrow aa \rightarrow b \bar{b} \tau \tau $~[pb]      & \fbox{$2.7\e{-5}$} & \fbox{$1.9\e{-5}$}  & \fbox{$3.2\e{-5}$} & \fbox{$1.5\e{-5}$} & \fbox{$3.3\e{-5}$} & \fbox{$2\e{-5}$} & \fbox{$1.8\e{-2}$} & \fbox{$1.6\e{-2}$}  \\
$s \rightarrow aa \rightarrow b \bar{b} \mu \mu $~[pb]      & $2.5\e{-8}$ & $1.7\e{-8}$  & $7.2\e{-7}$ & $3.3\e{-7}$ & $2.7\e{-7}$ & $1.5\e{-7}$ & \fbox{$1.6\e{-4}$} & \fbox{$1.3\e{-4}$}  \\
$s \rightarrow aa \rightarrow b \bar{b}  b \bar{b} $~[pb]     & $6.8\e{-5}$ & $5.4\e{-5}$  & $6.1\e{-5}$ & $3.7\e{-5}$ & $4.1\e{-5}$ & $2.9\e{-5}$ & $3.6\e{-2}$ & $3.7\e{-2}$  \\
\bottomrule
\end{tabular}
\caption{Benchmark points for different  $\mathcal{Z}_N \times \mathcal{Z}_{M}$  flavor symmetries for di-flavon production channels with low flavon mass ($m_a$) in case of the soft symmetry-breaking scenario at the 14 TeV HL-LHC, assuming $f = 500$ GeV.}
\label{tab:diflav_light14}
\end{table}

\begin{table}[H]
\setlength{\tabcolsep}{2.3pt} 
\renewcommand{\arraystretch}{1.1} 
\centering
\begin{tabular}{@{}l|rr|rr|rr|rr@{}}
\toprule
 & \multicolumn{2}{c|}{Benchmark} & \multicolumn{2}{c|}{Benchmark} & \multicolumn{2}{c|}{Benchmark} & \multicolumn{2}{c}{Benchmark}\\
 & \multicolumn{2}{c|}{$\mathcal{Z}_2 \times \mathcal{Z}_{5}$} & \multicolumn{2}{c|}{$\mathcal{Z}_2 \times \mathcal{Z}_{9}$} & \multicolumn{2}{c|}{$\mathcal{Z}_2 \times \mathcal{Z}_{11}$} & \multicolumn{2}{c}{$\mathcal{Z}_8 \times \mathcal{Z}_{22}$}\\
 $m_{a}$~[GeV] & \myalign{c}{20} & \myalign{c|}{60} & \myalign{c}{20} & \myalign{c|}{60} & \myalign{c}{20} & \myalign{c|}{60} &\myalign{c}{20} & \myalign{c}{60}\\
\midrule
$s \rightarrow aa \rightarrow b \bar{b} \tau \tau $~[pb]      & \fbox{$6.8\e{-5}$} & \fbox{$5.5\e{-5}$}  & \fbox{$9.5\e{-5}$} & \fbox{$5.8\e{-5}$} & \fbox{$8.9\e{-5}$} & \fbox{$6.1\e{-5}$} & \fbox{$8.8\e{-2}$} & \fbox{$8.1\e{-2}$}  \\
$s \rightarrow aa \rightarrow b \bar{b} \mu \mu $~[pb]      & $6.4\e{-8}$ & $5\e{-8}$  & $2.2\e{-6}$ & $1.3\e{-6}$ & $7.1\e{-7}$ & $4.7\e{-7}$ & \fbox{$8\e{-4}$} & \fbox{$7\e{-4}$}  \\
$s \rightarrow aa \rightarrow b \bar{b}  b \bar{b} $~[pb]     & $1.7\e{-4}$ & $1.6\e{-4}$  & $1.8\e{-4}$ & $1.4\e{-4}$ & $1.1\e{-4}$ & $9.2\e{-5}$ & \fbox{$0.18$} & \fbox{$0.19$}  \\
\bottomrule
\end{tabular}
\caption{Benchmark points for different  $\mathcal{Z}_N \times \mathcal{Z}_{M}$  flavor symmetries for di-flavon production channels with low flavon mass ($m_a$) in case of the soft symmetry-breaking scenario at the 27 TeV HE-LHC, assuming $f = 500$ GeV.}
\label{tab:diflav_light27}
\end{table}

\begin{table}[H]
\setlength{\tabcolsep}{2.3pt} 
\renewcommand{\arraystretch}{1.1} 
\centering
\begin{tabular}{@{}l|rr|rr|rr|rr@{}}
\toprule
 & \multicolumn{2}{c|}{Benchmark} & \multicolumn{2}{c|}{Benchmark} & \multicolumn{2}{c|}{Benchmark} & \multicolumn{2}{c}{Benchmark}\\
 & \multicolumn{2}{c|}{$\mathcal{Z}_2 \times \mathcal{Z}_{5}$} & \multicolumn{2}{c|}{$\mathcal{Z}_2 \times \mathcal{Z}_{9}$} & \multicolumn{2}{c|}{$\mathcal{Z}_2 \times \mathcal{Z}_{11}$} & \multicolumn{2}{c}{$\mathcal{Z}_8 \times \mathcal{Z}_{22}$}\\
 $m_{a}$~[GeV] & \myalign{c}{20} & \myalign{c|}{60} & \myalign{c}{20} & \myalign{c|}{60} & \myalign{c}{20} & \myalign{c|}{60} &\myalign{c}{20} & \myalign{c}{60}\\
\midrule
$s \rightarrow aa \rightarrow b \bar{b} \tau \tau $~[pb]      & \fbox{$3.5\e{-4}$} & \fbox{$3.2\e{-4}$}  & \fbox{$6.9\e{-4}$} & \fbox{$5.5\e{-4}$} & \fbox{$5.2\e{-4}$} & \fbox{$4.2\e{-4}$} & \fbox{$1.17$} & \fbox{$1.11$ } \\
$s \rightarrow aa \rightarrow b \bar{b} \mu \mu $~[pb]      & $3.3\e{-7}$ & $2.9\e{-7}$  & \fbox{$1.6\e{-5}$} & $1.2\e{-5}$ & $4.2\e{-6}$ & $3.2\e{-6}$ & \fbox{$1.1\e{-2}$} & \fbox{$9.7\e{-3}$}  \\
$s \rightarrow aa \rightarrow b \bar{b}  b \bar{b} $~[pb]     & $8.8\e{-4}$ & $9.2\e{-4}$  & $1.3\e{-3}$ & $1.3\e{-3}$ & $6.4\e{-4}$ & $6.3\e{-4}$ & \fbox{$2.35$} & \fbox{$2.65$}  \\
\bottomrule
\end{tabular}
\caption{Benchmark points for different  $\mathcal{Z}_N \times \mathcal{Z}_{M}$  flavor symmetries for di-flavon production channels with low flavon mass ($m_a$) in case of the soft symmetry-breaking scenario at a 100 TeV collider, assuming $f = 500$ GeV.}
\label{tab:diflav_light100}
\end{table}


\begin{table}[H]
\setlength{\tabcolsep}{6pt} 
\renewcommand{\arraystretch}{1.1} 
\centering
\begin{tabular}{@{}l|rr|rr|rr@{}}
\toprule
 & \multicolumn{2}{c|}{Benchmark} & \multicolumn{2}{c|}{Benchmark} & \multicolumn{2}{c}{Benchmark} \\
 & \multicolumn{2}{c|}{$\mathcal{Z}_2 \times \mathcal{Z}_{5}$} & \multicolumn{2}{c|}{$\mathcal{Z}_2 \times \mathcal{Z}_{9}$} & \multicolumn{2}{c}{$\mathcal{Z}_2 \times \mathcal{Z}_{11}$} \\
 $m_{a}$~[GeV] & \myalign{c}{20} & \myalign{c|}{60} & \myalign{c}{20} & \myalign{c|}{60} & \myalign{c}{20} & \myalign{c}{60} \\
\midrule
$s \rightarrow aa \rightarrow b \bar{b} \tau \tau $~[pb]      & $8.7\e{-10}$ & $3.5\e{-12}$  & $1.9\e{-14}$ & $5.4\e{-17}$ & $3.4\e{-16}$ & $1.1\e{-18}$   \\
$s \rightarrow aa \rightarrow b \bar{b} \mu \mu $~[pb]      & $8.2\e{-13}$ & $3.2\e{-15}$  & $4.2\e{-16}$ & $1.2\e{-18}$ & $2.7\e{-18}$ & $8.8\e{-21}$   \\
$s \rightarrow aa \rightarrow b \bar{b}  b \bar{b} $~[pb]     & $2.2\e{-9}$ & $1\e{-11}$  & $3.6\e{-14}$ & $1.3\e{-16}$ & $4.2\e{-16}$ & $1.7\e{-18}$   \\
\bottomrule
\end{tabular}
\caption{Benchmark points for different  $\mathcal{Z}_N \times \mathcal{Z}_{M}$  flavor symmetries for di-flavon production channels with low flavon mass ($m_a$) in case of the symmetry-conserving scenario at the 14 TeV HL-LHC.}
\label{tab:limits_bench_sc14}
\end{table}

\begin{table}[H]
\setlength{\tabcolsep}{6pt} 
\renewcommand{\arraystretch}{1.1} 
\centering
\begin{tabular}{@{}l|rr|rr|rr@{}}
\toprule
 & \multicolumn{2}{c|}{Benchmark} & \multicolumn{2}{c|}{Benchmark} & \multicolumn{2}{c}{Benchmark} \\
 & \multicolumn{2}{c|}{$\mathcal{Z}_2 \times \mathcal{Z}_{5}$} & \multicolumn{2}{c|}{$\mathcal{Z}_2 \times \mathcal{Z}_{9}$} & \multicolumn{2}{c}{$\mathcal{Z}_2 \times \mathcal{Z}_{11}$} \\
 $m_{a}$~[GeV] & \myalign{c}{20} & \myalign{c|}{60} & \myalign{c}{20} & \myalign{c|}{60} & \myalign{c}{20} & \myalign{c}{60} \\
\midrule
$s \rightarrow aa \rightarrow b \bar{b} \tau \tau $~[pb]      & $1.9\e{-9}$ & $8.8\e{-12}$  & $4.7\e{-14}$ & $1.6\e{-16}$ & $7.7\e{-16}$ & $3\e{-18}$   \\
$s \rightarrow aa \rightarrow b \bar{b} \mu \mu $~[pb]      & $1.8\e{-12}$ & $7.9\e{-15}$  & $1.1\e{-15}$ & $3.6\e{-18}$ & $6.2\e{-18}$ & $2.3\e{-20}$   \\
$s \rightarrow aa \rightarrow b \bar{b}  b \bar{b} $~[pb]     & $4.7\e{-9}$ & $2.5\e{-11}$  & $9.1\e{-14}$ & $4.1\e{-16}$ & $9.5\e{-16}$ & $3.1\e{-18}$   \\
\bottomrule
\end{tabular}
\caption{Benchmark points for different  $\mathcal{Z}_N \times \mathcal{Z}_{M}$  flavor symmetries for di-flavon production channels with low flavon mass ($m_a$) in case of the symmetry-conserving scenario at the 27 TeV HE-LHC.}
\label{tab:limits_bench_sc27}
\end{table}

\begin{table}[H]
\setlength{\tabcolsep}{6pt} 
\renewcommand{\arraystretch}{1.1} 
\centering
\begin{tabular}{@{}l|rr|rr|rr@{}}
\toprule
 & \multicolumn{2}{c|}{Benchmark} & \multicolumn{2}{c|}{Benchmark} & \multicolumn{2}{c}{Benchmark} \\
 & \multicolumn{2}{c|}{$\mathcal{Z}_2 \times \mathcal{Z}_{5}$} & \multicolumn{2}{c|}{$\mathcal{Z}_2 \times \mathcal{Z}_{9}$} & \multicolumn{2}{c}{$\mathcal{Z}_2 \times \mathcal{Z}_{11}$} \\
 $m_{a}$~[GeV] & \myalign{c}{20} & \myalign{c|}{60} & \myalign{c}{20} & \myalign{c|}{60} & \myalign{c}{20} & \myalign{c}{60} \\
\midrule
$s \rightarrow aa \rightarrow b \bar{b} \tau \tau $~[pb]      & $7.8\e{-9}$ & $4.5\e{-11}$  & $2.4\e{-13}$ & $1.2\e{-15}$ & $3.4\e{-15}$ & $1.7\e{-17}$   \\
$s \rightarrow aa \rightarrow b \bar{b} \mu \mu $~[pb]      & $7.4\e{-12}$ & $4\e{-14}$  & $5.5\e{-15}$ & $2.5\e{-17}$ & $2.8\e{-17}$ & $1.3\e{-19}$   \\
$s \rightarrow aa \rightarrow b \bar{b}  b \bar{b} $~[pb]     & $2\e{-8}$ & $1.3\e{-10}$  & $4.7\e{-13}$ & $2.8\e{-15}$ & $4.2\e{-15}$ & $2.6\e{-17}$   \\
\bottomrule
\end{tabular}
\caption{Benchmark points for different  $\mathcal{Z}_N \times \mathcal{Z}_{M}$  flavor symmetries for di-flavon production channels with low flavon mass ($m_a$) in case of the symmetry-conserving scenario at a 100 TeV collider.}
\label{tab:limits_bench_sc100}
\end{table}

 
\section{Summary}
\label{sum}
The  $\mathcal{Z}_{\rm N} \times \mathcal{Z}_{\rm M} $ flavor symmetry  is a unique and novel framework that may have its origin either in a $U(1) \times U(1)$ symmetry or in an anomalous global $U(1)_A \times U(1)_A$ symmetry.  This symmetry provides a unique way to realize the FN mechanism, resulting in an explanation to the flavor problem of the SM.  In this work, we have thoroughly investigated the flavor  phenomenology  and the collider signatures of  different realizations of the  $\mathcal{Z}_{\rm N} \times \mathcal{Z}_{\rm M} $ flavor symmetry  motivated by different theoretical demands.

In the case of quark flavor physics, the $K^0-\bar K^0$ mixing places the strongest bounds on the parameter space of the  flavon of the $\mathcal{Z}_2 \times \mathcal{Z}_9$ flavor symmetry in the case of the soft symmetry-breaking scenario.  The $\mathcal{Z}_2 \times \mathcal{Z}_5$ flavor symmetry  is subjected to the weakest constraints from the  $K^0-\bar K^0$ mixing.  This scenario continues to remain the same even for the symmetry-conserving case, where the allowed parameter space is only along a straight dashed line.  The  $B_s -\bar B_s$ mixing better constraints the parameter space of the  flavon of the $\mathcal{Z}_{2} \times \mathcal{Z}_{11}$  and $\mathcal{Z}_{8} \times \mathcal{Z}_{22}$ flavor symmetries for the soft symmetry-breaking scenario. In the case of the symmetry-conserving potential, the $B_s -\bar B_s$ mixing provides more stringent bounds on the $\mathcal{Z}_{2} \times \mathcal{Z}_{11}$ flavor symmetry.  The  $B_d -\bar B_d$ mixing places more tight bounds on the parameter space of the  flavon of the $\mathcal{Z}_{2} \times \mathcal{Z}_{11}$  and $\mathcal{Z}_{2} \times \mathcal{Z}_{5,9}$ flavor symmetries for the soft symmetry-breaking scenario.  This feature continues to hold even for the symmetry-conserving potential. The   $D^0 - \bar D^0$ mixing  works remarkably well in constraining the  parameter space of the  flavon of the $\mathcal{Z}_{2} \times \mathcal{Z}_{5}$ flavor symmetry for the soft symmetry-breaking scenario.  This is also true for the symmetry-conserving case.  The $B_{d,s} \rightarrow \mu^+ \mu^-$ decays impose relatively better bounds on the parameter space of the  flavon of the $\mathcal{Z}_{2} \times \mathcal{Z}_{11}$ symmetry.  On the other side, the $ K_L \rightarrow \mu^+ \mu^-$ decays are more sensitive to the parameter space of the  flavon of the $\mathcal{Z}_{2} \times \mathcal{Z}_{9}$ symmetry.   The $B_{d,s} \rightarrow \mu^+ \mu^-$  and $ K_L \rightarrow \mu^+ \mu^-$ decays  place tighter constraints on the parameter space of the  flavon of the $\mathcal{Z}_{2} \times \mathcal{Z}_{11}$ symmetry for the symmetry-conserving scenario.  The lifetime $\tau_{\mu\mu}$ is important in constraining the parameter space of the  flavon of the $\mathcal{Z}_{2} \times \mathcal{Z}_{5}$ symmetry.  For the soft symmetry-breaking scenario, the lifetime $\tau_{\mu\mu}$ places more significant bounds on the parameter space of the  flavon of the $\mathcal{Z}_{2} \times \mathcal{Z}_{11}$ symmetry.  The ratio $R_{\mu \mu}$ is a powerful observable to constrain the  parameter space of the  flavon of all  $\mathcal{Z}_{\rm N} \times \mathcal{Z}_{\rm M} $ flavor symmetries for the soft symmetry-breaking as well as symmetry-conserving scenarios, and will play a crucial role in determining the parameter space of the flavon of different $\mathcal{Z}_{\rm N} \times \mathcal{Z}_{\rm M} $ flavor symmetries.  A global fit to diverse data coming from the observables having $b \rightarrow s \mu^+ \mu^-$ transitions is performed, and allowed ranges of the Wilson coefficients are derived.  This results in  bounds on the parameter space of the  flavon of all  $\mathcal{Z}_{\rm N} \times \mathcal{Z}_{\rm M} $ flavor symmetries for the soft symmetry-breaking as well as symmetry-conserving scenarios. We also predict muon forward-backward asymmetry in $B \rightarrow K \mu^+ \mu^-$ decays for $f<200$ GeV.

On the leptonic flavor side, the decay $\mu \rightarrow e \gamma$ is able to constrain more the  parameter space of the  flavon of the $\mathcal{Z}_{2} \times \mathcal{Z}_{9}$  and the  $\mathcal{Z}_{8} \times \mathcal{Z}_{22}$ flavor symmetries for  the soft symmetry-breaking case.  However, the  parameter space of the  flavon of the $\mathcal{Z}_{2} \times \mathcal{Z}_{11}$ is more constrained for the symmetry-conserving scenario.  The $A~\mu\rightarrow A~e$ conversion is  more effective in constraining the  parameter space of the  flavon of the $\mathcal{Z}_{8} \times \mathcal{Z}_{22}$ flavor symmetry for the soft symmetry-breaking scenario.  In the case of symmetry-conserving scenario, the parameter space of the  flavon of the  $\mathcal{Z}_{2} \times \mathcal{Z}_{11}$ is more constrained.  Similar results are obtained for the $\mu\rightarrow 3e$ and $\tau \rightarrow 3 \mu$ decays.

In the study of collider physics of the flavon of different $\mathcal{Z}_{\rm N} \times \mathcal{Z}_{\rm M} $ flavor symmetries, we computed flavon decays as well as $t \rightarrow (c,u) a $ decays.  The  $t \rightarrow c a $ decays are within the reach of the HL-LHC, HE-LHC, and a 100 TeV collider only for the $\mathcal{Z}_{2} \times \mathcal{Z}_{11}$ and  the $\mathcal{Z}_{8} \times \mathcal{Z}_{22}$ flavor symmetries for the soft symmetry-breaking scenario. Furthermore,  we  have investigated inclusive, associative, and di-flavon production signatures of flavon  for the soft symmetry-breaking as well as symmetry-conserving scenarios.  We have predicted the sensitivities  of these channels for the HL-LHC, HE-LHC, and a 100 TeV collider in a model-independent way using the square root scaling of luminosity.   We compare our benchmark predictions for heavy and light flavon at the HL-LHC, HE-LHC, and a 100 TeV collider to estimated sensitivities of  the HL-LHC, HE-LHC and a 100 TeV collider for all the  $\mathcal{Z}_{\rm N} \times \mathcal{Z}_{\rm M} $ flavor symmetries discussed in this work.  In the case of inclusive production channels, a 100 TeV collider can probe the $\mathcal{Z}_{2} \times \mathcal{Z}_{5,9}$ and  the $\mathcal{Z}_{8} \times \mathcal{Z}_{22}$ flavor symmetries for the soft symmetry-breaking scenario for a heavy flavon. The HE-LHC will be able to probe the $\mathcal{Z}_{2} \times \mathcal{Z}_{5}$ and  the $\mathcal{Z}_{8} \times \mathcal{Z}_{22}$ flavor symmetries better than the HL-LHC for the soft symmetry-breaking scenario. In the case of a light flavon, the HE-LHC and a 100 TeV collider can probe all the $\mathcal{Z}_{\rm N} \times \mathcal{Z}_{\rm M} $ flavor symmetries in the $\tau \tau$ channel for the soft symmetry-breaking scenario. A few specific  inclusive signatures are within the reach of the 14 TeV high-luminosity LHC  only for the $\mathcal{Z}_{\rm 2} \times \mathcal{Z}_{\rm 5}$ and $\mathcal{Z}_{\rm 8} \times \mathcal{Z}_{\rm 22}$  flavor symmetries.  The symmetry-conserving scenario is beyond the reach of detection capabilities of any collider for the inclusive processes.

For associative production modes, only the  parameter space of the  flavon of the $\mathcal{Z}_{8} \times \mathcal{Z}_{22}$ flavor is accessible to  the HL-LHC, HE-LHC, and a 100 TeV collider  through the  $t  \bar{t} a \rightarrow t  \bar{t} t  \bar{t} $  channel in the soft symmetry-breaking scenario for  high flavon mass. The di-flavon production channels for heavy flavon are insensitive to  the HL-LHC, HE-LHC, and even to a 100 TeV collider for all the $\mathcal{Z}_{\rm N} \times \mathcal{Z}_{\rm M} $ flavor symmetries in the case of soft symmetry-breaking scenario.  However, a light flavon scenario can be probed for all the $\mathcal{Z}_{\rm N} \times \mathcal{Z}_{\rm M} $ flavor symmetries at the HL-LHC, HE-LHC, and a 100 TeV collider. In the case of the symmetry-conserving scenario, the di-flavon productions cross-sections are too small to be accessible at the HL-LHC, HE-LHC as well as a 100 TeV collider, and can be further constrained through the precision flavor physics in future.  Collider simulations for some particular signatures, such as $pp \rightarrow a \rightarrow t \bar{t}$,   could be an interesting future investigation, which is an expected sequel to the present work \cite{sequel}.

\section*{Acknowledgement}
We acknowledge the use of {\tt JaxoDraw} \cite{Binosi:2008ig} for creating the Feynman diagrams in this work.  This work is supported by the  Council of Science and Technology,  Govt. of Uttar Pradesh,  India through the  project ``   A new paradigm for flavor problem "  no.   CST/D-1301,  and Science and Engineering Research Board,  Department of Science and Technology, Government of India through the project `` Higgs Physics within and beyond the Standard Model" no. CRG/2022/003237. NS acknowledges the support through the INSPIRE fellowship by the Department of Science and Technology, Government of India.

\section*{Benchmark points for the Yukawa couplings}
\label{bench_appendix}

We perform a $\chi^2$ fit to the quark and charged-lepton masses, along with the quark mixing parameters, by defining,
\bea
\chi^2 &=& \dfrac{(m_q - m_q^{\rm{model}} )^2}{\sigma_{m_q}^2}+  \dfrac{(m_\ell - m_\ell^{\rm{model}} )^2}{\sigma_{m_\ell}^2}  + \dfrac{(\sin \theta_{ij} - \sin \theta_{ij}^{\rm{model}} )^2}{\sigma_{\sin \theta_{ij}}^2} + \dfrac{(\sin 2 \beta  - \sin 2 \beta^{\rm{model}} )^2}{\sigma_{\sin2\beta}^2} \nonumber \\ &+& \dfrac{( \alpha  - \alpha^{\rm{model}} )^2}{(\sigma_{\alpha})^2}
+ \dfrac{( \gamma   - \gamma^{\rm{model}} )^2}{(\sigma_{\gamma})^2},
\eea
where $q=\{u,d,c,s,t,b\}$, $\ell=\{e,\mu,\tau\}$, and $i,j=1,2,3$. The CKM matrix phases in the standard parametrization are given by,
\begin{eqnarray}
\beta^{\text{model}} =\text{arg} \left(- \dfrac{V_{cd} V_{cb}^*}{V_{td} V_{tb}^*}\right),~\alpha^{\text{model}} =\text{arg} \left(- \dfrac{V_{td} V_{tb}^*}{V_{ud} V_{ub}^*}\right),~\gamma^{\text{model}} =\text{arg} \left(- \dfrac{V_{ud} V_{ub}^*}{V_{cd} V_{cb}^*}\right).
\end{eqnarray}
To reproduce the fermion masses, we use the following values of the quark and charged-lepton masses at $ 1$TeV\cite{Xing:2007fb},
\begin{eqnarray}
\{m_t, m_c, m_u\} &\simeq& \{150.7 \pm 3.4,~ 0.532^{+0.074}_{-0.073},~ (1.10^{+0.43}_{-0.37}) \times 10^{-3}\}~{\rm GeV}, \nonumber \\
\{m_b, m_s, m_d\} &\simeq& \{2.43\pm 0.08,~ 4.7^{+1.4}_{-1.3} \times 10^{-2},~ 2.50^{+1.08}_{-1.03} \times 10^{-3}\}~{\rm GeV},
\nonumber \\
\{m_\tau, m_\mu, m_e\} &\simeq& \{1.78\pm 0.2,~ 0.105^{+9.4 \times 10^{-9}}_{-9.3 \times 10^{-9}},~ 4.96\pm 0.00000043 \times 10^{-4}\}~{\rm GeV}.
\end{eqnarray}
The magnitudes and phases  of the CKM mixing elements are \cite{Zyla:2021},
\bea
|V_{ud}| &=& 0.97370 \pm 0.00014,  |V_{cb}| = 0.0410 \pm 0.0014, |V_{ub}| = 0.00382 \pm 0.00024, \\ \nonumber
\sin 2 \beta &=& 0.699 \pm 0.017, ~ \alpha = (84.9^{+5.1}_{-4.5})^\circ,~  \gamma = (72.1^{+4.1}_{-4.5})^\circ, \delta = 1.196^{+0.045}_{-0.043}
\eea

\subsection*{The $\mathcal{Z}_2 \times \mathcal{Z}_{5}$ model }
The dimensionless coefficients $y_{ij}^{u,d,\ell,\nu}= |y_{ij}^{u,d,\ell,\nu}| e^{i \phi_{ij}^{q,\ell,\nu}}$  are scanned with $|y_{ij}^{u,d,\ell, \nu}| \in [0.1, 4 \pi]$ and $ \phi_{ij}^{q,\ell,\nu} \in [0,2\pi]$. The fit results  are,
\begin{equation*}
Y_u =\begin{pmatrix}
-1.68 - 3.37 i &  -0.09 + 0.03 i  &   -0.1 - 0.02 i    \\
1.53 + 4.95 i    &- 0.57 + 0.55 i  &  0.48 + 0.002 i    \\
0.76 + 0.18 i   &  -1.04 + 0.46 i   &  0.58 - 0.65 i
\end{pmatrix},  
Y_d = \begin{pmatrix}
-4.15 + 3.58 i &  2.20 - 0.89 i & 2.62 - 4.20 i   \\
-0.33 - 0.36 i  & 0.07 -0.075 i &  0.17 + 0.47 i  \\
- 0.24 -0.07 i &  -0.06 - 0.084 i   &  -0.07 - 0.12 i
\end{pmatrix},  \\
\end{equation*}

\begin{equation*}
Y_l = \begin{pmatrix}
-0.07 - 0.06 i &  0.099 - 0.004 i & 0.45 -0.32   i   \\
-0.14 - 0.09 i  & 0.08 - 0.06 i &  -0.63 + 0.24 i  \\
-0.04 + 0.09 i &  -0.09 + 0.06 i   &  0.10 - 0.0003 i
\end{pmatrix}
\end{equation*}
with $\epsilon = 0.1$, $\delta = 1.196$, and $\chi^2_{min} = 3.16$.

\subsection*{The $\mathcal{Z}_2 \times \mathcal{Z}_{9}$ model}
The dimensionless coefficients $y_{ij}^{u,d,\ell,\nu}= |y_{ij}^{u,d,\ell,\nu}| e^{i \phi_{ij}^{q,\ell,\nu}}$  are scanned with $|y_{ij}^{u,d,\ell, \nu}| \in [0.9, 2 \pi]$ and $ \phi_{ij}^{q,\ell,\nu} \in [0,2\pi]$. The fit results  are,
\begin{equation*}
Y_u =\begin{pmatrix}
1  &  0.87 - 0.49 i  &   -0.23 + 0.97 i    \\
-0.9 + 1.05 i    &- 0.7 - 0.72 i  &  1    \\
0.94 - 0.33 i   &  0.55 + 0.84 i   &  0.9 
\end{pmatrix}, 
Y_d = \begin{pmatrix}
0.99 - 0.09 i &  3.24 - 1.05 i & 1    \\
0.99 - 0.10 i  & 0.92 + 0.39 i &  0.9   \\
1  &  1    &  -1.04 + 0.54 i
\end{pmatrix},  \\
\end{equation*}

\begin{equation*}
Y_l = \begin{pmatrix}
0.9  &  0.9  & 1.5    \\
0.9   & 1.5  &  1.5  \\
1.5  &  1.5    &  0.9 
\end{pmatrix}
\end{equation*}
with $\epsilon = 0.23$, $\delta = 1.196$, and $\chi^2_{min} = 5.91$.

\subsection*{The $\mathcal{Z}_2 \times \mathcal{Z}_{11}$ model}
The dimensionless coefficients $y_{ij}^{u,d,\ell,\nu}= |y_{ij}^{u,d,\ell,\nu}| e^{i \phi_{ij}^{q,\ell,\nu}}$  are scanned with $|y_{ij}^{u,d,\ell, \nu}| \in [0.7, 2 \pi]$ and $ \phi_{ij}^{q,\ell,\nu} \in [0,2\pi]$. The fit results  are,
\begin{equation*}
Y_u =\begin{pmatrix}
2.19 - 0.18 i &  1.40 - 0.09 i  &   0.70 - 0.002 i    \\
2.12 + 0.02 i    &- 1.02 + 2.12 i  &  4.45 - 1.01 i    \\
0.33 - 0.71 i   &  0.92 + 0.42 i   &  0.82 - 0.26 i
\end{pmatrix}, 
Y_d = \begin{pmatrix}
0.40 + 0.59 i &  3.61 - 0.16 i & 3.34 - 4.06 i   \\
1.02 + 0.01 i  & -0.36 + 1.18 i &  1.29 + 0.36 i  \\
- 0.77 + 0.33 i &  0.82 + 0.02 i   &  -0.51 + 0.48 i
\end{pmatrix}  \\
\end{equation*}

\begin{equation*}
Y_l = \begin{pmatrix}
-0.73 + 0.001 i &  -0.58 + 0.53 i & 0.001 + 1.29  i   \\
6.24 + 0.37 i  & 3.48 - 2.44 i &  1.71 + 1.7 \times 10^{-5} i  \\
0.24 - 1.16 i &  -0.36 - 0.61 i   &  -0.60 + 0.35 i
\end{pmatrix}  \\
\end{equation*}
with $\epsilon = 0.28$, $\delta = 1.196$, and $\chi^2_{min} = 11.82$.

\subsection*{The $\mathcal{Z}_8 \times \mathcal{Z}_{22}$ model}
The dimensionless coefficients $y_{ij}^{u,d,\ell,\nu}= |y_{ij}^{u,d,\ell,\nu}| e^{i \phi_{ij}^{q,\ell,\nu}}$  are scanned with $|y_{ij}^{u,d,\ell, \nu}| \in [0.9, 2]$ and $ \phi_{ij}^{q,\ell,\nu} \in [0,2\pi]$. The fit results  are,
\begin{equation*}
Y_u =\begin{pmatrix}
0.12 + 1.44 i &  - 0.38 - 0.82 i  &   0.989 + 0.004 i    \\
- 1.27 - 1.38 i    & - 0.56 + 0.82 i &  - 1.22 - 0.24 i     \\
- 1.12 - 0.25 i    &  - 1.12 - 0.43 i   &  - 2.70 - 2.61 i
\end{pmatrix}, 
Y_d = \begin{pmatrix}
- 1.36 + 0.39 i &  0.31 - 1.28 i & 0.71 + 0.61 i   \\
- 1.05 + 0.27 i  & -0.44 + 0.85 i &  0.47 - 0.79 i  \\
- 1.11 - 0.37 i & - 0.81 + 0.42 i   &  0.80 + 0.82 i
\end{pmatrix}  \\
\end{equation*}

\begin{equation*}
Y_l = \begin{pmatrix}
-1.40 - 0.21 i &  1.14 - 0.004 i & - 0.78 + 0.45  i   \\
- 0.88 + 0.16 i  & - 0.54 + 0.78 i &  - 1.16 + 0.11 i  \\
0.85 + 0.34 i &  0.899 + 0.003 i   &  0.899 - 0.003 i
\end{pmatrix}  \\
\end{equation*}
with $\epsilon = 0.23$, $\delta = 1.196$, and $\chi^2_{min} = 1.62$.

\section*{The SM background}
\begin{table}[H]
    \setlength{\tabcolsep}{5.5pt} 
\renewcommand{\arraystretch}{0.8} 
    \begin{center}
    \begin{tabular}{|p{0.78in}||c|p{2.4in}|}
\hline 
\bf{Production mode} & \bf{Channel}  & \bf{SM Backgrounds} \\ \hline 
  & jet-jet & $p p \rightarrow j j$ \\
\cline{2-3}
 & $\ell^+ \ell^- (\ell = e, \mu, \tau)$ & $p p \rightarrow \ell^+ \ell^-$ \\
 \cline{2-3}
 & $\mu e $ & $p p \rightarrow t \bar{t}, p p \rightarrow V V (V= W^+, W^-, Z),$\\ & & $p p \rightarrow \ell^+ \ell^- , p p \rightarrow t W b$ \\
 \cline{2-3}
\centering{Inclusive}  & $\mu \tau, e \tau $ & $p p \rightarrow$ multijets, \hspace{0.6cm} $p p \rightarrow W +$ jets ,
\\ \centering{production} & & $p p \rightarrow t \bar{t}, p p \rightarrow V V (V= W^+, W^-, Z),$     \hspace{0.8cm}    $p p \rightarrow \ell^+ \ell^- , p p \rightarrow t W b$ \\
\cline{2-3}
 & $b \overline{b}$ & $p p \rightarrow b \bar{b}$ \\
 \cline{2-3}
 & $\gamma \gamma$ & $p p \rightarrow \gamma \gamma$ \\
 \cline{2-3}
 & $t \overline{t}$ & $p p \rightarrow t \bar{t}$ \\
 \hline
\centering{Associative} & $t  \bar{t} a \rightarrow t  \bar{t} t  \bar{t} $ & $p p \rightarrow t  \bar{t} t  \bar{t} $ \\
\cline{2-3}
\centering{production} & $a b \bar{b} \rightarrow \tau \tau b \bar{b}$ & $p p \rightarrow \tau \tau b \bar{b}$ \\
 \cline{2-3}
  & $a b  \rightarrow \tau \tau b $ &  $p p \rightarrow$ multijets, $p p \rightarrow W + $ jets, $p p \rightarrow t \bar{t} $ \\ & & $p p \rightarrow \ell^+ \ell^- , p p \rightarrow t W b$, $p p \rightarrow V V$  \\
 \hline
 \centering{Di-flavon} & $a  a \rightarrow b  \bar{b} \ell  \ell (\ell = \mu, \tau) $ & $p p \rightarrow  b  \bar{b} \ell  \ell $ \\
\cline{2-3}
\centering{production} & $a a \rightarrow b \bar{b} b \bar{b}$ & $p p \rightarrow b \bar{b} b \bar{b}$ \\
 \hline
    \end{tabular}
    \caption{SM Backgrounds for the flavon production through various channels.}
    \label{tab:BG}
    \end{center}
\end{table}

\section*{The dark-technicolor paradigm}
In this appendix, we show how the  $\mathcal{Z}_{\rm N} \times \mathcal{Z}_{\rm M} $ flavor symmetry and the FN mechanism of the  $\mathcal{Z}_8 \times \mathcal{Z}_{22}$ model naturally emerge from a dark-technicolour paradigm \cite{Abbas:2017vws,Abbas:2020frs,Abbas:2023bmm}. The technicolour paradigm is obtained by $SU(\rm N_{\rm TC}) \times SU(\rm N_{\rm DTC}) \times SU(\rm{N}_{\rm F})$, where TC denotes the standard  technicolour gauge group,  DTC is for the dark-technicolour, and the F stands for the  strong dynamics of vector-like fermions. 

The TC fermions  transform under the  symmetry $\rm SM \times \mathcal{G}$  as\cite{Abbas:2020frs},
\begin{eqnarray}
T_{q}^i  &\equiv&   \begin{pmatrix}
T  \\
B
\end{pmatrix}_L:(1,2,0,\rm{N}_{\rm TC},1,1),  \\ \nonumber
T_{R}^i &:& (1,1,1,\text{N}_{\rm{TC}},1,1), B_{R}^i : (1,1,-1,\rm{N}_{\rm TC},1,1), 
\end{eqnarray}
where $i=1,2,3 \cdots $,  and   $+\frac{1}{2}$ is the electric charge for $T$, and $-\frac{1}{2}$ is that of the  $B$.

The transformation of the DTC fermions under the  $\rm SM \times \mathcal{G}$  symmetry is \cite{Abbas:2020frs},
\begin{eqnarray}
 \mathcal{D}_{ q}^i &\equiv& \mathcal{C}_{L,R}^i  : (1,1, 1,1,\text{N}_{\text{DTC}},1),~\mathcal{S}_{L,R}^i  : (1,1,-1,1,\text{N}_{\text{DTC}},1), 
\end{eqnarray}
where  $i=1,2,3 \cdots $,  and  electric charges are $+\frac{1}{2}$ for $\mathcal C$ and $-\frac{1}{2}$ for $\mathcal S$.  

The fermions of the $SU(N_{\text{F}})$ symmetry have the following behavior under the $\rm SM \times \mathcal{G}$  symmetry \cite{Abbas:2020frs},
\begin{eqnarray}
F_{L,R} &\equiv &U_{L,R}^i \equiv  (3,1,\dfrac{4}{3},1,1,\text{N}_\text{F}),
D_{L,R}^{i} \equiv   (3,1,-\dfrac{2}{3},1,1,\text{N}_\text{F}),  \\ \nonumber 
N_{L,R}^i &\equiv&   (1,1,0,1,1,\text{N}_\text{F}), 
E_{L,R}^{i} \equiv   (1,1,-2,1,1,\text{N}_\text{F}),
\end{eqnarray}
where  $i=1,2,3 \cdots$.

We notice that there are  three axial $U(1)_A^{\rm TC, DTC,  F}$ symmetries in the dark-technicolor paradigm.  They are broken by instanton effects to a cyclic discrete group as $ U(1)_A^{\rm TC, DTC, F} \rightarrow \mathcal{Z}_{2 \rm K_{\rm TC, DTC, F}}$ \cite{Harari:1981bs},  where $\rm K_{\rm TC, DTC, F} $ show the number of   massless flavors of the TC, DTC and F gauge symmetries  in the $N$-dimensional representation of the gauge group $SU(\rm N)_{\rm TC, DTC, F}$.  Thus, we observe a natural emergence of a generic  $\mathcal{Z}_{\rm N} \times \mathcal{Z}_{\rm M} \times \mathcal{Z}_{\rm P}$ flavor symmetry, whose subset is the  $\mathcal{Z}_{\rm N} \times \mathcal{Z}_{\rm M} $ flavor symmetry.

\begin{figure}[h!]
	\centering
 \includegraphics[width=\linewidth]{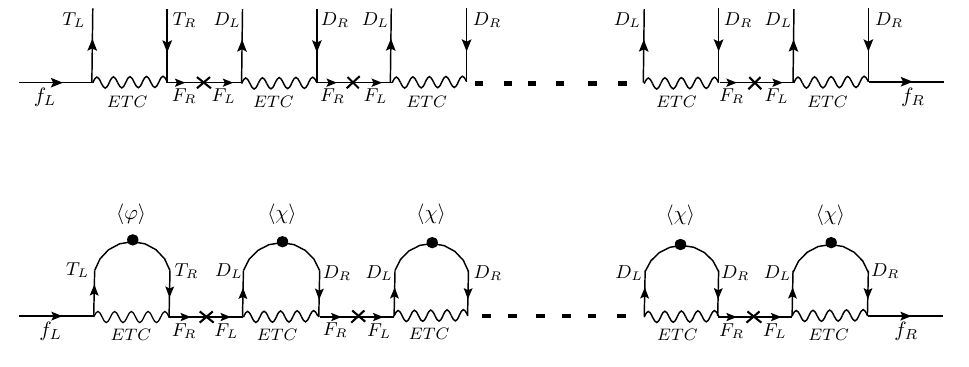}
    \caption{Feynman diagrams for the masses of the quarks and charged leptons in the  dark-technicolour paradigm.  On the top, the generic interactions of the SM, TC, F and DTC fermions are shown.  In the bottom, we see the formations of the on-shell  condensates $\langle \varphi \rangle$ and $\langle \chi \rangle$ (circular blobs),    and the resulting mass of the SM fermions.}
 \label{DTP}	
 \end{figure}

In the next step, we assume the existence of an extended technicolour  dynamics (ETC) which accommodates the SM, TC, and DTC sectors. This provides the required interactions for the masses of fermions in the FN mechanism.  These interactions are shown in figure \ref{DTP}, where $\langle \varphi \rangle$ and $\langle \chi \rangle$  show the Higgs and flavon VEVs, respectively,  in the form of an on-shell chiral condensate.

The multi-fermion condensate can be written as  \cite{Aoki:1983yy}, 
\be 
\label{VEV_h}
\langle  ( \bar{\psi}_L \psi_R )^n \rangle \sim \left(  \Lambda \exp(k \Delta \chi) \right)^{3n},
\ee
where $\Delta \chi$ is the chirality of an operator, $k$  is a constant, and $\Lambda$ represents the scale of the underlying gauge theory. 

Now, the masses of the SM fermions in the FN mechanism can be written as,
\bea
\label{TC_masses}
m_{f} & \approx &   \frac{\Lambda_{\text{TC}}^{3}}{\Lambda_{\text{ETC}}^2} \left( \dfrac{1}{\Lambda} \frac{\Lambda_{\text{DTC}}^3}{\Lambda_{\text{ETC}}^{2}} \exp(2 k) \right)^{n_{ij}^f},
\eea
where $f=u,d$, and $\Lambda_{\text{TC}}$, $\Lambda_{\text{DTC}}$ and  $\Lambda $ are the scale of the TC, DTC, and F dynamics, respectively.  We can now identify the order parameter $\epsilon$ as,

\begin{align}
    \epsilon = \frac{\langle \chi \rangle}{\Lambda}   \propto \dfrac{1}{\Lambda} \frac{\Lambda_{\text{DTC}}^3}{\Lambda_{\text{ETC}}^{2}} \exp(2 k),  
\end{align}

and the SM Higgs VEV is given by,
\begin{align}
  \langle \varphi \rangle \propto  \frac{\Lambda_{\text{TC}}^{3}}{\Lambda_{\text{ETC}}^2}. 
\end{align}

\end{document}